%% file: Main.tex
\documentclass[spanish, letterpaper, 12pt, twoside, openany]{book} 

\usepackage{textcomp}                           % Agrega muchos símbolos extra para utilizar en el texto
\usepackage[T1]{fontenc, url}                   % Para los enlaces url y las fuentes
\usepackage[utf8]{inputenc}                     % Introduce la codificación utf8 que permite el uso de tildes y otros.
\usepackage{titlesec}                           % Suplemento a fancyhdr
\usepackage{multirow}                           % Extensión para tablas
\usepackage{pdflscape}                          % usado para rotat pag (no estoy seguro)
\usepackage{booktabs}
\usepackage{enumitem}                           % Para modificar el interlineado entre los items
\usepackage{graphicx}                           % Para poner imágenes
\usepackage{amsmath,amssymb,amsfonts,dsfont,mathrsfs,amsthm,mathtools}          % paquetes matemáticos
\usepackage{color}                              % Agrega colores.
\usepackage{cancel}
\usepackage{caption}
\usepackage{subcaption}                         % Permite subtextos en las figuras. Es incompatible con el paquete subfig y subfigure
\usepackage[toc,page]{appendix}                 % Permite el uso de apéndices
\usepackage{chngcntr}                           % Necesario para enumerar las tablas
\usepackage{stackengine}                        % Necesario para enumerar las figuras

\counterwithin{table}{section}                  % Numeración de tablas 
\counterwithin{figure}{section}                 % Numeración de figuras
\numberwithin{equation}{section}                % Numeración de ecuaciones
\usepackage{xparse,nameref}                     % Ayuda en la creación de nuevos comandos
\usepackage[bottom,hang,flushmargin]{footmisc}  % notas al pie de página están fijas al fondo y sin sangría
\usepackage{lipsum}                             % Para crear texto de relleno.

\newcommand{\tongo}[1]{\hat{\mathbf{#1}}} % Ejemplo: \tongo{u}
\usepackage[spanish]{babel}          % Idioma
\decimalpoint                   %, a . en decimal

\graphicspath{{Images/}{../Images/}}                                % Dónde estarán las imágenes
\usepackage[left=2.3cm,top=3.5cm,bottom=2.5cm,right=2.3cm]{geometry}    % Márgenes del documento
\usepackage{setspace}                                               % Permite elegir el interlineado
\usepackage{microtype}                                              % Permite la modificación de los caracteres.
\usepackage{ragged2e}

\usepackage{fancyhdr}                           % Permite formatear las cabeceras, pies, enumeración, etc.
\usepackage{subfiles}                           % Para agregar los capítulos que se escriben aparte.
\usepackage{array}                              % Para ordenar texto y ecuaciones.
\usepackage[rightcaption]{sidecap}              % Permite agregar texto lateral
\usepackage{wrapfig}                            % Permite poner figuras con texto al rededor.
\usepackage{float}                              % Permite poner figuras en cualquier lugar.
\usepackage[labelfont=bf]{caption}              % Texto en negrita para descripciones (\caption)
\usepackage[para]{threeparttable}               % Tablas vistosas, mirar antes de utilizar.
\usepackage{url}                                % Permite el uso de enlaces URL.
\usepackage[table,xcdraw,dvipsnames]{xcolor}    % Agranda la cantidad de colores.
\usepackage{makecell}                           % Ayuda en la creación de tablas
\usepackage{hhline}                             % Agranda las opciones de las líneas
\usepackage{textcomp}                           % Símbolo de derechos de autor
\usepackage{graphicx}
\usepackage{epsfig}
\usepackage[export]{adjustbox} %this allows me to put a frame (as option to the includegraphics cmd) to images in figures

\usepackage[noadjust]{cite}
\def\biblio{\clearpage\bibliographystyle{apalike}\bibliography{References}} % Define el comando \biblio para referencias en subarchivos- NO CAMBIAR

\fancypagestyle{plain}{                     % Se redefine el estilo automático (plain) para que calce con el resto. En particular la 1ra página de cada capítulo
\fancyhf{}                                  % Elimina la cabecera y los pies
\fancyhead[RO,LE]{\thepage}                 % Número de página en la izquierda para par y derecha para impar
\fancyhead[RE,LO]{\nouppercase{\leftmark}}  % Nombre del capítulo en la derecha para par y la izquierda para impar en la cabecera
\fancyfoot{}                                % Elimina el número de la página abajo
}

\fancypagestyle{resumen}{                   % Se redefine el estilo resumen para que calce con el resto. 
\fancyhf{}                                  % Elimina la cabecera y los pies
\fancyhead[RO,LE]{\thepage}                 % Número de página en la izquierda para par y derecha para impar
\fancyhead[RE,LO]{\nouppercase{Resumen}}    % Nombre del capítulo en la derecha para par y la izquierda para impar en la cabecera
\fancyfoot{}                                % Elimina el número de la página abajo
}

\fancypagestyle{abstract}{                  % Se redefine el estilo resumen para que calce con el resto. 
\fancyhf{}                                  % Elimina la cabecera y los pies
\fancyhead[RO,LE]{\thepage}                 % Número de página en la izquierda para par y derecha para impar
\fancyhead[RE,LO]{\nouppercase{Abstract}}   % Nombre del capítulo en la derecha para par y la izquierda para impar en la cabecera
\fancyfoot{}                                % Elimina el número de la página abajo
}

\fancypagestyle{agradecimientos}{                   % Se redefine el estilo resumen para que calce con el resto. 
\fancyhf{}                                          % Elimina la cabecera y los pies
\fancyhead[RO,LE]{\thepage}                         % Número de página en la izquierda para par y derecha para impar
\fancyhead[RE,LO]{\nouppercase{Agradecimientos}}    % Nombre del capítulo en la derecha para par y la izquierda para impar en la cabecera
\fancyfoot{}                                        % Elimina el número de la página abajo
}

\fancypagestyle{frontpage}{             % Se define el estilo frontpage.
	\fancyhf{}                          % Elimina la cabecera y los pies
	\vspace*{1\baselineskip}
	
	\   % Línea 2 de la cabecera derecha, debe ir el nombre de la facultad en mayúsculas.
	\fancyhead{\centering \includegraphics[width=1.8in]{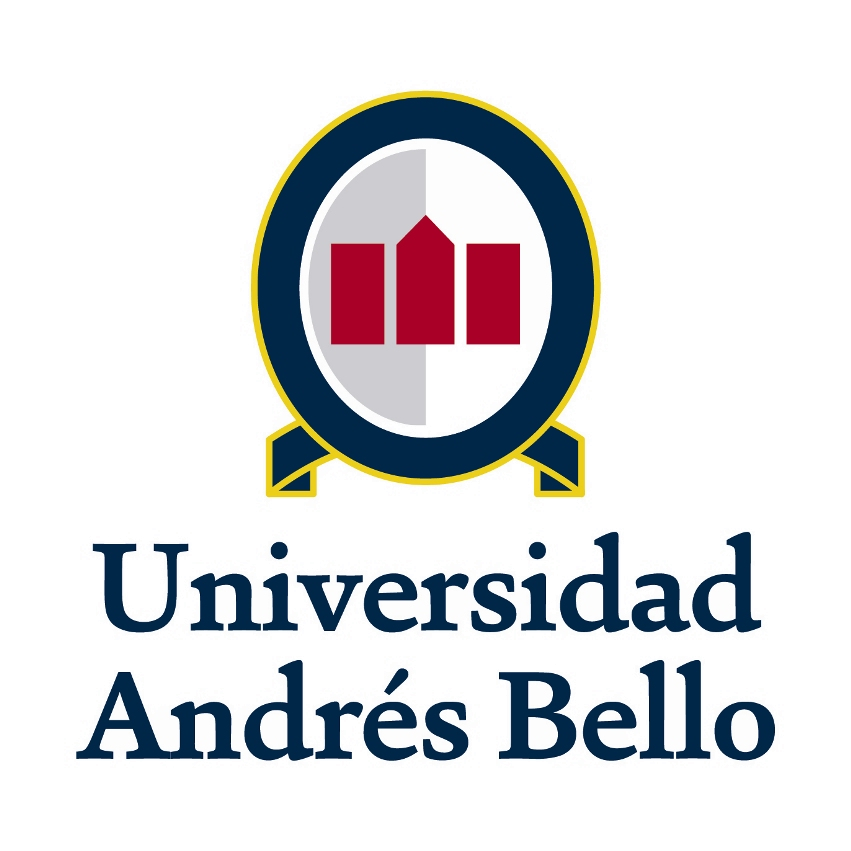}                       % Línea 1 de la cabecera derecha, debe ir el nombre de la UdeC en mayúsculas.
	\linebreak  Facultad de Ciencias Exactas\\
 Departamento de Ciencias Físicas} % LOGO que irá en la cabecera izquierda.
}

\usepackage{hyperref}   % Permite que todo el documento sea clickeable.
\newcommand\myshade{85} % Permite la redefinición de colores a gusto del usuario

\colorlet{mylinkcolor}{DarkOrchid}   % Define Color de los enlaces clickeables internos (índice, referencias cruzadas, etc.). En este caso DarkOrchid
\colorlet{mycitecolor}{YellowOrange} % Define Color de las citas ubicadas en las referencias. En este caso YellowOrange
\colorlet{myurlcolor}{Aquamarine}    % Define Color de los enlaces url. En este caso Aquamarine

\hypersetup{  %Define la forma en que se verán las cosas clickeables.
  	linkcolor  = mylinkcolor!\myshade!black,    % Aplica el color definido arriba. En este caso DarkOrchid
  	citecolor  = mycitecolor!\myshade!black,    % Aplica el color definido arriba. En este caso YellowOrange
  	urlcolor   = myurlcolor!\myshade!black,     % Aplica el color definido arriba. En este caso Aquamarine
  	colorlinks = true,                          % Elimina las cajas al rededor de lo clickeable y lo reemplaza por palabras a color.
}

\newcommand{\sgn}{\text{sgn}}

\begin{document}
\def\biblio{}   % Resetea el comando biblio, de lo contrario una lista de referencias será producida después de cada capítulo
                % resets the biblio command, if not here a new reference list will be produced after every chapter

\include{Otros/Portada}   % Genera la portada desde el archivo Portada.tex . Para editar este archivo ir a Otros -> Portada.tex
\vfill

%----------------Página de derechos de autor: elegir entre a) o b) y borrar/comentar la opción NO utilizada-----------------
\thispagestyle{empty}
\mbox{}                         % Ayuda a bajar el texto
\vfill                          % Deja el texto al fondo
\textcopyright\ 2024, Sebastián Filipini Parra  \\ % Derechos de autor
%a)
%Ninguna parte de esta tesis puede reproducirse o transmitirse bajo ninguna forma o por ningún medio o procedimiento, sin permiso por escrito del autor.\\\\
%b)
Se autoriza la reproducción total o parcial, con fines académicos, por cualquier medio o procedimiento, incluyendo la cita bibliográfica del documento.
\vspace{1cm}    % lo separa del fondo
\restoregeometry % Devuelve los márgenes después de la portada

\newpage
\phantomsection
\addcontentsline{toc}{chapter}{Resumen} % Agrega esta sección al índice
\section*{Resumen}                      % Con asterisco para que no sea numerada.

\subfile{Otros/Resumen}             % Llama al archivo Resumen en la carpeta Otros.

%-----------------Página de acronimos y convensiones-------------

\newpage
\phantomsection
\addcontentsline{toc}{chapter}{Acrónimos y Unidades} 
\section*{Acrónimos y Unidades}                   
\subfile{Otros/Acronimos}

%--------------Página de índice.  

%\nocite{*}     % Des-comentar si se desea que TODAS las referencias sean impresas en la lista de referencias, incluyendo las que no fueron finalmente citadas en el texto.

\newpage
{\setstretch{1.0}   % Interlineado de la lista.
\tableofcontents
}

% \newpage
% {\setstretch{1.0} 
% \listoftables}

% \newpage
% {\setstretch{1.0} 
% \listoffigures}

\newpage
\addtocontents{toc}{\protect\setcounter{tocdepth}{4}}   % La profundidad del índice queda en 4, 1.1.1.1
\pagenumbering{arabic}                                  % Comienza la numeración arábiga (números normales)
\setcounter{page}{1}                                    % Comienza el contador de páginas en 1

% A continuación se dejan nombres de diversos capítulos o secciones, para cambiar el nombre del archivo tan solo se debe hacer en la carpeta "capitulos" y luego llamarlos de la forma correcta en "\subfile{Capitulos/nuevonombre}".
% Los nombres de los archivos no pueden llevar tíldes ni espacios para el correcto funcionamiento del compilador, esto no tiene nada que ver con que tengan o no tilde en el documento final.

\chapter{Introducción y Motivación}                      % Nombre de section/chapter, este si usa tilde y es lo que se verá impreso

\subfile{Capitulos/01Introduccion}      % Incluye el subarchivo del section/chapter 
\clearpage                                  % limpia la página luego de que el capítulo concluya.

\chapter{Fases Topológicas}\label{Fases Topológicas}

\subfile{Capitulos/02FasesTopologicas}
\clearpage

\chapter{Efecto Magnetoeléctrico Topológico}\label{Efecto Magnetoeléctrico Topológico}
    \subfile{Capitulos/03TME}
\clearpage
  
\chapter{Guías de Ondas}\label{Guías de Ondas}
    \subfile{Capitulos/04Waveguide}
\clearpage

\chapter{Solución TEM en una Guía de Onda formada por Aislantes Topológicos}\label{TEM}

\subfile{Capitulos/05TEM}
\clearpage

\chapter{Guía de Onda en forma de slab: Solución Híbrida de Aislantes Topológicos}\label{Slab}

\subfile{Capitulos/06Slab}
\clearpage

\chapter{Conclusión}\label{Conclusión}

\subfile{Capitulos/07Conclusion}
\clearpage

\chapter{Apéndice}\label{Apéndice}

\subfile{Capitulos/Apendice}
\clearpage

\newpage
\phantomsection
\renewcommand\refname{Referencias}          % Nombre para la lista de referencias, también se utiliza "Bibliografía"
{\setstretch{0.8}                           % Interlineado de las referencias 
\addcontentsline{toc}{chapter}{Referencias} % Cambia el nombre de la lista de referencias en el índice 
\bibliography{Referencias.bib}              % Agrega las referencias al documento, estas se ubican en el archivo Referencias.bib
}

\newpage
\renewcommand{\appendixpagename}{Apéndices}     % Nombre al inicio.
\addcontentsline{toc}{chapter}{Apéndices}       % Agrega "Apéndices" al índice

% \appendix   % Empieza el ambiente de apéndices, desde ahora en adelante los capítulos, secciones, tablas, figuras, etc. vuelven a empezar su numeración

% \chapter{Test}                      % Nombre del apéndice, el capítulo será "Apéndice A" en vez de "Capítulo 1"
%     \subfile{Capitulos/Apendice}    % Llama al archivo apéndice
% \clearpage

% Este segundo apéndice tiene información general sobre LaTeX. Está comentado para que no aparezca en tu documento, pero si lo deseas puedes descomentarlo para ver su contenido.

%\chapter{Algunos consejos sobre \LaTeX{}}
%    \subfile{Otros/Consejos}
%\clearpage

% ÉXITO EN TU TESIS

\end{document}

%% file: Otros/Portada.tex
\begin{titlepage}
	
	\newgeometry{top=1 in, bottom=1 in, left=1 in, right= 1 in} 
	
	\thispagestyle{frontpage}
	
	\begin{center}
		
		\vspace*{8\baselineskip}

		{\LARGE \textbf{Ondas Electromagnéticas Confinadas en Medios Formados por Aislantes Topológicos\\}}%No abreviar, no subrayar, no usar comillas. Se escribe completamente en mayúsculas
	\vspace*{1.5\baselineskip}

		% \large{\textit{subtítulo}}\\ %No abreviar, no subrayar, no usar comillas. Solo la priemra letra usa mayúsculas
        
        \large{Tesis presentada para optar al grado de Doctor en Ciencias Físicas} %Cambiar a grado académico correspondiente. 

        \vspace*{2\baselineskip}
    
    \large{Autor:\linebreak}
  \Large{\textbf{Sebastián Ignacio Filipini Parra}}\\ %Nombre como aparece en registro académico
		\vspace{2\baselineskip}	        		\large{Director de Tesis: \\
        Dr. Mauro Cambiaso, Universidad Andrés Bello\\}

        \vspace{1cm}
        \large{Comisión de Evaluación: \\
        Dr. Cristian Villavicencio, Universidad del Bio-Bio \\
        Dr. Jerónimo Maze, Universidad Catolica de Chile\\
        Dr. Luis Foà-Torres, Universidad de Chile\\
        Dr. Sebastián Reyes, Universidad Andrés Bello}
        \end{center}

 \vspace{0.5\baselineskip}
  
	\begin{center}
        Santiago de Chile, Septiembre 2024 \end{center}
	
	%\vspace*{4\baselineskip}
	
	% La cabecera de esta página se cambia en el documento principal, en "cabecera portada"
	
\end{titlepage}

%% file: Otros/Resumen.tex
Los aislantes topológicos son una clase de materiales cuánticos que, al igual que los aislantes convencionales, presentan una brecha energética en las bandas electrónicas del \textit{bulk}. Sin embargo, a diferencia de los aislantes comunes, poseen estados conductores en sus bordes o superficies, los cuales están protegidos contra el desorden gracias a la simetría de inversión temporal. Estos estados conductores emergen en la interfaz entre materiales que exhiben diferentes propiedades topológicas en la función de onda cuántica de los electrones.

Los aislantes topológicos fueron predichos teóricamente por primera vez en grafeno por Kane y Mele en 2005 \cite{PhysRevLett.95.226801}, y en sistemas semiconductores bidimensionales por Bernevig y Zhang en 2006 \cite{Bernevig2006quantum}. El mismo año, Bernevig, Hughes y Zhang \cite{Bernevig2006HgTe} propusieron un modelo específico de aislante topológico bidimensional en estructuras de pozos cuánticos de HgCdTe, una familia de semiconductores con fuertes interacciones espín-órbita. Un año después, el grupo de Molenkamp \cite{König2007HgTe} presentó la primera evidencia experimental de un aislante topológico bidimensional. La generalización a los aislantes topológicos tridimensionales fue realizada independientemente en 2007 por los grupos teóricos de Fu, Kane y Mele \cite{Fu_Kane_Mele_PhysRevLett.98.106803} y de Moore y Balents \cite{More2007topological}. Finalmente, en 2008, el grupo de Hsieh \cite{hsieh_topological_2008} reportó el descubrimiento experimental del primer aislante topológico tridimensional en Bi$_{1-x}$Sb$_x$. Actualmente, los aislantes topológicos siguen siendo un área de investigación activa, con avances recientes que exploran nuevas fases topológicas y aplicaciones en tecnologías cuánticas emergentes \cite{Bernevig2022progress}.
%y cristales de Bi$_2$Te$_3$ y Bi$_2$Se$_3$.

En esta tesis, a modo de marco teórico, revisamos la clasificación topológica de las fases de la materia, y exploramos su respuesta electromagnética única. Destacamos el efecto magnetoeléctrico topológico, que surge cuando se rompe la simetría de inversión temporal en la superficie. Este efecto, presente en aislantes topológicos, se describe mediante un término efectivo que acopla los campos eléctricos y magnéticos, con un ``campo'' tipo axión $\theta$ que toma el valor $\pi\,(\textup{mod}\,2\pi)$ para los aislantes topológicos y $0$ para los aislantes triviales. El término de Maxwell, junto con el $\theta$-término, da lugar a una electrodinámica de Maxwell extendida, conocida como electrodinámica $\theta$, que describe la aparición de una polarización eléctrica al aplicar un campo magnético y una magnetización al aplicar un campo eléctrico. Uno de los efectos más sorprendentes de la electrodinámica $\theta$ es la modificación de las condiciones de borde que deben satisfacer los campos. Estas modificaciones se reflejan, por ejemplo, en cambios en las propiedades de propagación de las ondas electromagnéticas, como la rotación de Kerr y Faraday, y el ángulo de Brewster.

Las ecuaciones de Maxwell se modifican mediante términos proporcionales a $\boldsymbol{\nabla}\theta$ y $\partial_t \theta$. En un sistema que contiene un aislante topológico, el parámetro topológico $\theta$ es constante en el tiempo y cambia solo en la superficie del aislante, es decir, $\boldsymbol{\nabla}\theta$ tiene soporte solo en la superficie. Cuando ese es el caso (que es el considerado en esta tesis) el campo electromagnético fuera de la superficie satisface las ecuaciones convencionales de Maxwell. Por lo tanto, aunque las ecuaciones diferenciales que gobiernan la dinámica de los campos electromagnéticos en el interior del aislante topológico o fuera de él (no en la superficie) son idénticas a las ecuaciones convencionales de Maxwell, las soluciones son distintas debido a que las condiciones de borde, que determinan los coeficiente de la solución, son diferentes, lo que altera el espacio de soluciones.

El principal aporte en esta tesis radica en el estudio de los efectos del confinamiento de ondas electromagnéticas en guías de onda formadas por aislantes topológicos. Primero, realizamos un análisis de la energía electromagnética disipada y reactiva en una guía de onda, reportando cambios en ambas energías cuando la onda electromagnética tiene una polarización circular o elíptica. %Observamos que estos cambios se deben a la interacción entre la estructura topológica del TI y las propiedades de polarización de la onda electromagnética. 
Segundo, reportamos nuevas soluciones de ondas electromagnéticas transversales que, de otra manera, estarían prohibidas según el teorema de Earnshaw en la electrodinámica usual de Maxwell. Al aplicar una onda electromagnética externa a cilindros coaxiales de aislantes topológicos, observamos una rotación de la polarización debido a discontinuidades transversales del parámetro topológico $\theta$. Dicha rotación es distinta a las ya reportadas en \cite{tse_giant_2010,maciejko_topological_2010,PhysRevB.80.113304}. La transversalidad exacta de la onda electromagnética permite que se propaguen como una fibra óptica, sin reflexiones internas totales sucesivas, reduciendo las pérdidas en las curvas agudas de la fibra. Además, hemos logrado confinar ondas electromagnéticas transversales con menos de dos conductores en un rango de valores del ``campo'' $\theta$ donde es puramente imaginario. Hemos dado una explicación física de cómo se produce la rotación de la polarización a través de la generación, sucesiva e infinita, de fuentes topológicas en la superficie, que converge a las expresiones analíticas de los campos electromagnéticos totales. Estos resultados fueron reportados en \cite{Filipini2024polarization}. 

Finalmente, estudiamos la solución exacta en un guía de ondas en forma de \textit{slab}. Reportamos cambios en los modos de propagación y una rotación de la polarización dentro de la guía, observando que la polarización de la onda electromagnética no transmuta perfectamente entre un modo transversal eléctrico y transversal magnético, como se ha reportado en \cite{crosse_theory_2017}, si no que emergen modos híbridos como se sugiere en \cite{Talebi2016optical}, el cual corresponde a un estudio independiente de materiales no topológico, pero que presentan efectos magnetoeléctricos. Para analizar los nuevos modos permitidos de la onda electromagnética confinada, hemos acoplado un haz Gaussiano en el borde de la guía. Estos resultados, en preparación \cite{underprep2}, pueden abrir nuevas posibilidades en óptica y fotónica utilizando aislantes topológicos para manipular, por ejemplo, la polarización, el confinamiento y la perdidas energéticas de la luz.

\par\vspace*{\fill} % Mueve las palabras clave al final de la página
%\textbf{\textit{Keywords --}}  %Agregar todas las palabras claves asosciadas con la tesis aquí.

%-----------Si se desea poner el Abstract Des-comentar lo siguiente-----------
\newpage
\phantomsection
\addcontentsline{toc}{chapter}{Abstract} %Agrega esta sección al índice
\section*{Abstract}

Topological insulators are a class of quantum materials that, like conventional insulators, exhibit an energy gap in the electronic bands of the bulk material. However, unlike common insulators, they possess conductive states on their edges or surfaces, which are protected against disorder due to time-reversal symmetry. These conductive states emerge at the interface between materials with differing topological properties in the quantum wave function of electrons.

Topological insulators were first theoretically predicted in graphene by Kane and Mele in 2005 \cite{PhysRevLett.95.226801}, and in two-dimensional semiconductor systems by Bernevig and Zhang in 2006 \cite{Bernevig2006quantum}. That same year, Bernevig, Hughes, and Zhang \cite{Bernevig2006HgTe} proposed a specific model of a two-dimensional topological insulator in HgCdTe quantum well structures, a family of semiconductors with strong spin-orbit interactions. A year later, Molenkamp's group \cite{König2007HgTe} provided the first experimental evidence of a two-dimensional topological insulator. The generalization to three-dimensional topological insulators was independently accomplished in 2007 by the theoretical groups of Fu, Kane, and Mele \cite{Fu_Kane_Mele_PhysRevLett.98.106803} and of Moore and Balents \cite{More2007topological}. Finally, in 2008, Hsieh's group \cite{hsieh_topological_2008} reported the experimental discovery of the first three-dimensional topological insulator in Bi$_{1-x}$Sb$_x$. Currently, topological insulators remain an active area of research, with recent advances exploring new topological phases and applications in emerging quantum technologies \cite{Bernevig2022progress}.

In this thesis, as a theoretical framework, we review the topological classification of phases of matter and explore their unique electromagnetic response. We highlight the topological magnetoelectric effect, which arises when time-reversal symmetry is broken on the surface. This effect, present in topological insulators, is described by an effective term that couples electric and magnetic fields, featuring an axion-like "field" $\theta$ that takes the value $\pi,(\textup{mod},2\pi)$ for topological insulators and $0$ for trivial insulators. The Maxwell term, together with the $\theta$-term, leads to an extended form of Maxwell's electrodynamics, known as $\theta$ electrodynamics, which describes the emergence of an electric polarization when a magnetic field is applied and a magnetization when an electric field is applied. One of the most remarkable effects of $\theta$ electrodynamics is the modification of the boundary conditions that the fields must satisfy. These modifications are reflected, for example, in changes to the propagation properties of electromagnetic waves, such as Kerr and Faraday rotation, and the Brewster angle.

Maxwell's equations are modified by terms proportional to $\boldsymbol{\nabla}\theta$ and $\partial_t \theta$. In a system containing a topological insulator, typically the topological parameter $\theta$ is constant in time and changes only at the bounding surface of the insulator, i.e., $\boldsymbol{\nabla}\theta$ has support at the surface only. When that is the case (which is the one considered in this thesis) the electromagnetic field away from the surface satisfies conventional Maxwell equations. Thus, even though the differential equations governing the dynamics of the electromagnetic fields in the bulk of the topological insulator or away from it, (not at the surface) are identical to the conventional Maxwell equations, the solutions are different because the boundary conditions, which determine the coefficients of the solution, are altered, thus changing the solution space.

The main contribution of this thesis lies in the study of the effects of electromagnetic wave confinement in waveguides formed by topological insulators. First, we conducted an analysis of the dissipated and reactive electromagnetic energy in a waveguide, reporting changes in both energies when the electromagnetic wave has circular or elliptical polarization. Second, we report new solutions for transverse electromagnetic waves that would otherwise be forbidden according to Earnshaw's theorem in conventional Maxwell electrodynamics. By applying an external electromagnetic wave to coaxial cylinders of topological insulators, we observed a rotation of the polarization due to transverse discontinuities in the topological parameter $\theta$. This rotation is distinct from those previously reported in \cite{tse_giant_2010,maciejko_topological_2010,PhysRevB.80.113304}. The exact transversality of the electromagnetic wave allows it to propagate as an optical fiber without successive total internal reflections, reducing losses in sharp bends of the fiber. Additionally, we have succeeded in confining transverse electromagnetic waves with fewer than two conductors within a range of values of the $\theta$ "field" where it is purely imaginary. We provided a physical explanation of how polarization rotation occurs through the successive and infinite generation of topological sources on the surface, which converges to the analytical expressions of the total electromagnetic fields. These results were reported in \cite{Filipini2024polarization}.

Finally, we studied the exact solution in a slab waveguide. We report changes in the propagation modes and a rotation of the polarization within the guide, observing that the polarization of the electromagnetic wave does not perfectly transition between a transverse electric mode and a transverse magnetic mode, as reported in \cite{crosse_theory_2017}, but instead hybrid modes emerge, as suggested in \cite{Talebi2016optical}, which corresponds to an independent study of non-topological materials that exhibit magnetoelectric effects. To analyze the new allowed modes of the confined electromagnetic wave, we coupled a Gaussian beam at the edge of the guide. These results, currently in preparation \cite{underprep2}, could open new possibilities in optics and photonics by using topological insulators to manipulate, for example, polarization, confinement, and energy losses of light.

\par\vspace*{\fill} % Mueve las palabras clave al final de la página
\textbf{\textit{Keywords --}} Topological insulator, axion electrodynamics, electromagnetic waveguide, TEM wave, slab waveguide, polarization rotation.   % Agregar las palabras claves en inglés

\biblio %Se necesita para referenciar cuando se compilan subarchivos individuales - NO SACAR

%% file: Otros/Acronimos.tex
\small
\begin{itemize}[itemsep=0.1em]
    \item Las unidades Gaussianas se utilizan cuando el texto se refiere al efecto magnetoeléctrico topológico y a la propagación de ondas confinadas en aislantes topológicos. En este sistema de medidas, las constantes electromagnéticas $\epsilon$ y son adimensionales, la unidad de medida de la longitud es el $cm$ y los campos eléctrico y magnético tienen la misma unidad.
    \item Aislante topológico (TI, por sus siglas en ingles).
    \item Condiciones de borde (BC, por sus siglas en ingles).
    \item Bidimensional (2D).
    \item Bernevig-Hughes-Zhang (BHZ).
    \item Efecto Hall Anómalo cuántico (QAHE, por sus siglas en ingles).
    \item Efecto Hall cuántico entero (IQHE, por sus siglas en ingles).
    \item Efecto Hall de espín cuántico (QSHE, por sus siglas en ingles).
    \item Efecto magnético quiral (CME, por sus siglas en ingles).
    \item Efecto magnetoeléctrico topológico (TME, por sus siglas en ingles).
    \item Electrodinámica $\theta$ ($\theta$-ED).
    \item Electromagnético (EM).
    \item Estados superficiales (SS, por sus siglas en inglés).
    \item Haz Gaussiano (haz-G).
    \item Landau-Ginzburg (LG).
    \item Momentos invariantes en el tiempo (TRIM, por sus siglas en ingles).
    \item Ondas electromagnéticas (OEM).
    \item Polarizabilidad magnetoeléctrica topológica (TMEP, por sus siglas en ingles).
    \item Simetría de inversión temporal (TRS, por sus siglas en ingles).
    \item Teoría de campo topológico (TFT, por sus siglas en ingles).
    \item Teoría de bandas topológicas (TBT, por sus siglas en ingles).
    \item Transversales electromagnéticas (TEM).
    \item Zona de Brillouin (BZ, por sus siglas en ingles).
    \item Zona de Brillouin efectiva (EBZ, por sus siglas en ingles).
    \item Cualquier otra convención se explicará a lo largo del texto.
\end{itemize}

\biblio %Se necesita para referenciar cuando se compilan subarchivos individuales - NO SACAR

%% file: Capitulos/01Introduccion.tex
Las reglas que rigen los sistemas de materia condensada, compuestos por un número de partículas del orden del número de Avogadro ($\sim 6 \cdot 10^{23}$ partículas/mol), pueden diferir significativamente de aquellas que gobiernan la física de partículas elementales o sistemas con pocas partículas. Una diferencia fundamental es la aparición de excitaciones colectivas llamadas \textit{cuasipartículas}, que emergen como soluciones efectivas a bajas energías y describen el comportamiento de partículas en interacción dentro de un sólido. Estas cuasipartículas pueden comportarse como fermiones o bosones y comprenden fenómenos como huecos, fonones, plasmones, magnones, y otras más exóticas como fermiones de Weyl, anyones y fermiones compuestos. Las cuasipartículas que residen en los bordes o superficiales de los aislantes topológicos (TIs, por sus siglas en ingles) \cite{Hasan2010Colloquium} son el foco de esta tesis. Estos fenómenos emergentes, que se manifiestan a bajas energías (es decir, a energías mucho menores que las características de las interacciones fundamentales) dentro de sólidos con espacio físico finito, facilitan la comprobación experimental de predicciones teóricas y su aplicación tecnológica. 

\begin{SCfigure}[1][t]
\caption{Un sólido cristalino tiene una estructura periódica que es invariante bajo traslaciones discretas. En un líquido, los átomos están distribuidos de manera aleatoria e isotrópica, lo que hace que este estado sea invariante bajo traslaciones continuas. Esta transición entre un sólido y un líquido está caracterizada por el rompimiento de una simetría, que puede ser descrita utilizando la teoría de campo efectivo de Landau-Ginzburg.
\label{fig:SoLiquido}}
\includegraphics[scale=0.32]{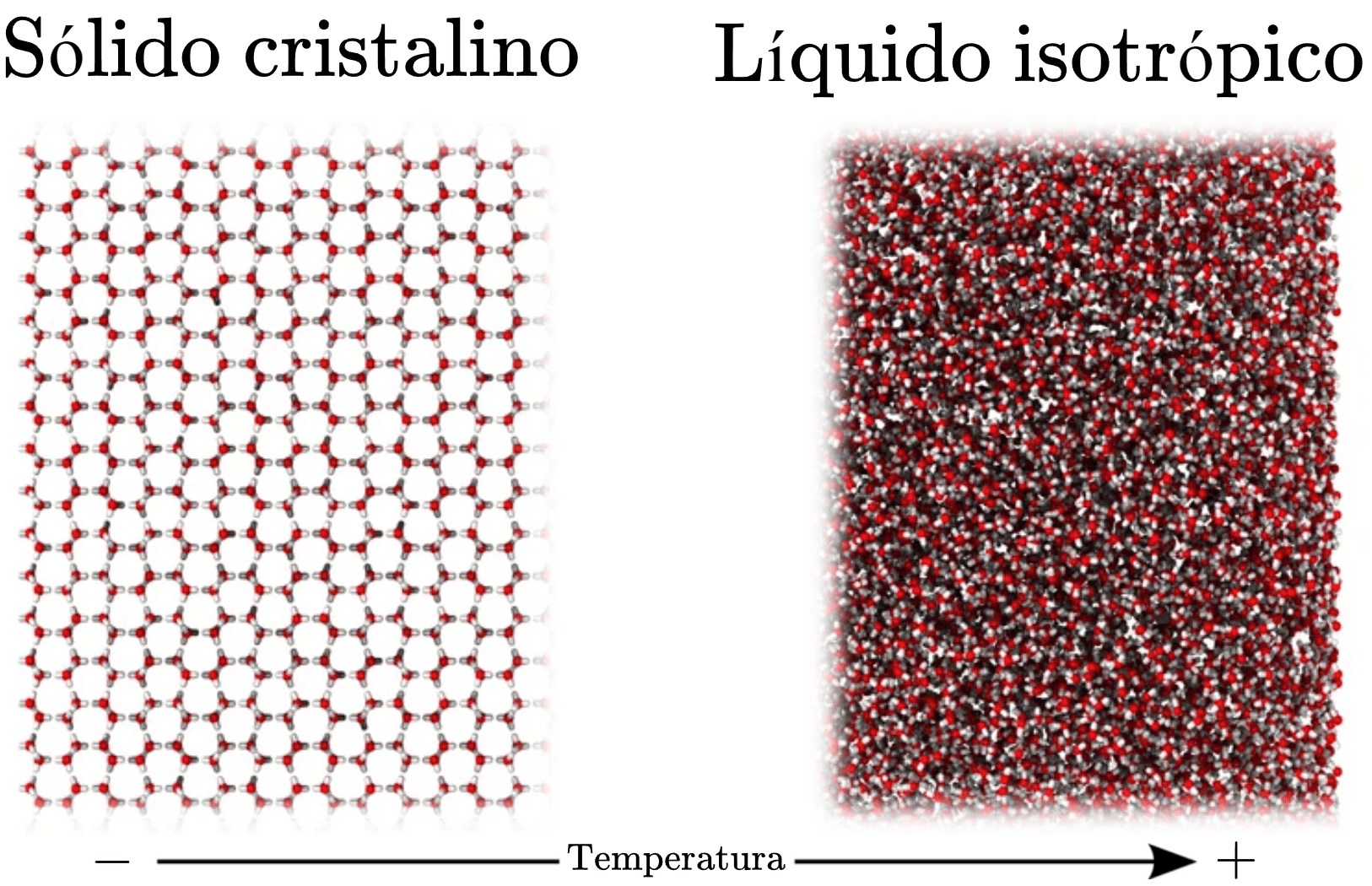}
\end{SCfigure}

%Las excitaciones elementales del sistema están influenciadas por la fase cuántica del sistema. Por lo tanto,
La necesidad de clasificar sin ambigüedades las fases cuánticas de la materia es crucial.
%, es decir, especificar los criterios formales que describen y distinguen fases como sólidos cristalinos, ferromagnetos y antiferromagnetos, superfluidos, superconductores, etc.
La teoría de campo efectivo de Landau-Ginzburg (LG) nos permite entender las fases de la materia mediante parámetros de orden que dependen del patrón de rompimiento espontáneo de la simetría del sistema \cite{wen2004quantum}. Por ejemplo, la transición de fase de líquido a sólido que se muestra en la Fig. (\ref{fig:SoLiquido}): En un liquido las moléculas se distribuyen de manera aleatoria, y decimos que este estado es invariante bajo traslaciones espaciales continuas. Al enfriar un líquido por debajo de una temperatura crítica, se convierte en sólido cristalino, rompiendo la simetría de traslación continua a una discreta. Al pasar del estado líquido al sólido, la materia se ordena y su simetría disminuye.

%---------------Figure------------
%---------------End Figure---------

\begin{figure}[t]
    \centering
    \includegraphics[width=0.8\linewidth]{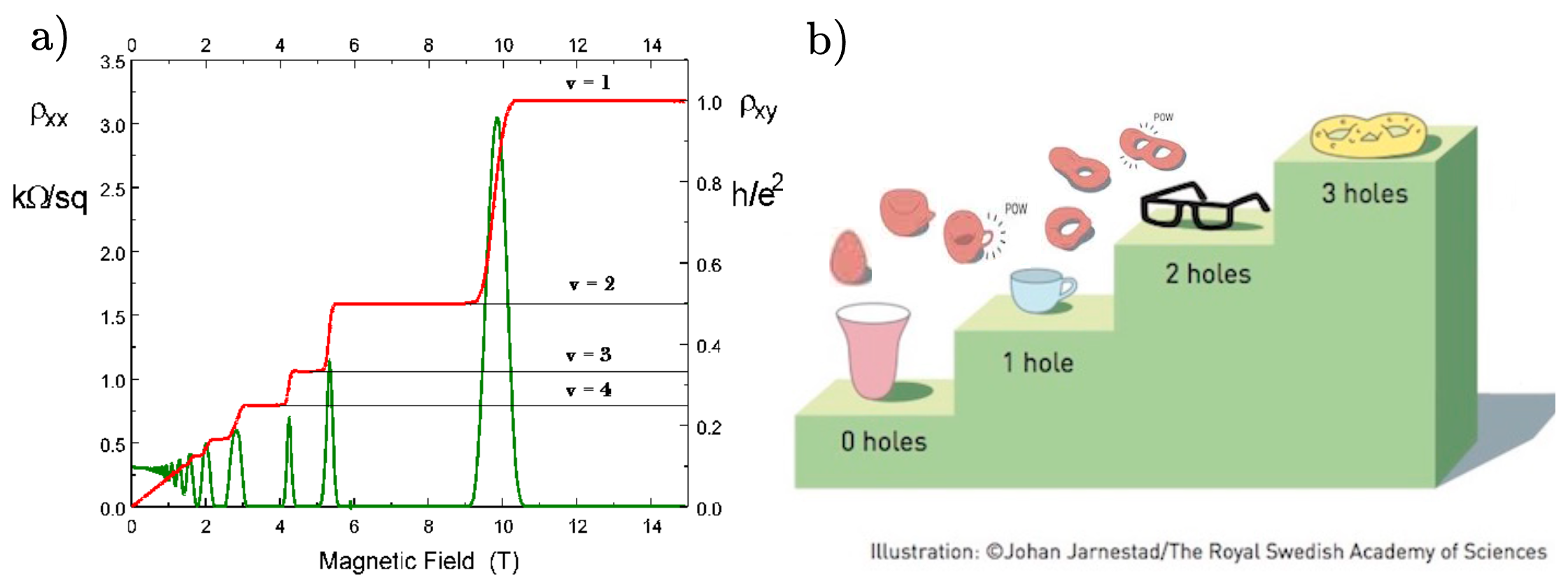}
    \caption{a) Efecto Hall cuántico entero: gráfico de la resistencia Hall (transversal) $\rho_{xy}$ y longitudinal $\rho_{xx}$ en función del campo magnético. Los números $\nu$ indican el factor de llenado de los niveles de Landau y están relacionados con el numero de canales quirales en el borde \cite{Hasan2010Colloquium}. Estas mesetas son robustas independientes de las imperfecciones de la muestra. b) Al igual que no podemos deformar continuamente una superficie de un agujero a dos agujeros sin romperla, tampoco podemos pasar de una meseta a otra sin cerrar la brecha energética. Esquema de transiciones de fase topológica relacionado con el Premio Nobel de Física de 2016.}
    \label{fig:topoQHE}
\end{figure}

Aunque la teoría de LG ha sido exitosa en describir las fases cuánticas en términos de simetrías, el descubrimiento de los TIs ha revelado que esta clasificación es incompleta. El orden que caracteriza a los TIs no puede explicarse simplemente como una ruptura de simetría, lo que ha llevado al desarrollo de una nueva clasificación basada en el concepto de orden topológico. Este concepto se introdujo para describir el efecto Hall cuántico entero (IQHE) \cite{thouless1982quantized}, %y el efecto Hall cuantico fraccionario (FQHE) \cite{wen1995topological}, 
el cual corresponde a una versión cuántica del efecto de Hall cuando el material bidimensional (2D) está sometido a intensos campos magnéticos y bajas temperaturas. Las propiedades fundamentales, como la conductividad de Hall, $\sigma_{H}=\nu(e^{2}/h)$ cuantizada en valores enteros de la constante fundamental $e^2/h=1/(25 812.807 572\,\Omega)$,  
%y los modos de conducción en el borde
no se ven afectadas por cambios suaves en los parámetros del sistema (como diferentes impurezas o geometrías de la muestra), a menos que ocurra una transición de fase topológica, ver Fig. (\ref{fig:topoQHE},a). Por este descubrimiento junto con otras investigaciones, Thouless, Haldane y Kosterlitz recibieron el premio Nobel el año 2016 por sus descubrimientos teóricos de las transiciones de fase topológicas y fases topológicas de la materia.  Estas propiedades, invariantes y bien definidas, son consecuencia del orden topológico y muestran la relación entre la física de materia condensada y la topología.

La topología es la rama de la matemática que estudia las propiedades de superficies, sean de cuerpos geométricos o abstractas, que permanecen invariantes bajo transformaciones continuas. Estas propiedades no dependen de la forma precisa del objeto, sino de su estructura fundamental. Un ejemplo notable de la conexión entre geometría y topología es la fórmula de Gauss-Bonnet \cite{Nakahara2018geometry}, que se expresa como $1/2\pi\int_{S}Kd^{2}x=2(1-g)$. Esta ecuación relaciona la curvatura Gaussiana $K$ de una superficie cerrada $S$ con una propiedad topológica global, el \textit{genus} $g$,  que corresponde al número de agujeros de la superficie. El genus es un invariante topológico, lo que significa que no cambia bajo deformaciones continuas de la superficie. Por ejemplo, un toro (o dónut) con $g=1$ es topológicamente equivalente a una taza con $g=1$, ya que ambos tienen un solo agujero y pueden transformarse entre sí mediante deformaciones continuas sin cortar ni pegar superficies , ver Fig. (\ref{fig:topoQHE},b). 
%Por ejemplo, según la fórmula de Gauss-Bonnet \cite{Nakahara2018geometry}, $1/2\pi\int_{S}Kd^{2}x=2(1-g)$, donde $K$ es la curvatura Gaussiana y $g$ es el número de agujeros o \textit{genus} y no depende de los detalles geométricos del objeto, más bien de una propiedad global como el número de agujeros. De esta manera, un toro (o dónut) con $g=1$ es topológicamente equivalente a una taza con $g=1$, en el sentido de que las deformaciones continuas pueden transformar un objeto en el otro sin romper ni unir superficies.

En el contexto de la física de la materia condensada, la topología no se refiere a la forma geométrica del sólido, sino a las propiedades globales de la función de onda de los electrones en un espacio más abstracto conocido como espacio de momentos\footnote{Los estados electrónicos adquieren una fase geométrica, conocida como fase de Berry, cuando el estado evoluciona a lo largo de una trayectoria en el tiempo, en el espacio de parámetros del Hamiltoniano o en el espacio del número de onda dentro de la primera zona de Brillouin.}.  En un sólido cristalino, los átomos se ordenan de forma periódica, lo que permite definir una celda unitaria que se repite en el espacio para formar el sólido. Esta repetición se describe mediante el vector de red $\mathbf{R}$, que indica las traslaciones necesarias para cubrir todo el cristal. Los electrones que se mueven en el potencial periódico de la red cristalina son descritos por la función de onda de Bloch,
\begin{align}
    |\psi_{n\mathbf{k}}(\mathbf{r})\rangle=e^{i\mathbf{k}\cdot \mathbf{r}} |u_{n\mathbf{k}}(\mathbf{r})\rangle
\end{align}
donde $|u_{n\mathbf{k}}(\mathbf{r})\rangle$ es una función periódica con la periodicidad de la red, $\mathbf{k}$ es el momento cristalino, y $n$ es el índice que etiqueta los niveles de energía. Las bandas de energía surgen de la disposición del gran número de átomos en el sólido y representan los valores de energía que puede tomar un electrón dentro de la red. Para un electrón que se propaga en el vacío, $\mathbf{k}$ puede tomar cualquier valor, pero en el interior de un cristal, la estructura periódica limita los posibles valores. La periodicidad de la red impone que $|\psi_{n\mathbf{k}}(\mathbf{r})\rangle = |\psi_{n\mathbf{k}}(\mathbf{r} + \mathbf{R})\rangle$. Esta propiedad implica que el momento $\mathbf{k}$ del electrón solo puede tomar valores dentro de una región finita, denominada zona de Brillouin (BZ, por sus siglas en inglés).
%, donde $\mathbf{k}$ es el momento cristalino de Bloch, $\mathbf{r}$ es la coordenada espacial del material y $n$ es la etiqueta de las bandas  de energía $\varepsilon_{n\mathbf{k}}$ del solido. 

Si permitimos que las funciones de onda de Bloch evolucionen adiabáticamente a lo largo de una curva cerrada en el espacio de parámetros del Hamiltoniano del sistema, por ejemplo, el vector de onda  $\mathbf{k}$ recorre un \textit{loop} en la BZ, estas funciones de onda acumulan una fase adicional conocida como la fase de Berry. Esta fase es intrínseca al estado cuántico y no puede eliminarse mediante una transformación de \textit{gauge}, lo que la convierte en una fase observable y físicamente significativa. Esta acumulación de fase nos permite definir una curvatura $F$ en el espacio de parámetros, conocida como la curvatura de Berry. De manera análoga a cómo las formas geométricas se clasifican topológicamente mediante el genus, la estructura de bandas de un material puede clasificarse utilizando la fórmula de Gauss-Bonnet-Chern,
\begin{align}
    \frac{1}{2\pi}\int_{S=BZ}Fd^{2}k=C_1,
\end{align}
donde la integral se realiza sobre la BZ, que tiene una geometría toroidal, y $C_1\in\mathbb{Z}$ es un invariante topológico denominado el primer número de Chern \cite{thouless1982quantized,thouless1998topological}. Este número caracteriza la estructura topológica de las bandas de un material. Así como es imposible transformar una esfera (con genus cero) en un toro (con genus uno) sin romper o unir superficies, tampoco es posible convertir una banda convencional (con número de Chern cero) en una banda topológica (con número de Chern distinto de cero) sin cerrar la brecha energética. 

%Las bandas de energía que poseen una brecha, separando el estado fundamental (banda de valencia) del excitado (banda de conducción), pueden modificarse continuamente, añadiendo pequeñas perturbaciones, sin cerrar esta brecha en el \textit{bulk} (volumen) del material. De forma similar a cómo se clasifica topológicamente las formas geométricas a través del genus, podemos clasificar la estructura de bandas de un material mediante la fórmula de Gauss-Bonnet-Chern, $1/2\pi\int_{S}Fd^{2}k=C_1$, donde $F$ es la \textit{curvatura de Berry} y $C_1$ es el primer \textit{número de Chern} \cite{thouless1982quantized,thouless1998topological}. Así como no podemos convertir una esfera (con genus cero) en un toro (con genus uno) sin romper ni unir superficies, no podemos convertir una banda convencional (con número de Chern cero) en una banda topológica (con número de Chern distinto de cero) sin cerrar la brecha energética.

El IQHE, descubierto por K. von Klitzing en 1980 \cite{PhysRevLett.45.494}, es el primer ejemplo de un estado cuántico topológicamente distinto de otros estados de la materia. En 1981, Laughlin \cite{laughlin1981quantized} 
presentó un argumento para explicar el IQHE. Laughlin consideró un gas de electrones 2D confinado en una cinta en forma de anillo con un fuerte campo magnético normal a su superficie. Además, introdujo un flujo magnético $\Phi(t)$ que enhebra al anillo y al variar en el tiempo simula el campo eléctrico angular que genera corriente. El aumento adiabático de $\Phi$ a $\Phi+\Phi_0$, con $\Phi_0=hc/e$ como el cuanto de flujo magnético, el sistema vuelve a su estado inicial. Durante el proceso de un ciclo, se ``bombeó'' un número entero de electrones, dando lugar a la conductancia de Hall cuantizada en unidades de $e^2/h$, el cuanto de conductancia Hall. La cuantización precisa de $\sigma_{H}$ se explica por ser un invariante topológico, el primer número de Chern $C_1$ \cite{thouless1982quantized}, entonces $\sigma_{H}=C_1 e^2/h$. 
%En la geometría moderna, la curvatura local de una superficie se define como el desajuste angular tras el recorrido de un bucle cerrado infinitesimal, dividido por el área del bucle.
Este concepto topológico identifica a dos aislantes cuyos Hamiltonianos tienen una brecha de energía que separa la banda de conducción de la de valencia. A diferencia de un aislante convencional, un aislante de Hall presenta estados de conducción en el borde quirales debido a que el campo magnético obliga a los portadores de carga a ``enroscarse'' alrededor del campo. Así, en el borde de la muestra, las partículas se propagan a lo largo de este. De esta manera, el orden topológico se relaciona físicamente con el número de canales unidimensionales quirales en el borde de la muestra, determinado por el número de Chern.

El interés en el orden topológico ha resurgido con el descubrimiento de los TIs \cite{PhysRevLett.95.226801, kane_z2_PhysRevLett.95.146802, Bernevig2006HgTe, Kane2007topological, hsieh_topological_2008}. Estos materiales tienen una brecha de energía en el bulk, similar a un aislante ordinario, pero presentan estados de conducción en el borde/superficie, al igual que un aislante de Hall. En los TIs 2D, el sistema se comporta como dos copias superpuestas del IQHE para cada proyección de espín. Debido al acoplamiento espín-órbita, el campo magnético efectivo se cancela, resultando en estados de conducción helicoidales en el borde, protegidos por simetría de inversión temporal (TRS). Este fenómeno es conocido como el efecto Hall de espín cuántico (QSHE) o TI 2D, teorizado en \cite{Bernevig2006quantum} y demostrado experimentalmente en pozos cuánticos de HgTe/CdTe \cite{Bernevig2006HgTe}. La extensión de estos sistemas a 3D, propuesta en \cite{Kane2007topological}, llevó al descubrimiento experimental de los TIs 3D en compuestos como Bi$_2$Se$_3$, Bi$_2$Te$_3$, y Sb$_2$Te$_3$ \cite{xia2009observation, zhang_topological_2009}.

La teoría de bandas topológicas (TBT) fue la primera en describir los TIs gracias al trabajo pionero de Kane y Mele \cite{kane_z2_PhysRevLett.95.146802}, quienes introdujeron el invariante topológico $\mathbb{Z}_{2}$ dentro de la teoría de bandas no interactiva. Este invariante se ha generalizado para los TIs 3D, donde hay cuatro invariantes $\mathbb{Z}_{2}$ para caracterizar completamente la topología. Otro enfoque es la teoría de campo topológico (TFT), introducida por Zhang y Hu \cite{doi:10.1126/science.294.5543.823}, que desarrollaron un modelo microscópico de un TI con TRS en (4+1)D, mostrando que la teoría efectiva está descrita por la acción de Chern-Simons en (4+1)D \cite{bernevig2002effective}. Mediante un procedimiento de reducción dimensional, se construyó la TFT para TIs 3D y 2D \cite{Qi2008topological}. La TFT es válida para sistemas con interacciones y desorden, y recientemente se demostró que se reduce exactamente a la TBT en el límite de no interacción \cite{wang2010equivalent}.

En los TIs, pueden surgir propiedades interesantes cuando la simetría de inversión temporal se conserva en el volumen pero se rompe en la superficie mediante perturbaciones magnéticas \cite{Essin2009magnetoelectric}. Esto separa 
%el cono de Dirac 
las bandas conductoras de los estados de borde en los extremos del material, dando lugar al efecto magnetoeléctrico topológico (TME). Este fenómeno se describe mediante un término topológico en la acción que acopla los campos eléctrico y magnético y es proporcional a un ``campo'' axión \cite{Wilczek1987two,Hasan2010Colloquium,Qi2008topological,Qi2011topological}. El término topológico, $S_\theta$, que se le añade a la acción de Maxwell es, 
\begin{equation} 
\label{eq:effaction}
    S_{\theta}=-\frac{\alpha}{4\pi^{2}}\int d^{4}x \theta(\mathbf{r},t)\mathbf{E}\cdot \mathbf{B}=-\frac{\alpha}{32\pi^{2}}\int d^{4}x \theta(\mathbf{r},t)\epsilon ^{\mu \nu \sigma \tau  }F_{\mu \nu}F_{\sigma \tau}.
\end{equation}
Llamamos a $\theta(\mathbf{r},t)$ un ``campo'' axiónico por analogía con su interpretación como potencial gauge fundamental en la física de altas energías. En el contexto de la materia condensada, $\theta(\mathbf{r},t)$ es llamado polarizabilidad magnetoeléctrica topológica (TMEP) \cite{Essin2009magnetoelectric}. Este término se puede derivar de un modelo general de \textit{tight-binding} acoplado al campo gauge electromagnético $A_{\mu}$ en el régimen no interactivo, donde $\alpha = e^2 / \hbar c\approx 1/137$ es la constante de estructura fina. Actualmente, la teoría considerada correcta para describir la respuesta electromagnética (EM) de materiales con fases topológicas es la electrodinámica axiónica, o simplemente $\theta$-ED, descrita por la acción efectiva de la Ec. \eqref{eq:effaction}. En este contexto, $\theta(\mathbf{r},t) = 0,\pi~(\textup{mod}~ 2\pi)$ para aislantes triviales y para TIs respectivamente.

%Revisar si en la accion poner \theta=\frac{\alpha}{\pi}\theta_{TI} las unidades de $\theta$ en los WSM cambian.
Otro material de interés con fase topológica son los semimetales de Weyl (WSM) \cite{wan2011topological,burkov2011weyl}. Para los WSM, el TMEP es $\theta(\mathbf{r},t) = 2(\mathbf{b} \cdot \mathbf{r} - b_{0}t)$, donde $2\mathbf{b}$ refleja la separación de los conos de Dirac en pares de nodos de Weyl en el espacio de momento, y $2b_{0}$ cuantifica la diferencia de energía.

En general, el parámetro $\theta$ puede considerarse como un caracterizador efectivo de los grados de libertad topológicos, similar a como la permitividad eléctrica $\epsilon$ y la permeabilidad magnética $\mu$ caracterizan las propiedades ópticas de los materiales. Debido al TME, algunas predicciones tradicionales del electromagnetismo se modifican significativamente. Estos cambios incluyen efectos como los de Kerr y Faraday \cite{tse_giant_2010}, la descripción de campos estáticos dentro y fuera de un TI \cite{qi_monopole_2009,martin2015green,martin2016electro,martin2016electromagnetic}, el efecto Casimir \cite{martin2016green}, y la fórmula de Fresnel para la reflexión de ondas electromagnéticas en interfaces entre aislantes triviales y topológicos \cite{maciejko_topological_2010,crosse_optical_2016,crosse_electromagnetic_2015,wu_quantized_2016}, el ángulo de Brewster \cite{PhysRevB.80.113304}, rotaciones de la polarización y cambios en la relación de dispersión en guías de ondas \cite{Gomes2010testing,melo_topological_2016,crosse_theory_2017}, entre otras.

%Además, estos materiales permiten realizar experimentos ópticos que cuantifican magnitudes fundamentales, como la constante de estructura fina, mediante las respuestas electromagnéticas de estos materiales, independientemente de sus detalles. Esto resalta la naturaleza topológica de los estados emergentes en estos materiales que dan lugar al TME.

Un tema central de esta tesis es la comprensión y caracterización de los TIs, materiales que permiten explorar y entender nuevos estados cuánticos de la materia. 
%Los aislantes topológicos se distinguen por poseer un orden topológico, una propiedad global que es robusta frente a perturbaciones locales y que no se puede describir mediante los parámetros de orden convencionales. 
Estos materiales son de gran relevancia teórica tanto para físicos de partículas como para físicos de materia condensada, ya que ofrecen un marco para estudiar fenómenos exóticos como los estados de borde protegidos topológicamente y las cuasipartículas no convencionales. Los TIs tienen aplicaciones prometedoras en diversos campos, incluyendo óptica, fotónica, espintrónica y metrología. En óptica y fotónica, la capacidad de manipular la respuesta EM de los TIs abre nuevas posibilidades para el control de la luz a nano y micro escala, lo que es crucial para el desarrollo de dispositivos optoelectrónicos avanzados. 
%En espintrónica, los estados de borde protegidos pueden permitir la transmisión de corriente de espín sin disipación, lo que es fundamental para el desarrollo de dispositivos de almacenamiento y procesamiento de información de alta eficiencia. 
Además, los TIs tienen un potencial para aportar a nuevos entendimientos en la física. En particular, el estudio de su respuesta EM puede elucidar aspectos no comprendidos de estos materiales, como la naturaleza de las interacciones entre los estados de borde y las excitaciones colectivas en el volumen del material. Esto, a su vez, puede conducir a la concepción de configuraciones experimentales novedosas para medir el TME. Explorar nuevas posibilidades de transporte eléctrico, de espín o de luz en los TIs no solo enriquece nuestro conocimiento fundamental, sino que también tiene implicaciones prácticas. Por ejemplo, el control preciso de la luz en dispositivos fotónicos podría llevar a avances en la comunicación óptica y en la metrología, donde la precisión y estabilidad son cruciales.

En los primeros capítulos de este trabajo, revisaremos el marco teórico que condujo al desarrollo de los TIs. Específicamente, en el capítulo (\ref{Fases Topológicas}) examinaremos de forma general los aspectos cuánticos de estos materiales y cómo surgen y se caracterizan las fases topológicas. En el capítulo (\ref{Efecto Magnetoeléctrico Topológico}) describiremos la teoría efectiva que explica la respuesta electromagnética de los TIs cuando se abre la brecha en la superficie del material, además de explorar algunos efectos peculiares que se manifiestan en ellos. En el capítulo (\ref{Guías de Ondas}), abordaremos la teoría general de la propagación de ondas electromagnéticas en medios confinados que incluyen TIs en el sistema, así como la respuesta de las corrientes superficiales y cómo estas modifican las leyes de conservación de la energía, presentando un primer resultado de esta tesis. En el capítulo (\ref{TEM}), presentaremos nuestro estudio sobre la propagación de ondas electromagnéticas transversales (TEM) confinadas por TIs. En este estudio, analizamos la dinámica de los campos eléctricos y magnéticos en una guía de onda cilíndrica, logrando obtener soluciones TEM exactas que serían imposibles sin el efecto magnetoeléctrico, que eluden el teorema de Earnshaw como se menciono por primera vez en \cite{martin2016electro}. Los resultados de este estudio han sido reportados en \cite{Filipini2024polarization}. Además, conseguimos confinar los modos TEM reduciendo el número de conductores a uno o incluso a cero, minimizando así la pérdida de potencia por calentamiento Joule \cite{underprep}. En el capítulo (\ref{Slab}), exploraremos la solución completa de la $\theta$-ED en una guía de onda en forma de slab (o losa), donde observamos cambios significativos en comparación con trabajos de la literatura, como los reportados en \cite{crosse_theory_2017}. Reportamos variaciones en los modos de propagación, la relación de dispersión, la potencia transmitida y las amplitudes de acoplamiento \cite{underprep2}. Finalmente, en el capítulo (\ref{Conclusión}), presentaremos algunas conclusiones sobre los problemas estudiados y proponemos futuras líneas de investigación en medios y problemas de confinamiento más elaborados.

\biblio %Se necesita para referenciar cuando se compilan subarchivos individuales - NO SACAR

%% file: Capitulos/02FasesTopologicas.tex
La estructura de bandas energéticas, que describe los rangos de energía permitidos y prohibidos para los electrones en sólidos cristalinos, proporciona información detallada sobre sus propiedades electrónicas. Cada material presenta un espectro único de autoenergías que define sus bandas, y este espectro depende de la estructura cristalina, la composición química, las interacciones y las simetrías presentes, entre otros factores. A pesar de las diferencias en los espectros, los materiales pueden clasificarse según sus propiedades electrónicas en categorías como metales, semiconductores y aislantes. En los aislantes, las bandas de valencia están completamente ocupadas por electrones, mientras que las bandas de conducción están vacías. Estos materiales se caracterizan por una brecha de energía entre las bandas de valencia y de conducción. En contraste, los metales tienen bandas de conducción parcialmente ocupadas. Esta diferencia en la ocupación de las bandas de energía da lugar a distintas respuestas de los materiales frente a campos eléctricos y magnéticos, por ejemplo, diferencias en las propiedades de conducción eléctrica.

El Hamiltoniano que describe cada sistema puede modificarse continuamente en sus parámetros, permitiendo una interpolación entre dos sistemas sin que se cierre la brecha energética en ambos. En el contexto de la física de la materia condensada, modificar el Hamiltoniano significa ajustar parámetros o añadir perturbaciones al modelo, como variar la magnitud de un campo externo o cambiar la interacción entre partículas, sin que se produzca una transición de fase cuántica, es decir, sin que la estructura fundamental del sistema cambie drásticamente \cite{Qi2011topological}. Así, dos materiales con brechas energéticas pueden ser ambos aislantes, pero con una clasificación topológica distinta: triviales o no triviales, como los TIs. A primera vista, es imposible distinguir la topología observando únicamente las bandas energéticas en el \textit{bulk} del material; para esto, es necesario estudiar los estados de Bloch y cómo cambian cuando se varía adiabáticamente el Hamiltoniano.

En este capítulo, presentamos algunos de los desarrollos teóricos más importantes en la clasificación topológica de los materiales. Siguiendo el trabajo de Ando \cite{Ando2013topological}, en la Sec. (\ref{2.1}) revisaremos brevemente el concepto de estado de Bloch en un potencial periódico y definiremos la fase de Berry que el estado de Bloch adquiere cuando el Hamiltoniano depende del tiempo a través de cambios adiabáticos en los parámetros. Estos conceptos son fundamentales en la teoría de bandas topológicas. En la Sec. (\ref{2.2}), discutiremos el cálculo de la conductividad de Hall en 2D y cómo esta se cuantiza al aplicar un campo magnético externo que rompe la simetría de inversión temporal. Además, exploraremos la relación de esta cuantización con la topología de los materiales, basándonos en el trabajo pionero de Thouless, Kohmoto, Nightingale y Den Nijs \cite{thouless1982quantized} (TKNN). En la Sec. (\ref{2.3}), analizaremos cómo algunos materiales magnéticos bidimensionales presentan números cuánticos topológicos en la conductividad de Hall sin necesidad de un campo magnético externo. En la Sec. (\ref{2.4}), introduciremos un nuevo número topológico para materiales 2D que preservan la simetría de inversión temporal, lo que da lugar a corrientes de espín cuantizadas en el borde de la muestra. Para ello, revisaremos el concepto de pares de Kramers y su conexión con la polarización del material. Presentaremos el modelo de Bernevig-Hughes-Zhang \cite{Bernevig2006HgTe}, que proporciona la primera propuesta experimental de un aislante topológico 2D. Finalmente, en la Sec. (\ref{2.5}), mostraremos cómo calcular la topología de un material 3D que preserva la simetría de inversión temporal, conocido como aislante topológico 3D. Se comunica al lector que, aunque este primer capítulo está centrado en la teoría de bandas, en esta tesis no realizaremos cálculos detallados en este ámbito. Nos interesaremos particularmente en cómo estos aislantes topológicos 3D, definidos en la última sección, responden a campos electromagnéticos externos.

%Finalmente, examinaremos la teoría de campos topológica que describe las respuestas electromagnéticas de los aislantes topológicos, proporcionando un marco unificado para entender estos fascinantes fenómenos.
%
\section{Fase de Berry de los estados de Bloch}\label{2.1}
Desde el punto de vista microscópico, el concepto de fase de Berry \cite{Berry1984quantal} es crucial para comprender el orden topológico en la materia. En esta tesis, consideraremos brevemente las ideas de Berry en un cristal perfecto con simetría de traslación discreta, \textit{i.e.}, $H(\mathbf{r}) = H(\mathbf{r} + \mathbf{R})$, donde $\mathbf{R}$ es el vector de traslación de la red. El estado de Bloch \cite{Bloch1929elektronen} se expresa como $|\psi_{n\mathbf{k}}(\mathbf{r})\rangle = e^{i\mathbf{k} \cdot \mathbf{r}} |u_{n\mathbf{k}}(\mathbf{r})\rangle$, donde $n$ es el índice de la banda de energía, $\mathbf{k}$ es el momento cristalino y $|u_{n\mathbf{k}}(\mathbf{r})\rangle$ es una función periódica con la misma periodicidad que la red, es decir, $|u_{n\mathbf{k}}(\mathbf{r})\rangle = |u_{n\mathbf{k}}(\mathbf{r} + \mathbf{R})\rangle$. Esta última propiedad, restringe los valores de $\mathbf{k}$ y solo puede tomar valores dentro de la BZ. La ecuación de Schrödinger para los estados $|u_{n\mathbf{k}}\rangle$ es,
\begin{align}
    H_{\mathbf{k}}|u_{n\mathbf{k}}\rangle=\varepsilon_{n\mathbf{k}}|u_{n\mathbf{k}}\rangle, ~~~~~~~~~~~~~~~~~  \langle u_{n\mathbf{k}}|u_{m\mathbf{k}}\rangle=\delta_{nm},
\end{align}
donde $H_{\mathbf{k}} = e^{-i\mathbf{k} \cdot \mathbf{r}} H(\mathbf{r}) e^{i\mathbf{k} \cdot \mathbf{r}}$, y $\varepsilon_{n\mathbf{k}}$ son las autoenergías que definen las bandas del cristal. Supongamos que el sólido está sometido a variaciones adiabáticas, i.e., cambios lentos de los parámetros del sistema $\boldsymbol{\lambda}(t)=(\lambda_1,\lambda_2,...)$, ordenados como vector. De esta forma, el Hamiltoniano del sistema cambia en el tiempo, y por consecuencia, sus autoestados y autoenergías satisfacen, 
\begin{align}
    H_{\mathbf{k}}(t)|u_{n\mathbf{k}}(t)\rangle=\varepsilon_{n\mathbf{k}}(t)|u_{n\mathbf{k}}(t)\rangle, ~~~~~~~~~~~~~~~~~  \langle u_{n\mathbf{k}}(t)|u_{m\mathbf{k}}(t)\rangle=\delta_{nm}.
\end{align}
En el sólido, los parámetros del sistema por excelencia son los momentos de Bloch, $\boldsymbol{\lambda}(t)=\mathbf{k}(t)$, pero también pueden incluir, por ejemplo, campos magnéticos externos $\boldsymbol{\lambda}(t)=(\mathbf{k}(t),\mathbf{B}(t))$. Supongamos que el nivel de energía $\varepsilon_{n\mathbf{k}}(t)$ está separado de otros niveles por una separación mínima $\Delta_{0}(t)$ durante toda la evolución, y que la frecuencia característica del parámetro cambiante $\Omega_0 \ll \Delta_0 / \hbar$, es adiabático. De acuerdo con el teorema adiabático, un estado $|u_{n\mathbf{k}}(0)\rangle$ permanecerá en el mismo nivel de energía $n$, y después del tiempo $t$, evoluciona al estado $|\Psi_{n\mathbf{k}}(t)\rangle$ \cite{Sakurai2020modern},
\begin{align}
    |u_{n\mathbf{k}}(0)\rangle~~\to~~|\Psi_{n\mathbf{k}}(t)\rangle=e^{i \gamma_{n}(t)}e^{i\vartheta_{n}(t)}|u_{n\mathbf{k}}(t)\rangle,
\end{align}
donde,
\begin{align}
    \vartheta_{n}(t)=-\frac{1}{\hbar}\int_{0}^{t}dt'\varepsilon_{n\mathbf{k}}(t'), ~~~~~~\textup{y,}~~~~~~\gamma_{n}(t)=i\int_{\mathbf{k}(0)}^{\mathbf{k}(t)}d\mathbf{k}\cdot\langle u_{n\mathbf{k}}|\partial_{\mathbf{k}}|u_{n\mathbf{k}}\rangle
\end{align}
corresponden a la \textit{fase dinámica} y a la \textit{fase de Berry}, respectivamente. Cuando los parámetros se mueven sobre una trayectoria cerrada $C$ en el tiempo $t = T$, es decir, $\mathbf{k}(T) = \mathbf{k}(0)$, la fase de Berry $\gamma_{n}[C]$ para esta trayectoria cerrada $C$ en el espacio de parámetros es,
\begin{align}\label{eq:FasedeBerryC}
    \gamma_{n}[C]&=i\oint_{C} d\mathbf{k}\cdot\langle u_{n\mathbf{k}}|\partial_{\mathbf{k}}|u_{n\mathbf{k}}\rangle\\
    &=\oint_{C} d\mathbf{k}\cdot\mathbf{A}_{n}(\mathbf{k})\\
    &=\int_{S} d\mathbf{S}\cdot\mathbf{F}_{n}(\mathbf{k})
\end{align}
donde hemos usado el teorema de Stokes en la última igualdad \footnote{Donde hemos asumido un espacio vectorial de parámetros 3D. En dimensiones superiores, el teorema de Stokes sigue siendo válido, pero es necesario escribirlo en el lenguaje de formas diferenciales.}, con $S$ la superficie abierta de borde $C$. Hemos definido la \textit{conexión de Berry},
\begin{align}
    \mathbf{A}_{n}(\mathbf{k})\equiv i\langle u_{n\mathbf{k}}|\partial_{\mathbf{k}}|u_{n\mathbf{k}}\rangle
\end{align}
y su rotación es la \textit{curvatura de Berry},
\begin{align}
    \mathbf{F}_{n}(\mathbf{k})\equiv \boldsymbol{\nabla}_{k}\times \mathbf{A}_{n}(\mathbf{k})
\end{align}
La fase de Berry solo depende de la geometría de la trayectoria $C$, y no de la velocidad de $\boldsymbol{\lambda}$, siempre que sea lo suficientemente lenta como para que el teorema adiabático siga siendo válido.
%despreciar la transición entre niveles. 
%La integral superficial en la Ec. \eqref{eq:FasedeBerryC} se realiza sobre la primera zona de Brillouin (BZ); si tenemos mas parámetros que cambian adiabaticamente, la integral se realiza sobre la superficie generada por el vector de parámetros $\boldsymbol{\lambda}(t)$. 
La conexión de Berry corresponde al campo gauge definido en ese espacio de parámetros, similar al potencial vectorial para campos electromagnéticos en el espacio real. Esta afirmación se puede ver al hacer una transformación de gauge a la parte periódica de los estados de Bloch,
\begin{align}
    |u_{n\mathbf{k}}\rangle\to |u_{n\mathbf{k}}\rangle'&=e^{i\chi_{n}(\mathbf{k})}|u_{n\mathbf{k}}\rangle,
\end{align}
donde $e^{i\chi_{n}(\mathbf{k})}$ es una función monovaluada, entonces,
\begin{align}
    \mathbf{A}_{n}(\mathbf{k})\to\mathbf{A}_{n}'(\mathbf{k})&=\mathbf{A}_{n}(\mathbf{k})-\partial_{\mathbf{k}}\chi(\mathbf{k}), &&\textup{y,} & \mathbf{F}_{n}(\mathbf{k})\to\mathbf{F}_{n}'(\mathbf{k})&=\mathbf{F}_{n}(\mathbf{k})
\end{align}
Esto es similar a la transformación gauge de un potencial vectorial y un campo magnético en electromagnetismo, en este caso, vemos que la curvatura de Berry es invariante ante una transformación de gauge, y que la conexión de Berry cambia. Además, podemos demostrar que la exponencial de la fase de Berry es invariante de gauge debido a que,
\begin{align}
    \gamma_{n}[C]\to\gamma_{n}[C]'&=\gamma_{n}[C]+2\pi\times\textup{entero,}
\end{align}
lo que implica que, $e^{i \gamma_{n}[C]}\to e^{i \gamma_{n}[C]'}=e^{i \gamma_{n}[C]}$. 

La noción de fase de Berry ha permeado todas las ramas de la física, teniendo un impacto profundo en las propiedades de los materiales. Esta fase es responsable de una variedad de fenómenos, incluyendo la polarización, el magnetismo orbital, diversos efectos Hall cuánticos, anómalos y de espín, así como el bombeo cuántico de carga \cite{Xiao2010berryphase}. 
\section{Invariante TKNN}\label{2.2}

El invariante topológico definido para el sistema Hall cuántico, 
%entero, 
el invariante TKNN \cite{thouless1982quantized}, está estrechamente relacionado con la fase de Berry. Para ver esto, derivamos el invariante TKNN calculando la conductividad Hall de un sistema de electrones 2D de tamaño $L\times L$ en campos magnéticos perpendiculares, donde el campo eléctrico $\mathbf{E}$ y el campo magnético $\mathbf{B}$ se aplican a lo largo de los ejes $\tongo{y}$ y $\tongo{z}$, respectivamente. El Hamiltoniano del sistema, que incorpora el \textit{acoplamiento mínimo} entre los campos electromagnéticos y las partículas cargadas, se expresa como,
\begin{align}
    H_{\mathbf{k}}=\frac{1}{2m}(\mathbf{p}+\hbar\mathbf{k}-\frac{q}{c}\mathbf{A})^{2}+q\Phi+V_{L}(\mathbf{r})
\end{align}
donde $V_{L}(\mathbf{r})$ es el potencial de la red atómica, $\mathbf{E}=-\boldsymbol{\nabla}\Phi$ es la contribución del campo eléctrico y $q$ es la carga eléctrica. El campo magnético, $\mathbf{B}=\boldsymbol{\nabla}\times\mathbf{A}$, genera niveles de energía discretos conocidos como niveles de Landau \cite{Landau1930diamagnetismus}. Tratando el efecto del campo eléctrico $\mathbf{E}$ como un potencial de perturbación $\Phi=-Ey$, se puede utilizar la teoría de perturbaciones \cite{Ando2013topological} para aproximar el estado propio perturbado $|u_{n\mathbf{k}}\rangle$ a primer orden $|u_{n\mathbf{k}}\rangle_{E}$ como sigue,
\begin{align}
    |u_{n\mathbf{k}}\rangle_{E}=|u_{n\mathbf{k}}\rangle+\sum_{m(\neq n)}\frac{\langle u_{m\mathbf{k}}|(-qEy)|u_{n\mathbf{k}}\rangle}{\varepsilon_{n\mathbf{k}}-\varepsilon_{m\mathbf{k}}}|u_{m\mathbf{k}}\rangle
\end{align}
donde $n$ y $m$  Utilizando este estado propio perturbado, se puede obtener el valor esperado de la densidad de corriente superficial a lo largo del eje $\hat{x}$, $K_x$, en presencia del campo $E$ como,
\begin{align}\notag
    \langle K_{x} \rangle_{E}&=\sum_{n\mathbf{k}}f(\varepsilon_{n\mathbf{k}})_{E}\langle u_{n\mathbf{k}}|\Big{(}\frac{qv_{x}}{L^{2}}\Big{)}|u_{n\mathbf{k}}\rangle_{E}
\end{align}
donde $v_{x}$ es la velocidad de la carga a lo largo del eje $\hat{x}$ y $f(\varepsilon_{n\mathbf{k}})$ es la función de distribución de Fermi. Al reemplazar en $\sigma_{xy}=\langle K_{x} \rangle_{E}/E$ obtenemos la corriente de Hall (ver apéndice \ref{ConductividadHall} para la derivación),
\begin{align}\label{eq:ConductividadHallconFermi}
    \sigma_{xy}&=-\frac{q^{2}}{\hbar L^{2}}\sum_{n\mathbf{k}}f(\varepsilon_{n\mathbf{k}})F_{nz}.
\end{align}
Para un electrón $q=-e$, a temperatura $T = 0$, es decir, donde la distribución de Fermi es una función escalón ($f(\varepsilon_{n\mathbf{k}})=1$ para $E_{\textup{Fermi}}<\varepsilon_{n\mathbf{k}}$ y $f(\varepsilon_{n\mathbf{k}})=0$ para $E_{\textup{Fermi}}>\varepsilon_{n\mathbf{k}}$) con $N$ sub-bandas de Landau llenas, entonces la conductividad Hall es,
\begin{align}\label{eq:HallConductance}
    \sigma_{xy}&=-\frac{e^{2}}{h}\sum_{n=1}^{N}\Big(\frac{1}{2\pi}\int_{BZ=T^2}d^{2}kF_{nz}(\mathbf{k})\Big)
\end{align}
donde $BZ=T^2=S^1\times S^1$ es un toro en 2D 
%y corresponde al producto cartesiano de dos círculo. 
debido a la periodicidad de la red. Además, hemos considerado un cristal ideal e infinito, que nos permite pasar de $1/L^{2}\sum_{\mathbf{k}}\to 1/(2\pi)^{2}\int d^2k$. Así, la integral sobre la BZ dentro del paréntesis es un número entero\footnote{El número entero está relacionado con el número de singularidades de la conexión de Berry en la BZ, por lo tanto, estamos contando el número de ``cargas de Berry'' similares a monopolos magnéticos pero en el espacio de parámetros.} y se trata de una cantidad topológica \cite{thouless1982quantized} denominada invariante TKNN o primer número de Chern,
\begin{align}
    C_{1}^{(n)}=\frac{1}{2\pi}\int_{BZ}d^{2}kF_{nz}(\mathbf{k})~\in~\mathbb{Z}.
\end{align}
Por lo tanto, en un aislante 2D con $N$ bandas llenas donde se rompe la simetría de inversión temporal con un campo magnético externo, se observaría un IQHE, $\sigma_{xy}=-\frac{e^{2}}{h}\sum_{n=1}^{N}C_{1}^{(n)}$, independiente de los detalles de las bandas de energía.

La conductancia Hall de la Ec. \eqref{eq:HallConductance} también puede obtenerse a partir de la teoría de respuesta lineal utilizando la formula de Kubo \cite{Tkachov2015topological}, que puede generalizarse para incluir la interacción de electrones y el desorden. Se puede demostrar que, a pesar de estas complicaciones, la conductancia Hall permanece cuantizada, siempre y cuando la brecha de energía permanezca abierta \cite{Niu1985quantized,Hastings2015quantization}.
%
%Escribir una oracion del FQHE
%
\section{Efecto Hall anomalo cuántico 2D}\label{2.3}
Antes de adentrarnos en la discusión sobre los TIs 2D, es crucial comprender el efecto Hall anómalo cuántico (QAHE) \cite{Nagaosa2010anomalous}. En el IQHE, las cargas se desvían debido a la Fuerza de Lorentz en presencia de un campo magnético externo. Sin embargo, en un ferromagneto, los electrones se desvían debido a un campo magnético efectivo generado por la interacción de intercambio, que corresponde a la interacción entre las funciones de ondas de los electrones cercanos y es uno de los procesos que alinea los momentos magnéticos de los electrones en un material magnético. Para que el movimiento de los electrones esté acoplado a la dirección de la magnetización, es necesario un acoplamiento espín-órbita, que describe cómo el espín de un electrón interactúa con su movimiento orbital dentro de un material. Estos efectos intrínsecos son los que pueden producir el QAHE. 

\begin{figure}
\centering
\stackinset{l}{20pt}{t}{4pt}{(a)}{\includegraphics[width=0.24\textwidth]{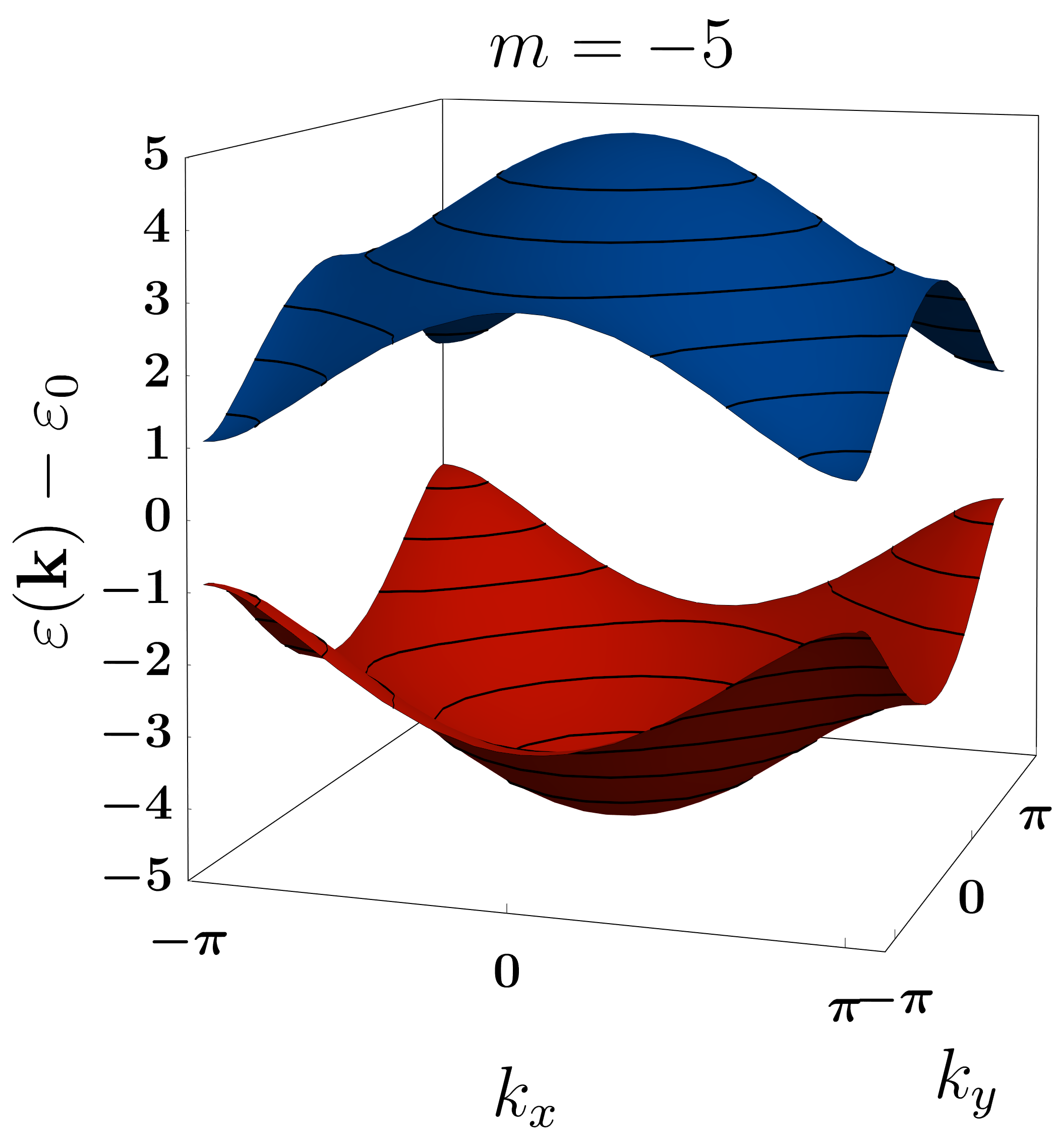}}
\stackinset{l}{20pt}{t}{4pt}{(b)}{\includegraphics[width=0.24\textwidth]{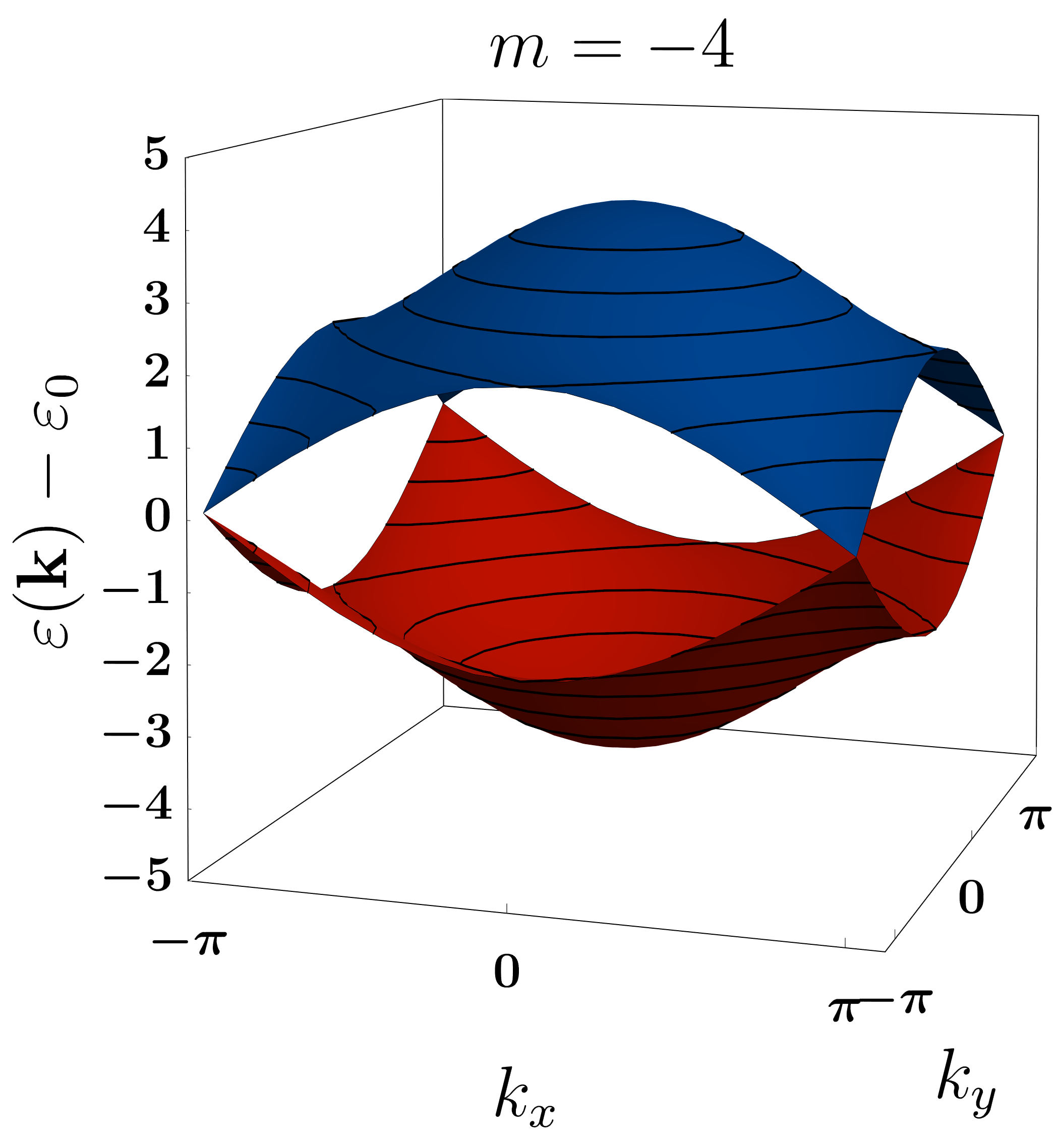}}
\stackinset{l}{20pt}{t}{4pt}{(c)}{\includegraphics[width=0.24\textwidth]{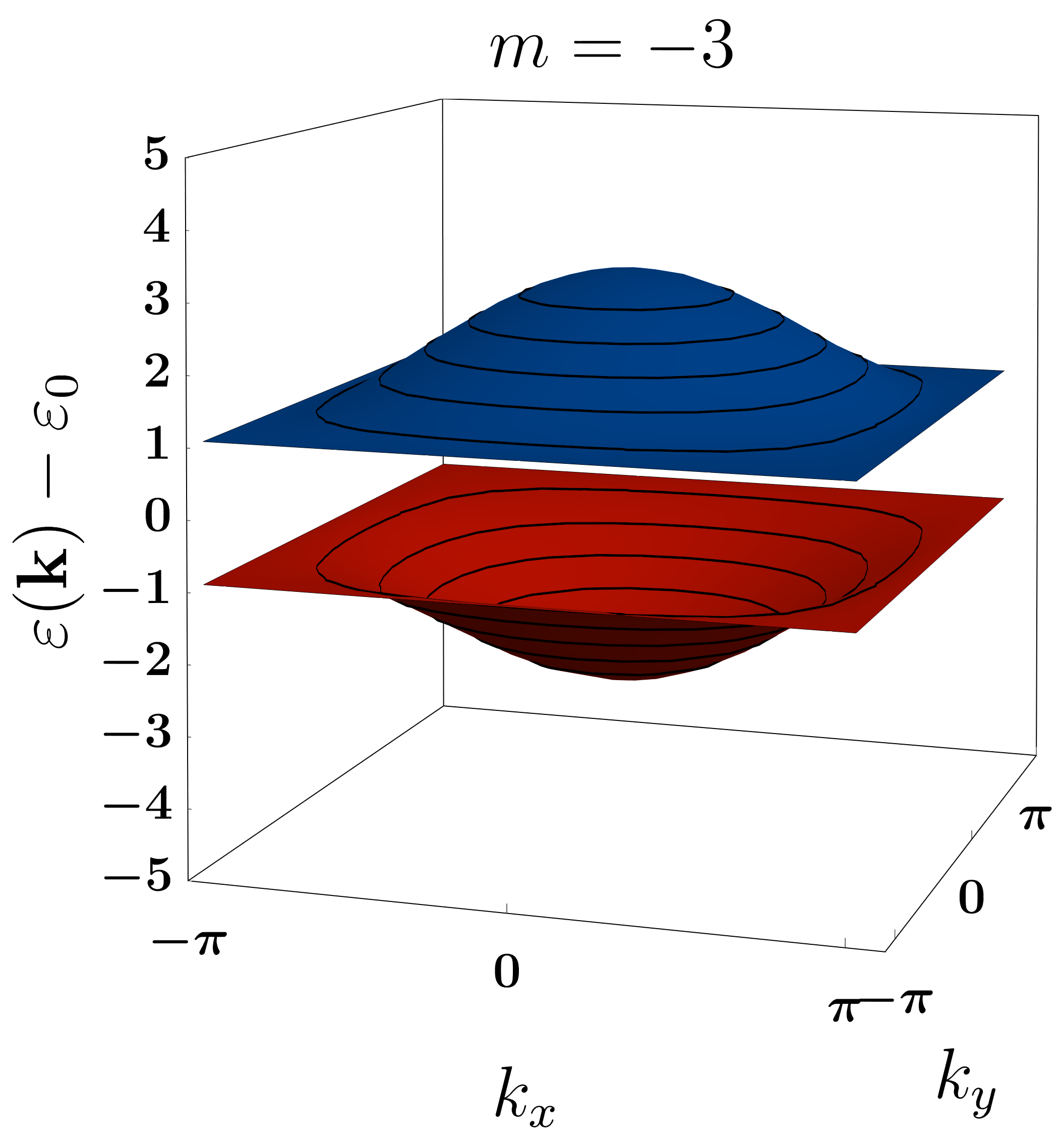}}
\stackinset{l}{20pt}{t}{4pt}{(d)}{\includegraphics[width=0.24\textwidth]{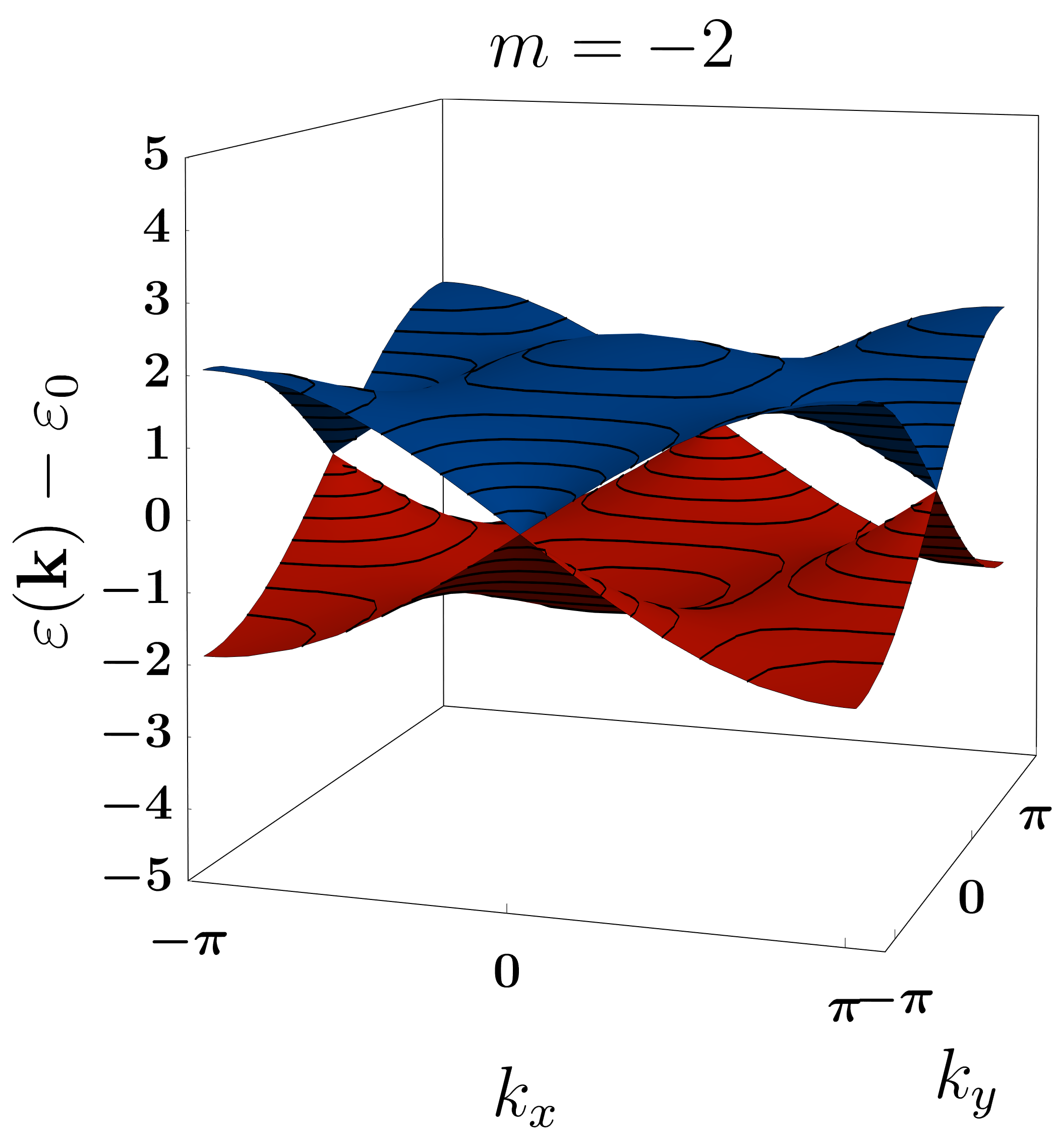}}
\stackinset{l}{20pt}{t}{4pt}{(e)}{\includegraphics[width=0.24\textwidth]{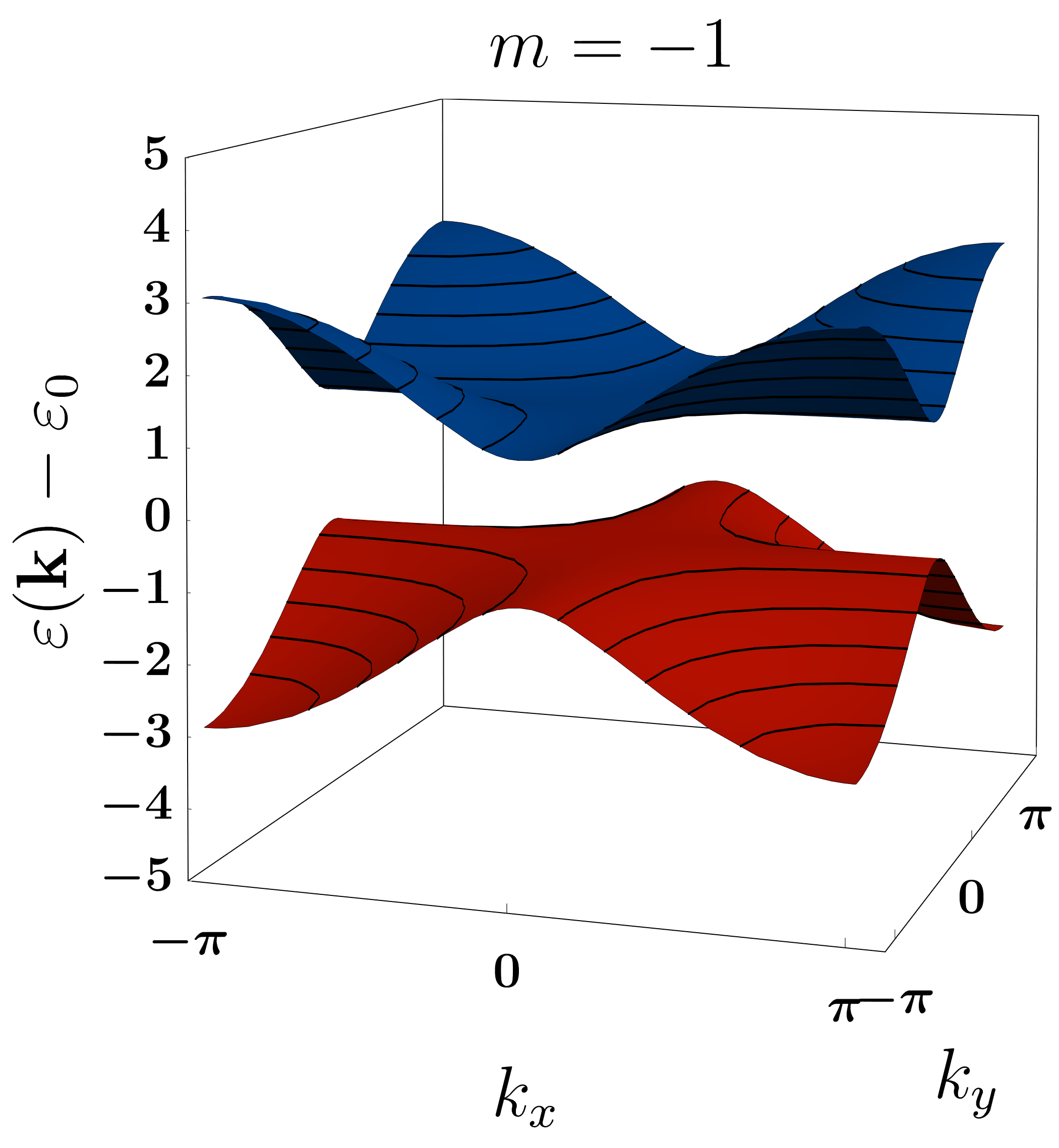}}
\stackinset{l}{20pt}{t}{4pt}{(f)}{\includegraphics[width=0.24\textwidth]{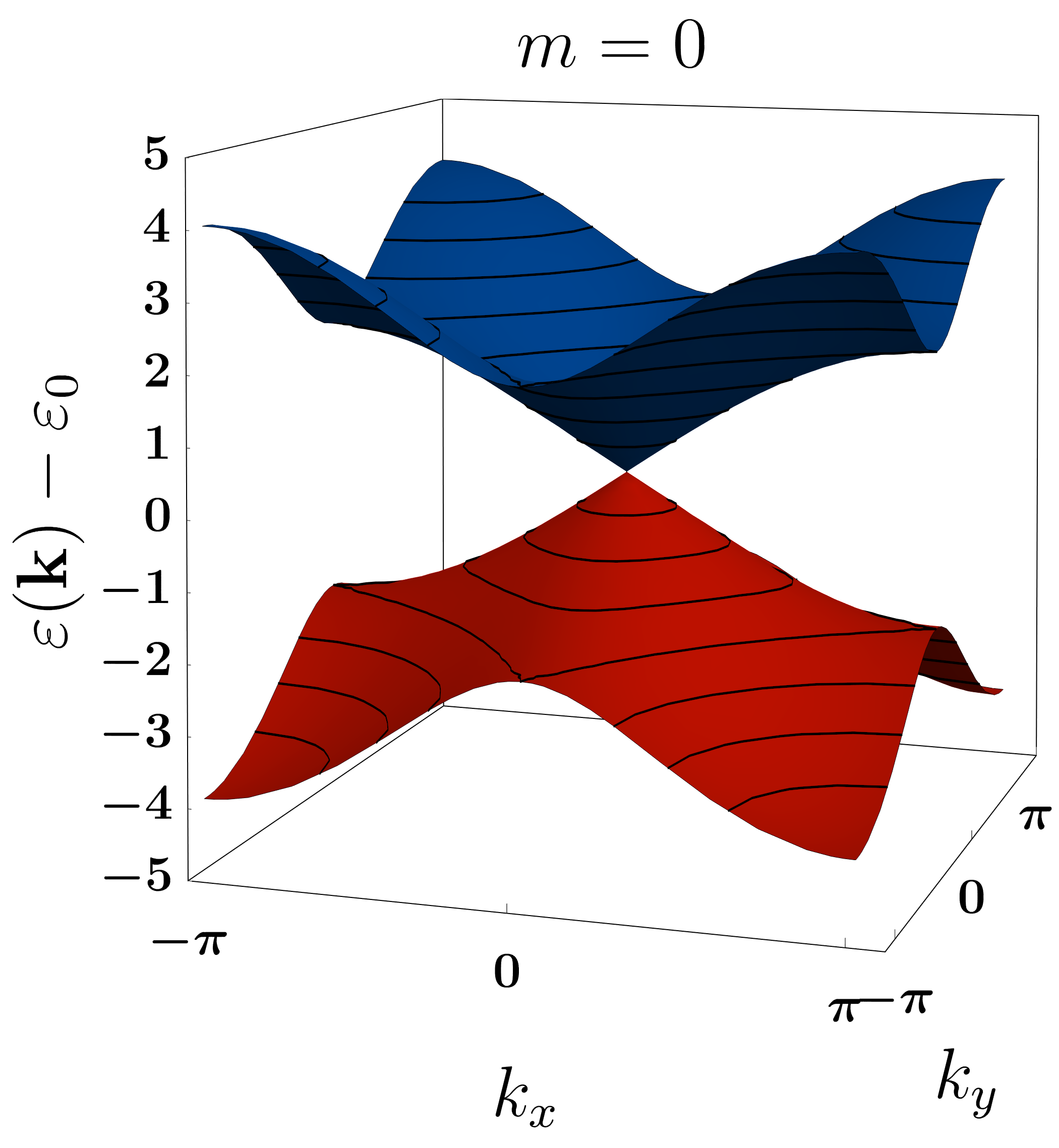}}
\stackinset{l}{20pt}{t}{4pt}{(g)}{\includegraphics[width=0.24\textwidth]{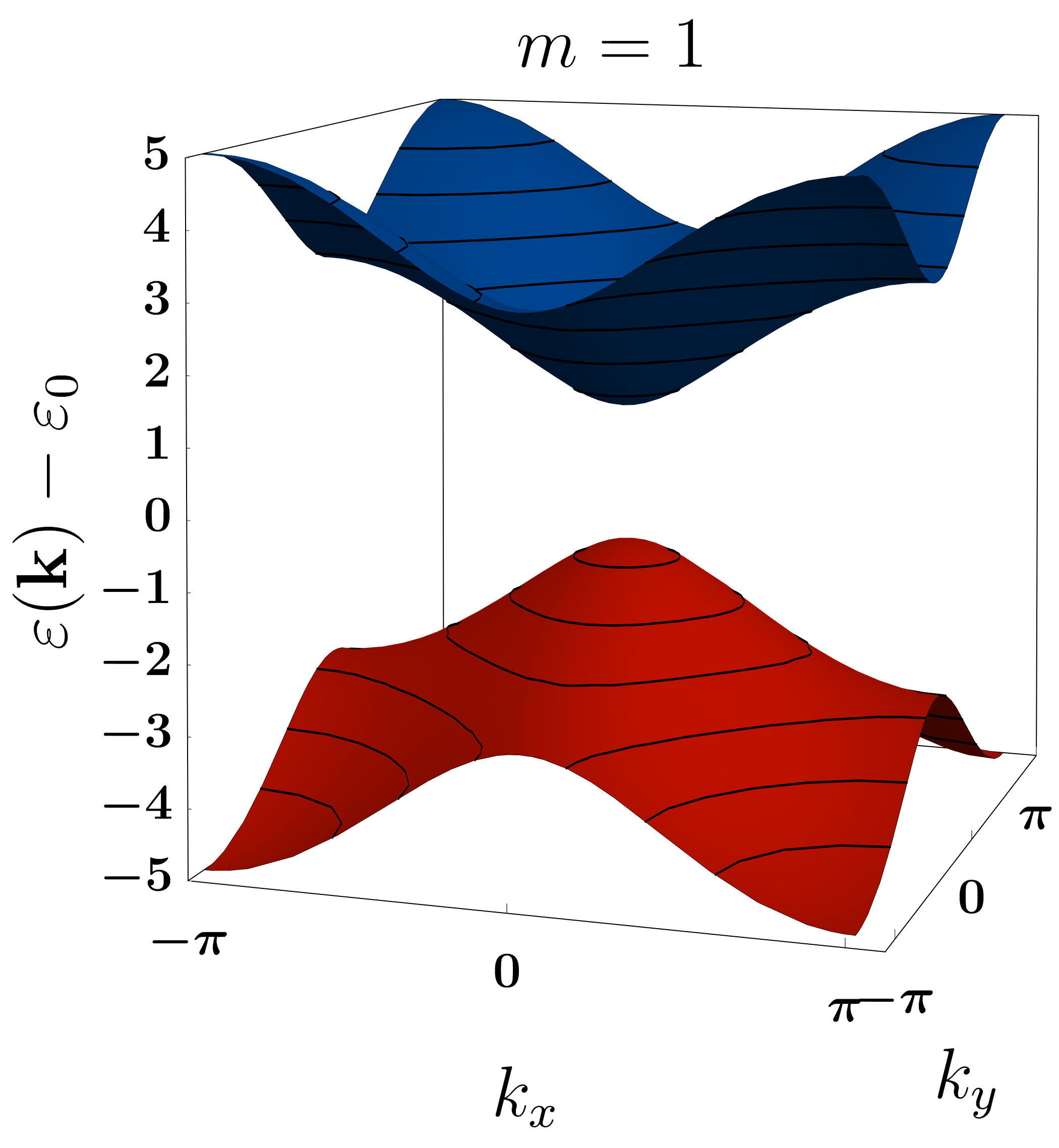}}
\stackinset{l}{20pt}{t}{-7pt}{(h)}{\includegraphics[width=0.24\textwidth]{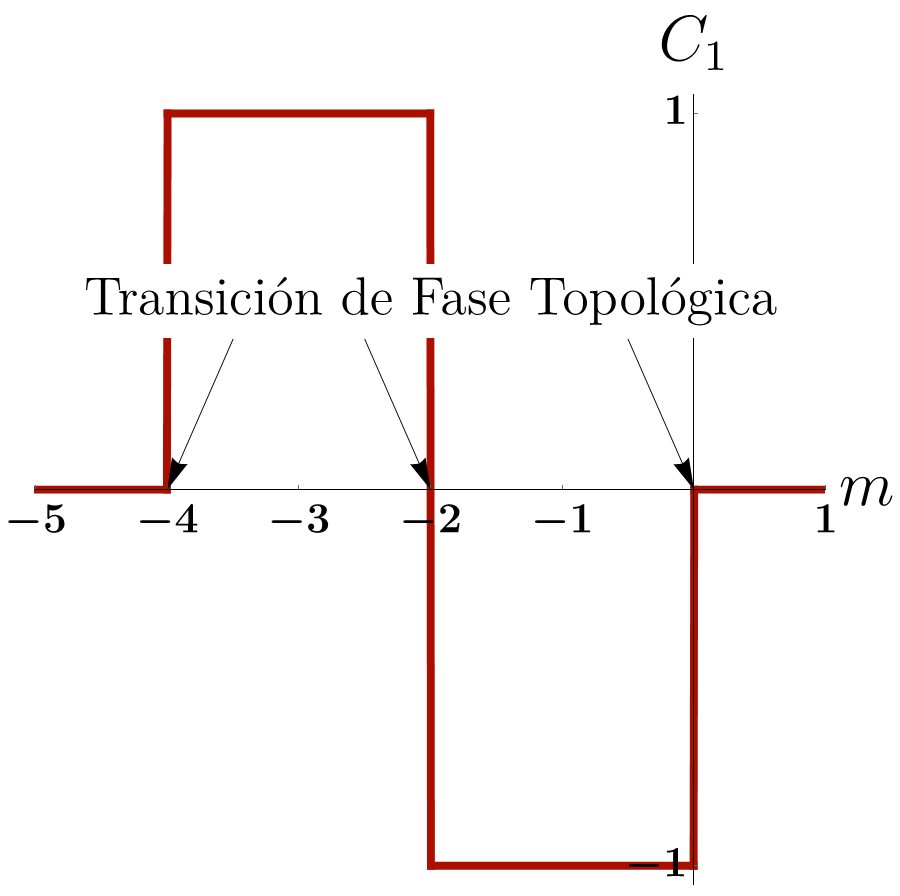}}
    \caption{(a-g) Dispersión de las bandas de energía para diferentes valores de $m$. Para $m=-4,-2,0$ la brecha energética se cierra y el material es un metal. Cuando $m\neq -4,-2,0$ el material es un aislante. En (h) mostramos el primer número de Chern $C_1$ de la banda de valencia (roja) en función de $m$. Cuando el material es un metal $C_1$ no esta definido y ocurre una transición de fase topológica. Cuando es un aislante, este puede ser trivial ($C_1=0$) o topológico ($C_1\neq 0$).}
    \label{fig:BandasyC1}
\end{figure}

Como ejemplo, consideremos brevemente el modelo propuesto por Qi, Wu y Zhang (QWZ) \cite{Qi2006topological}. En este modelo, los fermiones se encuentran en una red cuadrada 2D, donde cada punto de la red puede tener dos grados de libertad, que se pueden asociar con el espín-1/2 del electrón o a los grados de libertad orbitales. El Hamiltoniano QWZ es,
%
% \begin{align}
%     H_{\mathbf{k}}=\frac{1}{2m}(\mathbf{p}+\hbar\mathbf{k})^{2}+V_{L}(\mathbf{r})+\mathbf{h}(\mathbf{k})\cdot\boldsymbol{\sigma}\equiv \varepsilon_{0}+\mathbf{b}(\mathbf{k})\cdot\boldsymbol{\sigma}
% \end{align}
\begin{align}\label{eq:HamiltonianoQWZ}
    H_{\mathbf{k}}^{\textup{QWZ}}= \varepsilon_{0}+\mathbf{b}(\mathbf{k})\cdot\boldsymbol{\sigma},
\end{align}
el cual puede derivarse de un modelo general de \textit{tight binding} 2D. Como ejemplo de QWZ, consideraremos,
\begin{align}
    \mathbf{b}(\mathbf{k})=\Big(\lambda\sin k_{x}a,\lambda\sin k_{y}a,m+t(2-\cos k_{x}a-\cos k_{y}a)\Big)
\end{align}
donde $\boldsymbol{\sigma}=(\sigma_x,\sigma_y,\sigma_z)$ corresponde a las matrices de Pauli, $a$ es la constante de red, $m$ caracteriza la interacción de intercambio, $t$ es el término de \textit{hopping} de la red cuadrada y $\lambda$ caracteriza el acoplamiento espín-órbita. Las auto-energías de \eqref{eq:HamiltonianoQWZ} son,
\begin{align}\label{eq:BandasQWZ}
    \varepsilon_{\pm}(\mathbf{k})=\varepsilon_{0}\pm |\mathbf{b}(\mathbf{k})|.
\end{align}
Para simplificar $t=a=\lambda=1$. La brecha energética se cierra cuando $|\mathbf{b}(\mathbf{k})|=0$ y el sistema pasa por una transición de fase topológica. Notamos que la brecha energética se cierra en ciertos puntos de la BZ:
\begin{enumerate}
    \item $\mathbf{k}_{0}=(0,0)\to \varepsilon_{\pm}(\mathbf{k}_{0})=\varepsilon_{0}\pm m$
    \item $\mathbf{k}_{0}=(\pi,0),(0,\pi)\to \varepsilon_{\pm}(\mathbf{k}_{0})=\varepsilon_{0}\pm |m+2|$
    \item $\mathbf{k}_{0}=(\pi,\pi)\to \varepsilon_{\pm}(\mathbf{k}_{0})=\varepsilon_{0}\pm |m+4|$
\end{enumerate}
cuando $m=0,-2$ y $-4$, como se puede ver en la Fig. (\ref{fig:BandasyC1}). Al igual que el IQHE podemos calcular la conductividad de Hall y notamos que,
\begin{enumerate}
    \item cuando $m>0$ y $m<-4$ $\Rightarrow  \sigma_{xy}=0$.
    \item cuando $-4<m<-2$  $\Rightarrow  \sigma_{xy}=-\frac{e^2}{h}$
    \item cuando $-2<m<0$ $\Rightarrow  \sigma_{xy}=+\frac{e^2}{h}$
\end{enumerate}
En general, cuando la energía de Fermi está entre la brecha energética, la conductividad de Hall es $\sigma_{xy}=C_{1}e^2/h$, donde $C_1~\in~\mathbb{Z}$ es el primer número de Chern y corresponde al número de veces que la superficie $\mathbf{b}(\mathbf{k})$ se envuelve sobre el origen cuando recorre la BZ. Si tenemos un material topológico junto con otro no topológico (por ejemplo el aire) deben existir estados de borde conductores quirales debido a que espacialmente la topología cambia y por consecuencia debe cerrarse la brecha en la interfase para que ocurra la transición de fase. Los estados de borde del modelo QWZ son estables frente a las perturbaciones y sólo pueden destruirse si la brecha energética de las bandas del bulk se cierra de forma que la topología de los estados electrónicos se trivialice. En general, la interfaz entre dos materiales con topologías diferentes tendría estados de interfaz robustos.
\section{Aislante topológico 2D}\label{2.4}
Las simetrías juegan un rol importante en la clasificación de las fases topológicas de la materia, y para entender a los TIs debemos estudiar la simetría de inversión temporal. 
%Para un estudio completo de un conjunto mas amplio de simetrías puede consultar en \cite{murakami a y b}.  
En la introducción de los TIs 2D y 3D seguiremos los argumentos de \cite{Ando2013topological}. 

Si un sólido cristalino es simétrico ante una transformación de inversión temporal (TRS, por sus siglas en ingles), el Hamiltoniano de Bloch satisface $\Theta H_{\mathbf{k}}\Theta^{-1}=H_{-\mathbf{k}}$, donde $\Theta$ es el operador de inversión temporal $\Theta: t\to -t$. Este operador es antiunitario y en general se escribe como $\Theta=UK$, donde $U$ es un operado unitario que depende del espín del estado cuántico (por ejemplo, $U=1$ para partículas sin espín y $U=-i\sigma_{y}$ para partículas con espín $1/2$) y $K$ es el operador complejo conjugado, \textit{i.e.}, $Ki=-iK$. Un sistema de partículas donde el espín es ignorado (sistema sin espín) o es entero, el operador $\Theta^2=1$, a diferencia de sistemas con espín semi-entero, donde $\Theta^2=-1$. Esta última propiedad es importante para el teorema de Kramers que explicaremos a continuación.

Hasta 2005-2006, se pensaba que sólo los sistemas que rompían la TRS presentaban fenómenos topológicos interesantes. Un campo magnético o una magnetización rompe la TRS como en el QHE y QAHE respectivamente, así en ambos casos se observa la conductividad de Hall $\sigma_{xy}$ debido a la topología del primer número de Chern $C_1$. En 2005, Kane y Mele, mostraron que sistemas con TRS, aunque no pueden mostrar una conductividad de Hall $\sigma_{xy}=0$, pueden, sin embargo, mostrar otros fenómenos topológicos igualmente interesantes, como la clasificación topológica $\mathbb{Z}_2$ no trivial.

Para un sistema con espín semi-entero que presenta TRS el teorema de Kramers afirma que cada estado propio $|\psi\rangle$ y su invertido temporal $|\Theta\psi\rangle$, son ortogonales $\langle \psi |\Theta\psi\rangle=0$ y degenerados. Para Hamiltonianos de Bloch, donde los estados se indexan con $\mathbf{k}$, la degeneración ocurre para dos estados, uno en $\mathbf{k}$ con espín $\mathbf{s}$ y otro en $-\mathbf{k}$ con espín $-\mathbf{s}$, es decir $\varepsilon_{n\mathbf{k}\mathbf{s}}=\varepsilon_{n\mathbf{-k}\mathbf{-s}}$;
%(recordemos que el operador $\Theta$ invierte el signo del espín como si fuera un momento angular) 
este par de estados relacionados son llamados \textit{pares de Kramers}. Sólo los estados en momentos especiales ($\mathbf{k}=-\mathbf{k}\equiv \boldsymbol{\Lambda}$) mantienen una doble degeneración en el mismo $\mathbf{k}$. Estos puntos especiales de la BZ se denominan momentos invariantes en el tiempo (TRIM, por sus siglas en ingles). Un modelo mínimo para un TI con TRS de electrones de espín semi-entero tiene que tener al menos cuatro bandas: un modelo de dos bandas siempre sería un metal porque debe existir una doble degeneración entre las bandas en los TRIM.

% Se puede demostrar que la conductancia Hall es cero para fermiones sin espín (o espín entero) con TRS. Tambien es posible demostrar  (aunque técnicamente es más engorroso) una afirmación idéntica para fermiones de espín semi-entero con TRS. Esto ocurre porque la curvatura de Berry para los puntos $\mathbf{k}$ y $-\mathbf{k}$ en la BZ son opuestos entre sí y se cancela cuando se integra en toda la BZ.

% Consideremos un Hamiltoniano de Bloch con acoplamiento de espín-orbita,
% %
% \begin{align}
%     H_{\mathbf{k}}=d_{0}(\mathbf{k})+\mathbf{d}(\mathbf{k})\cdot\boldsymbol{\sigma}
% \end{align}
% %
% el cual es TRS. Para entender la topologia de los TIs seguiremos el argumento de More y Balent \cite{More2007topological}.  Debido a la TRS, los estados degenerados de Bloch para $\mathbf{k}$ y $-\mathbf{k}$ en una BZ son conjugados en inversión temporal sus curvaturas de Berry se cancelan entre sí, y asi, el primer número de Chern $C_{1}$ para una banda llena es cero. Dado que el dominio de los estados de Bloch \textit{independientes} cubre sólo la mitad de la BZ (llamada zona de Brillouin efectiva, o EBZ), cabe preguntarse si la integral de la curvatura de Berry sobre la EBZ podría cuantizarse. Moore y Balents demostraron que, $C_1~\textup{mod}~2$, es una propiedad intrínseca de la propia EBZ. Así pues, tenemos dos clases topológicas: $0$ que es el aislante habitual, y $1$ que es el aislante topológico. Por lo tanto, un TI 2D se caracteriza por un número topológico $\mathbb{Z}_2$.

Para determinar el número topológico $\mathbb{Z}_2$ debemos definir la matriz que relaciona un estado de Bloch con su invertido temporal,
\begin{align}
    w_{n\alpha\beta}(\mathbf{k})\equiv\langle u_{n-\mathbf{k}\alpha}|\Theta u_{n\mathbf{k}\beta}\rangle,
\end{align}
donde $\alpha,\beta=+,-$ son índices de ``espín''. Se puede demostrar que la matriz $[\textup{w}_n(\mathbf{k})]_{\alpha\beta}=w_{n\alpha\beta}(\mathbf{k})$ es unitaria, $\textup{w}_n\textup{w}_n^{\dagger}=1$, y que satisface $w_{n\beta\alpha}(-\mathbf{k})=-w_{n\alpha\beta}(\mathbf{k})$. Esta última propiedad implica que, en los TRIM la matriz $\textup{w}_{n}(\boldsymbol{\Lambda})$ es antisimétrica, \textit{i.e.},
\begin{align}
    \textup{w}_{n}(\boldsymbol{\Lambda})=\begin{pmatrix}
0 & w_{n+-}(\boldsymbol{\Lambda })\\ 
-w_{n+-}(\boldsymbol{\Lambda }) & 0 
\end{pmatrix}=w_{n+-}(\boldsymbol{\Lambda })i\sigma_{y}.
\end{align}
Las conexión de Berry entre dos pares de Kramers satisface,
\begin{align}\label{eq:ATR}
    \textup{A}_{n}(\mathbf{-k})=\textup{w}_{n}(\mathbf{k})\textup{A}_{n}^{*}(\mathbf{k})\textup{w}_{n}^{\dagger}(\mathbf{k})-i\textup{w}_{n}(\mathbf{k})\partial_{\mathbf{k}}\textup{w}_{n}^{\dagger}(\mathbf{k})
\end{align}
donde $[\textup{A}_{n}(\mathbf{k})]_{\alpha\beta}=\mathbf{A}_{n\alpha\beta}(\mathbf{k})=i\langle u_{n\mathbf{k}\alpha}|\partial_{\mathbf{k}} u_{n\mathbf{k}\beta}\rangle$, es un vector de matrices. Por otro lado, la curvatura de Berry  para un par de Kramers es,
\begin{align}
    \textup{F}_{n}(-\mathbf{k})=-\textup{w}_{n}(\mathbf{k})\textup{F}^{*}_{n}(\mathbf{k})\textup{w}_{n}^{\dagger}(\mathbf{k}),
\end{align}
donde $[\textup{F}_{n}(\mathbf{k})]_{\alpha\beta}=\mathbf{F}_{n\alpha\beta}(\mathbf{k})$. Por lo tanto, debido al TRS, los estados de Bloch en $\mathbf{k}$ y $-\mathbf{k}$ están relacionados. También lo están las conexiones de Berry y las curvaturas de Berry. El número de Chern total de una banda de energía con un par de Kramers desaparece debido a la integral de la curvatura de Berry se cancela de a pares. Sin embargo, se puede denotar un número topológico alternativo sobre la mitad de la BZ, llamada zona de Brillouin efectiva (EBZ, por sus siglas en ingles), que es la mínima que debemos conocer para saber el estado completo del par de Kramers.

Consideremos dos bandas que forman un par de Kramers, y denotamos los estados de Bloch como $|u_{n\mathbf{k}+}\rangle$ y $|u_{n\mathbf{k}-}\rangle$. En la teoría moderna de la polarización la \textit{polarización de carga} en una dimensión (1D) se define en función de la fase de Berry como sigue \cite{KingSmith1993Theoryofpolarization},
\begin{align}
    P_{x}(k_y)&=\sum_{n~\textup{ocupados}}\int_{-\pi}^{\pi}\frac{dk_{x}}{2\pi}\textup{tr}[\textup{A}_{x,n}(k_x,k_y)]\\
    &=\sum_{n~\textup{ocupados}}\int_{-\pi}^{\pi}\frac{dk_{x}}{2\pi}[A_{x,n++}(k_x,k_y)+A_{x,n--}(k_x,k_y)]\\
    &\equiv P_{x,+}(k_y)+P_{x,-}(k_y)
\end{align}
donde $\textup{tr}[~]$ es la traza sobres los índices de espín $(\alpha,\beta)$ y $P_{x,i}(k_y)$ es la contribución parcial de cada banda llamada polarización parcial. El punto clave, es definir la \textit{polarización inversa en el tiempo} como sigue,
\begin{align}\label{eq:Polarizacionxtheta}
    P_{x,\theta}(k_y)\equiv P_{x,+}(k_y)-P_{x,-}(k_y).
\end{align}
Intuitivamente, $P_{x,\theta}(k_y)$ es la diferencia de polarización de carga entre las bandas de espín arriba y espín abajo, del par de Kramers para un determinado $k_y$. Al reemplazar en la Ec. \eqref{eq:Polarizacionxtheta} obtenemos,
\begin{align}\label{eq:Polarizacionxtheta2}
     P_{x,\theta}&=\sum_{n~\textup{ocupados}}\int_{0}^{\pi}\frac{dk_{x}}{2\pi}[A_{x,n++}(k_x)+A_{x,n++}(-k_x)-A_{x,n--}(k_x)-A_{x,n--}(-k_x)],
\end{align}
donde hemos omitido la dependencia en $k_y$. Usando la transformación de la Ec. \eqref{eq:ATR}, que relaciona las conexiones de Berry de los momentos conjugados, \textit{i.e.} $A_{x,n++}(-k_x)=A^{*}_{x,n--}(k_x)-iw_{n+-}(k_x)\partial_{k_x}w_{n+-}(k_x)$ y los mismo para $A_{x,n--}(-k_x)$. Reemplazando en la Ec. \eqref{eq:Polarizacionxtheta2} y con un poco de álgebra se obtiene,
\begin{align}\label{eq:Polarizationtheta}
    P_{x,\theta}(k_y)&=\frac{1}{i\pi}\log\Big{(} \prod_{n}\frac{w_{n12}(\pi,k_y)}{\sqrt{w_{n12}^{2}(\pi,k_y)}}\frac{w_{n12}(0,k_y)}{\sqrt{w_{n12}^2(0,k_y)}}\Big{)}+2m,~~~m~\in~\mathbb{Z}.
\end{align}
El argumento del logaritmo puede ser $+1$ o $-1$, por lo tanto $P_{x,\theta}(k_y)=0,1 + 2m$, donde $2m$ viene de una transformación de gauge que queda libre y hace que la polarización no esté bien definida. Al igual que la polarización eléctrica, esta ambigüedad puede eliminarse si se calcula la diferencia, $\Delta P_{x,\theta}\equiv P_{x,\theta}(k_{y}=\pi)-P_{x,\theta}(k_{y}=0)$ el cual se obtiene,
\begin{align}
    (-1)^{\Delta P_{x,\theta}}=\prod_{n}\frac{w_{n12}(\Lambda_{1})}{\sqrt{w_{n12}^{2}(\Lambda_{1})}}\frac{w_{n12}(\Lambda_{2})}{\sqrt{w_{n12}^{2}(\Lambda_{2})}}\frac{w_{n12}(\Lambda_{3})}{\sqrt{w_{n12}^{2}(\Lambda_{3})}}\frac{w_{n12}(\Lambda_{4})}{\sqrt{w_{n12}^{2}(\Lambda_{4})}}=\pm 1
\end{align}
donde $\Lambda_{i}$ son los TRIM de la EBZ, \textit{i.e.}, $\Lambda_{1}=(0,0)$, $\Lambda_{2}=(0,\pi)$, $\Lambda_{3}=(\pi,0)$ y $\Lambda_{4}=(\pi,\pi)$. Así, $\Delta P_{x,\theta}= 1,0~\textup{mod}~2~\in~\mathbb{Z}_{2}$ es gauge invariante y caracteriza una fase topológica $\Delta P_{x,\theta}=1$ de una no topológica $\Delta P_{x,\theta}=0$.

Para $N$ pares de Kramers  la matriz $w_{n}$ se generaliza y sera una matriz antisimétrica de $2N\times 2N$. Así podemos escribir $\Delta P_{x,\theta}$ como,
\begin{align}
    (-1)^{\Delta P_{x,\theta}}=\prod_{i=1}^{4}\delta(\Lambda_{i}),~~~~\textup{con,}~~~\delta(\Lambda_{i})\equiv \prod _{n~\textup{ocupados}}\frac{w_{n}(\Lambda_{i})}{\sqrt{w_{n}^{2}(\Lambda_{i})}}=\frac{\textup{pf}[\textup{w}(\Lambda_{i})]}{\sqrt{\textup{det}[\textup{w}(\Lambda_{i})]}}
\end{align}
%Aquii GPT
donde $\textup{pf}[~]=$ es el Pfaffiano. Un aislante con $\Delta P_{x,\theta}=1,~0$ es llamado un aislante trivial y un TI respectivamente. Físicamente estamos viendo un material donde se ha polarizado por espín en la dirección $x$. Por simetría, podemos hacer este mismo ejercicio para una polarización en $y$, lo cual implica que se ha transportado el espín por el bordes de la muestra, donde las cargas con \textit{espín-up} y \textit{espín-down} se mueven en direcciones opuestas hasta llegar al borde. Estos estados de borde del TI, polarizados por espín, conocido como el efecto Hall cuántico de espín (QSHE, por sus siglas en ingles), son robustos a impurezas no magnéticas.

Como ejemplo ilustrativo del primer TI 2D confirmado experimentalmente \cite{König2007HgTe}, consideraremos un modelo simple de Bernevig-Hughes-Zhang (BHZ) \cite{Bernevig2006HgTe} que describe la inversión de bandas para un pozo cuántico de CdTe/HgTe/CdTe que da lugar al QSHE. La inversión de la banda en el bulk de HgTe ocurre por el acoplamiento espín-órbita y tiene una estrecha relación con el número entero $\mathbb{Z}_2$. El modelo simplificado de BHZ, que tiene TRS, sólo se consideran la banda de conducción originada del orbital-s y una banda de valencia del orbital-p (cada una de ellas es doblemente degenerada). En Hamiltoniano en la base de espín es,
\begin{align}
    H^{\textup{BHZ}}=\sum_{\mathbf{k}}(\langle\mathbf{k}s\uparrow|,\langle\mathbf{k}p\uparrow|,\langle\mathbf{k}s\downarrow|,\langle\mathbf{k}p\downarrow|)\begin{pmatrix}
 h(\mathbf{k})& 0\\ 
0 & h^{*}(-\mathbf{k})
\end{pmatrix}\begin{pmatrix}
|\mathbf{k}s\uparrow\rangle\\ 
|\mathbf{k}p\uparrow\rangle\\ 
|\mathbf{k}s\downarrow\rangle\\ 
|\mathbf{k}p\downarrow\rangle
\end{pmatrix}
\end{align}
donde,
\begin{align}
    h(\mathbf{k})&=\varepsilon_0+\mathbf{b}(\mathbf{k})\cdot\boldsymbol{\tau}
\end{align}
con, 
\begin{align}
    \mathbf{b}(\mathbf{k})=(2t_{sp}\sin k_y,2t_{sp}\sin k_x,\frac{\varepsilon_s-\varepsilon_p}{2}-(t_{ss}+t_{pp})(\cos k_x+\cos k_y)).
\end{align}

\begin{figure}[t]
    \centering
    \stackinset{l}{20pt}{t}{0pt}{(a)}{\includegraphics[width=0.4\textwidth]{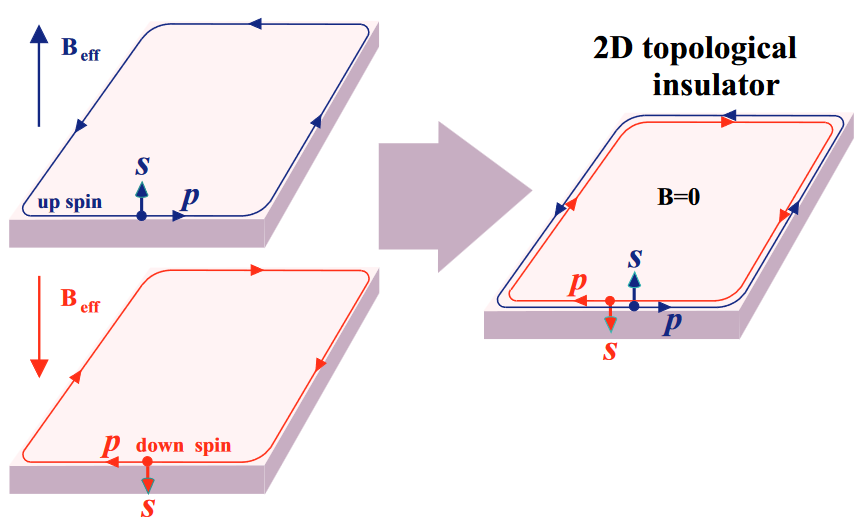}}
    \stackinset{l}{20pt}{t}{0pt}{(b)}{\includegraphics[width=0.31\textwidth]{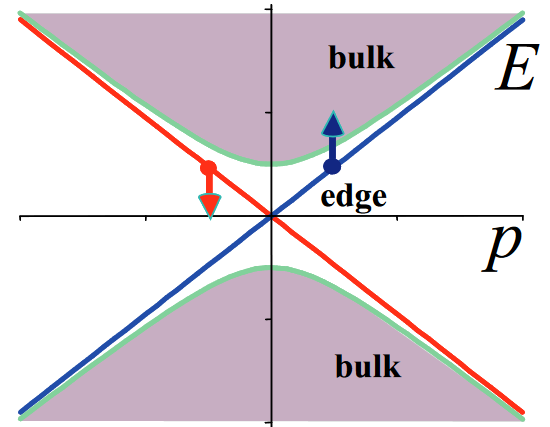}}
    \caption{(a) En el espacio real, hay un estado de borde helicoidal a lo largo del límite de una TI 2D, compuesto de dos estados de borde quirales, uno para cada espín. (b) En el espacio del momento ($k\to p$), las curvas de dispersión de energía ($\varepsilon\to E$) de los estados de borde se cruzan en un TRIM. Los estados con espín-up se mueven en la dirección contraria a los estados con espín-down. Imagen de \cite{Tkachov2015topological}.}
    \label{fig:2DTI}
\end{figure}

Aquí $\boldsymbol{\tau}$ son las matrices de Pauli del cuasi-espín relacionado con los dos orbitales, $\varepsilon_i$ es la energía de sitio del orbital $i$ y $t_{ij}$ son las amplitudes de \textit{hopping} del orbital $i$ al $j$.  Debido a la diagonalización de bloques, el espín-up y el espín-down están explícitamente desacoplados. Podemos notar que el Hamiltoniano BHZ se compone de dos subsistemas del Hamiltoniano de QWZ independientes $h(\mathbf{k})$ y $h^{*}(-\mathbf{k})$, esta afirmación se representa en la Fig. (\ref{fig:2DTI}.a). Para un subsistema, uno puede comprobar que si $\varepsilon_s-\varepsilon_p>4(t_{ss}+t_{pp})$, la conductividad de Hall es $\sigma_{xy}=0$ y para $\varepsilon_s-\varepsilon_p<4(t_{ss}+t_{pp})$, la conductividad de Hall es $\sigma_{xy}=e^2/h$. En el punto crítico, $\varepsilon_s-\varepsilon_p=4(t_{ss}+t_{pp})$, se cierra la brecha energética. La transición de la fase trivial a la fase topológica va acompañada de una inversión de las bandas de energía, cuando $\varepsilon_{+}$ y $\varepsilon_{-}$ cambian de posición.

Cuando el subsistema se encuentra en la fase QAHE, de acuerdo con la discusión al final de la Sec. (\ref{2.3}), tiene estados de borde quirales. Puesto que este subsistema consta sólo de electrones de espín-up, los electrones de los estados de borde son de espín-up. Por otra parte, el subsistema conjugado $h^{*}(-\mathbf{k})$ tiene $\sigma_{xy}=-e^2/h$. Sus electrones del borde con espín-down se transportan en la dirección opuesta (ver Fig. (\ref{fig:2DTI}.a)). En el espacio de momento, la dispersión de energía del estado de borde es lineal en el límite $\mathbf{k}$ pequeño. Uno tiene pendiente positiva (velocidad positiva), y el otro tiene pendiente negativa (velocidad negativa). Debido a la degeneración de Kramers, estas dos curvas de dispersión tienen que cruzarse en un TRIM (ver Fig. (\ref{fig:2DTI}.b)). Esta degeneración puntual sólo puede eliminarse si se rompe el TRS.
 
\section{Aislante topológico 3D}\label{2.5}
\begin{figure}[t]
    \centering
    \includegraphics[width=0.9\textwidth]{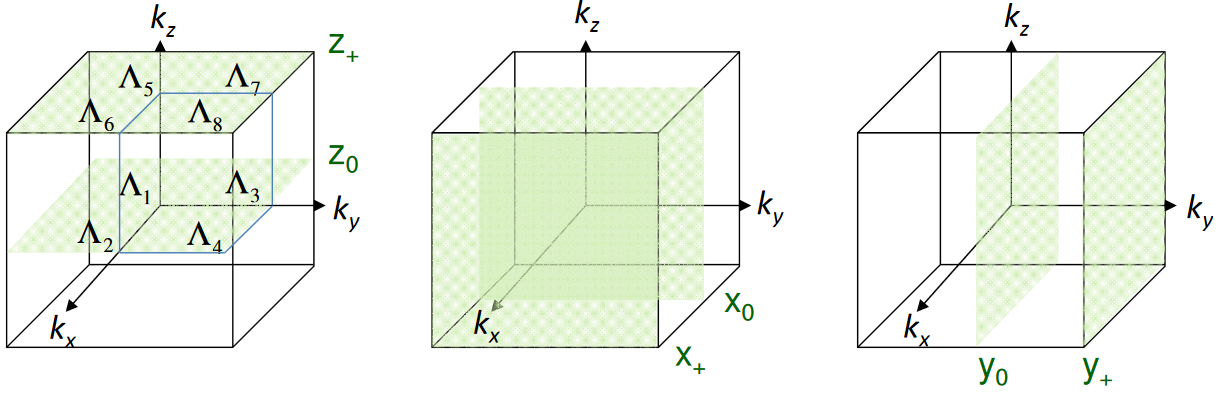}
    \caption{BZ 3D de una red cristalina cubica. Los planos 2D coloreados en la BZ 3D son invariante bajo TR. Es decir, bajo una transformación TR, se mapean a sí mismos. Podemos definir seis de estos planos. A cada plano se le puede asignar un número topológico $\mathbb{Z}_2$. Los ocho puntos $\boldsymbol{\Lambda}_i$ corresponden a los TRIM, donde las bandas de energía de los pares de Kramers se degeneran. Imagen de \cite{chang2021lecture}.}
    \label{fig:BZ3D}
\end{figure}

Basándose en argumentos de teoría de homotopía, que describen clases de equivalencia bajo deformaciones suaves de mapeos entre variedades, Moore y Balents demostraron \cite{More2007topological} que existen cuatro invariantes $\mathbb{Z}_2$ para los sistemas 3D. Aunque la construcción matemática de la homotopía es compleja, el origen físico de estos cuatro invariantes puede entenderse de manera clara. Para simplificar, consideremos un sistema cúbico y tomemos la constante de red $a=1$. En la BZ 3D de este sistema, hay ocho TRIMs denotados como $\Lambda_{i}$,
%$\Lambda_{0;0;0}$, $\Lambda_{\pi;0;0}$, $\Lambda_{0;\pi;0}$, $\Lambda_{0;0;\pi}$, $\Lambda_{\pi;0;\pi}$, $\Lambda_{0;\pi;\pi}$, $\Lambda_{\pi;\pi;0}$, y $\Lambda_{\pi;\pi;\pi}$
como se ve en la Fig. (\ref{fig:BZ3D}), donde $\Lambda_{8}=(\pi,\pi,\pi)$ por ejemplo. En estos puntos de la BZ 3D, el Hamiltoniano de Bloch es TRS, \textit{i.e.}, $\Theta H(\Lambda_{i}) \Theta^{-1}=H(\Lambda_{i})$, y cada par de bandas de Kramers se degenera.

Nótese que los seis planos en la BZ 3D, $k_x=0$, $k_x=\pi$, $k_y=0$, y $k_y=\pi$ , $k_z=0$, y $k_z=\pi$ poseen las simetrías de la BZ 2D, y por tanto cada uno de ellos tiene un invariante $\mathbb{Z}_2$. Las seis invariantes pueden denotarse como $x_0$, $x_+$, $y_0$, $y_+$, $z_0$ y $z_+$, pero no todas ellas son independientes \cite{More2007topological}. Esto se debe a que los productos de los invariantes $x_0x_+$, $y_0y_+$ y $z_0z_+$ son redundantes, debido a que estos tres productos dan el mismo resultado, es decir,  $x_0x_+=y_0y_+=z_0z_+=\prod_{i=1}^{8}\delta(\Lambda_{i})$; la multiplicación de los ocho TRIM y, por tanto, son iguales. Esto significa que hay dos relaciones de restricción $x_0x_+=y_0y_+=z_0z_+$, lo que dicta que sólo pueden determinarse independientemente cuatro invariantes en un sistema 3D.

La construcción concreta de los cuatro invariantes $\mathbb{Z}_2$ fue dada por Fu y Kane \cite{Fu_Kane_Mele_PhysRevLett.98.106803}. 
%Para cada TRIM $\Lambda_i$, definimos,
%
% \begin{align}
%     \delta(\Lambda_i)\equiv\frac{\textup{pf}[w(\Lambda_i)]}{\sqrt{\textup{det}[w(\Lambda_i)]}}.
% \end{align}
%
%Usando este $\delta(\Lambda_i)$
, donde los cuatro invariantes $\mathbb{Z}_2$, $\nu_0$, $\nu_1$, $\nu_2$, $\nu_3$ se definen como,
%
% \begin{align}
%     (-1)^{\nu_0}&=\prod_{n_{j}=0,\pi}\delta(\Delta_{n_1;n_2;n_3}),\\
%     (-1)^{\nu_1}&=\prod_{n_{j\neq i}=0,\pi;n_{i}=\pi}\delta(\Delta_{n_1;n_2;n_3}).
% \end{align}
\begin{align}
    (-1)^{\nu_0}&=\prod_{i=1}^{8}\delta(\Delta_{i}),\\
    (-1)^{\nu_1}&=\delta(\Delta_{2})\delta(\Delta_{4})\delta(\Delta_{6})\delta(\Delta_{8}),\\
    (-1)^{\nu_2}&=\delta(\Delta_{3})\delta(\Delta_{4})\delta(\Delta_{7})\delta(\Delta_{8}),\\
    (-1)^{\nu_3}&=\delta(\Delta_{5})\delta(\Delta_{6})\delta(\Delta_{7})\delta(\Delta_{8}).
\end{align}
\begin{figure}[t]
    \centering
    \includegraphics[width=0.7\linewidth]{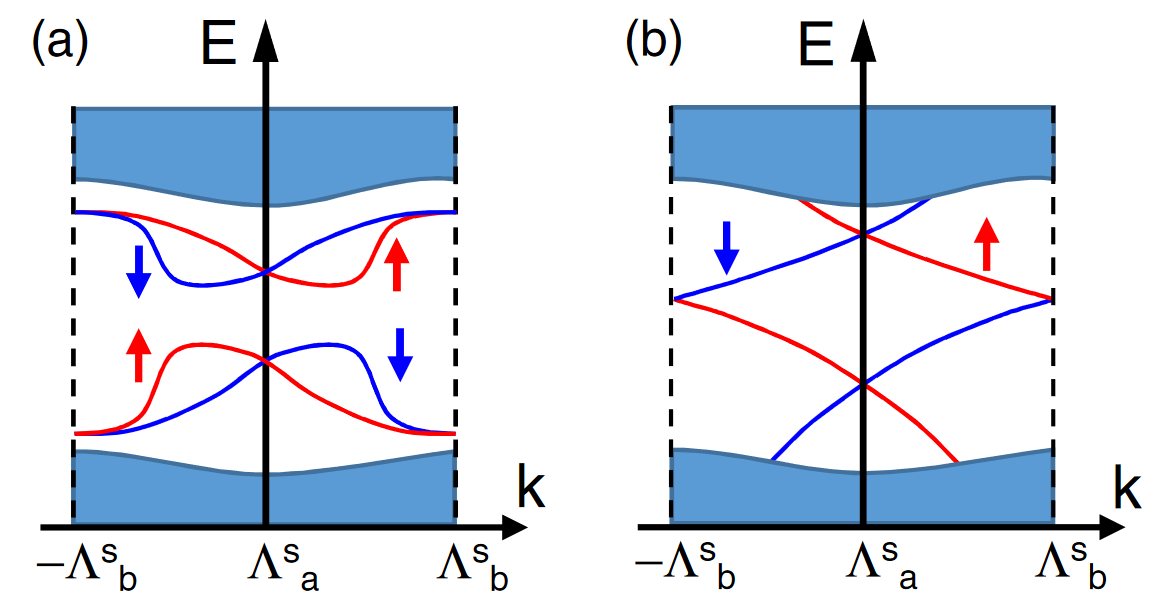}
    \includegraphics[width=0.32\linewidth]{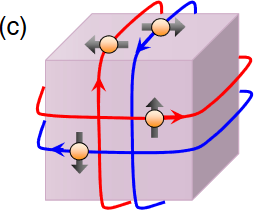}
    \includegraphics[width=0.33\linewidth]{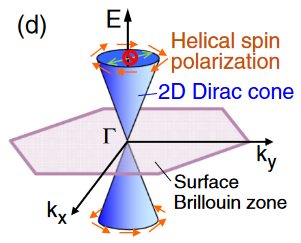}
    \caption{Esquema de la formación de los estados de superficie (s) entre dos TRIM de superficie, $\Lambda_{a}^{s}$ y $\Lambda_{b}^{s}$, para (a) casos topológicamente triviales y (b) casos topológicamente no triviales. En el último, la característica de ``cambian de pareja'' es un reflejo del cambio en la polarización TR. Las regiones sombreadas representan los estados del bulk. (c) Imagen esquemática en espacio real del estado de superficie helicoidal 2D de un TI 3D. (d) Dispersión de energía del estado superficial no degenerado de espín de un TI 3D que forma un cono de Dirac 2D; debido a la polarización helicoidal de espín, la retrodispersión de $\mathbf{k}$ a $-\mathbf{k}$ está prohibida. Imagen de \cite{Ando2013topological}.}
    \label{fig:surfacestates3d}
\end{figure}

El invariante $\nu_0$ viene dado por un producto de las ocho $\delta(\Lambda_{i})$, por lo que es único para un sistema 3D. Por otro lado, $\nu_i$ son un producto de cuatro $\delta(\Lambda_{i})$ y es similar al invariante $\mathbb{Z}_2$ en el caso 2D. Por ejemplo, $\nu_3$
% %
% \begin{align}
%     (-1)^{\nu_3}=\delta(\Delta_{0,0,\pi})\delta(\Delta_{\pi,0,\pi})\delta(\Delta_{0,\pi,\pi})\delta(\Delta_{\pi,\pi,\pi})
% \end{align}
% %
corresponde a la invariante $\mathbb{Z}_2$ en el plano $z=\pi$, que puede verse considerando la polarización TR definida en este plano, $P_{x,\theta}(k_y=0,k_z=\pi)=\frac{1}{i\pi}\log [\delta(\Lambda_{5})\delta(\Lambda_{6})]$ y $P_{x,\theta}(k_y=\pi,k_z=\pi)=\frac{1}{i\pi}\log [\delta(\Lambda_{7})\delta(\Lambda_{8})]$. Cuando la polarización TR cambia entre $k_y=0$ y $k_y =\pi$ [es decir, $P_{x,\theta}(k_y=0,k_z=\pi)$ y $P_{x,\theta}(k_y=\pi,k_z=\pi)$ son diferentes], entonces la topología $\mathbb{Z}_2$ es no trivial y $\nu_3$ se convierte en 1. 

La consecuencia física de una invariante $\mathbb{Z}_2$ no trivial es la aparición de estados superficiales topológicamente protegidos. 
Esto se muestra gráficamente en la Fig. (\ref{fig:surfacestates3d},a,b), en la que se comparan estados superficiales topológicamente triviales y no triviales. En el caso no trivial, los pares de Kramers en el estado de superficie ``cambian de pareja'', y como resultado, el estado de superficie está garantizado para cruzar cualquier energía de Fermi dentro de la brecha de energía del bulk. Esta característica de cambio de pareja es un reflejo del cambio en la polarización TR comentado anteriormente. Así, estos estados superficiales son robustos ante perturbaciones que no destruyan los pares de Krames, \textit{i.e.}, perturbaciones no magnéticas. 

%Es habitual escribir la combinación de las cuatro invariantes en la forma $(\nu_0; \nu_1, \nu_2, \nu_3)$, porque $(\nu_1, \nu_2, \nu_3)$ puede interpretarse como índices de Miller para especificar la dirección del vector i en el espacio recíproco. 
Un TI 3D se denomina ``fuerte'' cuando $\nu_0 =1$, mientras que se denomina ``débil'' cuando $\nu_i=1$ para algún $i=1,2,3$. El índice fuerte es intrínseco al TI 3D, mientras que los otros 3 están relacionados, a grandes rasgos, con el apilamiento de TIs 2D a lo largo de las dirección $i$. Para un 3D TI, las bandas energéticas de los estados de superficie se tocan formando números impares de cruces (o conos de Dirac para un TI ideal) donde se mantiene la polarización de espín y existe un bloqueo espín-momento, \textit{i.e.}, el estado superficial es de espín no degenerado y la dirección del espín es perpendicular al vector momento y está confinado principalmente en el plano superficial (ver Fig. (\ref{fig:surfacestates3d},c,d)).

El BiSb es el primer aislante topológico 3D confirmado experimentalmente \cite{hsieh_topological_2008}. Posteriormente, se han predicho y verificado muchos más \cite{Bansil2016topological}. Un TI ideal sería un aislante con una gran brecha energética en el bulk. Pero esto es difícil de conseguir, porque la inversión de la brecha energética suele ser resultado del acoplamiento espín-órbita, que no es fácil de potenciar. 

Hemos dedicado un esfuerzo considerable a describir cómo emergen las fases topológicas de la materia a partir de una teoría microscópica de los sólidos. Este esfuerzo ha sido fundamental para contextualizar al lector sobre los primeros desarrollos en el estudio de estas fases. Sin embargo, en el resto de esta tesis, nos centraremos principalmente en el efecto magnetoeléctrico en los TIs. Para abordar este fenómeno, es esencial que consideremos la interacción de un TI con un campo electromagnético externo arbitrario. La descripción precisa de un sistema con interacción requiere el uso de una teoría de campos, ya que la teoría de bandas, en su forma estándar, no incorpora interacciones. En el próximo capítulo, profundizaremos en esta teoría de campos y, de manera análoga a los desarrollos de TKNN, mostraremos cómo se llega a la cuantización de la conductividad Hall mediante un enfoque basado en la teoría de campos.

\biblio %Se necesita para referenciar cuando se compilan subarchivos individuales - NO SACAR

%% file: Capitulos/03TME.tex
%

% En el campo de la física de la materia condensada, la \textit{Teoría de Bandas Topológicas} (TBT) \cite{Bansil2016topological} se ha destacado como un marco fundamental para el estudio y la descripción de la estructura microscópica de los TIs, tal como exploramos previamente. La TBT ha introducido conceptos fundamentales como la fase de Berry y la curvatura de Berry, que son aspectos cruciales de los estados cuánticos de Bloch en la primera BZ. Además, el invariante topológico $\mathbb{Z}_{2}$, que es crucial para distinguir  entre un aislante trivial y uno topológico, se ha vuelto indispensable, especialmente en el estudio de TIs (3+1)D. %En estos sistemas, se observa una conductividad de Hall cuantizada en la superficie del material, lo que da lugar a propiedades electromagnéticas (EMs) notables.

En el ámbito de la física de la materia condensada, la \textit{Teoría de Bandas Topológicas} (TBT) \cite{Bansil2016topological} se ha consolidado como un marco esencial para el estudio y la descripción de las estructuras microscópicas de las fases topológicas, tal como se ha explorado anteriormente. La TBT ha introducido conceptos fundamentales como la fase de Berry y la curvatura de Berry, los cuales son componentes cruciales de los estados cuánticos de Bloch en la primera BZ para definir la topología del material. Se ha demostrado que al aplicar campos electromagnéticos (EMs) externos a un sistema 2D, se puede observar el IQHE. Este fenómeno topológico se manifiesta al calcular la conductividad de Hall $\sigma_{xy}$ y expresarla en función de la fase de Berry, que revela la curvatura en el espacio de parámetros. Además, se ha comprobado que cuando el sistema preserva la simetría de inversión temporal, el invariante topológico $\mathbb{Z}_{2}$ es fundamental para distinguir entre un aislante trivial y uno topológico, dando lugar al QSHE. Estos resultados, revisados brevemente en el capítulo anterior, se han vuelto indispensables para el estudio de las fases topológicas.

% Otro enfoque relevante en la descripción de las respuestas EMs de materiales con fases topológicas es la \textit{Teoría de Campo Topológica} (TFT), propuesta inicialmente por Zhang y Hu \cite{doi:10.1126/science.294.5543.823}. En este marco, se ha desarrollado un modelo microscópico para un TI con simetría de inversión temporal (TRS) en (4+1)D. La teoría efectiva de este modelo puede describirse mediante la acción de Chern-Simons en (4+1)D \cite{bernevig2002effective}. A través de un procedimiento de reducción dimensional, este modelo se ha extendido para incluir TIs con TRS en (3+1)D y (2+1)D, incluso en presencia de interacciones y desorden \cite{Qi2008topological}. Un avance significativo radica en la reciente observación de que la TFT se reduce de manera exacta a la TBT en el límite de no interacción \cite{wang2010equivalent}. 

Otro enfoque relevante en la descripción de las respuestas EMs de materiales con fases topológicas es la \textit{Teoría de Campo Topológica} (TFT, por sus siglas en ingles). Inicialmente Zhang, Hansson y Kivelson en 1989 \cite{Zhang1989effective} mostraron que tanto el IQHE como el efecto Hall cuántico fraccionario pueden describirse mediante la TFT de Chern-Simons en (2+1)D. Esta TFT captura todos los efectos topológicos de los sistemas 2D donde la TRS está rota, incluyendo la cuantización de la conductancia Hall, la carga fraccionaria y las estadísticas de las cuasipartículas \cite{Zhang1992ChernSimons}. Mas tarde, Zhang y Hu \cite{doi:10.1126/science.294.5543.823} desarrollaron un modelo microscópico para un TI con TRS en (4+1)D. La teoría efectiva de este modelo puede describirse mediante la acción de Chern-Simons en (4+1)D \cite{bernevig2002effective}. A través de un procedimiento de reducción dimensional, este modelo se ha extendido para incluir TIs con TRS en (3+1)D y (2+1)D, incluso en presencia de interacciones y desorden \cite{Qi2008topological}. Un avance significativo radica en la observación de que la TFT se reduce de manera exacta a la TBT en el límite de no interacción \cite{wang2010equivalent}. 

La existencia de estados metálicos en los bordes (o superficies) de los TIs, que están protegidos topológicamente por 
%la correspondencia volumen-límite y 
la TRS, da lugar a una respuesta única a un campo EM. Cuando se aplica un campo eléctrico, se induce una corriente Hall en la superficie del TI, y esta corriente a su vez genera un campo magnético que magnetiza el material. Esta magnetización es una respuesta de la fase topológica de la materia al campo eléctrico aplicado y resulta independiente de los detalles microscópicos del sistema. Esta peculiar respuesta a los campos EMs da lugar al fenómeno conocido como \textit{Efecto Magnetoeléctrico Topológico} (TME). La magnetoelectricidad es un fenómeno en el cual se produce una interacción entre el campo magnético y el campo eléctrico en un material, \textit{i.e.}, un campo magnético puede inducir una polarización eléctrica y un campo eléctrico puede inducir una magnetización. La magnetoelectricidad se ha estudiado mucho antes de los TIs \cite{landau2013electrodynamics} y es más general que las respuestas magnetoeléctricas de los TI, véase \cite{lindell_1994_electromagnetic}.
%\textcolor{red}{Poner demostración sobre porque es la ED axionica y no otra. Ver Essin 2009. Tambien ver Zhang. A lo mejor poner el la Sec siguiente}

El TME puede describirse mediante una teoría de campos llamada \textit{Electrodinámica Axiónica} \cite{Wilczek1987two}. Esta teoría incluye un término nuevo al Lagrangiano, $\mathcal{L}_{\theta}$,  proveniente de la física de partículas, que se añade al Lagrangiano EM convencional de Maxwell, $\mathcal{L}_{0}$. Este término acopla los campos eléctrico y magnético con un factor $\theta(x)$. El Lagrangiano efectivo que describe el TME en (3+1)D, en unidades Gaussianas, viene dado por,
\begin{equation} 
\label{eq:Leff}
    \mathcal{L}_{\textup{eff}}^{3+1}=\mathcal{L}_{0}+\mathcal{L}_{\theta}=-\frac{1}{16\pi} F_{\mu\nu}F^{\mu\nu}-\frac{1}{16\pi} \theta(x)F_{\mu\nu}\tilde{F}^{\mu \nu}-\frac{1}{c}A_{\mu}J^{\mu},
\end{equation}
donde $\tilde{F}^{\mu\nu}=\frac{1}{2}\epsilon^{\mu\nu\delta\sigma}F_{\delta\sigma}$ es el tensor electromagnético dual, $\epsilon^{\mu\nu\delta\sigma}$ es el símbolo de Levi-Civita ($\epsilon^{0123}=+1$), $A_{\mu}$ es el campo de gauge electromagnético y $J^{\mu}$ es la corriente externa conservada. En este contexto, el factor $\theta(x)$ no tiene dinámica, a diferencia del campo axiónico . En lenguaje vectorial, la ecuación \eqref{eq:Leff} se escribe como,
\begin{align}
\mathcal{L}_{\textup{eff}}^{3+1}=\frac{1}{8\pi}\left (\mathbf{E}^{2}-\mathbf{B}^{2} \right )+\frac{1}{4\pi}\theta(\mathbf{x}) \mathbf{E}\cdot \mathbf{B} -\rho \mathbf{\Phi}+\frac{1}{c}\mathbf{J}\cdot\mathbf{A} 
\end{align}
donde la métrica la tomamos como $(+,-,-,-)$. Aquí $F^{i0}=E^{i}$, $F^{ij}=-\epsilon^{ijk}B_{k}$ y $\tilde{F}^{i0}=B^{i}$, $\tilde{F}^{ij}=\epsilon^{ijk}E_{k}$.

En la Sec. (\ref{3.1}), presentaremos algunas características de la electrodinámica axiónica y cómo esta describe la respuesta EM de los TIs 3D. Además, definiremos el parámetro topológico que se acopla a los campos EMs desde una perspectiva microscópica. En la Sección (\ref{3.2}), abordaremos la electrodinámica $\theta$ y sus diferencias con la electrodinámica de Maxwell, destacando, por ejemplo, los cambios en las condiciones de contorno. En la Sección (\ref{3.3}), revisaremos algunas aplicaciones de la electrodinámica axiónica en el contexto de los TIs y sus efectos, como los campos magnéticos inducidos por un monopolo magnético imagen y las rotaciones de la polarización de las ondas electromagnéticas que se reflejan y transmiten en un TI. También explicaremos las nuevas contribuciones de esta tesis a la literatura. Finalmente, en la Sección (\ref{3.4}), demostraremos que la electrodinámica axiónica puede reescribirse para satisfacer las ecuaciones de Maxwell, pero con modificaciones en las relaciones constitutivas. Argumentaremos que la magnetoelectricidad de los TIs no es el caso más general, ya que ha sido estudiada previamente en otros materiales. Sin embargo, medir la magnetoelectricidad en un TI es un enfoque para explorar las corrientes superficiales robustas, algo que no se logra de la misma manera en un medio magnetoeléctrico trivial.
\section{Electrodinámica axiónica en aislantes topológicos}\label{3.1}
%
% El concepto del campo de axiones y su excitación fundamental, el axión, se introdujo hace más de 40 años para explicar la ausencia de violación de la paridad de carga en la cromodinámica cuántica 
% %interacción fuerte entre quarks 
% \cite{Peccei1977CP,Wilczek1978problem,Weinberg1978boson}. Desde entonces, este concepto ha sido adoptado en diversos campos, desde la teoría de cuerdas hasta la cosmología, y el axión se ha considerado un candidato para la materia oscura \cite{PRESKILL1983127}. Sin embargo, hasta la fecha, no se han observado pruebas directas de la existencia del axión.

El axión, fue introducido hace más de 40 años en la física de partículas para resolver el problema de la violación de la simetría CP fuerte, en la cromodinámica cuántica (QCD, por sus siglas en inglés). En la QCD las interacciones nucleares fuertes se describen mediante el intercambio de gluones entre los quarks que están al interior de los hadrones. La evidencia experimental indica que la física de las interacciones fuertes es invariante bajo la transformación de CP, la aplicación conjunta de una transformación de carga C (el intercambio de cada partícula por su anti-partícula) y paridad P (el intercambio de las partículas de quiralidad izquierda por las de quiralidad derecha), i.e., la naturaleza parece indicar que la transformación CP es una simetría. No obstante, la QCD predice que dicha simetría puede violarse y no en menor medida. El mecanismo de Peccei-Quinn \cite{Peccei1977CP,Wilczek1978problem,Weinberg1978boson} introduce una nueva simetría global, la cual producto de un rompimiento espontáneo, deja como ``remanente'' un pseudo bosón de Goldstone, el axión de la QCD. Y pues la propia dinámica del axión impone que la teoría sea CP invariante. Así entonces el axión permite explicar que dicha violación sea nula. Desde entonces, este concepto ha sido adoptado en diversos campos de la física, \textit{e.g.}, en cosmología, los axiones se han propuesto como candidatos para la materia oscura fría \cite{PRESKILL1983127}, en teoría de cuerdas \cite{Mariño2005strings,PeterSvrcek_2006}, 
%los campos de axiones aparecen naturalmente en compactificaciones y son fundamentales para entender ciertos mecanismos de estabilización de moduli y la resolución de anomalías \cite{Witten1984dimensional}.
en teorías de gravedad cuántica \cite{Freese1990natural}, 
%Además, en teorías de gravedad cuántica y teoría de cuerdas, los axiones desempeñan un papel crucial en la construcción de modelos de inflación y en la posible unificación de las fuerzas fundamentales \cite{Freese1990natural}.
y más recientemente, en la física de la materia condensada \cite{Qi2008topological,Nenno2020axion}. A pesar del amplio interés teórico y experimental, hasta la fecha no se han obtenido pruebas directas de la existencia del axión, lo que sigue motivando investigaciones intensivas en múltiples frentes.

El concepto de ``electrodinámica axiónica'' fue impulsado por la idea de que el campo de axiones podría afectar las interacciones EMs, introduciendo un término adicional en las ecuaciones de Maxwell que daría lugar a efectos físicos observables. Uno de los primeros trabajos importantes en este contexto fue realizado por Frank Wilczek, quien en 1987 \cite{Wilczek1987two} introdujo el término ``\textit{axion electrodynamic'}' para describir cómo un campo de axiones acoplado al electromagnetismo podría modificar las leyes de la electrodinámica de Maxwell. Este trabajo fue clave para establecer la idea de que los axiones, si existen, podrían tener efectos observables en los campos EM y, por lo tanto, en fenómenos como la óptica y la magnetoóptica en materiales.
%\textcolor{red}{Revisar}

El término topológico $\mathcal{L}_{\theta}$ del Lagrangeano \eqref{eq:Leff} es una de las formas de Chern-Simons \cite{CS1974} y a diferencia de los primero trabajos, en el contexto de la materia condensada $\theta(x)$ no tiene dinámica y puede interpretarse como una contribución a la polarización magnetoeléctrica de los orbitales extendidos \cite{Qi2008topological,Essin2009magnetoelectric}. Nos referiremos a este modelo en particular como $\theta$-electrodinámica ($\theta$-ED). Es importante notar que esta extensión, al ser un término topológico, es una derivada total $\mathcal{L}_{\theta}=-\frac{1}{8\pi}\theta(x)\partial_{\mu}(A_{\nu}\tilde{F}^{\mu\nu})$, y no produce ninguna contribución a las ecuaciones de campo si $\theta(x)$ es uniforme en todo el espacio. Así, la acción efectiva es,
\begin{align}
 \label{eq:Seff}
%S_{\textup{eff}}^{3+1}&=\int_{\mathcal{M}} d^{4}x\left (-\frac{1}{16\pi} F_{\mu\nu}F^{\mu\nu}-\frac{1}{16\pi} \theta(x)F_{\mu\nu}\tilde{F}^{\mu \nu}-A_{\mu}J^{\mu}\right ),\\
    S_{\textup{eff}}^{3+1}&=\int d^{4}x\left (-\frac{1}{16\pi} F_{\mu\nu}F^{\mu\nu}+\frac{1}{8\pi} \partial_{\mu}\theta(x)A_{\nu}\tilde{F}^{\mu \nu}-A_{\mu}J^{\mu}\right ).
\end{align} 
con,
\begin{align}
    \theta(x)=-\frac{1}{4\pi}\int_{BZ}d^3k\epsilon^{\mu\nu\sigma}\textup{tr}\Big{[}a_\mu\partial_\nu a_\sigma -i\frac{2}{3}a_\mu a_\nu a_\sigma\Big{]}
\end{align}

donde la traza corre sobre el índice de las bandas de la conexión de Berry, $a$, entre las bandas ocupadas.  El campo $\theta(x)$ se puede derivar de un modelo de tight-binding acoplado al campo $A_{\mu}$ en el rango no interactuante, y toma diferentes valores dependiendo de la clase topológica del material.

Actualmente, se acepta que la teoría más adecuada para describir la respuesta EM de materiales con fases topológicas es la $\theta$-ED, representada por la acción efectiva \eqref{eq:Seff}. Sin embargo, para que la respuesta EM de un material pueda describirse correctamente mediante este formalismo, el material debe cumplir ciertas restricciones y consideraciones. Hasan y Kane \cite{Hasan2010Colloquium} discuten diversos fenómenos exóticos y propiedades únicas que pueden surgir en la superficie de los TIs cuando se induce una brecha energética $\Delta$ en las bandas de los estados superficiales (SS, por sus siglas en inglés), como es el caso del TME. 

Si la TRS se rompe mediante la aplicación de un campo magnético externo \cite{Kane2007topological}, se abrirá una brecha en los SS, lo que puede conducir al IQHE. Alternativamente, esta brecha también puede abrirse depositando una película magnética en la superficie del TI, lo que da lugar al QAHE \cite{Qi2008topological}. La conductividad Hall en la superficie de un TI puede ser investigada mediante métodos ópticos \cite{Essin2009magnetoelectric}, siempre y cuando la energía de los fotones incidentes sea mucho menor que la brecha energética del bulk y de los SS, asegurando así la validez de la descripción mediante la $\theta$-ED en el límite de baja frecuencia, $\hbar\omega\ll\Delta,E_g$, donde $E_g$ es la brecha de energía en los estados electronicos del bulk. Para un valor típico de $E_{g}=0.3~\textup{eV}$ \cite{Qi2008topological}, se obtiene que $f=\omega/2\pi\ll 73~\textup{THz}$, lo que corresponde a la región infrarroja o de microondas del espectro EM. En principio, es posible encontrar TIs con una brecha de energía mayor, lo cual permitiría explorar una ventana más amplia del espectro EM \cite{Ando2013topological}.

La respuesta ante un campo EM externo $A_{\mu}$ de los TIs, se obtiene de las ecuaciones de movimiento Euler-Lagrange para los campos EMs,
\begin{align}
\label{eq:E-L}
    \partial_{\mu}\left ( \frac{\partial \mathcal{L}_{\textup{eff}}}{\partial(\partial_{\mu}A_{\nu})} \right )-\frac{\partial\mathcal{L}_{\textup{eff}}}{\partial A_{\nu}}=0,
\end{align}
reemplazando el Lagrangeano de la Ec. \eqref{eq:Leff} obtenemos,
\begin{align}
\label{eq:Maxwelltheta}
\partial_{\mu}F^{\mu\nu}+\partial_{\mu}\theta\tilde{F}^{\mu\nu}=\frac{4\pi}{c}J^{\nu},
\end{align}
el cual corresponde a las dos ecuaciones de Maxwell no homogéneas. El primer término del lado izquierdo corresponde a las derivadas usuales del campo EM y el segundo término del lado izquierdo está asociado a la respuesta EM de los TIs. Para ``separar'' las contribuciones de las corrientes topologicas de la superficie del TI, podemos asociar en el lado derecho todo lo que tiene que ver con la nueva 4-corriente y asi definir una 4-corriente de Hall,
\begin{align}
\label{eq:4corrientetheta}
    J^{\nu}_{\theta}=-\frac{c}{4\pi}\partial_{\mu}\theta\tilde{F}^{\mu\nu}.
\end{align}
En el lenguaje vectorial obtenemos,
\begin{align}
    \rho_{\theta}&=-\frac{1}{4\pi}\boldsymbol{\nabla}\theta\cdot\mathbf{B}, &
    \mathbf{J}_{\theta}&=\frac{1}{4\pi}(c\boldsymbol{\nabla}\theta\times\mathbf{E}+\partial_{t}\theta\,\mathbf{B}),\label{eq:corrientetopologica}
\end{align}
donde vemos que un cambio espacial de $\theta=\theta(\mathbf{r})$, induce una corriente de Hall, $\mathbf{J}_{\textup{Hall}}=\frac{c}{4\pi}\boldsymbol{\nabla}\theta\times\mathbf{E}$ al aplicar un campo eléctrico externo $\mathbf{E}$. Además, cuando hay presente un campo magnético $\mathbf{B}$, éste induce una densidad de carga extra $\rho_{\theta}$ \cite{Wilczek1987two}. Por otro lado, cuando hay un cambio temporal de $\theta=\theta(t)$, se observa un efecto magnético quiral (CME, por sus siglas en ingles), $\mathbf{J}_{\textup{CME}}=\frac{1}{4\pi}\partial_{t}\theta\,\mathbf{B}$, que corresponde a la generación de corriente eléctrica inducida por un campo magnético aplicado, ver \cite{Urrutia2020EMpropagation,Urrutia2023CherenkovCM,Mizher2018aspects}. Podemos verificar que la 4-corriente inducida \eqref{eq:4corrientetheta} en un medio con fases topológica se conserva,
\begin{align}
    \partial_{\mu}J^{\mu}_{\theta}=0,
\end{align}
o, en lenguaje vectorial, $\partial_{t}\rho_{\theta}+\boldsymbol{\nabla}\cdot\mathbf{J}_{\theta}=0$. Para demostrar la continuidad de la 4-corriente hemos usado la identidad de Bianchi, $\partial_{\mu}\tilde{F}^{\mu\nu}=0$, que nos proporciona las dos ecuaciones homogéneas usuales de Maxwell. 

Como hemos mencionado anteriormente, el campo $\theta(x)$ toma diferentes valores dependiendo del material considerado.
%de la clase topológica del material. 
En el contexto de los TIs, $\theta(x)=\frac{\alpha}{\pi}\theta_{\textup{TI}}$, donde $\alpha=e^2/\hbar c \approx 1/137$ es la constante de estructura fina y $\theta_{\textup{TI}}$ se denomina Polarizabilidad Magnetoeléctrica Topológica (TMEP). Su origen es mecánico-cuántico y codifica las propiedades microscópicas que caracterizan a los TIs. Al romper la TRS para abrir la brecha energética de los SS, $\theta_{\textup{TI}}$ puede tomar los siguiente valores,
\begin{align}
\theta_{\textup{TI}}=\left\{\begin{matrix}
\pm(2n+1)\pi, & \textup{Para TI}.\\ 
0, & \textup{Para aislantes triviales}.
\end{matrix}\right.
\end{align}
donde $n\in\mathbb{Z}$ y el signo $\pm$ está determinado por la dirección de ruptura de la TRS en la superficie \cite{Qi2008topological,maciejko_topological_2010}. 

\begin{figure}[t]
\stackinset{l}{5pt}{t}{11pt}{(a)}{\includegraphics[width=0.3\textwidth,height=3.25cm]{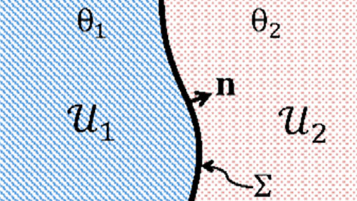}}
\hspace{-0.1cm}  
\stackinset{l}{5pt}{t}{25pt}{(b)}{\includegraphics[scale=0.46]{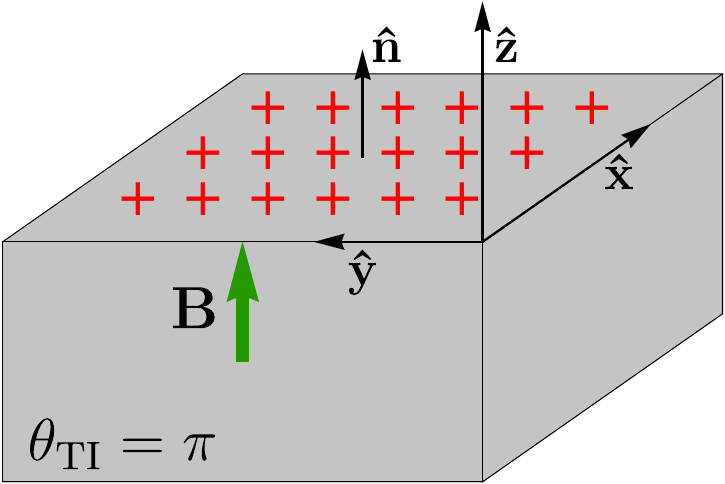}} 
\hspace{-0.1cm}  
\stackinset{l}{5pt}{t}{25pt}{(c)}{\includegraphics[scale=0.46]{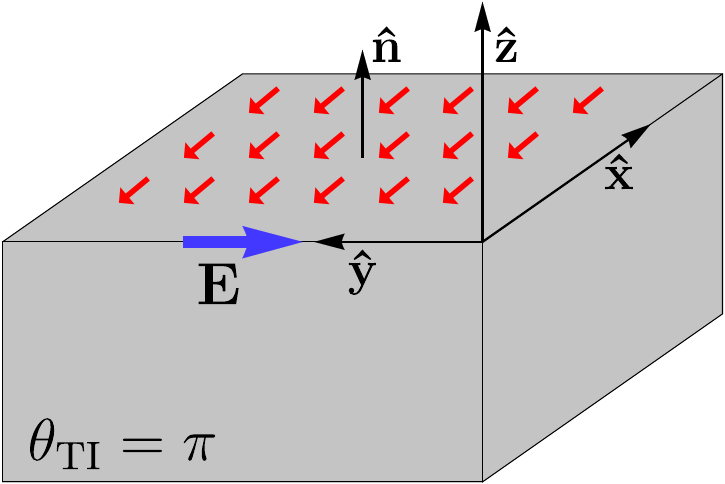}}
\caption{(a) Interfase entre dos materiales con diferentes propiedades topológicas. En la superficie $\boldsymbol{\Sigma}$ es donde se inducen las cargas y corrientes topológicas proporcionales a $(\theta_2-\theta_1)$ cuando aplicamos campos EM externos. (b) Un campo magnético induce cargas eléctricas en la superficie de un TI. (c) Un campo eléctrico induce una corriente Hall en la superficie de un TI. Solo los campos EM en $\delta(z)$ inducen cargas y corrientes topológicas, que a su vez son fuentes de campos EM en todo el espacio. La medición de estos campos EM inducidos será un complemento importante de los experimentos ARPES.}
\label{FIG:HallChargeAndCurrent}
\end{figure}

%Para observar la respuesta EM de los TI, 
Para observar el TME que inducen los SS de los TIs, es necesario que el parámetro $\theta$ cambie de un medio a otro como se ve en la Fig. (\ref{FIG:HallChargeAndCurrent},a). Así, un TI cubierto por un material con distinta topología, por ejemplo el vació, inducirá un cambio espacial de $\theta$, \textit{i.e.}, para dos medios $\mathcal{U}_{1}$ y $\mathcal{U}_{2}$ tenemos $\boldsymbol{\nabla}\theta=\tilde{\theta}\delta(\mathbf{r}-\boldsymbol{\Sigma})\mathbf{\hat{n}}$ y $\partial_{t}\theta=0$, 
%es necesario tener una interfaz que separe dos regiones con valores de $\theta_{\textup{TI}}$ distintos. Esto implica que para un TI, $\partial_{t}\theta=0$ y $\boldsymbol{\nabla}\theta=\tilde{\theta}_{i}\delta(\Sigma_{i})\boldsymbol{\hat{n}}$, 
donde $\tilde{\theta}\equiv\theta_{2}-\theta_{1}$ y $\mathbf{\hat{n}}$ es el vector unitario normal que va desde el medio $1$ al medio $2$ definido en la interfaz $\Sigma$. Por lo tanto, las densidades de cargas y corrientes,
\begin{align}
    \sigma_{\theta}^{\textup{TI}}&=\int \rho_{\theta}dl=-\frac{1}{4\pi}\int(\boldsymbol{\nabla}\theta\cdot\mathbf{B})dl=-\frac{\tilde{\theta}}{4\pi}\int \delta(\mathbf{r}-\boldsymbol{\Sigma})(\mathbf{B}\cdot\mathbf{\hat{n}})dl=-\frac{\tilde{\theta}}{4\pi}(\mathbf{B}\cdot\mathbf{\hat{n}})|_{\boldsymbol{\Sigma}}\\
    \mathbf{K}_{\theta}^{\textup{TI}}&=\int \mathbf{J}_{\theta}dl=\frac{c}{4\pi}\int(\boldsymbol{\nabla}\theta\times\mathbf{E})dl=\frac{c}{4\pi}\tilde{\theta}\int\delta(\mathbf{r}-\boldsymbol{\Sigma})(\mathbf{\hat{n}}\times \mathbf{E})dl=\frac{c}{4\pi}\tilde{\theta}(\mathbf{\hat{n}}\times \mathbf{E})|_{\boldsymbol{\Sigma}}
\end{align}
se inducen en la interfaz. Donde $dl$ es el elemento de línea que recorre la dirección de la normal a la superficie. 

Supongamos que el espacio semiinfinito debajo del plano $z=0$ está ocupado por un TI, tal que $\theta_{\textup{TI}} = \pi$ para $z<0$, y $\theta_{\textup{TI}} = 0$ para $z>0$. Si se aplica un campo magnético uniforme a lo largo del eje $z$ (ver Fig. (\ref{FIG:HallChargeAndCurrent} .b)), entonces la densidad de carga efectiva inducida por $\theta$ es,
\begin{align}\label{eq:cargatheta}
    \sigma_{\theta}^{\textup{TI}}=-\frac{\tilde{\theta}}{4\pi}(\mathbf{B}\cdot\mathbf{\hat{n}})|_{\boldsymbol{\Sigma}}=\frac{\alpha}{4\pi}B_{z}(x,y,0),
\end{align}
Es decir, hay una fina capa de cargas positivas en la superficie del TI. Por otro lado, si se aplica un campo eléctrico uniforme paralelo a la superficie (ver Fig. (\ref{FIG:HallChargeAndCurrent} .c)), la densidad de corriente efectiva inducida por $\theta$ es,
\begin{align}\label{eq:corrientetheta}
    \mathbf{K}_{\theta}^{\textup{TI}}=\frac{c}{4\pi}\tilde{\theta}(\mathbf{\hat{n}}\times \mathbf{E})|_{\boldsymbol{\Sigma}}=-\frac{\alpha c}{4\pi}(\mathbf{\hat{z}}\times \mathbf{E}(x,y,0)).
\end{align}
Es decir, hay una fina capa de corriente en la superficie del TI, perpendicular al campo $\mathbf{E}$. Esta es una corriente Hall con conductividad Hall semi-cuantizada, $\sigma_{\textup{Hall}} = \alpha c/4\pi = e^{2}/2h$. Por tanto, el término axión produce respuestas EMs correctas del estado de la superficie de TI, a saber, la conductividad de Hall. 

La observación de las corrientes superficiales asociadas a este efecto magnetoeléctrico será un complemento importante de los experimentos ARPES. Sin embargo, debe enfatizarse que a pesar del estado topológicamente cuantizado de $\theta$, las corrientes superficiales deben distinguirse de otras corrientes ligadas no cuantizadas que puedan estar presentes \cite{Hasan2010Colloquium}.

% Como consecuencia directa de la respuesta EM es el efecto topológico magneto-eléctrico (TME) \cite{qi2008topological}, el cual significa que la polarización eléctrica obtiene una nueva contribución proporcional a $\theta \mathbf{B}$ y la magnetización obtiene una contribución proporcional a $\theta \mathbf{E}$. 
%---------------Figure------------
%---------------End Figure------------
\section{Ecuaciones de la $\theta$-ED}\label{3.2}
En un sistema EM donde hay presente un medio con fases topológicas, la electrodinámica está dada por la Ec. \eqref{eq:Maxwelltheta}, junto con la identidad de Bianchi. Además, si consideramos que el medio a consideración responde con las propiedades ópticas usuales, como la permitividad $\epsilon$ y permeabilidad $\mu$, tenemos que las ecuaciones de la $\theta$-ED, escritas en lenguaje vectorial son,
\begin{align}
\mathbf{\nabla}\cdot (\epsilon \mathbf{E}) &=4\pi\rho-\boldsymbol{\nabla}\theta \cdot \mathbf{B},  \label{eq:MCS1}\\ 
\mathbf{\nabla}\times  \mathbf{E}&=-\frac{1}{c}\frac{\partial }{\partial t}\mathbf{B},\label{eq:MCS2}\\
\mathbf{\nabla}\cdot \mathbf{B}&=0,\label{eq:MCS3}\\ 
\mathbf{\nabla}\times (\frac{1}{\mu}\mathbf{B})&=\frac{4\pi}{c}\mathbf{J}+\frac{1}{c}\frac{\partial}{\partial t} (\epsilon \mathbf{E})+\boldsymbol{\nabla}\theta \times \mathbf{E}+\frac{1}{c}\frac{\partial \theta }{\partial t}\mathbf{B},\label{eq:MCS4}
\end{align}
donde $\rho$ y $\mathbf{J}$ son las densidades de carga y corrientes eléctricas libres convencionales y son las fuentes de los campos eléctricos $\mathbf{E}$ y magnéticos $\mathbf{B}$ respectivamente. En general, $\theta$ puede interpretarse como un parámetro macroscópico efectivo determinado por la topología de la estructura de bandas del material, aparte de la permitividad eléctrica $\epsilon$ y la permeabilidad magnética $\mu$ usuales del electromagnetismo de Maxwell. 
%Estas ecuaciones proporcionan una ilustración física de la respuesta EM de materiales con fases topológica.

%Dado que las ecuaciones de la electrodinámica son modificadas en materiales con fases topológicas se puede verificar fácilmente que las condiciones de borde (BC, por sus siglas en ingles) cambian. 
Los efectos de la interfaz $\theta$, es añadir discontinuidades adicionales en los campos producidas por las densidades de carga y corrientes superficiales. Si consideramos que no depositamos fuentes externas en la interfaz que separa dos medios, \textit{i.e.}, $\rho=0$ y $\mathbf{J}=0$, vemos que de las ecuaciones no homogéneas, \eqref{eq:MCS1} y \eqref{eq:MCS4}, las condiciones de borde (BCs, por sus siglas en ingles) se modifican como siguen, 
\begin{align}
\Delta [\epsilon\mathbf{E}_{\perp }]|_{\Sigma_{i}}=-\tilde{\theta}_{i}\mathbf{B}_{\perp }|_{\Sigma_{i}}, \quad \textrm{y} \quad   \Delta \left[\mu^{-1}\mathbf{B}_{\parallel} \right]|_{\Sigma_{i} }=\tilde{\theta}_{i}\mathbf{E}_{\parallel}|_{\Sigma_{i}}, 
\label{EQ:BCsdiscontinuas}
\end{align}
donde $\Delta [A]\equiv A_{i+1}-A_{i}$, $\tilde{\theta}_{i}=\theta_{i+1}-\theta_{i}$, y los sub-símbolos $\perp$ y $\parallel$ son las componentes perpendiculares y paralelas a la superficie $\Sigma_{i}$. Las Ecs. (\ref{EQ:BCsdiscontinuas}) conducen a diferentes soluciones tanto en las interfaces como en el volumen del sistema. Por otro lado, de las ecuaciones homogéneas, \eqref{eq:MCS2} y \eqref{eq:MCS3}, las otras BCs quedan sin modificaciones,
\begin{align}
\Delta [\mathbf{E}_{\parallel}]|_{\Sigma_{i}}=0, \quad \textrm{y} \quad   \Delta \left[\mathbf{B}_{\perp} \right]|_{\Sigma_{i} }=0, 
\label{EQ:BCscontinuas}
\end{align}
manteniendo la continuidad de las componentes paralelas del campo eléctrico y la componente perpendicular del campo magnético. La Ec. \eqref{EQ:BCsdiscontinuas} es una forma explícita de ver como se manifiesta el TME, es decir, la inter-relación entre el campo el eléctrico y magnético, aún en el caso estático\footnote{Sabemos de la ley de Faraday que un campo magnético variable en el tiempo induce un campo eléctrico y de la ecuación de Amp\`ere-Maxwell sabemos lo contrario, es decir, existe una inter-relación entre estos dos campos en el caso dinámico. Pero en una interfaz TI-aislante trivial, vemos que un campo magnético estático induce un campo eléctrico y viceversa.}.

Es crucial recalcar el siguiente punto: Aunque las ecuaciones diferenciales que rigen la dinámica de los campos EM lejos de las interfaces de los TIs son idénticas a las ecuaciones de Maxwell convencionales, las soluciones varían debido a que las BCs, que determinan el conjunto de amplitudes de la solución, son diferentes y, por lo tanto, alteran el espacio de soluciones.

Otro material con fase topológica de gran interés son los semi-metales de Weyl (WSM) \cite{wan2011topological,burkov2011weyl}, los cuales exhiben un efecto Hall en el volumen y un CME. La respuesta electromagnética de los WSM se describe igualmente por la acción \eqref{eq:Seff}. A diferencia de los TIs, los WSM no tienen el ángulo axiónico constante por partes; en cambio, el campo $\theta(x)$ es $\theta(\mathbf{r},t)=2(\mathbf{b} \cdot\mathbf{r}-b_{0}t)$, donde $2\mathbf{b}$ refleja la separación de los conos de Dirac en pares de nodos de Weyl en el espacio de momento, y $2b_{0}$ está relacionado con su diferencia de energía. 
\section{Aplicaciones de la $\theta$-ED en materiales con fases topológicas}\label{3.3}
En el contexto de la $\theta$-ED, se ha investigado la respuesta EM de materiales con fases topológicas. Revisaremos brevemente algunos trabajos, por ejemplo, lo que ocurre al colocar una carga puntual estática sobre la superficie plana de un TI. También revisaremos cómo una onda electromagnética (OEM) se comporta al incidir de manera normal en la superficie de un TI, y cómo los ángulos de rotación de las ondas reflejadas y transmitidas se ven afectados. Por último, mostraremos las modificaciones en las ecuaciones de Fresnel cuando una OEM incide oblicuamente en una superficie con diferentes propiedades topológicas, lo cual es crucial en este trabajo, ya que las OEMs que se propagan en un medio confinado sufren continuas reflexiones debido a su incidencia oblicua en las paredes de la guía de ondas formada por TIs.

\begin{SCfigure}[1][t]
\caption{Un TI con constante dieléctrica $\varepsilon_1$ y permeabilidad magnética $\mu_1$ ocupa el semi-espacio inferior ($z<0$), mientras que el semi-espacio superior ($z>0$) es ocupado por un dieléctrico convencional con constante dieléctrica $\varepsilon_2$ y permeabilidad magnética $\mu_2$. Una carga eléctrica puntual $q$ ubicada en $(0,0,d)$ sobre un TI induce monopolos imágenes. Cuando se ve el campo desde el medio superior se observa un \textit{dyon} en $(0,0,-d)$ con carga $q_{2}$ y $g_{2}$. Lo mismo si se observa desde el medio inferior. Imagen de \cite{qi_monopole_2009}.
\label{FIG:MonopoleMagnetic}}
\includegraphics[scale=0.6]{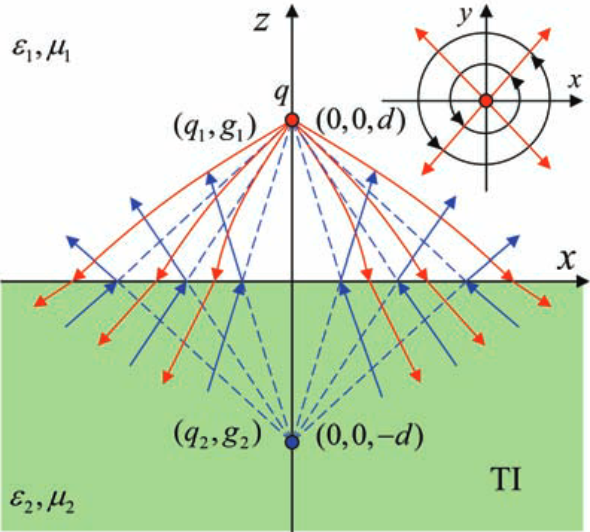}
\end{SCfigure}

Un fenómeno peculiar originado por la TME, es la inducción de un monopolo magnético imagen debido al acercamiento de una carga eléctrica a un TI semi-infinito estudiado primero en \cite{qi_monopole_2009}. La configuración de tal sistema se muestra en la Fig. (\ref{FIG:MonopoleMagnetic}).
%---------------Figure------------
%---------------End Figure---------
Debido al campo eléctrico originado por la carga $q$ y de las BCs modificadas, se induce una corriente Hall tipo vórtice sobre la superficie del TI. El campo magnético creado por esta corriente Hall puede describirse en términos de una carga magnética imagen dentro del TI, la cual se puede calcular utilizando el método de las imágenes. Para resolver este problema, suponemos que en el semi-espacio inferior el campo eléctrico está generado por una carga puntual efectiva $q/\varepsilon_1$ y una carga imagen $q_1$ en $(0, 0, d)$, mientras que el campo magnético proviene de un monopolo magnético imagen $g_1$ en $(0, 0, d)$. En el semi-espacio superior, el campo eléctrico está dado por una carga eléctrica $q/\varepsilon_1$ en $(0, 0, d)$ y una carga imagen $q_2$ en $(0, 0, -d)$, y  el campo magnético por un monopolo magnético imagen $g_2$ en $(0, 0, -d)$. Este enfoque satisface las ecuaciones de Maxwell a cada lado de la frontera. En la frontera $z=0$, se igualan las soluciones, lo que, según las BCs, conduce al siguiente resultado,
\begin{align}
    g_{1}=-g_{2}=-\frac{4\alpha P_{3}}{(\epsilon_{1}+\epsilon_{2})(1/\mu_{1}+1/\mu_{2})+4\alpha^{2}P_{3}^{2}}q
\end{align}
Donde $\alpha$ es la constante de estructura fina, $P_{3}=\theta_{\textup{TI}}/2\pi$ es la polarización EM introducida relacionada con el TMEP, donde $\theta_{\textup{TI}}=\pm \pi$ con el signo determinado por la dirección de la magnetización superficial. Las cargas eléctricas se relacionan de la siguiente manera,
\begin{align}
q_{1}=q_{2}=\frac{1}{\epsilon_{1}}\frac{(\epsilon_{1}-\epsilon_{2})(1/\mu_{1}+1/\mu_{2})-4\alpha^{2}P_{3}^{2}}{(\epsilon_{1}+\epsilon_{2})(1/\mu_{1}+1/\mu_{2})+4\alpha^{2}P_{3}^{2}}q
\end{align}
Las partículas imágenes anteriores son los \textit{dyons}, cuasipartículas que poseen carga eléctrica y magnética. Al considerar solo la respuesta $\theta$, \textit{i.e.}, $\varepsilon_{1}=\varepsilon_{2}=\mu_{1}=\mu_{2}=1$, se observa que las cargas de los dyones están relacionadas por $q_{1,2}=\pm(\theta/2)g_{1,2}=\pm(\alpha\theta_{\textup{TI}}/2\pi)g_{1,2}$, a través del término topológico $\theta$. Los monopolos imagen que emergen en estos sistemas pueden investigarse mediante microscopía de fuerza magnética, la cual permite mapear el flujo magnético generado por estas cargas \cite{qi_monopole_2009,Hasan2010Colloquium,RevModPhys.83.1057}. Es importante señalar que estos monopolos no son reales; es decir, sus campos inducidos se comportan como si existiera un monopolo en el dominio de la imagen, aunque no haya un monopolo físico presente.

En \cite{PhysRevB.99.155142}, se investigó el campo electromagnético generado por una carga eléctrica en proximidad a un WSM. A diferencia de lo que ocurre en un TI, se observó que las líneas del campo magnético inducido por la carga emergen desde la superficie, similar a lo que sucede con un monopolo magnético imagen en el semiespacio inferior. Sin embargo, estas líneas de campo magnético se curvan y terminan nuevamente en la superficie, aunque en un punto diferente al de su origen.

Como vimos, la respuesta topológica de un TI puede ser detectada mediante experimentos que involucren mediciones sin contacto a través de la interacción con el campo EM y/o fuentes. Por ejemplo, mediante sondas que tengan una carga de prueba cerca de la superficie del TI. En este contexto, en \cite{martin2015green,martin2016electro,martin2016electromagnetic} se sistematizó el uso de la formulación de Green para problemas EMs con fuentes arbitrarias en la vecindad de un TI. Además, en \cite{martin2016green} se analizó la interacción del TI con el campo EM cuantizado en uno de los escenarios más emblemáticos de la \textit{Quantum electrodynamics}, a saber, en el contexto del efecto Casimir. Aspectos formales del planteamiento y solución del problema EM fueron reportados en \cite{martin2019magnetoelectric2}.

De igual manera, se ha explorado la interacción entre OEMs y TIs, revelando notables respuestas, en función del parámetro topológico $\theta$. Entre estas respuestas, destaca la rotación del plano de polarización de la OEM al incidir sobre la superficie de un TI, manifestada a través de los efectos de Kerr y Faraday. El efecto Faraday se refiere a la rotación del plano de polarización de la onda EM transmitida al atravesar un material magnetizado, inducida por la birrefringencia que surge de la presencia de un campo magnético alineado con la dirección de propagación de la onda. Por otro lado, el efecto Kerr se observa en la onda EM reflejada, donde el plano de polarización se rota debido a la interacción magneto-óptica en un material sometido a un campo magnético perpendicular a la superficie. Ambos efectos surgen debido a los términos fuera de la diagonal en el tensor de permitividad, $\epsilon_{ij}$, que son responsables de la anisotropía inducida por la magnetización del material. Estos términos rompen TRS, haciendo que la permitividad dependa no solo del campo eléctrico aplicado, sino también de la magnetización interna del material. Así, los efectos magneto-ópticos de Kerr y Faraday ofrecen una poderosa herramienta para investigar directamente la ruptura de TRS en sólidos \cite{tse_giant_2010}. Para un TI, las rotaciones son causadas por el TMEP al romper la TRS en la superficie. 

\begin{SCfigure}[1][t]
\caption{Una OEM incidente se propaga en la dirección $\tongo{z}$ con una polarización inicial en la dirección $\tongo{x}$. Los ángulos de Kerr y Faraday dependen de la amplitud del campo EM incidente, como se muestra en la Ec. \eqref{eq:Kerr&Faraday}. La OEM incide perpendicularmente (los rayos de luz se dibujan con ángulos simplemente para mayor claridad) sobre una película gruesa de TI con un espesor $\ell$, que se encuentra sobre un sustrato aislante trivial, en presencia de un campo magnético perpendicular $\mathbf{B}$. Imagen de \cite{maciejko_topological_2010}.
\label{FIG:KerryFaraday}}
\includegraphics[scale=0.47]{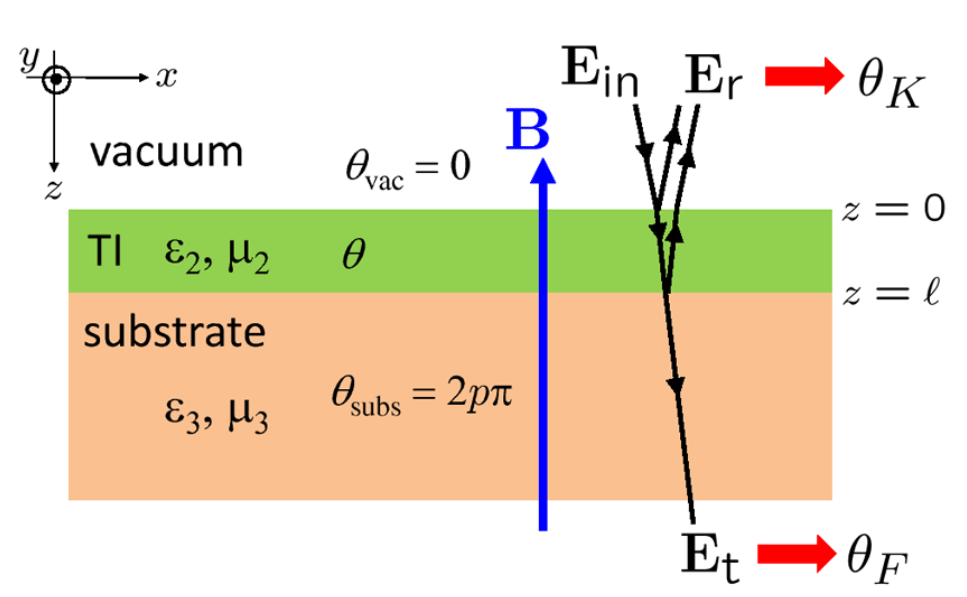}
\end{SCfigure}

Consideremos la propagación de una OEM en la dirección $\tongo{z}$, asi la polarizacion de la luz se encuentra en el plano $z=constante$ como se muestra en la Fig. (\ref{FIG:KerryFaraday}). Los ángulos de Kerr ($\phi_{\textup{K}}$) y Faraday ($\phi_{\textup{F}}$) se definen en términos de las amplitudes de los campos eléctricos de la OEM incidente. Cuando la luz está en $\tongo{x}$, las rotaciones se definen como,
\begin{align}\label{eq:Kerr&Faraday}
    \tan \phi_{\textup{K}}&=\frac{E_{r}^{y}}{E_{r}^{x}}, & \tan \phi_{\textup{F}}&=\frac{E_{t}^{y}}{E_{t}^{x}}
\end{align}
donde $E_{r}^{x}$ y $E_{r}^{y}$ son las componentes $x$ y $y$ del campo eléctrico reflejado y $E_{t}^{x}$ y $E_{t}^{y}$ del campo eléctrico transmitido. Una onda con esta polarización, cuando incide sobre un TI depositado sobre un sustrato trivial y en presencia de un campo magnético $\mathbf{B}$, como se ilustra en la Fig (\ref{FIG:KerryFaraday}), presenta señales de los estados topológicos que residen en la superficie del TI \cite{maciejko_topological_2010}.
%---------------Figure------------
%---------------End Figure---------
%
En la disposición anterior, el sustrato es un aislante trivial, pero dado que la conductividad Hall en la superficie depende de los detalles de la interfaz, puede diferir entre las dos interfaces del TI. En el caso general, considerar $\theta_{\textup{subs}}=2p\pi$ donde $p \in \mathbb{Z}$ implica diferentes valores para la conductancia Hall en las distintas interfaces,
\begin{align}
    \sigma_{H}^{z=0}&=\frac{\theta}{2\pi}\frac{e^{2}}{h}, & \sigma_{H}^{z=\ell}&=(p-\frac{\theta}{2\pi})\frac{e^{2}}{h},
\end{align}
En el sistema ilustrado en la Fig. (\ref{FIG:KerryFaraday}), la reflectividad está definida por,
\begin{align}
    R\equiv \frac{|\mathbf{E}_{r}|^{2}}{|\mathbf{E}_{\textup{in}}|^{2}}
\end{align}
La cual alcanza un mínimo cuando el espesor del aislante topológico $\ell$ es un múltiplo semi entero de la longitud de onda de la radiación electromagnética dentro del TI. Cuando la frecuencia de operación de la onda incidente se ajusta a este mínimo, se encuentra una relación entre los ángulos de Kerr y Faraday y la constante de estructura fina,
\begin{align}
    \frac{\cot \phi'_{\textup{F}}+\cot \phi'_{\textup{K}}}{1+\cot^{2} \phi'_{\textup{F}}}=\alpha p
    \label{eq:KerrandFaradayanlge}
\end{align}
La relación anterior resalta la naturaleza topológica de los estados de la superficie, ya que la combinación de los ángulos de Kerr y Faraday es independiente de las propiedades ópticas del TI y del sustrato ($\varepsilon_{1}, \mu_1, \varepsilon_{2}$ y $\mu_2$). Además, esta respuesta está cuantificada en términos de la constante de estructura fina \cite{maciejko_topological_2010,RevModPhys.83.1057}. En otros trabajos sobre mediciones de los ángulos de Kerr y Faraday, demostraron una rotación universal del ángulo de Faraday $\phi_{\textup{F}}=\tan^{-1}\alpha$, mientras que una rotación gigante para el ángulo de Kerr $\phi_{\textup{K}}=\pi/2$, en ondas de bajas frecuencias \cite{tse_giant_2010,RevModPhys.83.1057}. 

\begin{SCfigure}[1][t]
\caption{Reflexión y refracción de una OEM en la interfaz entre dos materiales con índices de refracción $n$ y $n'$ y TMEP de $\Theta$ y $\Theta'$. En la polarización paralela o polarización TM, el campo eléctrico $E_{\parallel}$ se encuentran en el plano de incidencia (plano $y\!=\!0$) y los campos magnéticos son perpendiculares a dicho plano (a lo largo de la dirección $\tongo{y}$). En la polarización perpendicular o polarización TE, el campos eléctrico $E_{\perp}$ es perpendiculares al plano de incidencia, y los campos magnéticos se encuentran en ese plano. Imagen de \cite{PhysRevB.80.113304}.
\label{fig:FresnelEcs}}
\includegraphics[scale=0.47]{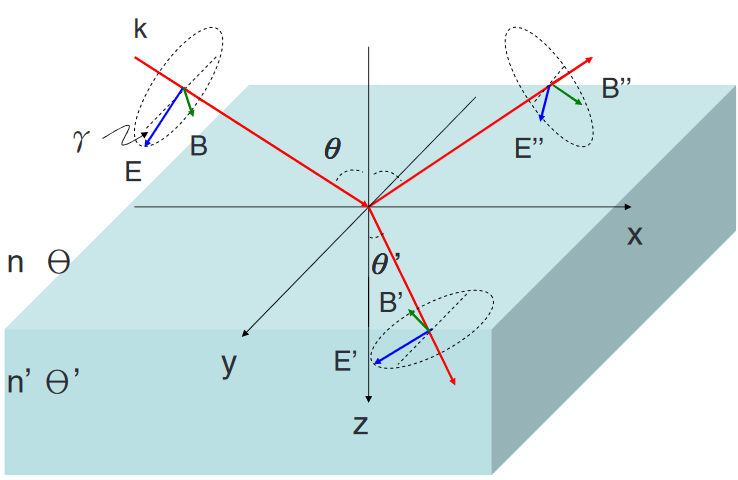}
\end{SCfigure}

Otra consecuencia del efecto TME se observa en la modificación de las ecuaciones de Fresnel para las componentes del campo eléctrico de la OEM reflejada en presencia de un TI, lo que resulta, \textit{e.g.}, en un cambio medible en el ángulo de Brewster. En \cite{PhysRevB.80.113304}, se consideró una OEM que se propaga desde un medio con índice de refracción $n$ y TMEP de $\Theta$, e incide oblicuamente con un ángulo $\theta$ sobre un TI con índice de refracción $n'$ y TMEP de $\Theta'$, como se muestra en la Fig. (\ref{fig:FresnelEcs}). La polarización de la OEM sufre una rotación, que es descrita por una matriz de las amplitudes del campo,
%---------------Figure------------
%---------------End Figure---------
\begin{align}
    \begin{pmatrix}
E''_{\perp}\\ 
E''_{\parallel}
\end{pmatrix}=\frac{1}{\Delta}\begin{pmatrix}
(n^{2}-n'^{2}-\bar{\alpha}^{2})+nn'\xi_{-}  & 2\bar{\alpha}n\\ 
2\bar{\alpha}n & -(n^{2}-n'^{2}-\bar{\alpha}^{2})+nn'\xi_{-} 
\end{pmatrix}\begin{pmatrix}
E_{\perp}\\ 
E_{\parallel}\end{pmatrix}
\end{align}
donde $\perp$ y $\parallel$ corresponde a las direcciones perpendiculares y paralelas al plano de incidencia respectivamente (los modos TE y TM respectivamente), $\bar{\alpha}=\alpha(\Theta-\Theta')/\pi$, $\Delta=n^{2}+n'^{2}+\bar{\alpha}^{2}+nn'\xi_{+}$, y 
\begin{align}
    \xi_{\pm}\equiv\frac{\cos\theta}{\cos\theta'}\pm\frac{\cos\theta'}{\cos\theta}
\end{align}

con $\theta'$ determinado por la Ley de Snell $n\sin\theta=n'\sin\theta'$, que no es modificada cuando $\Theta \neq 0$. Esta \textit{mezcla} de las componentes perpendiculares y paralelas desencadena una modificación en los modos de polarización TE y TM de la onda incidente \cite{PhysRevB.80.113304}. 

Este hallazgo cobra una relevancia significativa al abordar la problemática del confinamiento de ondas en TIs, que constituye un aspecto clave de esta tesis. Se destaca que los modos que son puramente TE o TM no se propagan a lo largo de la guía de ondas definida por interfaces TI, ya que esta mezcla se manifiesta en cada incidencia de la OEM en las paredes guía. Esta característica particular, originada por el efecto TME en la interfaz, distingue las guías de ondas examinadas en este trabajo de las guías convencionales.
\section{Relaciones constitutivas electromagnéticas: medios magneto-eléctricos}\label{3.4}

Las ecuaciones de la $\theta$-ED se pueden reorganizar del tal forma que la respuesta EM de los materiales con fases topológicas quede encapsulada en ecuaciones constitutivas  que relacionan los parámetros efectivos de los medios con los campo eléctricos y magnéticos. Para entender esto, consideremos las ecuaciones de Maxwell usuales en la materia, 
\begin{align}
\mathbf{\nabla}\cdot\mathbf{D} &=4\pi\rho,  & & & 
\mathbf{\nabla}\times\mathbf{H}&=\frac{4\pi}{c}\mathbf{J}+\frac{1}{c}\frac{\partial}{\partial t}\mathbf{D},\label{eq:MnoHomogeneas}
\end{align}
donde $\mathbf{D}$ es el vector de desplazamiento eléctrico y $\mathbf{H}$ es el vector auxiliar ``H''. Estos campos describen cómo la materia se polariza y/o magnetiza cuando se somete a campos EM externos, y es necesario especificar la relación constitutiva para cada material antes de realizar cálculos.  En general, $\mathbf{D} = \mathbf{E} + 4\pi\mathbf{P}$, tiene en cuenta los efectos de la polarización $\mathbf{P}$ del medio, mientras que $\mathbf{H} = \mathbf{B} - 4\pi\mathbf{M}$, tiene en cuenta los efectos de la magnetización $\mathbf{M}$ del medio. 

La mayoría de los materiales exhiben una respuesta lineal a campos EMs externos, y es lo que consideraremos de aquí en adelante. Las cargas ligadas a los átomos y/o moléculas del material responden al campo eléctrico separándose en direcciones opuestas y alineándolos a lo largo de la dirección del campo, resultando en la polarización del material. Este fenómeno se describe comúnmente mediante la relación $\mathbf{P}_{0}=\chi_{e}\mathbf{E}$, donde $\chi_{e}$ representa la susceptibilidad eléctrica del medio. Además, algunos materiales pueden responder a un campo magnético, donde los dipolos magnéticos atómicos en el material tienden a alinearse con el campo magnético externo. Como resultado, algunos materiales adquieren una magnetización paralela a $\mathbf{B}$ (paramagnetismo), otros una magnetización opuesta a $\mathbf{B}$ (diamagnetismo), y unas pocas sustancias conservan su magnetización incluso después de que se elimina el campo externo (ferromagnetismo). Esta respuesta magnética se describe mediante la relación $\mathbf{M}_{0}=\chi_{m}\mathbf{H}$, donde $\chi_{m}$ representa la susceptibilidad magnética del medio. En general, la respuesta EM de los materiales puede ser más compleja y es necesario considerar que $\chi_{e}$ y $\chi_{m}$, puedan ser tensores de rango 2 para considerar las anisotropías del medio.

En materiales que presentan fases topológicas, se observan contribuciones adicionales a la polarización y magnetización (porque son materiales magnetoeléctricos). Esta contribución adicional, junto con la respuesta usual, se representa en las siguientes relaciones constitutivas, 
\begin{align}
\label{eq:ConstitutiveRelations}
\mathbf{D}&=\epsilon\mathbf{E}+\theta\mathbf{B}, &&&
\mathbf{H}&=\frac{1}{\mu}\mathbf{B}-\theta\mathbf{E},
\end{align}
donde $\epsilon = 1 + 4\pi\chi_{e}$ y $\mu = 1 + 4\pi\chi_{m}$ son la permitividad y permeabilidad relativa del material respectivamente.  Estas ecuaciones dan cuenta de la polarización $\mathbf{P}_{\theta}=\frac{1}{4\pi}\theta\mathbf{B}$, inducida por un campo magnético y la magnetización $\mathbf{M}_{\theta}=\frac{1}{4\pi}\mu\theta\mathbf{E}$, inducida por un campo eléctrico. Estas son consecuencias del efecto TME, ya que un $\mathbf{B}_{\textup{ext}}$ induce un $\mathbf{E}_{\theta}$ y un $\mathbf{E}_{\textup{ext}}$ induce un $\mathbf{B}_{\theta}$\footnote{Uno podría pensar en el siguiente procedimiento $\mathbf{B}_{ext}\to \mathcal{O}(\mathbf{E}_{\theta})\to\mathcal{O}^{2}(\mathbf{B}_{\theta})\to\mathcal{O}^{3}(\mathbf{E}_{\theta})$ y así suponer una divergencia de los campos y simplemente quedarse con los términos lineales en $\theta$ debido a que $\theta\ll 1$. De esta forma, los términos más altos son despreciables. En la sección (\ref{Sec:CargasTopoInducidas}) demostramos explícitamente la convergencia de este procedimiento a todos los órdenes sin aproximar al término lineal.}. El TME es una propiedad fundamental de los TIs tridimensionales, similar a la conductividad de Hall semi-cuantizada. 

Para entender el TME intuitivamente, podemos imaginar un material magnetoeléctrico con $\theta\neq 0$, donde las unidades básicas (los constituyentes de la materia) consisten en dipolos eléctricos permanentes que están conectados por fuerzas no electromagnéticas a dipolos magnéticos permanentes \cite{tellegen1948gyrator}. 
%Esta imagen semi-microscópica, propuesta inicialmente por Tellegen \cite{tellegen1948gyrator}, nos ayuda a comprender cómo funciona el efecto TME.
Cuando se aplica un campo magnético, los dipolos magnéticos giran alineándose al campo externo. Como resultado, los dipolos eléctricos también giran, lo que da como resultado la polarización electrica. Un efecto similar para explicar la magnetización ocurre cuando se expone el material a un campo eléctrico externo. 

En la comunidad de electrodinámica macroscópica, los materiales que responden de acuerdo con las ecuaciones constitutivas como en \eqref{eq:ConstitutiveRelations} se conocen como \textit{medios Tellegen}. Los medios Tellegen y una clase más amplia de materiales bi-anisotrópicos han sido objeto de intensa investigación \cite{serdyukov2001electromagnetics}. Los materiales bi-anisotrópicos son aquellos cuya respuesta es anisotrópica tanto ante un campo eléctrico como ante un campo magnético. Se han propuesto ejemplos de metaátomos que presentan una respuesta Tellegen efectiva \cite{tretyakov2003artificial}, y los metamateriales, que son medios estructurados artificialmente con una EM menor que la longitud de onda, permitiendo el diseño y control de propiedades electromagnéticas avanzadas no convencionales diseñadas \cite{ViktorGVeselago_1968,caloz2005electromagnetic}, también pueden describirse como materiales bi-anisotrópicos.

En el ámbito de la física de la materia condensada, las relaciones constitutivas expresadas por la Ec. \eqref{eq:ConstitutiveRelations} caracterizan diversos tipos de materiales no topológicos que exhiben un fenómeno magnetoeléctrico inherente, como se observa en los materiales multiferroicos \cite{eerenstein2006multiferroic,Pyatakov_2012}. Estos materiales, inicialmente predichos de manera teórica \cite{dzyaloshinskii1960magneto}, fueron posteriormente identificados en la naturaleza, siendo el Cr$_{2}$O$_{3}$ el primer ejemplo, seguido por una amplia variedad de otros compuestos magnetoeléctricos \cite{eerenstein2006multiferroic,Pyatakov_2012}. 
%La comunidad de la materia condensada ha dirigido una atención significativa hacia los materiales multiferroicos, los cuales han encontrado aplicaciones prácticas, como en la tecnología E-Ink utilizada en libros electrónicos. 
A pesar de esto, la respuesta $\theta$ de tales estructuras tiende a ser relativamente tenue $(\theta_{\textup{natural}}\approx 10^{-3} \div 10^{-2})$ y, en algunos casos, requiere bajas temperaturas para manifestarse \cite{Pyatakov_2012}.

Somos conscientes de que las relaciones constitutivas \eqref{eq:ConstitutiveRelations} no abarcan todo el espectro de respuestas EMs. El modelo de los medios bi-anisotrópicos mencionados anteriormente presentan dos efectos magnetoeléctricos físicamente diferentes. La magnetoelectricidad se ha estudiado mucho antes del advenimiento de los TIs \cite{dzyaloshinskii_1960} y es mucho más general que las respuestas magnetoeléctricas de los TIs, ver, por ejemplo, \cite{lindell_1994_electromagnetic}. Para los medios lineales, las respuestas magnetoeléctricas se pueden codificar en relaciones constitutivas modificadas del tipo
\begin{align}
\mathbf{D}&=\epsilon\mathbf{E}+(\chi+i\kappa)\mathbf{B}, & \mathbf{H}&=\frac{1}{\mu}\mathbf{B}-(\chi-i\kappa)\mathbf{E}. \label{eq:GeneralContitutiveRelation}
\end{align}
Para los medios bi-anisotrópicos generales, todos los parámetros ópticos, $\epsilon$, $\mu$, $\chi$ y $\kappa$, son tensores de rango 2 con elementos fuera de la diagonal. Los términos $ \chi \mp i \kappa $ son responsables de la magnetoelectricidad. Si $ \chi $ y $ \kappa $ desaparecen, no hay respuesta magnetoeléctrica, pero es suficiente que uno de ellos sea diferente de cero para tener un efecto magnetoeléctrico. 

En \cite{lindell_1994_electromagnetic}, el parámetro $\kappa$ se define para tener en cuenta la quiralidad, la cual se refiere a la propiedad de un material que presenta respuestas diferentes para OEMs de polarización circular izquierda y derecha, debido a su asimetría estructural. Por otro lado, el parámetro $\chi$ se utiliza para describir la no reciprocidad, una característica de los medios que responden de manera diferente a ondas que viajan en direcciones opuestas, rompiendo así la TRS. Los medios para los cuales $\kappa = 0$ y $\chi \neq 0$ se conocen como medios no quirales y no recíprocos (o Tellegen), mientras que aquellos con $\chi = 0$ y $\kappa \neq 0$ se conocen como medios quirales y recíprocos (o Pasteur). Si ambos parámetros son cero ($\chi = 0$ y $\kappa = 0$), el medio se clasifica como no quiral y recíproco. En el caso de que ambos parámetros sean distintos de cero ($\chi \neq 0$ y $\kappa \neq 0$), el medio se denomina medio bi-anisotrópico general. 

En \cite{serdyukov_1996} se muestra que los parámetros individuales $\chi$ y $\kappa$ dan lugar a respuestas magnetoeléctricas intrínsecamente diferentes; es decir, los efectos asociados a quiralidad y no reciprocidad no se pueden transformar uno en otro mediante transformaciones de dualidad. Estas transformaciones intercambian los campos eléctricos y magnéticos en las ecuaciones de Maxwell, pero solo son aplicables en medios dual-simétricos donde las respuestas son invariantes bajo tal intercambio. Esta distinción es relevante porque la quiralidad, aunque importante en la electrodinámica, no será considerada en esta tesis. Por ejemplo, en \cite{PhysRevB.79.035407} se discute una aplicación de la quiralidad como efecto magnetoeléctrico utilizando OEMs en meta-materiales.

En resumen, si algún material se describe mediante las relaciones constitutivas de la Ec. \eqref{eq:ConstitutiveRelations}, sus propiedades EMs son capturadas con precisión por las ecuaciones de la $\theta$-ED \cite{shaposhnikov2023emergent}. Aun así, medir la respuesta magnetoeléctrica de materiales con fases topológicas no es lo mismo que medir la magnetoelectricidad de materiales que no son topológicos. Para materiales topológicos estamos accediendo a estudiar cómo las corrientes robustas superficiales responden a campos EM externos.

%Campos EM en funcion de los potenciales debido a las ecuaciones homogeneas. 

%Energia EM (Ver mejor en OEM)

\biblio %Se necesita para referenciar cuando se compilan subarchivos individuales - NO SACAR

%% file: Capitulos/04Waveguide.tex
El estudio y la caracterización de las propiedades electrónicas que emergen en las fases topológicas de los materiales son fundamentales tanto para la teoría de campos de alta energía como para la física de la materia condensada. Este enfoque es especialmente relevante debido a las posibles aplicaciones en áreas como la espintrónica y la computación cuántica topológica \cite{Hasan2010Colloquium,mellnik_2014_spin}. Un aspecto clave de este estudio es comprender el TME, el cual es esencial para revelar las propiedades y el comportamiento de los estados superficiales de los TIs. Medir el TME permite detectar directamente la cuantización de las respuestas EMs en la superficie, lo cual es una firma inequívoca de los estados topológicos protegidos por la simetría en los TIs.

La medición precisa del efecto TME, sin embargo, presenta desafíos significativos debido a que la respuesta EM se ve suprimida por la constante de estructura fina $\alpha\approx 1/137$, la cual es tres órdenes de magnitud más pequeña que los valores típicos de la permitividad $\epsilon$. 
La propagación de ondas electromagnéticas (OEM) en medios confinados, como guías de ondas o fibras ópticas, se fundamenta en las propiedades de reflexión del campo EM en las interfaces, lo que la convierte en un escenario idóneo para estudiar la respuesta del efecto TME en los TIs, dada su fuerte dependencia de las BCs.

Muchas de las tecnologías contemporáneas aprovechan el hecho de que el campo EM puede confinarse y propagarse de manera controlada de un punto del espacio a otro. Esta tecnología, conocida como \textit{guided-wave optics}, inicialmente desarrollada para transmitir luz a largas distancias sin la necesidad de lentes de retransmisión, ha evolucionado para desempeñar un papel crucial en diversas aplicaciones fundamentales y prácticas, como la comunicación por fibra óptica, imágenes biomédicas de difícil acceso y la interconexión de componentes en sistemas ópticos y optoelectrónicos miniaturizados. En este contexto, la \textit{integrated photonics} emerge como una disciplina que fusiona varios dispositivos y componentes ópticos en un único sustrato, conocido como ``chip''. Estos dispositivos son esenciales para la generación, enfoque, división, combinación, aislamiento, polarización, acoplamiento, conmutación, modulación y detección controlada de la luz. De manera crucial, las \textit{optical waveguide} actúan como conexiones entre estos componentes, creando así circuitos integrados fotónicos, análogos ópticos de los circuitos integrados electrónicos. La fotónica integrada permite la miniaturización de la óptica, de manera similar a cómo los circuitos integrados de silicio han revolucionado la miniaturización en la electrónica. 

En este escenario tecnológico y teórico de las guías de ondas, surge la necesidad de explorar nuevos materiales que puedan mejorar y expandir sus capacidades. Aquí es donde los TIs juegan un papel fundamental. Debido a sus propiedades electrónicas únicas y a la robustez de sus estados superficiales protegidos por la simetría, los TIs presentan una respuesta EM excepcional que puede ser descrita de manera efectiva mediante la $\theta$-ED. Este formalismo no solo proporciona un marco teórico para entender la interacción de los TIs con campos EMs, sino que también abre la puerta a innovadoras aplicaciones en fotónica integrada. Mi contribución en esta área se centra en el estudio de las respuestas EMs en guías de ondas formadas por TIs, utilizando la $\theta$-ED para investigar cómo estas propiedades pueden ser aprovechadas para desarrollar nuevas tecnologías ópticas y optoelectrónicas. Este enfoque permite no solo el confinamiento y la propagación de la luz en las guías de ondas topológicas, sino también la posibilidad de manipular y controlar las propiedades de la luz de maneras que no son posibles con materiales convencionales.

% Además de la aplicaciones tecnológicas, las guías de ondas tienen diversas aplicaciones científicas que abarcan el estudio de propiedades electromagnéticas, la investigación en óptica, la simulación de condiciones específicas en entornos controlados, la exploración de fenómenos cuánticos, la caracterización de materiales, el estudio de fenómenos EMs avanzados, y la realización de pruebas de teorías de campo, incluyendo aquellas relacionadas con la ruptura de la simetría de Lorentz. Estas herramientas se utilizan para investigar y comprender desde la dispersión y reflexión de ondas hasta fenómenos cuánticos y avanzados de la teoría electromagnética. Su versatilidad contribuye a diversas disciplinas, incluyendo nanotecnología y campos cuánticos {\color{red}Citar documentos con aplicaciones científicas y tecnologías}.

\begin{SCfigure}[1][t]
\caption{En una geometría plana (\textit{slab}) con índices de refracción $n_{1} > n_{2}$, un rayo de luz que incide con un ángulo $\theta$ puede ser guiado por las paredes de la guía de ondas. Si el ángulo de incidencia cumple con la condición $\theta < \cos^{-1}(n_1/n_2)$, el rayo es confinado mediante reflexiones internas totales. Sin embargo, si $\theta \ge \cos^{-1}(n_1/n_2)$, los rayos no pueden ser confinados y se irradian fuera de la guía de ondas.}
\includegraphics[width=0.53\linewidth]{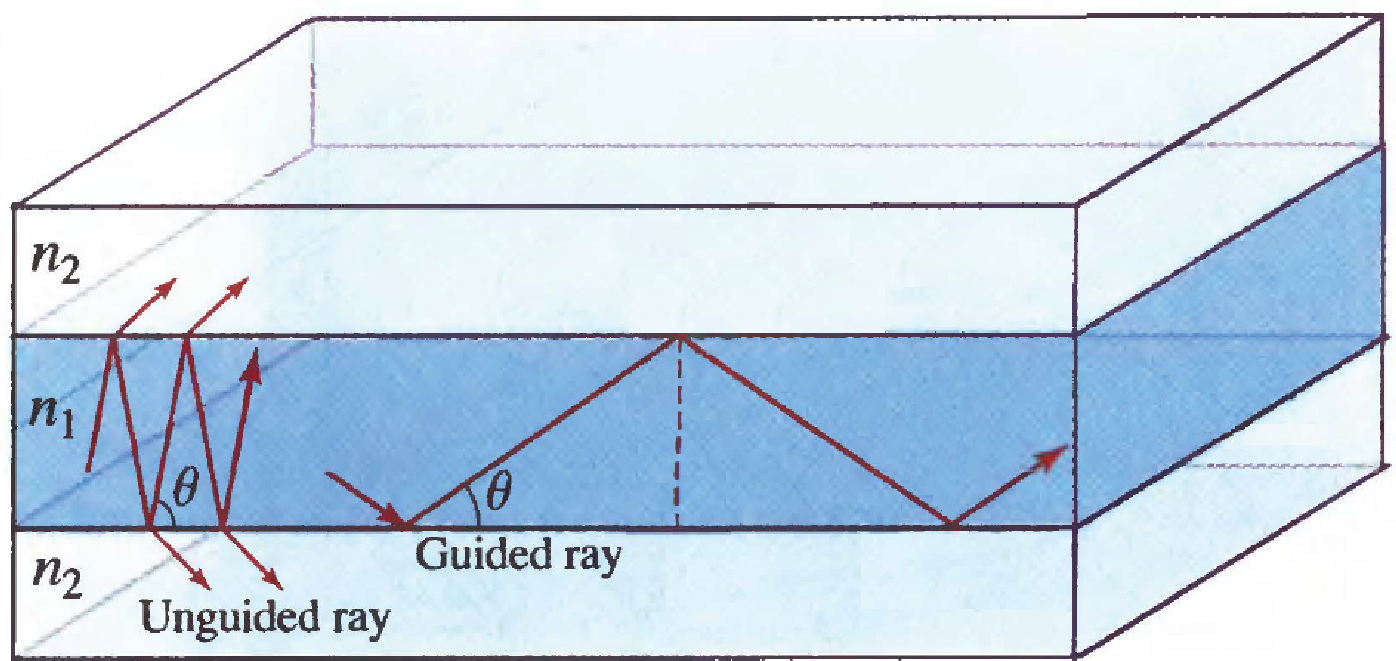}
\label{fig:IRtotal}
\end{SCfigure}

El principio subyacente del confinamiento óptico es sencillo. Un medio de índice de refracción $n_{1}$, incrustado en un medio de índice de refracción más bajo $n_{1}>n_{2}$, actúa como una ``trampa'' de luz dentro de la cual los rayos ópticos permanecen confinados por múltiples reflexiones internas totales en las interfaces, ver Fig. (\ref{fig:IRtotal}). 
%---------------Figure------------
%---------------End Figure---------
En la Sec. (\ref{4.1}), argumentaremos que los materiales deben cumplir con el principio de causalidad cuando el campo EM cambia en el tiempo dentro de ellos. Esta hipótesis nos lleva a formular las ecuaciones de la $\theta$-ED en el dominio de la frecuencia, donde tanto los campos como los parámetros topo-ópticos dependen de la frecuencia $\omega$ del campo EM. Revisaremos brevemente el modelo de Lorentz para un dieléctrico sometido a campos oscilantes y veremos cómo cambia la permitividad en función de la frecuencia. En la Sec. (\ref{4.2}), discutiremos porque el parámetro topológico $\theta$ adquiere una dependencia en la frecuencia y presenta tanto una parte real como una imaginaria, similar a la permitividad. Esta sección plantea una pregunta clave: ¿por qué la parte real de la conductividad de Hall contribuye a la energía reactiva y la parte imaginaria a la energía disipada? En la Sec. (\ref{4.3}), analizaremos la ley de conservación de la energía EM en sistemas donde el campo EM oscila a una frecuencia $\omega$. Observaremos términos adicionales en las ecuaciones que describen la energía disipada y reactiva, los cuales son fundamentales para responder la pregunta anterior. Esta ley de conservación de la energía EM modificada en el dominio de la frecuencia constituye la primera contribución relevante de esta tesis. Concluiremos que en medios donde la OEM tiene una polarización circular o elíptica, hay un intercambio de energía con el medio topológico, mientras que las OEM con polarización lineal no lo hacen. Finalmente, en la Sec. (\ref{4.4}), realizaremos la descomposición longitudinal-transversal de la electrodinámica $\theta$-ED, que servirá como punto de partida para los cálculos del siguiente capítulo. En esta última sección, describiremos algunos cálculos realizados por otros autores y delinearemos nuestras propias contribuciones para abordar las preguntas que han quedado abiertas.
\section{Ecuaciones de la $\theta$-ED causales en el espacio de frecuencia} \label{4.1}
Consideremos un sistema electrodinámico en (3+1)D formado por $N$ materiales distintos, cada uno con propiedades ópticas que definen el material. Para mantener la generalidad, asumiremos que todos los medios satisfacen las siguientes relaciones constitutivas,
\begin{align}
    \mathbf{D}&=\epsilon\mathbf{E}+\theta\mathbf{B}, & \textup{y,}& & \mathbf{H}&=\frac{1}{\mu}\mathbf{B}-\theta\mathbf{E}.
\end{align}

Más adelante, consideraremos casos específicos, como $\theta_{\textup{aire}}=0$ o $\theta_{\textup{Bi}{2}\textup{Se}{3}}=\pi$. Además, está bien establecido fenomenológicamente que la densidad de corriente en muchos materiales obedece la \textit{ley de Ohm}, $\mathbf{J}=\sigma\mathbf{E}$, donde $\sigma$ es la conductividad longitudinal del material. Asumimos que todos los medios que constituyen el sistema son totalmente isotrópicos. En este sentido, las propiedades como la permitividad $\epsilon_{ij}=\epsilon(\mathbf{r})\delta_{ij}$, permeabilidad $\mu_{ij}=\mu(\mathbf{r})\delta_{ij}$, conductividad $\sigma_{ij}=\sigma(\mathbf{r})\delta_{ij}$, y TMEP $\theta_{ij}=\theta(\mathbf{r},t) \delta_{ij}$ son todas funciones escalares.

Las ecuaciones de Maxwell se simplifican considerablemente en el caso de dependencia armónica del tiempo. A través de la transformada inversa de Fourier, las soluciones generales de las ecuaciones de Maxwell se pueden construir haciendo uso del principio de superposición, es decir, como una combinaciones lineal de las soluciones de una sola frecuencia,
\begin{align}\label{TransFourier}
    \mathbf{E}(\mathbf{r},t)=\frac{1}{2\pi}\int_{-\infty}^{\infty}\mathbf{E}(\mathbf{r};\omega)e^{-i\omega t}d\omega,
\end{align}
donde,
\begin{align}\label{eq;TransFourier}
    \mathbf{E}(\mathbf{r};\omega)=\int_{-\infty}^{\infty}\mathbf{E}(\mathbf{r},t)e^{i\omega t}dt
\end{align}
es la trasformada de Fourier de $\mathbf{E}(\mathbf{r},t)$. Por lo tanto, asumimos que todos los campos dependen del tiempo de forma armónica $e^{-i\omega t}$,
\begin{align}
    \mathbf{E}(\mathbf{r},t)&=\mathbf{E}(\mathbf{r;\omega})e^{-i\omega t}, & \mathbf{B}(\mathbf{r},t)&=\mathbf{B}(\mathbf{r};\omega)e^{-i\omega t}
\end{align}
donde las amplitudes fasoriales $\mathbf{E}(\mathbf{r;\omega})$, $\mathbf{B}(\mathbf{r;\omega})$ tienen valores complejos. Recordemos que el campo y todas las cantidades físicas corresponden a la parte real de tal cantidad, \textit{i.e.}, $\textup{Re}[\mathbf{E}(\mathbf{r},t)]$.

Consideraremos un material en el que el campo EM oscila con una frecuencia $\omega$. En este caso, el comportamiento del material difiere notablemente de su respuesta a un campo EM estático. Específicamente, las propiedades ópticas del material, como la permitividad dieléctrica y la permeabilidad magnética, dependen de la frecuencia del campo EM aplicado. Esta dependencia se debe a que el material no puede responder de manera instantánea a perturbaciones externas, lo que implica la necesidad de satisfacer la causalidad. 

Un ejemplo clásico que revisaremos es el modelo de Lorentz, que describe la permitividad efectiva de un dieléctrico considerando el movimiento de un electrón ligado en presencia de un campo eléctrico aplicado. Este modelo simple para la dinámica del desplazamiento $\mathbf{r}$ del electrón ligado se expresa mediante la siguiente ecuación de movimiento:
\begin{align}\label{eq:dielectricmodel1}
    m\ddot{\mathbf{r}}&=-e\mathbf{E}(t)-k\mathbf{r}-m\gamma\dot{\mathbf{r}}, &\rightarrow& & \mathbf{r}&=\frac{q}{m}\frac{1}{(\omega_{0}^{2}-\omega^2-i\omega\gamma)}\mathbf{E},
\end{align}
donde se asume un campo eléctrico oscilante $\mathbf{E}(t)=\mathbf{E}e^{-i\omega t}$ con frecuencia $\omega$, que induce una respuesta del electrón a la misma frecuencia, $\mathbf{r}(t)=\mathbf{r}e^{-i\omega t}$. En esta ecuación, la fuerza de restauración $-k\mathbf{r}$ se debe al enlace del electrón con el núcleo, similar a un resorte, y está caracterizada por la constante $k$, donde 
$\omega_0^2=k/m$  es la frecuencia de resonancia del oscilador. También se incluye una fuerza de fricción proporcional a la velocidad del electrón, representada por el término de amortiguamiento $-m\gamma\dot{\mathbf{r}}$, donde $\gamma$ es la tasa de colisiones por unidad de tiempo.

El desplazamiento $\mathbf{r}$ induce un momento dipolar $\mathbf{p}=-e\mathbf{r}$, lo que a su vez genera una polarización $\mathbf{P}=N\mathbf{p}$, donde $N$ es la densidad de electrones. La relación entre el desplazamiento eléctrico $\mathbf{D}$y el campo eléctrico $\mathbf{E}$ esta dada por,
\begin{align}
    \mathbf{D}&=\mathbf{E}+4\pi\mathbf{P}, &\rightarrow& & \mathbf{D}=&\epsilon(\omega)\mathbf{E}
\end{align}
donde la permitividad dependiente de la frecuencia se define como,
\begin{align}\label{eq:dielectricmodel2}
    \epsilon(\omega)=1+\frac{\omega_{p}^{2}}{(\omega_{0}^{2}-\omega^2-i\omega\gamma)},
\end{align}
y $\omega_{p}^{2}=4\pi Ne^2/m$, es la frecuencia de plasma del material. Las partes real e imaginaria de $\epsilon(\omega)=\epsilon'(\omega)+i\epsilon''(\omega)$ caracterizan las propiedades de refracción y absorción del material. De la Ec. \eqref{eq:dielectricmodel2} se encuentra que,
\begin{align}\label{eq:dielectricmodel3}
    \epsilon'(\omega)&=1+\frac{\omega_{p}^{2}(\omega_{0}^{2}-\omega^{2})}{(\omega^{2}-\omega_{0}^{2})^{2}+\gamma^{2}\omega^{2}}, & \textup{y,}&& \epsilon''(\omega)&=\frac{\omega_{p}^{2}\omega\gamma}{(\omega^{2}-\omega_{0}^{2})^{2}+\gamma^{2}\omega^{2}}
\end{align}
\begin{figure}[t]
    \centering
    \includegraphics[width=0.95\linewidth]{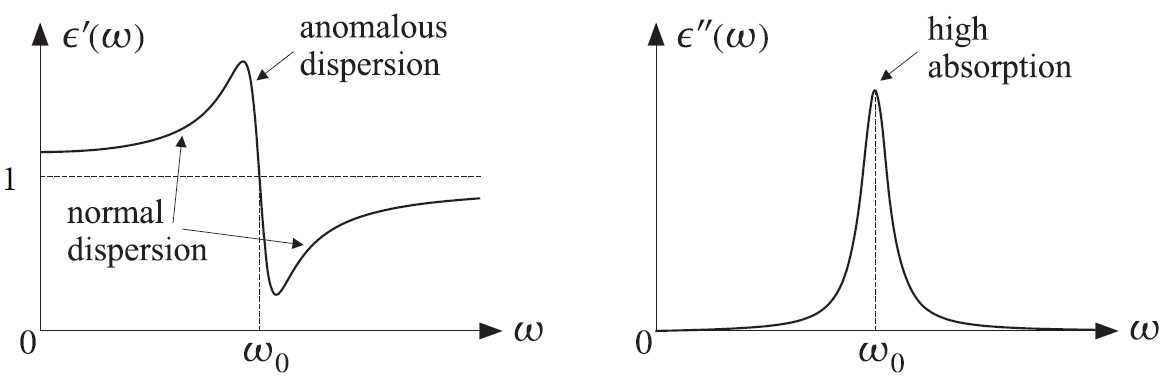}
    \caption{Partes reales e imaginarias de la permitividad efectiva $\epsilon(\omega)=\epsilon'(\omega)+i\epsilon''(\omega)$, donde $\epsilon'(\omega)$ y $\epsilon''(\omega)$ están definidas en la Ec. \eqref{eq:dielectricmodel3}. Imagen de \cite{Orfanidis2016electromagnetic}.}
    \label{fig:ReImepsilon}
\end{figure}
La Fig. (\ref{fig:ReImepsilon}) muestra un gráfico de $\epsilon'(\omega)$ y $\epsilon''(\omega)$. Alrededor de la frecuencia de resonancia $\omega_0$, la parte real $\epsilon'(\omega)$ se comporta de manera anómala, es decir, cae rápidamente con la frecuencia a valores inferiores a $1$ y el material presenta una fuerte absorción. El término ``dispersión normal'' se refiere a una $\epsilon'(\omega)$ que es una función creciente de $\omega$, como es el caso en los extremos izquierdo y derecho de la frecuencia resonante.

% {\color{red} Citas para el modelo de la permitividad para los TT: Jenkins PRB 82 (2010), Valdés Aguilar PRL 108 (2012), Sharma Materials Science and Engineering: B 272 (2021), Wenjie Nie PRB 88 (2013), Grushin \& Cortijo, PRL 106 (2011).}

Los materiales dieléctricos reales exhiben, por supuesto, varias frecuencias resonantes correspondientes a varios modos de vibración y mecanismos de polarización (por ejemplo, electrónicos, iónicos, etc.). Además, es necesario considerar correcciones cuánticas, donde las frecuencias de oscilación características $\omega_0$ se reemplazan por las frecuencias de transición de los átomos en el medio (frecuencias de resonancia) $\omega_{ij}=(E_{i}-E_{j})/\hbar$, que dependen de las diferencias de energía entre los niveles electrónicos. Además, la densidad de electrones $N$ se ajusta con las poblaciones de los niveles de energía, reemplazándola por $f_{ij}(N_{i}-N_{j})$, donde $f_{ij}$ es conocido como la "fuerza del oscilador". Finalmente, el término de amortiguamiento $\gamma$ se ajusta a $\gamma_{ij}$, que puede variar dependiendo de las transiciones entre diferentes niveles de energía. Así la permitividad cambia a \cite{Orfanidis2016electromagnetic}, 
\begin{align}
    \epsilon(\omega)=1+\frac{4\pi e^2}{m}\sum_{j>i}\frac{f_{ij}(N_i-N_j)}{(\omega_{ij}^{2}-\omega^2-i\omega\gamma_{ij})}.
\end{align}
Los medios en los que estas propiedades, como la permitividad dieléctrica y la permeabilidad magnética, varían con la frecuencia se denominan \textit{dispersivos}. Esta respuesta dinámica y causal puede describirse mediante el \textit{Teorema de Convolución}, el cual debe aplicarse a todas las relaciones constitutivas del material (incluida la ley de Ohm),  \cite{Toll1956causality},
\begin{align}\label{eq:causal1}
    \mathbf{D}(\mathbf{r},t)&=\int_{-\infty}^{\infty}\left [ \epsilon(\mathbf{r},t-t')\mathbf{E}(\mathbf{r},t')+\theta(\mathbf{r},t-t')\mathbf{B}(\mathbf{r},t') \right ]dt',\\\label{eq:causal2}
    \mathbf{H}(\mathbf{r},t)&=\int_{-\infty}^{\infty}\left [ \mu(\mathbf{r},t-t')^{-1}\mathbf{B}(\mathbf{r},t')-\theta(\mathbf{r},t-t')\mathbf{E}(\mathbf{r},t') \right ]dt',\\\label{eq:causal3}
\mathbf{J}(\mathbf{r},t)&=\int_{-\infty}^{\infty}\sigma(\mathbf{r},t-t')\mathbf{E}(\mathbf{r},t')dt'.
\end{align}
La forma convolucional de las Ecs. \eqref{eq:causal1}-\eqref{eq:causal3} implica causalidad, es decir que el valor de $\mathbf{D}(\mathbf{r},t)$ en el tiempo $t$ depende únicamente de los valores pasados de $\mathbf{E}(\mathbf{r},t')$, $t'\leq t$, así con todos los parámetros del material. Esta condición es equivalente a requerir que la respuesta dieléctrica $\epsilon(t)$ sea una función del tiempo tal que, $\epsilon(t)=0$ para $t<0$. Al reemplazar la transformada de Fourier inversa a todas las funciones de las Ecs. \eqref{eq:causal1}-\eqref{eq:causal3}, se obtiene una multiplicación en el dominio de la frecuencia,
\begin{align}\label{eq:Constitutiveomega1}
    \mathbf{D}(\mathbf{r},t)&=\frac{1}{2\pi}\int_{-\infty}^{\infty}\left [\epsilon(\mathbf{r};\omega)\mathbf{E}(\mathbf{r};\omega)+\theta(\mathbf{r};\omega)\mathbf{B}(\mathbf{r};\omega)\right ]e^{-i\omega t}d\omega,\\\label{eq:Constitutiveomega2}
    \mathbf{H}(\mathbf{r},t)&=\frac{1}{2\pi}\int_{-\infty}^{\infty}\left [\mu(\mathbf{r};\omega)^{-1}\mathbf{B}(\mathbf{r};\omega)-\theta(\mathbf{r};\omega)\mathbf{E}(\mathbf{r};\omega)\right ]e^{-i\omega t}d\omega,\\\label{eq:Constitutiveomega3}
    \mathbf{J}(\mathbf{r},t)&=\frac{1}{2\pi}\int_{-\infty}^{\infty}\sigma(\mathbf{r};\omega)\mathbf{E}(\mathbf{r};\omega)e^{-i\omega t}d\omega.
\end{align}
% \begin{align}
%     \Upsilon(\mathbf{r},t)&=\frac{1}{2\pi}\int_{-\infty}^{\infty}\Upsilon(\mathbf{r};\omega)e^{-i\omega t}d\omega, &\textup{y}& & \Upsilon(\mathbf{r};\omega)&=\int_{-\infty}^{\infty}\Upsilon(\mathbf{r},t)e^{i\omega t}dt,
% \end{align}
% %
% con $\Upsilon = (\epsilon,~\mu^{-1},~\theta)$. 
Equivalentemente, una multiplicación en el espacio de frecuencia implica causalidad en el espacio temporal. Remplazando las Ecs. \eqref{eq:Constitutiveomega1}-\eqref{eq:Constitutiveomega3} en las ecuaciones de la $\theta$-ED \eqref{eq:MCS1}-\eqref{eq:MCS4} sin fuentes libres ($\rho=0$) obtenemos \cite{Qi2008topological,Obukhov2005piecewise},
\begin{align}
\mathbf{\nabla}\cdot (\epsilon \mathbf{E}) &=-\boldsymbol{\nabla}\theta \cdot \mathbf{B},  \label{eq:MCS1time}\\ 
\mathbf{\nabla}\times  \mathbf{E}&=ik_{0}\mathbf{B},\label{eq:MCS2time}\\
\mathbf{\nabla}\cdot \mathbf{B}&=0,\label{eq:MCS3time}\\ 
\mathbf{\nabla}\times (\frac{1}{\mu}\mathbf{B})&=-ik_{0}\epsilon \mathbf{E}+\boldsymbol{\nabla}\theta \times \mathbf{E},\label{eq:MCS4time}
\end{align}
donde $k_{0}=\omega/c$ corresponde a la relación de dispersión de una OEM en el vacío y $\epsilon\equiv\epsilon(\omega)+ i\frac{4\pi}{\omega}\sigma(\omega)$. Hemos suprimido la dependencia espacial y de la frecuencia por simplicidad.

\section{Respuesta electromagnética de una superficie metálica topológica}\label{4.2}
Notemos la introducción de una dependencia de la frecuencia angular $\omega$ en el término axiónico $\theta=\theta(\mathbf{r};\omega)$. Aunque inicialmente hemos considerado que $\theta$ para un TI es constante por partes y no depende del tiempo, recientemente se ha explorado la respuesta de la conductividad Hall y longitudinal en la superficie de un TI en función de la frecuencia \cite{PhysRevLett.97.106804,tse_giant_2010,PhysRevB.87.205424,PhysRevB.88.045442}, revelando comportamientos no triviales de estas cantidades físicas. Esto implica la existencia de un $\theta_{\textup{TI}}(t)$, relacionado con la respuesta causal de los estados metálicos en la superficie del TI, sin influir inicialmente en las ecuaciones de Maxwell planteadas.

En \cite{tse_giant_2010}, 
%se aborda la respuesta EM de la superficie de un TI cuando la TRS se rompe débilmente abriendo una brecha energética $\Delta$, dando lugar a efectos magnetoeléctricos intensos en los estados de superficie \cite{qi2008topological,PhysRevLett.102.146805}. 
se evaluó la respuesta eléctrica de la superficie del TI utilizando una ecuación cinética cuántica linealizada sin colisiones. Se descubrió que el campo eléctrico oscilante induce corrientes en las superficies del TI,
\begin{align}
    K_{i}&=\sigma_{ij}^{\textup{sup}}E_{j}, & \textup{con}~~~~&\{i,j\}=\{x,y\},
\end{align}
donde $\mathbf{K}$ es la corriente superficial inducida en la superficie del TI y $\sigma_{ij}^{\textup{sup}}$ es la conductividad superficial tensorial. Al considerar un campo eléctrico oscilando con frecuencia $\omega$ en la dirección $x$, se obtiene $J_{x}=\sigma_{xx}(\omega)E_{x}$ y $J_{y}=\sigma_{xy}(\omega)E_{x}$. Esto significa que el campo eléctrico induce una respuesta tanto longitudinal como de Hall en las superficies del TI, ambas con partes real e imaginaria debido a la dependencia en $\omega$,
\begin{align}
 \sigma_{xx}(\omega)&=\sigma_{xx}'+i\sigma_{xx}'' & \sigma_{xy}(\omega)&=\sigma_{xy}'+i\sigma_{xy}''
\end{align}
%frecuencias de transición de los átomos en el medio

Las partes imaginarias están relacionadas con comportamientos anómalos de las cargas cuando oscilan cerca de las frecuencias de resonancia $\omega_{ij}$, similar a como se expresa $\epsilon(\omega)=\epsilon'+i\epsilon''$, ver Fig. \ref{fig:ReImepsilon}. 

En el estudio \cite{tse_giant_2010}, se menciona que $\sigma_{xx}'$ y $\sigma_{xy}''$ son las componentes disipativas, mientras que $\sigma_{xx}''$ y $\sigma_{xy}'$ son las componentes reactivas. Las componentes disipativas representan la capacidad del material para absorber energía EM y convertirla en calor, lo que se traduce en pérdidas por efecto Joule. Por otro lado, las componentes reactivas describen cómo el material almacena y devuelve energía EM sin disiparla en forma de calor. 

Aunque varios estudios \cite{PhysRevLett.97.106804,tse_giant_2010,PhysRevB.87.205424,PhysRevB.88.045442} no explican por qué la parte real de la conductividad de Hall se considera reactiva y la parte imaginaria disipativa, en esta tesis se presenta un análisis que intenta esclarecer esta clasificación utilizando cálculos clásicos. La conclusión derivada de este análisis sugiere que hay contribuciones adicionales a la energía electromagnética cuando un campo EM oscila a una frecuencia $\omega$. Este análisis, presentado en la siguiente sección, es el primer resultado novedoso que mostraré de la tesis.

% que se propaga en medios confinados, lo cual permite la posibilidad de configurar el sistema para que la energía se propague a mayor distancia con menores pérdidas por efecto Joule.

En el límite estático $\omega\rightarrow 0$, los artículos de la literatura mencionan que $\sigma_{xx}'(0)=\sigma_{xx}''(0)=\sigma_{xy}''(0)=0$ y $\sigma_{xy}'(0)=(\alpha c/4\pi)(\tilde{\theta}_{\textup{TI}}/\pi)$, que corresponde al límite correcto y concuerda con la ecuación \eqref{eq:corrientetheta}. Bajo estas consideraciones, las BCs para la componente paralela del campo magnético experimenta modificaciones a causa de las corrientes en la superficie \cite{PhysRevB.88.045442},
\begin{align}
\Delta \left[\mu^{-1}\mathbf{B}_{\parallel} \right]|_{\Sigma_{i} }=\tilde{\theta}_{i}(\omega)\mathbf{E}_{\parallel}|_{\Sigma_{i}}+\frac{4\pi}{c}\sigma^{\textup{sup}}(\omega)\mathbf{E}_{\parallel}|_{\Sigma_{i}}\times \hat{\mathbf{n}}.
\label{EQ:BC_Bdiscontinua}
\end{align}
%
%\textcolor{red}{Este término $\sigma^{sup}$ puede esconderse en los $\epsilon(\omega)?$}
En el lado derecho de la expresión, el primer término está asociado con la corriente de Hall, mientras que el segundo término se relaciona con la corriente longitudinal. Si hacemos oscilar el campo EM a una frecuencia $\nu$ tal que la energía no supere la brecha energética de los estados de superficie, el término $\sigma^{\textup{sup}}(\omega)=0$, recuperando la BC de la Ec. \eqref{EQ:BCsdiscontinuas} con la diferencia de que $\tilde{\theta}_{i}$ pueda tener parte imaginaria. Este hecho sera relevante para los resultados obtenidos en la Sec. (\ref{5.6}).
%
% {\color{blue} No estoy seguro de poner explicitamente las ecuaciones de de $\sigma_{ij}(\omega)$ mostradas en Tse y MacDonald porque no estoy seguro si es la respuesta que todos los TIs tienen con la frecuencia o es un caso muy particular de un cierto material. O incluso la respuesta depende del modelo (obviamente asumiendo linealidad)}
%
\section{Energía electromagnética disipada y reactiva en un sistema con aislantes topológicos} \label{4.3}

En esta sección, exploraremos cómo los TIs modifican el teorema de conservación de la energía en sistemas electromagnéticos donde los campos oscilan de forma armónica, introduciendo nuevos términos asociados con la disipación y la reactividad de la energía electromagnética. La discusión siguiente presenta un enfoque que, hasta donde sabemos, no ha sido tratado de manera similar en la literatura, representando así un primer resultado de esta tesis.

El balance energético en un sistema electrodinámico donde los campos eléctricos y magnéticos cambian a través del espacio y del tiempo de manera arbitraria está dado por el teorema de Poynting,
\begin{align}\label{eq:TeoremaPoynting}
\partial_{t}u_{\textup{EM}}+\boldsymbol{\nabla}\cdot\mathbf{S}=-\mathbf{J}\cdot\mathbf{E}
\end{align}
donde $u_{\textup{EM}}$ corresponde a la densidad de energía EM y  $\mathbf{S}$ es el flujo de energía o vector de Poynting, definidos como sigue,
\begin{align}
\partial_{t}u_{\textup{EM}}&=\frac{1}{4\pi}(\mathbf{E}\cdot\partial_{t}\mathbf{D}+\mathbf{H}\cdot\partial_{t}\mathbf{B}),&&&
\mathbf{S}&=\frac{c}{4\pi}(\mathbf{E}\times\mathbf{H}).
\end{align}
La cantidad $\mathbf{J}\cdot\mathbf{E}$ representa las pérdidas Óhmicas, es decir, la potencia por unidad de volumen perdida en forma de calor debido a los campos. La ley de conservación de la energía EM se obtiene al integrar el teorema de Poynting sobre un volumen  $V$ encerrado por una superficie $\partial V$, y haciendo uso del teorema de Gauss para transformar el término de divergencia en una integral de superficie. En un sistema con materiales topológicos y no topológicos sin dispersión, es decir, cuando $\epsilon=\epsilon(\mathbf{r})$, $\mu=\mu(\mathbf{r})$ y $\theta=\theta(\mathbf{r})$ son constantes a pedazos, reales e isotrópicos, el teorema de Poynting está dado por \eqref{eq:TeoremaPoynting}, donde,
\begin{align}
u_{\textup{EM}}^{\textup{TI}}&=\frac{1}{8\pi}\left(\epsilon\mathbf{E}^{2}+\frac{1}{\mu}\mathbf{B}^{2}\right), &&\textup{y},&
\mathbf{S}^{\textup{TI}}&=\frac{c}{4\pi}\left(\frac{1}{\mu}\mathbf{E}\times\mathbf{B}\right).
\end{align}
Por lo tanto, en un sistema sin dispersión donde hay presente un TI, la ley de conservación de la energía EM es igual a la de un medio no topológico, como se reportó en \cite{martin2015green}. 

Los campos monocromáticos, que oscilan a una frecuencia angular $\omega$ en la materia, pierden energía irreversiblemente si la permitividad $\epsilon(\omega)$ o permeabilidad $\mu(\omega)$ de la materia huésped tiene una parte imaginaria, es decir, en presencia de dispersión tenemos una disipación de energía. La pregunta natural es: ¿La dispersión de $\theta(\omega)$ es responsable de añadir un nuevo término en la ley de conservación de energía, a diferencia de un sistema sin dispersión? Recordemos que para campos que dependen arbitrariamente del tiempo, pueden tratarse como una superposición de campos monocromáticos mediante el análisis de Fourier con la respectiva modulación. Por lo tanto, nos enfocaremos de ahora en adelante en campos monocromáticos. 

En una representación fasorial monocromática, los campos y fuentes dependen del tiempo como $e^{-i\omega t}$, por lo que escribimos,
\begin{align}
    \mathbf{E}(\mathbf{r},t)=\textup{Re}[\mathbf{E}(\mathbf{r})e^{-i\omega t }]=\frac{1}{2}[\mathbf{E}(\mathbf{r})e^{-i\omega t }+\mathbf{E}^{*}(\mathbf{r})e^{i\omega t }]
\end{align}
El campo $\mathbf{E}(\mathbf{r})$ es en general complejo, con magnitud y fase que cambian con la posición. Para productos escalares, como $\mathbf{J}(\mathbf{r},t)\cdot\mathbf{E}(\mathbf{r},t)$, tenemos,
\begin{align}
    \mathbf{J}(\mathbf{r},t)\cdot\mathbf{E}(\mathbf{r},t)=\frac{1}{2}\textup{Re}[\mathbf{J}^{*}(\mathbf{r})\cdot\mathbf{E}(\mathbf{r})+\mathbf{J}(\mathbf{r})\cdot\mathbf{E}(\mathbf{r})e^{-2i\omega t}].
\end{align}
Este producto tiene un término constante en el tiempo y otro que varía con una frecuencia de $2\omega$. Al aplicar el promedio temporal del producto anterior sobre un período $T$, obtenemos,
\begin{align}
    \left \langle \mathbf{J}(\mathbf{r},t)\cdot\mathbf{E}(\mathbf{r},t) \right \rangle\equiv\frac{1}{T}\int_{0}^{T}\mathbf{J}(\mathbf{r},t)\cdot\mathbf{E}(\mathbf{r},t)dt=\frac{1}{2}\textup{Re}[\mathbf{J}^{*}(\mathbf{r})\cdot\mathbf{E}(\mathbf{r})]
\end{align}
Por lo tanto, para promedios temporales de campos monocromáticos, la convención es tomar la mitad de la parte real del producto complejo conjugado. Así, podemos determinar el teorema de Poynting complejo en un material arbitrario,
\begin{align}\label{eq:ConservacionEnergiaComplejo}
    \frac{c}{4\pi}\boldsymbol{\nabla}\cdot(\mathbf{E}\times\mathbf{H}^{*})+\frac{i\omega}{4\pi}(\mathbf{E}\cdot\mathbf{D}^{*}-\mathbf{B}\cdot\mathbf{H}^{*})=-\mathbf{J}^{*}\cdot\mathbf{E}
\end{align}
donde todas las cantidades son funciones complejas de $\mathbf{r}$. Para TIs, donde las relaciones constitutivas están dadas por las ecuaciones \eqref{eq:ConstitutiveRelations}, tenemos el siguiente teorema de Poynting complejo,
%
% \begin{align}\label{eq:ConservacionEnergiaComplejoTI}
% \frac{c}{4\pi}\boldsymbol{\nabla}\cdot(\frac{1}{\mu^{*}}\mathbf{E}\times\mathbf{B}^{*})+\frac{i\omega}{4\pi}(\epsilon^{*}|\mathbf{E}|^{2}-\frac{1}{\mu^{*}}|\mathbf{B}|^{2})-\frac{c}{4\pi}\boldsymbol{\nabla}\theta^{*}\cdot(\mathbf{E}\times\mathbf{E}^{*})&=-\mathbf{J}^{*}\cdot\mathbf{E}
% \end{align}
%
%donde hemos usado el complejo conjugado de la ecuación de Faraday, $\boldsymbol{\nabla}\times\mathbf{E}^{*}=-i\omega\mathbf{B}^{*}/c$.
%
\begin{align}\label{eq:TeoremaPoyntingComplejo}
\boldsymbol{\nabla}\cdot\tilde{\mathbf{S}}+2i\omega \tilde{u}_{\textup{EM}}-P_{\theta}&=-\frac{1}{2}\mathbf{J}^{*}\cdot\mathbf{E}
\end{align}
donde,
\begin{align}\label{eq:PowerTheta}
    P_{\theta} &= \frac{c}{8\pi}\boldsymbol{\nabla}\theta^{*}\cdot(\mathbf{E}\times\mathbf{E}^{*}),\\\label{eq:Poyntingcomplejo}
    \tilde{\mathbf{S}} &=\frac{c}{4\pi}\left(\frac{1}{2\mu^{*}}\mathbf{E}\times\mathbf{B}^{*}\right),\\\label{eq:EnergiaCompleja}
    \tilde{u}_{\textup{EM}}&=\frac{1}{8\pi}\frac{1}{2}\left(\epsilon^{*}|\mathbf{E}|^{2}-\frac{1}{\mu^{*}}|\mathbf{B}|^{2}\right)
\end{align}
y $\boldsymbol{\nabla}\theta=\tilde{\theta}\delta(\mathbf{r}-\mathbf{\Sigma})\tongo{n}$, donde $\tongo{n}$ es el vector normal que sale de la superficie $\mathbf{\Sigma}$. El término $P_{\theta}$ representa la contribución de $\theta(\mathbf{r})$ a la conservación de la energía EM y tiene unidades de densidad de potencia. Además, el vector $\tilde{\mathbf{S}}$
%
% \begin{align}
%     \tilde{\mathbf{S}} &=\frac{c}{4\pi}\left(\frac{1}{2\mu^{*}}\mathbf{E}\times\mathbf{B}^{*}\right), &
%     \tilde{u}_{\textup{EM}}&=\frac{1}{8\pi}\frac{1}{2}\left(\epsilon^{*}|\mathbf{E}|^{2}-\frac{1}{\mu^{*}}|\mathbf{B}|^{2}\right)
% \end{align}
%
corresponde al vector de Poynting complejo y $\tilde{u}_{\textup{EM}}$ a la densidad de energía EM compleja. Al integrar la expresión anterior en un cierto volumen $V$ con superficie $\partial V$, obtenemos,
\begin{align}\label{eq:IntTeoremaPoyntingComplejo}
\oint_{\partial V}\tilde{\mathbf{S}}\cdot d\mathbf{a}+2i\omega \int_{V}\tilde{u}_{\textup{EM}}d^3x-\frac{c}{8\pi}\tilde{\theta}^{*}\int_{\Sigma}(\mathbf{E}\times\mathbf{E}^{*})\cdot d\mathbf{\Sigma}&=-\frac{1}{2}\int_{V}\mathbf{J}^{*}\cdot\mathbf{E}d^{3}x.
\end{align}
En la Ec. \eqref{eq:IntTeoremaPoyntingComplejo}, se puede observar un término proporcional a $\tilde{\theta}$ debido a que en la representación compleja, el término  $\mathbf{E}\times\mathbf{E}^{*}$ no es automáticamente igual a cero, a diferencia del caso sin dispersión. Para una polarización en el plano $z=constante$ general: $\mathbf{E}=|E_x|\tongo{x}+|E_y|e^{i\delta}\tongo{y}$, tenemos que $\mathbf{E}\times\mathbf{E}^{*}=-2i|E_x||E_y|\sin\delta\tongo{z}$. El término es puramente imaginario y se hace cero cuando la polarización es lineal, $\delta=n \pi$, y es distinto de cero cuando la polarización es circular o elíptica con $\delta\neq n \pi$.  

Por ejemplo, consideremos una OEM que se propaga en la dirección $\tongo{z}$ con polarización circular $\mathbf{E}(\mathbf{r})=E_{0}(\tongo{x}\pm i \tongo{y})$, cerca de una $\theta$-interfaz ubicada en el plano  $z=0$, por lo que $\tongo{n}=\tongo{z}$, y así el término $\tilde{\theta}^{*}(\mathbf{E}\times\mathbf{E}^{*})\cdot\tongo{z}=\mp 2i\tilde{\theta}^{*}|E_{0}|^{2}$. Vemos que este término solamente influye cuando la OEM \textit{incide} en la $\theta$-interfaz, en cualquier ángulo, con polarización circular o elíptica.

La Ec. \eqref{eq:TeoremaPoyntingComplejo} es análoga a la Ec. \eqref{eq:TeoremaPoynting} para campos armónicos. Esta es una ecuación compleja cuya parte real entrega la conservación de energía para cantidades promediadas en un período de tiempo, y cuya parte imaginaria se relaciona con la energía reactiva o almacenada y su flujo alterno. Al dividir la Ec. \eqref{eq:TeoremaPoyntingComplejo} en su parte real e imaginaria, obtenemos,
%
% \begin{align}
% \boldsymbol{\nabla}\cdot\tilde{\mathbf{S}}+2\omega(i\textup{Re}\tilde{u}_{\textup{EM}}-\textup{Im}\tilde{u}_{\textup{EM}})-\frac{c}{4\pi}(i\textup{Re}\boldsymbol{\nabla}\theta+\textup{Im}\boldsymbol{\nabla}\theta)\cdot(\textup{Im}\mathbf{E}\times\textup{Re}\mathbf{E}))&=-\frac{1}{2}\mathbf{J}^{*}\cdot\mathbf{E}
% \end{align}
%
\begin{align}\label{eq:TeoremaPoyntingComplejoRe}
\boldsymbol{\nabla}\cdot\textup{Re}(\tilde{\mathbf{S}})-2\omega\textup{Im}(\tilde{u}_{\textup{EM}})-\frac{c}{4\pi}\textup{Im}\boldsymbol{\nabla}\theta\cdot(\textup{Im}\mathbf{E}\times\textup{Re}\mathbf{E})+\frac{1}{2}\textup{Re}(\mathbf{J}^{*}\cdot\mathbf{E})&=0,\\\label{eq:TeoremaPoyntingComplejoIm}
\boldsymbol{\nabla}\cdot\textup{Im}(\tilde{\mathbf{S}})+2\omega\textup{Re}(\tilde{u}_{\textup{EM}})-\frac{c}{4\pi}\textup{Re}\boldsymbol{\nabla}\theta\cdot(\textup{Im}\mathbf{E}\times\textup{Re}\mathbf{E})+\frac{1}{2}\textup{Im}(\mathbf{J}^{*}\cdot\mathbf{E})&=0.
\end{align}
La Ec. \eqref{eq:TeoremaPoyntingComplejoIm} no puede obtenerse simplemente a partir de la forma dependiente del tiempo del teorema de Poynting. Estas relaciones dan cuenta de dos términos proporcionales al TMEP, donde su parte real $\textup{Re}(\theta(\mathbf{r}))=\theta'$ pertenece a la energía almacenada en un cierto volumen y su parte imaginaria $\textup{Im}(\theta(\mathbf{r}))=\theta''$ está relacionada con la pérdida o ganancia de energía que entra o sale de un volumen $V$ a través de su superficie $\partial V$, dando lugar a comportamientos no triviales. Como se mencionó anteriormente, las conclusiones obtenidas en este primer cálculo de la tesis sobre la influencia de $\theta'$ y $\theta''$ en la energía de un sistema, derivadas de cálculos electromagnéticos clásicos, son consistentes con los resultados obtenidos mediante cálculos cuánticos en estudios como el de \cite{tse_giant_2010}. 

Para medios topológicos no dispersivos, lineales, isotrópicos, óhmicos, con campos armónicos, donde los parámetros topo-ópticos ($\epsilon$, $\mu$ y $\theta$) son reales, tenemos,
\begin{align}\label{eq:RePartNoDisipativa}
\frac{c}{4\pi\mu}\boldsymbol{\nabla}\cdot\textup{Re}(\mathbf{E}\times\mathbf{B}^{*})+\sigma|\mathbf{E}|^{2}&=0,\\\label{eq:ImPartNoDisipativa}
\frac{c}{2\mu}\boldsymbol{\nabla}\cdot\textup{Im}(\mathbf{E}\times\mathbf{B}^{*})+\frac{\omega}{2}\left(\epsilon|\mathbf{E}|^{2}-\frac{1}{\mu}|\mathbf{B}|^{2}\right)-c\boldsymbol{\nabla}\theta\cdot(\textup{Im}\mathbf{E}\times\textup{Re}\mathbf{E})&=0.
\end{align}
La Ec. \eqref{eq:RePartNoDisipativa} revela que la energía electromagnética que ingresa al sistema se disipa en forma de calor al realizar trabajo en las fuentes, como es usual. Esto implica una conversión de energía EM a energía térmica dentro del sistema. Por otro lado, la ecuación \eqref{eq:ImPartNoDisipativa} describe cómo la energía se almacena en el sistema y cómo fluye dentro y fuera del mismo a lo largo del tiempo. Esta ecuación refleja la dinámica de la energía reactiva y su interacción con el medio, incluyendo los estados de superficie del TI. 

En resumen, la inclusión de TIs en un sistema EM introduce nuevos términos en el teorema de conservación de la energía. Las propiedades únicas de los estados de borde pueden afectar la disipación, distribución y la reactividad de la energía EM en el sistema, dando lugar a comportamientos no triviales. Estudiar estos efectos es crucial para comprender plenamente el comportamiento de los sistemas con TI y puede tener implicaciones significativas en el diseño de dispositivos electrónicos y ópticos avanzados.

\section{Descomposición Transversal-Longitudinal de la $\theta$-ED} \label{4.4}
\begin{SCfigure}[1][t]
\caption{Geometría arbitraria de una guía de onda formada por dos medios con diferentes propiedades topo-ópticas, \textit{i.e.}, diferentes $\{\epsilon,\mu,\theta\}$. El campo EM se propaga en la dirección $\tongo{z}$ y están confinadas en la dirección $\tongo{x}$ e $\tongo{y}$. Imagen modificada de \cite{Griffiths2023introduction}}.
\includegraphics[width=0.45\linewidth]{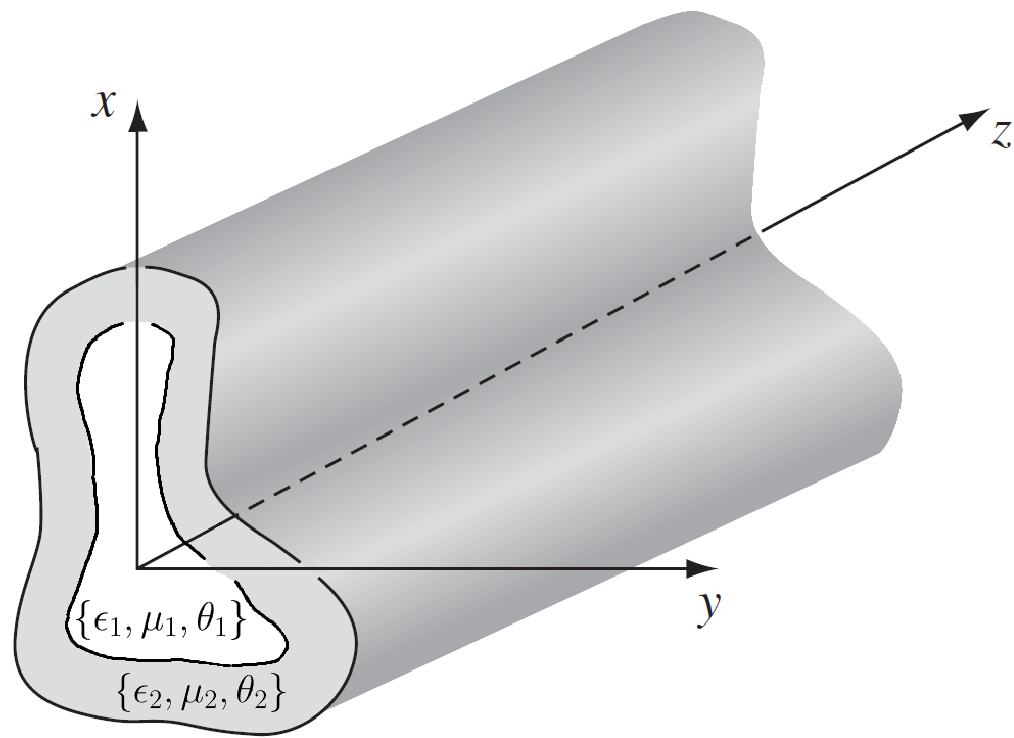}
\label{fig:Waveguide1}
\end{SCfigure}

En un sistema de guía de ondas, nos enfocamos en encontrar soluciones de las ecuaciones de Maxwell que se propaguen a lo largo de la dirección de la guía, es decir, en la dirección $\tongo{z}$, y que estén confinadas en las proximidades de la estructura de la guía, tal como se ilustra en la Fig. (\ref{fig:Waveguide1}). 
%---------------Figure------------
%---------------End Figure---------
Supondremos que la guía está compuesta, al menos en parte, por materiales que exhiben el efecto magnetoeléctrico, tales como TIs, WSM, meta-materiales o medios Tellegen (medios descritos por las relaciones constitutivas de la Ec. \eqref{eq:ConstitutiveRelations}), de manera que las OEM se ven afectadas por el TMEP. Por lo tanto, asumimos que los campos eléctrico y magnético tienen la siguiente forma,
\begin{align} \label{E_y_B_tot1} 
    \mathbf{E}(\mathbf{r},t)&=\frac{1}{2\pi}\int_{-\infty}^{\infty} \mathbf{E}_{0}(\mathbf{r}_{\perp};\omega)e^{i(k_{z}z-wt)}d\omega,\\ \label{E_y_B_tot2} 
    \mathbf{B}(\mathbf{r},t)&=\frac{1}{2\pi}\int_{-\infty}^{\infty}\mathbf{B}_{0}(\mathbf{r}_{\perp};\omega)e^{i(k_{z}z-wt)}d\omega,
\end{align}
donde $k_{z}=2\pi/\lambda_{g}$ es el número de onda, $\lambda_{g}$ la longitud de onda a lo largo de la dirección de la guía, $\omega=2\pi f$ con $f$ siendo la frecuencia de operación de la guía. Las coordenadas $\mathbf{r}_{\perp}$ se refieren a las coordenadas perpendiculares a la dirección de propagación de la onda, de manera que $\mathbf{r}_{\perp}\cdot\tongo{z}=0$. Las variables $\omega$ y $k_{z}$ no son independientes, debido a la condición impuesta por la relación de dispersión que satisface la OEM en la guía, la cual depende de la geometría de la misma. Como es habitual \cite{jackson1999classical}, descomponemos los campos y operadores diferenciales en sus componentes transversales y longitudinales con respecto a la dirección de propagación de la siguiente manera,
\begin{align}
\mathbf{C}_{0}&=\mathbf{C}_{\perp}+\tongo{z}C_{z}, &
\boldsymbol{\nabla}&=\boldsymbol{\nabla}_{\perp}+\tongo{z}\partial _{z} \label{DescomposicionVec}
\end{align}
donde $\mathbf{C}_{0}=\{\mathbf{E}_{0},\mathbf{B}_{0}\}$. Al sustituir en las Ecs. \eqref{eq:MCS1time}-\eqref{eq:MCS4time} de la $\theta$-ED, obtenemos,
\begin{align}
\boldsymbol{\nabla}_{\perp}\cdot  (\epsilon\mathbf{E}_{\perp})+\boldsymbol{\nabla}_{\perp}\theta\cdot \mathbf{B}_{\perp}&=-ik_{z}\epsilon E_{z}-\partial_{z}\theta B_{z}, 
\label{EQ:T-L1}\\
\boldsymbol{\nabla}_{\perp}\cdot \mathbf{B}_{\perp}&=-ik_{z}B_{z}, 
\label{EQ:T-L2}\\
ik_{z}\mathbf{E}_{\perp}+ik_{0}\tongo{z}\times \mathbf{B}_{\perp}&=\boldsymbol{\nabla}_{\perp}E_{z},
\label{EQ:T-L3}\\
\tongo{z}\cdot (\boldsymbol{\nabla}_{\perp}\times \mathbf{E}_{\perp})&=ik_{0}B_{z}, 
\label{EQ:T-L4}\\
ik_{z}\mu^{-1}\mathbf{B}_{\perp}- \partial_{z}\theta\mathbf{E}_{\perp}-ik_{0}\epsilon\tongo{z}\times\mathbf{E}_{\perp}&=\boldsymbol{\nabla}_{\perp}(\mu^{-1}B_{z})-E_{z}\boldsymbol{\nabla}_{\perp}\theta, 
\label{EQ:T-L5}\\
\tongo{z}\cdot (\boldsymbol{\nabla}_{\perp}\times (\mu^{-1}\mathbf{B}_{\perp})-\boldsymbol{\nabla}_{\perp}\theta\times \mathbf{E}_{\perp})&=-ik_{0}\epsilon E_{z}. 
\label{EQ:T-L6}
\end{align}
%
% \textcolor{red}{Revisar el limite de onda plana en un medio infinito. Aun tenemos el término $\partial_{z}\theta$ que afecta en las relaciones de dipersion ?. Si es asi, porque no observamos polarizaciones circulares R o L ?. Que pasa con las Interfazs en las direccion $\tongo{z}$? no es lo mismo que incidencia normal ?}\textcolor{blue}{Re: En el limite de onda plana tenemos que cambia la relacion de dispersion y tiene dependencia de $\partial_{z}\theta$ igual que los modos TEM en la guia de onda. Este término hace que las polarizaciones R y L se propaguen a diferentes velocidades. Ademas hace que una componente se desvanezca cuando se hacerca a la Interfaz, es decir, aparece una parte imaginaria en $k_{z}$ para una polarizacion. Si tenemos Interfazs en $\tongo{z}$ tendriamos el mismo efecto que incidencias normales.}
Hemos reemplazado $\partial_{z}\to ik_{z}$ y utilizado la propiedad vectorial $(\tongo{z}\times\mathbf{C}_{\perp})\times\tongo{z}=\mathbf{C}_{\perp}$. Además, hemos asumido que los parámetros ópticos solo dependen de las coordenadas transversales, es decir, $\epsilon=\epsilon(\mathbf{r}_{\perp};\omega)$ y $\mu=\mu(\mathbf{r}_{\perp};\omega)$. Por ahora, $\theta=\theta(\mathbf{r};\omega)$ puede variar espacialmente en cualquier dirección.

Por lo tanto, el problema consiste en encontrar $\mathbf{E}_{0}(\mathbf{r}_{\perp};\omega)$ y $\mathbf{B}_{0}(\mathbf{r}_{\perp};\omega)$ de manera que satisfagan el conjunto de ecuaciones de la $\theta$-ED \eqref{EQ:T-L1}-\eqref{EQ:T-L6}, sujeto a las condiciones de contorno \eqref{EQ:BCsdiscontinuas}, \eqref{EQ:BCscontinuas} y \eqref{EQ:BC_Bdiscontinua}. Dependiendo de si ambas, una o ninguna de las componentes longitudinales son nulas, podemos clasificar las soluciones como transversal electromagnética (TEM), transversal eléctrica (TE), transversal magnética (TM) o híbrida.  

\section{Hipótesis y relación con trabajos de la literatura}\label{4.5}
Las propiedades exóticas de los materiales con fases topológicas resultan fascinantes tanto en la teoría de campos de alta energía como en la física de la materia condensada, debido a sus posibles aplicaciones en tecnologías como la espintrónica y la computación cuántica topológica \cite{Hasan2010Colloquium,mellnik_2014_spin}. Como un nuevo estado de la materia, comprender sus interacciones con otras formas de radiación y/o materia es un tema de gran interés científico.

\begin{itemize}
    \item \textbf{Objetivo general:} El objetivo principal de esta tesis es avanzar en la descripción y caracterización de la interacción entre campos y fuentes electromagnéticas con materiales que presentan fases topológicas, como los TIs.
    \item \textbf{Hipótesis general:} La hipótesis principal de esta tesis sostiene que la teoría de campo efectivo de la $\theta$-ED es adecuada para describir la respuesta electromagnética de sistemas tridimensionales con fases topológicas.
    \item \textbf{Hipótesis específica 1:} Los sistemas en los que las BCs juegan un papel fundamental, como en la propagación de ondas en medios confinados, constituyen escenarios ideales para observar efectos significativos de los estados de borde de los TIs.
    \item \textbf{Objetivo específico 1:} Estudiar la propagación de OEM en medios confinados construidos con TIs, dieléctricos y/o conductores.
\end{itemize}

Los sistema de OEM confinadas en guías de ondas formadas por TIs han sido revisado por algunos trabajos como  \cite{Gomes2010testing,melo_topological_2016,crosse_theory_2017}.

En el estudio \cite{melo_topological_2016}, se analiza la propagación de OEM en una guía de ondas tipo \textit{slab} cuyas paredes están formadas por TIs. A partir de las Ecs. \eqref{EQ:T-L1}-\eqref{EQ:T-L6} los autores asumen que los estados superficiales tienen una longitud de penetración en el \textit{bulk} del TI, denotada como $l$, \textit{i.e.}, $\theta(x)=\frac{\alpha x}{\pi l}\theta_{\textup{TI}}$. Esto conduce a una modificación en la relación de dispersión de las ondas confinadas, donde el parámetro topológico $\theta$ introduce una frecuencia de corte $\omega_{\theta}=\frac{2c}{\epsilon}\frac{\alpha\theta_{\textup{TI}}}{\pi l}$. En este escenario, las frecuencias por debajo de $\omega_\theta$ ($\omega < \omega_\theta$) son reflejadas por las paredes de TI, mientras que para frecuencias superiores ($\omega > \omega_\theta$), una porción significativa de la luz incidente penetra en el bulk del TI, comprometiendo significativamente la propagación de la onda a través de la guía de onda. Para materiales como el Bi$_2$Se$_{3}$, se estima que $\omega_{\theta}\sim 10~\textup{THz}$.

Una limitación importante de este modelo es la suposición de que el campo EM dentro del TI se desvanece de manera similar a un conductor, sin considerar adecuadamente la distribución del campo dentro del TI ni los modos normales permitidos al incluir las BCs modificadas por la $\theta$-ED. 

En \cite{crosse_theory_2017}, se analiza la propagación de OEM en una guía de ondas \textit{slab} formada por capas de TIs. El autor asume que el campo EM dentro de la guía es una combinación lineal de los modos permitidos en la electrodinámica convencional, debido a que el TMEP es pequeño en comparación con los valores típicos de la permitividad eléctrica. Esta simplificación omite los términos cuadráticos en $\theta$ y se basa en el método de modos acoplados. Los resultados muestran una rotación de la polarización de $\approx 100~\textup{mrad}$, significativamente mayor que los valores observados cuando una OEM incide una sola vez en una interfaz TI-dieléctrico. Esta rotación se atribuye a una ``transmutación'' exacta de los modos de propagación, en la que los modos TE y TM se mezclan a lo largo de la guía. Sin embargo, el modelo no considera los cambios significativos en los modos de propagación.

\begin{itemize}
    \item \textbf{Hipótesis específica 2:} Los modos de propagación en guías de ondas formadas por TIs cambian significativamente al resolver el sistema de forma exacta, incluyendo todos los órdenes en $\theta$.
    \item \textbf{Objetivo específico 2:} Analizar y caracterizar los modos de propagación permitidos en guías de ondas de tipo \textit{slab} y su dependencia en las BCs y las propiedades del material.
\end{itemize}

En esta tesis, se demostrará que al resolver el sistema de forma exacta, los modos de propagación cambian de forma considerable. En particular, la transmutación exacta entre modos TE y TM deja de ser válida, y la propagación de la OEM se vuelve inherentemente híbrida. Además, el modelo incluirá las variaciones en las constantes de propagación, como el número de onda transversal, lo que permitirá un mejor confinamiento de las OEM gracias a las propiedades de las paredes de TI.

El trabajo de Ibanescu et al. \cite{ibanescu_all-dielectric_2000} estudia modos cuasi-TEM en una guía coaxial dieléctrica, destacando sus bajas pérdidas de energía y ausencia de frecuencia de corte. Sin embargo, debido al teorema de Earnshaw, verdaderos modos TEM no pueden existir en sistemas puramente dieléctricos, requiriéndose al menos dos conductores. Este estudio abre la posibilidad de investigar configuraciones donde el uso de TIs permita superar estas restricciones.

\begin{itemize}
    \item \textbf{Hipótesis específica 3:} La no aplicabilidad del teorema de Earnshaw para TIs sugiere que guías de ondas huecas con paredes topológicas pueden soportar modos TEM.
    \item \textbf{Objetivo específico 3:} Estudiar soluciones TEM en guías de ondas formadas por TIs y evaluar su potencial para minimizar pérdidas y mejorar el confinamiento de las OEM.
\end{itemize}

En esta tesis, aprovechamos las propiedades únicas de los TIs para superar estas limitaciones. Al utilizar el TME, demostramos que es posible obtener soluciones exactas para modos TEM en guías de ondas cilíndricas sin necesidad de dos conductores, eludiendo así el teorema de Earnshaw \cite{Filipini2024polarization}. Esto permite reducir el número de conductores a uno o incluso a cero, minimizando las pérdidas de potencia por calentamiento Joule y mejorando el confinamiento de los modos. Estos avances no solo resuelven los problemas identificados en el trabajo de Ibanescu, sino que también abren nuevas posibilidades para el diseño de guías de ondas en sistemas fotónicos avanzados.

\biblio %Se necesita para referenciar cuando se compilan subarchivos individuales - NO SACAR

%% file: Capitulos/05TEM.tex
La propagación en modo TEM es reconocida como el modo dominante en guías de ondas y es ampliamente utilizada para guiar energía EM en aplicaciones terrestres, como la propagación de ondas de radio a través de la atmósfera en comunicaciones por satélite, telefonía móvil, radio FM, transmisión de televisión y radar, entre otras. Su principal aplicación se encuentra en estructuras conocidas como líneas de transmisión de dos conductores, como los cables coaxiales y las líneas de placas paralelas. La existencia de ondas TEM en estas estructuras requiere necesariamente la presencia de al menos dos conductores, lo cual está relacionado con el teorema de Earnshaw \cite{landau2013electrodynamics}. Debido a este teorema, es imposible propagar modos TEM en guías de onda dieléctricas o huecas. Sin embargo, en el ámbito de la $\theta$-ED, se ha demostrado que es posible eludir las restricciones impuestas por el teorema de Earnshaw \cite{martin2016electro}. En un estudio preliminar, se investigó la posible existencia de modos TEM en una guía de onda rectangular hueca \cite{Cancino2019TEM}, aunque no se alcanzó una conclusión definitiva sobre su viabilidad. En esta tesis, hemos logrado desarrollar configuraciones concretas que eluden el teorema de Earnshaw, permitiendo así la existencia de soluciones TEM incluso sin la presencia de conductores, algo que no es permitido en el electromagnetismo clásico de Maxwell.

La propagación de ondas TEM es altamente valorada en ciencia y tecnología \cite{ibanescu_all-dielectric_2000} debido a varias características que la distinguen de otros modos. Una de las ventajas principales del modo TEM es que no tiene una frecuencia de corte, lo que significa que puede propagarse a cualquier frecuencia y siempre está presente en estructuras con dos conductores. Además, los campos electromagnéticos en el modo TEM se distribuyen uniformemente, garantizando una distribución homogénea de la energía a lo largo de la sección transversal del cable, lo que minimiza las pérdidas por radiación. Otra ventaja clave es la ausencia de dispersión modal, es decir, la diferencia en la velocidad de propagación entre diferentes modos; esto se debe a que el modo TEM puede ser el único que se propaga cuando la frecuencia se ajusta adecuadamente. Además, el modo TEM es robusto frente a variaciones geométricas en la línea de transmisión, como curvas pronunciadas o cambios en las propiedades del dieléctrico, lo que contribuye a una señal más estable y menos afectada por imperfecciones estructurales.

En cualquier línea de transmisión que propague modos TEM, las pérdidas resistivas en los conductores y en los materiales dieléctricos utilizados son significativas en la atenuación de la señal. Sin embargo, nuestro trabajo contribuye a reducir estas pérdidas al disminuir, e incluso eliminar, la necesidad de conductores para la propagación de modos TEM.

En este capítulo, mostraré varios cálculos desarrollados en esta tesis con el objetivo de llenar los vacíos en el estudio de ondas TEM confinadas en TIs, lo cual dio lugar a un artículo \cite{Filipini2024polarization}. En la Sec. (\ref{5.1}), presentaremos las ecuaciones diferenciales generales y la relación de dispersión que satisfacen una onda TEM. Demostraremos que el sistema de ecuaciones de la $\theta$-ED se reduce a un problema estático en 2D. En la Sec. (\ref{5.2}), mostraremos que una onda TEM confinada en un TI presenta una rotación de la polarización. Argumentaremos que esta rotación es diferente de los efectos conocidos como rotación de Faraday y Kerr de una OEM. En la Sec. (\ref{5.3}), resolveremos el sistema de ecuaciones en una geometría cilíndrica. Aun así, la solución no estará completamente determinada hasta que se apliquen las BCs. En la Sec. (\ref{5.4}), consideraremos un cilindro TI inmerso en una OEM externa. Al aplicar las BCs de la $\theta$-ED en la superficie del TI, encontraremos una solución TEM. En esta misma sección, presentaremos un procedimiento iterativo convergente que genera cargas topológicas en la superficie del TI. En la Sec. (\ref{5.5}), aplicamos las BCs para encontrar la solución en un sistema con un cilindro hueco de cierto grosor inmerso en una OEM externa. Obtendremos la solución completa de la onda TEM, el cual depende de un parámetro $\chi = R_1/R_2$ que cuantifica la razón geométrica de los cilindros. Encontramos una razón geométrica tal que maximiza el transporte de energía EM a lo largo de la dirección de la guía. Los cálculos de las Secs. (\ref{5.4}) y (\ref{5.5}) dieron lugar a un artículo \cite{Filipini2024polarization}. En la Sec. (\ref{5.6}), encontraremos la onda TEM en un sistema formado por un TI y un conductor que confina el campo EM en la región interna. Esta solución es aplicable cuando el TI tiene un término imaginario en el parámetro topológico, lo cual ocurre en materiales con dispersión. Los cálculos de la Sec. (\ref{5.6}) están en preparación para un artículo \cite{underprep}.
\section{Modelo de ondas electromagnéticas transversales confinadas en aislantes topologicos}\label{5.1}
En nuestro análisis, buscaremos condiciones bajo las cuales pueda existir la propagación TEM a lo largo del eje de la guía, donde $\mathbf{E}_{\perp}\neq 0$ y $\mathbf{B}_{\perp}\neq 0$, con $E_{z}=B_{z}=0$. Sustituyendo estas condiciones en las ecuaciones de la $\theta$-ED \eqref{EQ:T-L1}-\eqref{EQ:T-L6}, obtenemos,
\begin{align}
\boldsymbol{\nabla}_{\perp}\cdot  (\epsilon\mathbf{E}_{\perp})+\boldsymbol{\nabla}_{\perp}\theta\cdot \mathbf{B}_{\perp}&=0, 
\label{EQ:T-L1_tem}\\
\boldsymbol{\nabla}_{\perp}\cdot \mathbf{B}_{\perp}&=0, 
\label{EQ:T-L2_tem}\\
ik_{z}\mathbf{E}_{\perp}+ik_{0}\tongo{z}\times \mathbf{B}_{\perp}&=0,
\label{EQ:T-L3_tem}\\
\tongo{z}\cdot (\boldsymbol{\nabla}_{\perp}\times \mathbf{E}_{\perp})&=0, 
\label{EQ:T-L4_tem}\\
ik_{z}\mu^{-1}\mathbf{B}_{\perp}- \partial_{z}\theta\mathbf{E}_{\perp}-ik_{0}\epsilon\tongo{z}\times\mathbf{E}_{\perp}&=0, 
\label{EQ:T-L5_tem}\\
\tongo{z}\cdot (\boldsymbol{\nabla}_{\perp}\times (\mu^{-1}\mathbf{B}_{\perp})-\boldsymbol{\nabla}_{\perp}\theta\times \mathbf{E}_{\perp})&=0. 
\label{EQ:T-L6_tem}
\end{align}
De la Ec. \eqref{EQ:T-L3_tem}, observamos que $\mathbf{B}_{\perp}$ es transversal tanto a $\mathbf{E}_{\perp}$ como a $\tongo{z}$,
\begin{align}\label{eq:B_transverse}
    \mathbf{B}_{\perp}=\frac{k_{z}}{k_{0}}\tongo{z}\times\mathbf{E}_{\perp},
\end{align}
La relación entre $\mathbf{B}_{\perp}$ y $\mathbf{E}_{\perp}$ es análoga a la que se encuentra para las ondas planas en un medio infinito. Consideraremos que el TMEP a lo largo de la guía de ondas es homogéneo, \textit{i.e.}, $\theta(\mathbf{r}_{\perp},z)\equiv\theta(\mathbf{r}_{\perp})$, cambiando solo en las direcciones transversales de la guía. Al sustituir $\mathbf{B}_{\perp}$ en la Ec. \eqref{EQ:T-L5_tem}, determinamos la relación de dispersión para la OEM,
%
% \begin{align}
%     %ik_{z}^{2}\mu^{-1}\mathbf{E}_{\perp}- k_{0}\partial_{z}\theta\mathbf{E}_{\perp}\times\tongo{z}-ik_{0}^{2}\epsilon\mathbf{E}_{\perp}&=0\\
%     %(k_{z}^{2}-k_{0}^{2}\epsilon\mu)\mathbf{E}_{\perp}&=ik_{0}\mu\partial_{z}\theta\tongo{z}\times\mathbf{E}_{\perp}\\
%     %((k_{z}^{2}-k_{0}^{2}\epsilon\mu)^{2}-k_{0}^{2}\mu^{2}(\partial_{z}\theta)^{2})\mathbf{E}_{\perp}&=0\\
%     k_{z}^{2}-k_{0}^{2}\epsilon\mu&=\pm k_{0}\mu\partial_{z}\theta
% \end{align}
%
%Esta relación de dispersión es idéntica a la obtenida para una onda plana que se propaga en un medio infinito e incide en una superficie $z=cte$ con discontinuidad en $\theta(z)$ y desvela una birrefringencia circular. Este término tiene el mismo efecto que una incidencia normal, pero ahora, dentro de una guía con $\theta$-interfaces a lo largo de la dirección $\tongo{z}$. 

%De esta forma, la relación de dispersión corresponde a la de una onda plana en un medio homogéneo,
%
\begin{align}
    k_{z}=k_{0}n
\end{align}
donde $n=\sqrt{\epsilon\mu}$ es el índice de refracción del medio. Esta relación implica que la OEM puede propagarse a una frecuencia arbitraria, \textit{i.e.}, que no hay una frecuencia de corte, a diferencia de los modos TE, TM o híbrido. Además, la velocidad de fase y la velocidad de grupo son iguales, $v_{f}=v_{g}=c/n$.

Antes de continuar y reemplazar en las ecuaciones de Maxwell restantes, revisemos un hecho importante en la propagación de modos TEM. Considerando que el campo magnético es transversal al campo eléctrico, como se muestra en la Ec. \eqref{eq:B_transverse}, y que el campo magnético normal $B_{n}$ a las paredes de la guía debe ser continuo,
\begin{align*}
    B_{n 2}e^{i(k_{z2}z-\omega t)}&=B_{n 1}e^{i(k_{z1}z-\omega t)}\\
    \frac{k_{z2}}{k_{0}}(\tongo{z}\times\mathbf{E}_{\perp2})\cdot\tongo{n}e^{ik_{z2}z}&=\frac{k_{z1}}{k_{0}}(\tongo{z}\times\mathbf{E}_{\perp1})\cdot\tongo{n}e^{ik_{z1}z},
\end{align*}
además, el campo eléctrico paralelo a la superficie es continuo, $(\tongo{z}\times\mathbf{E}_{\perp2})\cdot\tongo{n}=(\tongo{z}\times\mathbf{E}_{\perp1})\cdot\tongo{n}$, implica la \textit{refractive index matching} (RIM),
\begin{align}\label{eq:RIM}
    n_{2}=n_{1}
\end{align}
La relación dada por la Ec. \eqref{eq:RIM}, es fundamental para la existencia de ondas TEM en guías de onda. Esto implica que las ondas TEM que se propagan a lo largo del eje $\tongo{z}$ en diferentes medios transversales, deben tener el mismo índice de refracción. Esta condición es crucial para garantizar la satisfacción de las BCs a lo largo de toda la guía y en todo momento en los modos TEM. Esta condición no aplica cuando buscamos soluciones TE o TM, como veremos en el siguiente capitulo.

La RIM es factible y tiene aplicaciones en campos como \cite{wiederseiner2011refractive}. Por ejemplo, se utiliza para superar los problemas de refracción en flujos líquidos \cite{budwig1994refractive}, donde la igualdad de índices de refracción facilita la transmisión de luz a través de interfaces líquido-líquido. Asimismo, la coincidencia de índices se emplea en la fabricación de recubrimientos anti-reflectantes de banda ancha \cite{xi2007optical}, donde garantiza la mínima reflectividad de la luz al atravesar diversas capas de material.

Finalmente de la Ecs. \eqref{EQ:T-L2_tem} y \eqref{EQ:T-L4_tem}, vemos que $\mathbf{E}_{\perp}=-\boldsymbol{\nabla}_{\perp}\Phi$, y de las Ecs. \eqref{EQ:T-L1_tem} y \eqref{EQ:T-L6_tem} obtenemos la siguiente ecuación diferencial,
\begin{align}\notag
%\nabla_{\perp}^{2}\Phi=Z~\tongo{z}\cdot(\boldsymbol{\nabla}_{\perp} \theta\times\boldsymbol{\nabla}_{\perp}\Phi),
\nabla_{\perp}^{2}\Phi&=Z~\boldsymbol{\nabla}_{\perp}\Phi\cdot(\tongo{z}\times\boldsymbol{\nabla}_{\perp} \theta),\\
&=Z\tilde{\theta}\delta(\mathbf{r}_{\perp}-\boldsymbol{\Sigma}_{\perp})\boldsymbol{\nabla}_{\perp}\Phi\cdot\tongo{t},
\label{eq:TEM_equation_potential}
\end{align}
donde $Z=\sqrt{\mu/\epsilon}$ es la impedancia del medio, $\Sigma_{\perp}$ es la $\theta$-interfaz que separa dos medios con diferentes TMEP, $\tilde{\theta}$ es la diferencia de los TMEP de ambos medios adyacentes y $\tongo{t}=\tongo{z}\times\tongo{n}$, es la dirección a lo largo del borde de la $\theta$-interfaz. Por lo tanto, el conjunto de la $\theta$-ED se reduce a un problema electrostático bidimensional equivalente con diferentes BCs. 

En un sistema de guía de ondas compuesto por materiales topológicamente triviales, donde $\theta=0$, la propagación TEM sigue la ecuación de Laplace 2D, $\boldsymbol{\nabla}_{\perp}\Phi=0$, lo que implica que el modo TEM no puede existir dentro de un solo conductor hueco de conductividad infinita debido al teorema de Earnshaw \cite{landau2013electrodynamics}. El conductor hueco actúa como un equipotencial, lo que hace que el campo eléctrico se anule en el interior. Es necesario contar con al menos dos superficies conductoras para que $\Phi$ no sea constante y así sostener el modo TEM.

En el contexto de la $\theta$-ED, se planteó por primera vez en \cite{martin-ruiz_electro-_2016} la posibilidad de eludir el teorema de Earnshaw con $\theta\neq 0$, lo que permitiría la propagación TEM con menos de dos conductores. La Ec. \eqref{eq:TEM_equation_potential} es un caso particular de la ecuación (19) del artículo \cite{martin-ruiz_electro-_2016}.

En regiones lejos de la superficie, donde $\theta$ es constante, vemos que de las Ecs. \eqref{EQ:T-L1} y \eqref{EQ:T-L4}, el campo eléctrico satisface,
\begin{align}
    \boldsymbol{\nabla}_{\perp}\cdot \mathbf{E}_{\perp}&=0, & &\textup{y,} & \boldsymbol{\nabla}_{\perp}\times \mathbf{E}_{\perp}&=0.
\end{align}
Esto significa que $\mathbf{E}_{\perp}$ es una solución de un problema electrostático en dos dimensiones, y de la Ec. \eqref{eq:TEM_equation_potential} vemos que el potencial eléctrico $\Phi$, satisface la ecuación de Laplace. Por lo tanto, la metodología consistirá en resolver la ecuación de Laplace en cada región transversal lejos de las superficie donde cambia el parámetro topológico $\theta$ y luego aplicar las BCs modificadas por $\theta$ para obtener las condiciones de existencia para la propagación TEM.

%\textcolor{blue}{hablar de las consideraciones generales de la energia EM en propagacion TEM, para mencionar que no hay perdidas.}

% \vspace{10cm}

% \begin{align}
% \boldsymbol{\nabla}\cdot\textup{Re}(\tilde{\mathbf{S}})-2\omega\textup{Im}(\tilde{u}_{\textup{EM}})-\frac{c}{4\pi}\textup{Im}\boldsymbol{\nabla}\theta\cdot(\textup{Im}\mathbf{E}\times\textup{Re}\mathbf{E})&=0,\\
% \boldsymbol{\nabla}\cdot\textup{Im}(\tilde{\mathbf{S}})+2\omega\textup{Re}(\tilde{u}_{\textup{EM}})-\frac{c}{4\pi}\textup{Re}\boldsymbol{\nabla}\theta\cdot(\textup{Im}\mathbf{E}\times\textup{Re}\mathbf{E})&=0.
% \end{align}

% \begin{align}
%     P_{\theta} &= \frac{c}{8\pi}\boldsymbol{\nabla}\theta^{*}\cdot(\mathbf{E}_{\perp}\times\mathbf{E}_{\perp}^{*}),\\
%     \tilde{\mathbf{S}} &=\frac{c}{8\pi}\frac{n^{*}}{\mu^{*}}\tongo{z}|\mathbf{E}_{\perp}|^{2},\\
%     \tilde{u}_{\textup{EM}}&=\frac{1}{8\pi}\frac{1}{2}\left(\epsilon^{*}|\mathbf{E}_{\perp}|^{2}-\frac{|n|^2}{\mu^{*}}|\mathbf{E}_{\perp}|^{2}\right)
% \end{align}

% \begin{align}
%     \mathbf{B}_{\perp}=n\tongo{z}\times\mathbf{E}_{\perp},
% \end{align}
%
\section{Rotación de la polarización debido a una interfaz $\theta$ transversal}\label{5.2}
Un efecto interesante que ocurre debido a una $\theta$-interfaz en la propagación TEM, es un nuevo tipo de rotación de la polarización no reportada anteriormente. La $\theta$-interfaz produce una discontinuidad de $\mathbf{E}_{\perp}$ que resulta en una rotación de la polarización de la OEM. En cualquier punto dado en la $\theta$-interfaz, las direcciones de $\mathbf{E}_{\perp}$ satisfacen,
\begin{align}
\label{EQ:refracangle}
\tan\gamma_{m+1}&=\frac{\tan\gamma_{m}}{1+Z(\theta_{m+1}-\theta_{m})\tan\gamma_m}~, %& o,& & \tan\bar{\gamma}_{m+1}&=Z(\theta_{m+1}-\theta_{m})+\tan\bar{\gamma}_m~,
%\tan \gamma_{m+1}= \tan \gamma_{m} \left(1+2 Z_\theta \tan \gamma_i \right)^{-1},
\end{align}
%
%---------------------------------------
\begin{figure}[t]
%\stackinset{l}{2pt}{t}{3pt}{(a)}{\includegraphics[scale=0.53]{Capitulos/imagenes/TEM_rot_2.pdf}}
%\raisebox{0.845cm}{\makebox[0pt][l]{\stackinset{l}{2pt}{t}{7pt}{(a)}{\includegraphics[scale=0.53]{Capitulos/imagenes/TEM_rot_2.pdf}}}}
\raisebox{-1.06cm}{\stackinset{l}{2pt}{t}{7pt}{(a)}{\includegraphics[scale=0.55]{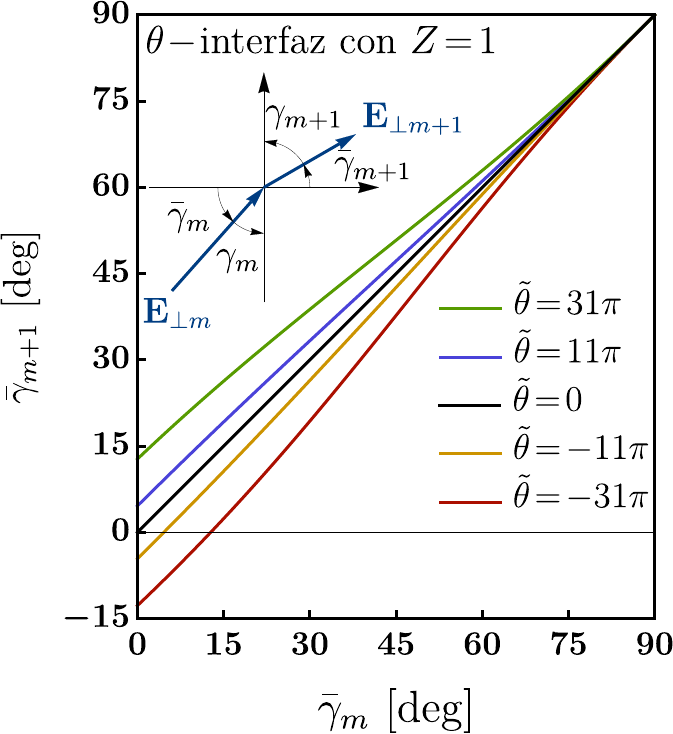}}}
\stackinset{l}{2pt}{t}{3pt}{(b)}{\includegraphics[scale=0.36, frame]{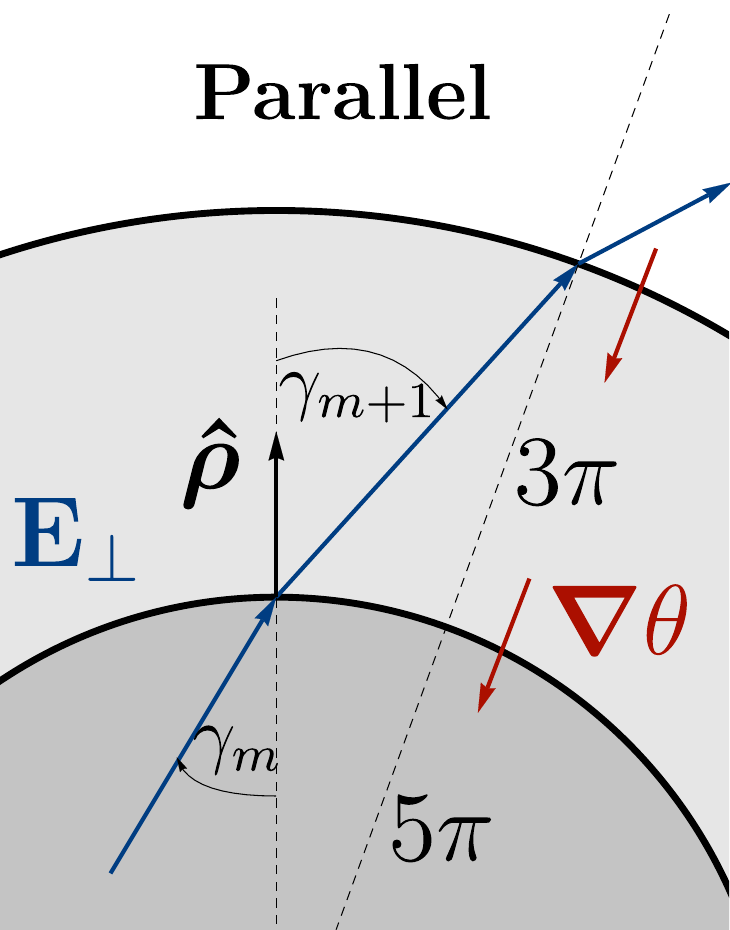}}
\hspace{0.1cm} 
\stackinset{l}{2pt}{t}{3pt}{(c)}{\includegraphics[scale=0.36, frame]{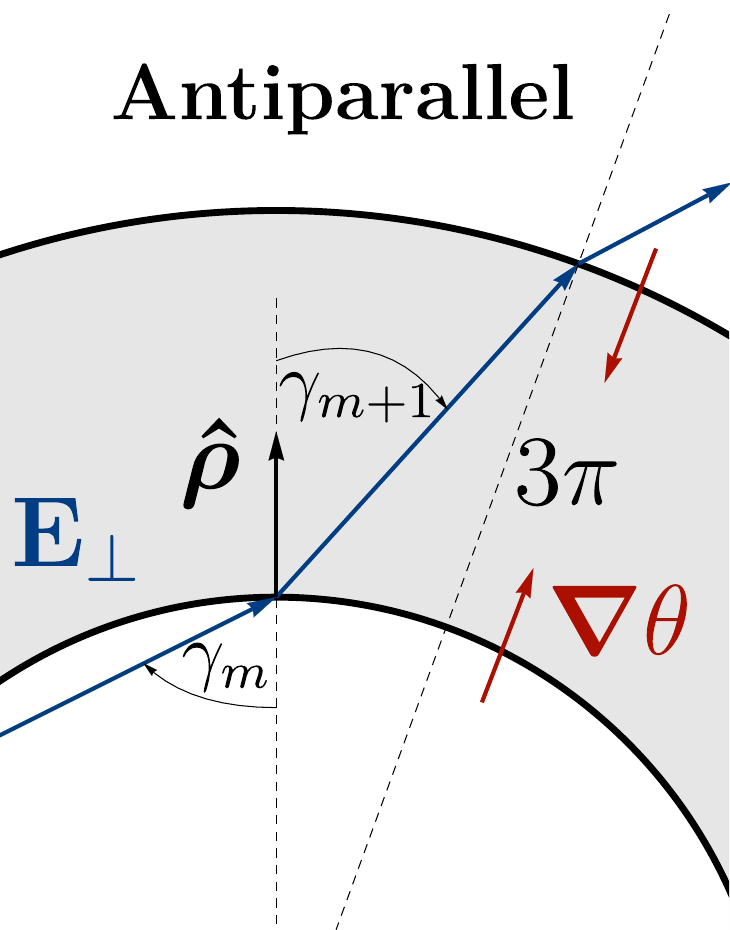}}
\caption{En (a) mostramos el ángulo $\bar{\gamma}_{m+1}$ en función de $\bar{\gamma}_{m}$ para diferentes valores y orientaciones de $\tilde{\theta}$ representado en colores. Los ángulos están definidos en el inset. En (b,c) mostramos una geometría cilíndrica dividida en tres medios en diferentes configuraciones de $\boldsymbol{\nabla}\theta$. El paralelo (b) y el antiparalelo (c) se refieren a las direcciones de $\boldsymbol{\nabla}\theta$ (rojo). Los campos transversales $\mathbf{E}_{\perp}$ se muestran en azul. Cada medio $m$ esta caracterizado por $\epsilon$, $\mu$ y $\theta_m$, por los cuales un OEM se propaga en la dirección $\tongo{z}$.} %.}
\label{FIG:rotacionTEM}
\end{figure}

%------------------------------------------------------
% \begin{figure}[t]
% \stackinset{l}{2pt}{t}{2pt}{(a)}{\includegraphics[scale=0.365,frame]{Capitulos/imagenes/AntiParallelCircular_3.pdf}}
% \stackinset{l}{2pt}{t}{2pt}{(b)}{\includegraphics[scale=0.365,frame]{Capitulos/imagenes/ParallelCircular_3.pdf}}
% \hspace{0.5em}% Añade un espacio horizontal entre las imágenes y la leyenda
% \begin{minipage}[b]{0.4\textwidth} % Define el ancho de la minipage y la alineación
%   \caption{Mostramos una geometría cilíndrica dividida en tres medios, cada uno caracterizado por $\epsilon$, $\mu$ y $\theta_m$. En (a,b) diferentes configuraciones de $\boldsymbol{\nabla}\theta$ se muestran, en los que una OEM se propaga en la dirección $\tongo{z}$. El antiparalelo (a) y el paralelo (b) se refieren a las direcciones de $\boldsymbol{\nabla}\theta$ (rojo). Los campos transversales $\mathbf{E}_{\perp}$ se muestran en azul.}
% \end{minipage}
% \label{FIG:rotacionTEM}
% \end{figure}
%--------------------------------
%
%\textcolor{red}{Revisar signos y hacer algún gráfico}\textcolor{blue}{Signo OK, grafico no estoy seguro de hacerlo}
donde $\gamma_m$ y $\gamma_{m+1}$ son los ángulos entre la normal de la $m$-ésima interfaz y $\mathbf{E}_{\perp}$ a cada lado de ella, y $\bar{\gamma}=\pi/2-\gamma$, es el ángulo complementario, que crece desde el eje horizontal, anti-horario. Esta situación se representa en la Fig. (\ref{FIG:rotacionTEM}.a) donde mostramos $\bar{\gamma}_{f}$ en función de $\bar{\gamma}_{i}$ en grados. Si hay un campo eléctrico $\mathbf{E}_{\perp}$ en el mismo sentido que el $\boldsymbol{\nabla}_{\perp}\theta$, es decir $\mathbf{E}_{\perp}\cdot\boldsymbol{\nabla}_{\perp}\theta>0$, el campo eléctrico en el medio adyacente se \textit{acerca} a la normal. Por el contrario, si $\mathbf{E}_{\perp}\cdot\boldsymbol{\nabla}_{\perp}\theta<0$, se \textit{aleja} de la normal. 

Existe un ángulo critico $\bar{\gamma}_{i}^{*}$ en el cual la componente normal del campo eléctrico del medio adyacente se anula, este efecto ocurre cuando $\tan\bar{\gamma}_{i}^{*}=-Z\tilde{\theta}$, es decir, $\bar{\gamma}_{f}(\bar{\gamma}_{i}^{*})=0$ y el campo eléctrico es puramente tangencial. La relación anterior se puede escribir con las componentes del campo eléctrico, \textit{i.e.} cuando $E_{n,m}=-\tilde{\theta}Z E_{t}\to E_{n,m+1}=0$, donde $E_{n,m}$ es la componente normal del campo eléctrico en el medio $m$ y $E_{t}$ es la componente tangencial continua del campo eléctrico. Cuando $\bar{\gamma}_{f}<0$, corresponde a un cambio de sentido en la componente normal del campo eléctrico al cruzar la $\theta$-interfaz. 

En las Figs. (\ref{FIG:rotacionTEM}.b,c) mostramos la rotación en una geometría cilíndrica para dos $\theta$-interfaces en configuración paralela y antiparalela. Añadir interfaces transversales permite un control de la rotación, en la configuración paralela el efecto es acumulable mientras que en la configuración antiparalela el efecto se anula al cruzar las interfaces.

Esta rotación es una señal interesante del TME que, debido a la naturaleza TEM del campo, difiere radicalmente de los efectos de rotación de Faraday y Kerr usuales, y de las que se han predicho en el contexto de los TI como señales del TME \cite{ahn_theory_2022,shuvaev_giant_2011,shuvaev_room_2013, shuvaev_universal_2022,crosse_theory_2017,crosse_optical_2016, wu_quantized_2016,tse_giant_2010,yang_transmission_2018}. Las diferencias se deben a los siguientes hechos: 
\begin{itemize}
    \item[(i)] La rotación no requiere un campo magnético $\mathbf{B}$ en la dirección de propagación de la OEM.
    \item[(ii)] No ocurre debido a la birrefringencia en la relación de dispersión, ya que esta es idéntica a la de una OEM en un medio homogéneo, $k_{z}=k_{0}n$.
    \item[(iii)] Tampoco ocurre porque la OEM incide desde un medio con, digamos $\theta_1$ sobre otro medio con $\theta_2 \neq \theta_1$. Observe que la propagación es a lo largo de $\boldsymbol{\hat z}$ y los diferentes medios ($\theta$) están separados por interfaces $\rho = R_i= \textrm{const}_i$, donde $z, \rho$ son coordenadas cilíndricas. Todas estas características, (i)-(iii), son típicas de los experimentos/configuraciones de rotación tipo Faraday.
    \item[(iv)] Por otro lado, no se aplica a la onda reflejada en un sentido similar al (iii) anterior, aquí la OEM no se refleja en la interfaz entre dos medios $\theta$ diferentes, de ahí la diferencia con la rotación de Kerr.
\end{itemize}
\section{Propagación TEM en geometría cilíndrica: consideraciones generales}\label{5.3}

Consideremos propagación TEM en una geometría cilíndrica orientado en la dirección $\tongo{z}$ como se muestra en la Fig. (\ref{FIG:rotacionTEM}.b,c). En esta geometría, $\boldsymbol{\nabla}_{\perp}\theta=\sum_{m=1}^{N}\tilde{\theta}_{m}\delta(\rho-R_{m})\tongo{\rho}$, donde $\tilde{\theta}_{m}\equiv\theta_{m+1}-\theta_{m}$ y $\theta_{m}$ siendo el valor del TMEP en el $m$-ésimo medio y $R_{m}$ define la interfaz $\Sigma_{m}$, $N$ es el número de interfaces cilíndricas trasversales. 

Utilizando el método de separación de variable se obtiene la solución de la ecuación de Laplace 2D en coordenadas polares para cada medio separado por una $\theta$-interfaz,
\begin{align}\notag
\Phi_{m}(\rho,\phi)&=A_{0m}+B_{0m}\ln\rho+C_{0m}\phi+D_{0m}\phi\ln\rho\\
+\sum _{\nu=1}^{\infty}&\left (\cos\nu\phi[A_{\nu m}\rho^{\nu}+B_{\nu m}\rho^{-\nu}]+\sin\nu\phi[C_{\nu m}\rho^{\nu}+D_{\nu m}\rho^{-\nu}]\right )\label{eq:SolLaplacePolar},
\end{align}
donde $\nu$ es la constante de separación de variable. Dado que el campo magnético depende del campo eléctrico, $\mathbf{B}_{\perp}=\sqrt{\epsilon\mu}\tongo{z}\times\mathbf{E}_{\perp}$, solo nos enfocaremos en el campo eléctrico transversal. Como $\mathbf{E}_{\perp}=-\boldsymbol{\nabla}_{\perp}\Phi$, tenemos,
\begin{align}\notag
\mathbf{E}_{\perp m}=&-\frac{B_{0m}}{\rho}\boldsymbol{\hat{\rho}}-\sum _{\nu=1}^{\infty}\nu\left (\cos\nu\phi[A_{\nu m}\rho^{\nu-1}-B_{\nu m}\rho^{-\nu-1}]+\sin\nu\phi[C_{\nu m}\rho^{\nu-1}-D_{\nu m}\rho^{-\nu-1}]\right )\boldsymbol{\hat{\rho}}\\
-&\sum_{\nu=1}^{\infty}\nu\left (\cos\nu\phi[C_{\nu m}\rho^{\nu-1}+D_{\nu m}\rho^{-\nu-1}]-\sin\nu\phi[A_{\nu m}\rho^{\nu-1}+B_{\nu m}\rho^{-\nu-1}]\right )\boldsymbol{\hat{\phi}}.
\label{EQ:GeneralElectricFieldcilindricas}
\end{align}
En general, hemos impuesto $D_{0m}=0$, debido a que no produce campos físicos por la dependencia lineal de $\phi$ en los campos. Por otro lado, $C_{0m}=0$, debido a que la componente angular del campo eléctrico tiene que ser continua $C_{0m+1}=C_{0m}$ y finito en $\rho=0$. Para determinar el resto de constantes tenemos que imponer las BCs en cada interfaz cilíndrica que consideremos. Específicamente, en una interfaz arbitraria definida en $\rho=R_{m}$ tenemos que el campo eléctrico satisface,
\begin{align}\label{eq:CBcilindricas}
    E_{\rho m+1}-E_{\rho m}&=Z\tilde{\theta}_{m}E_{\phi m}, & & & E_{\phi m+1}&=E_{\phi m}.
\end{align}
donde $E_{\rho m}$ y $E_{\phi m}$ son las componentes radial y angular del campo eléctrico en el medio $m$ respectivamente. A continuación, estudiaremos diferentes situaciones donde aplicaremos las BCs \eqref{eq:CBcilindricas} y veremos cuales son las restricciones que se deben satisfacer para que exista solución. Si la solución existe, significa que hemos encontrado una solución TEM, lo que constituye un caso concreto donde se ha eludido el teorema de Earnshaw como se ha mencionado en \cite{martin2016electro}. Para estudiar estas soluciones consideraremos los siguientes casos,
\begin{itemize}
    \item[(1)] Cilindro TI de radio $R$ en una OEM de fondo.
    \item[(2)] Cilindro TI hueco de radio interno $R_{1}$ y externo $R_{2}$ en una OEM de fondo.
    \item[(3)] Cilindro TI de radio $R_{1}$ cubierta por una capa cilíndrica coaxial conductora en $R_{2}$.
    %\item Cilindro TI hueco de radio interno $R_{1}$ y externo $R_{2}$.
    %\item Cilindro TI de radio $R$. 
\end{itemize}
Solamente detallaremos el procedimiento para el caso (1), en el resto de los casos, como el procedimiento es el mismo, pondremos los puntos claves y resultados del cálculo. Los casos (1) y (2) los hemos reportados en \cite{Filipini2024polarization}.
\section{Cilindro TI en un onda electromagnética de fondo}\label{5.4}
\begin{SCfigure}[1][t]
\caption{Guía de ondas compuesta por un único TI cilíndrico de radio $R$, en presencia de una OEM externa (o de fondo) con el campo eléctrico orientado en la dirección $y$, \textit{i.e.}, $\mathbf{E}{\textup{ext}}=E{0}e^{i(k_{z}z-\omega t)}\hat{\mathbf{y}}$. El medio circundante es topológicamente trivial, con $\theta_2=0$. Para mantener las BCs a lo largo de toda la guía de ondas, el TI y el medio circundante tienen el mismo índice de refracción, según la Ec.(\ref{eq:RIM}).
\label{FIG:GeoTEM1}}
\includegraphics[scale=0.7]{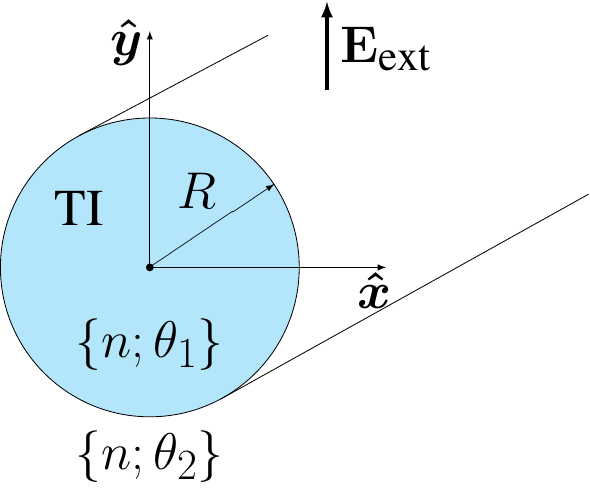}
\end{SCfigure}

Consideremos una OEM de fondo con una polarización definida, por ejemplo,
\begin{align}\label{eq:Efondo}
\mathbf{E}_{\textup{fondo}}&=E_{0}e^{i(k_{z}z-\omega t)}\tongo{y}, & \textup{y}& & \mathbf{B}_{\textup{fondo}}=-\sqrt{\epsilon\mu}E_{0}e^{i(k_{z}z-\omega t)}\tongo{x},
\end{align}
que polariza y magnetiza un cilindro TI de radio $R$ como se muestra en la Fig. (\ref{FIG:GeoTEM1}). En este sistema tenemos una única $\theta$-interfaz, es decir $N=1$, por lo tanto, existen dos soluciones del campo eléctrico, una interior $\mathbf{E}_{\perp 1}$ y otra exterior $\mathbf{E}_{\perp 2}$. La OEM de fondo define el comportamiento asintótico del campo exterior como sigue,
\begin{align}
    \lim_{\rho\to\infty}\mathbf{E}_{\perp 2}(\rho,\phi)=E_{0}\tongo{y}=E_{0}(\sin\phi\boldsymbol{\hat{\rho}}+\cos\phi\boldsymbol{\hat{\phi}}).
    \label{eq:comportamientoasintotico}
\end{align}
%---------------Figure------------
%---------------End Figure---------
De ahora en adelante eliminaremos el subíndice $\perp$. La única solución finita lejos del cilindro TI es para $\nu=1$. Comparando la solución general \eqref{EQ:GeneralElectricFieldcilindricas} con \eqref{eq:comportamientoasintotico} se observa que $A_{12}=0$ y $C_{12}=-E_{0}$ (para una polarización arbitraria las constantes son $A_{12}=-E_{0x}$ y $C_{12}=-E_{0y}$, pero siempre podemos orientar el sistema de referencia de tal manera que la OEM de fondo tenga una polarización en $\tongo{y}$). Cuando $\nu\neq 1$, $A_{\nu2}=C_{\nu2}=0$, que elimina las soluciones no físicas del campo eléctrico en el infinito. Para el cilindro interior, imponemos que sea finito en $\rho=0$ para obtener soluciones físicas, esto implica que $B_{01}=0$, $B_{\nu 1}=0$ y $D_{\nu 1}=0$, $\forall\nu$. 

Aplicando las BCs \eqref{eq:CBcilindricas}, en $\rho=R$ y explotando la ortogononalidad de las funciones trigonométricas, se obtiene el siguiente sistema de ecuaciones algebraicas,
\begin{align}\label{eq:Sistecuaciones1}
    \left(
\begin{array}{ccccc}
 1 & 2 Z_{\theta} & R^{-2} & 0 & 0 \\
 -2 Z_{\theta} & 1 & 0 & R^{-2} & 0 \\
 -R^2 & 0 & 1 & 0 & 0 \\
 0 & -R^2 & 0 & 1 & 0 \\
 0 & 0 & 0 & 0 & 1 \\
\end{array}
\right).\left(
\begin{array}{c}
 A_{11} \\
 C_{11} \\
 B_{12} \\
 D_{12} \\
 B_{02} \\
\end{array}
\right)=E_0\left(
\begin{array}{c}
 0 \\
 -1 \\
 0 \\
 R^2  \\
 0 \\
\end{array}
\right),
\end{align}
donde,
\begin{align}\label{eq:Zetatheta}
    Z_{\theta}\equiv& \frac{Z}{2}\tilde{\theta}, & &\textup{con,} & Z&=\sqrt{\frac{\mu}{\epsilon}} ~~~~\textup{y}~~~~\tilde{\theta}=(\theta_{2}-\theta_{1}).
\end{align}
El medio exterior lo consideramos como un medio trivial, es decir $\theta_{2}=0$. La solución de este sistema algebraico es,
\begin{align}\label{eq:solu1}
    A_{11}=\frac{Z_{\theta}E_{0}}{Z_{\theta}^2+1},~~~C_{11}=-\frac{E_{0}}{Z_{\theta}^2+1},~~~B_{12}=\frac{R^2 Z_{\theta} E_{0}}{Z_{\theta}^2+1},~~~D_{12}=\frac{R^2 Z_{\theta}^2 E_{0}}{Z_{\theta}^2+1}~~~\textup{y}~~~B_{02}=0
\end{align}
Para $\nu\neq 1$ tenemos el siguiente sistema algebraico,
\begin{align}\label{eq:Sistecuaciones2}
    \mathbb{M}.\left(
\begin{array}{c}
 A_{\nu 1} \\
 C_{\nu 1} \\
 B_{\nu 2} \\
 D_{\nu 2} \\
\end{array}
\right)=\left(
\begin{array}{c}
 0 \\
 0 \\
 0 \\
0  \\
\end{array}
\right), & & \textup{con,}&& \mathbb{M}&=\left(
\begin{array}{ccccc}
 1 & 2 Z_{\theta} & R^{-2\nu} & 0  \\
 -2 Z_{\theta} & 1 & 0 &R^{-2\nu}  \\
 -R^{2\nu} & 0 & 1 & 0  \\
 0 & -R^{2\nu} & 0 & 1  \\
\end{array}
\right)
\end{align}
donde una posible solución ocurre cuando el determinante de \eqref{eq:Sistecuaciones2} es igual a cero, \textit{i.e.} $\det(\mathbb{M})=0$. Se verifica que $\det(\mathbb{M})=Z_{\theta}^{2}+1$, por lo tanto, no es posible debido a que las constantes de las Ecs. \eqref{eq:solu1} se indefinen. Entonces, la única solución es $A_{\nu 1}=C_{\nu 1}=B_{\nu 2}=D_{\nu 2}=0$. De esta forma, los campos TEM totales que resuelven las ecuaciones de la $\theta$-ED junto con las BCs en cada medio, para $m=1,2$  son,
\begin{align}
\mathbf{E}_{m}^{\textup{tot}}=E_{0}\tongo{y}+E_{0}\mathbf{E}_{m}^{\theta},
\end{align}
donde,
\begin{align}
\mathbf{E}_{1}^{\theta}&=-\frac{Z_{\theta}}{(1+Z_{\theta}^{2})}[\tongo{x}+ Z_{\theta}\tongo{y}],
\label{eq:Etheta1}\\\
\mathbf{E}_{2}^{\theta}&=\frac{Z_{\theta}}{(1+Z_{\theta}^{2})}\left ( \frac{R}{\rho} \right )^{2} [ ( Z_\theta  \sin\phi+\cos\phi)\boldsymbol{\hat{\rho}}+\!(\sin\phi-\!Z_\theta  \cos\phi)\boldsymbol{\hat{\phi}}].
\label{eq:Etheta2}
\end{align}
Notemos que si $\tilde \theta = 0$, significa que no hay interfaz y las soluciones interior y exterior son idénticas a $E_{0}e^{i(kz-\omega t)}\, \tongo{y}$ como deben ser. Si $\tilde \theta \neq 0$ y $E_0\neq 0$ el campo total no es trivial, pero si $E_0 = 0$ no hay solución alguna. Por lo tanto, nuestra solución depende del campo de fondo. Pero, si $\tilde \theta \neq 0$ y $E_0\neq 0$, existe una solución TEM total en todo el espacio que no se puede obtener con materiales totalmente dieléctricos o de otro modo, y adquiere características nuevas y no triviales que son atribuibles solo a $\theta$, lo que constituye a nuevas señales del efecto TME. Es interesante notar que nuestra solucione TEM son independientes de la frecuencia de oscilación $\omega$.
%---------------------------------------------------------
\begin{figure}[t!]
\begin{minipage}{0.43\columnwidth}
\begin{subfigure}{\textwidth}
\stackinset{l}{20pt}{t}{12pt}{(a)}{\includegraphics[width=0.9\textwidth]{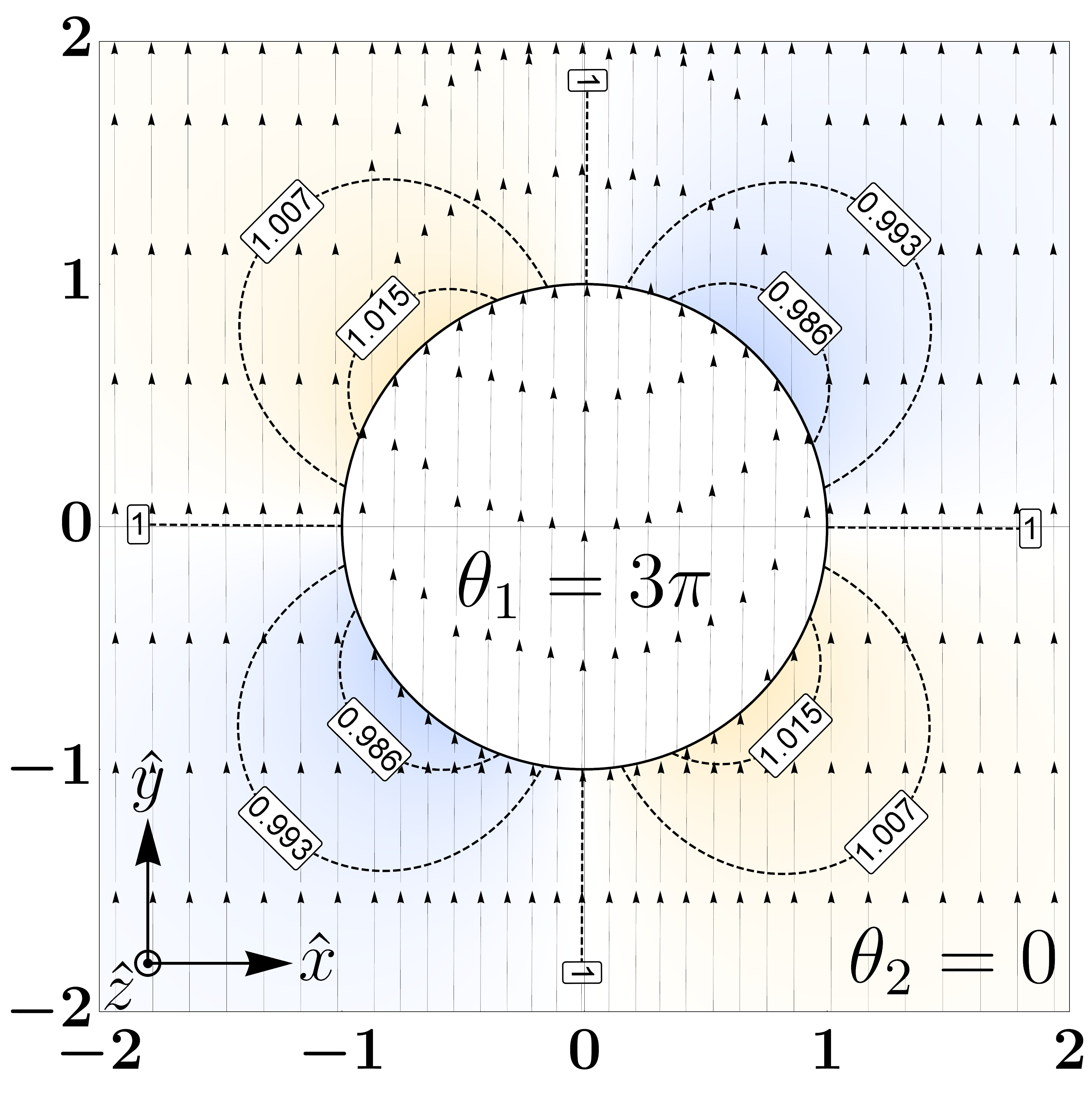}}
%\caption{}
\end{subfigure}
\begin{subfigure}{\textwidth}
\stackinset{l}{20pt}{t}{12pt}{(b)}{\includegraphics[width=0.9\textwidth]{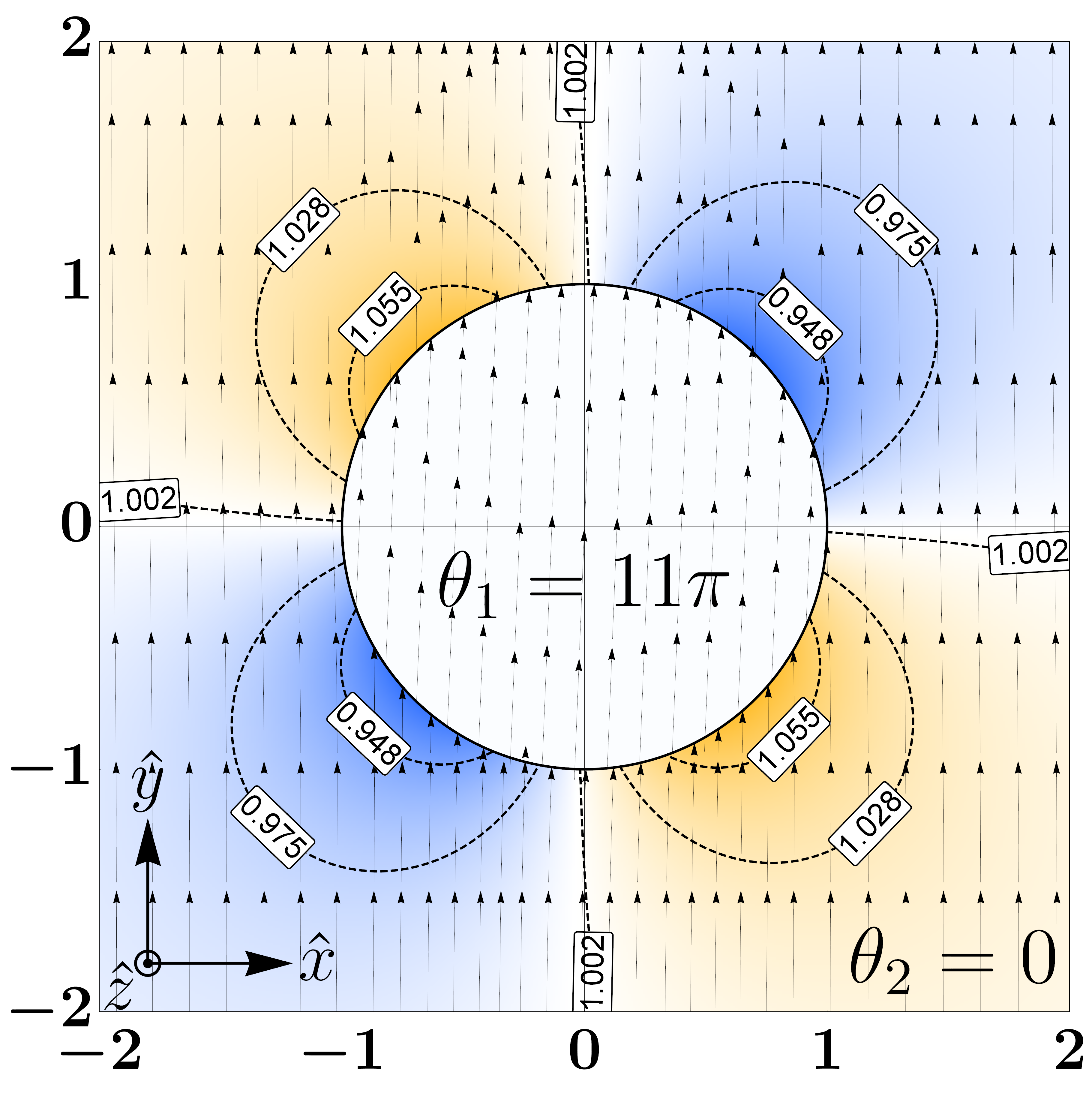}}
%\caption{}
\end{subfigure}
\end{minipage}
\hspace{0.1mm}
% \hfill
\begin{minipage}{0.43\columnwidth}
\begin{subfigure}{\textwidth}
\stackinset{l}{20pt}{t}{12pt}{(c)}{\includegraphics[width=0.9\textwidth]{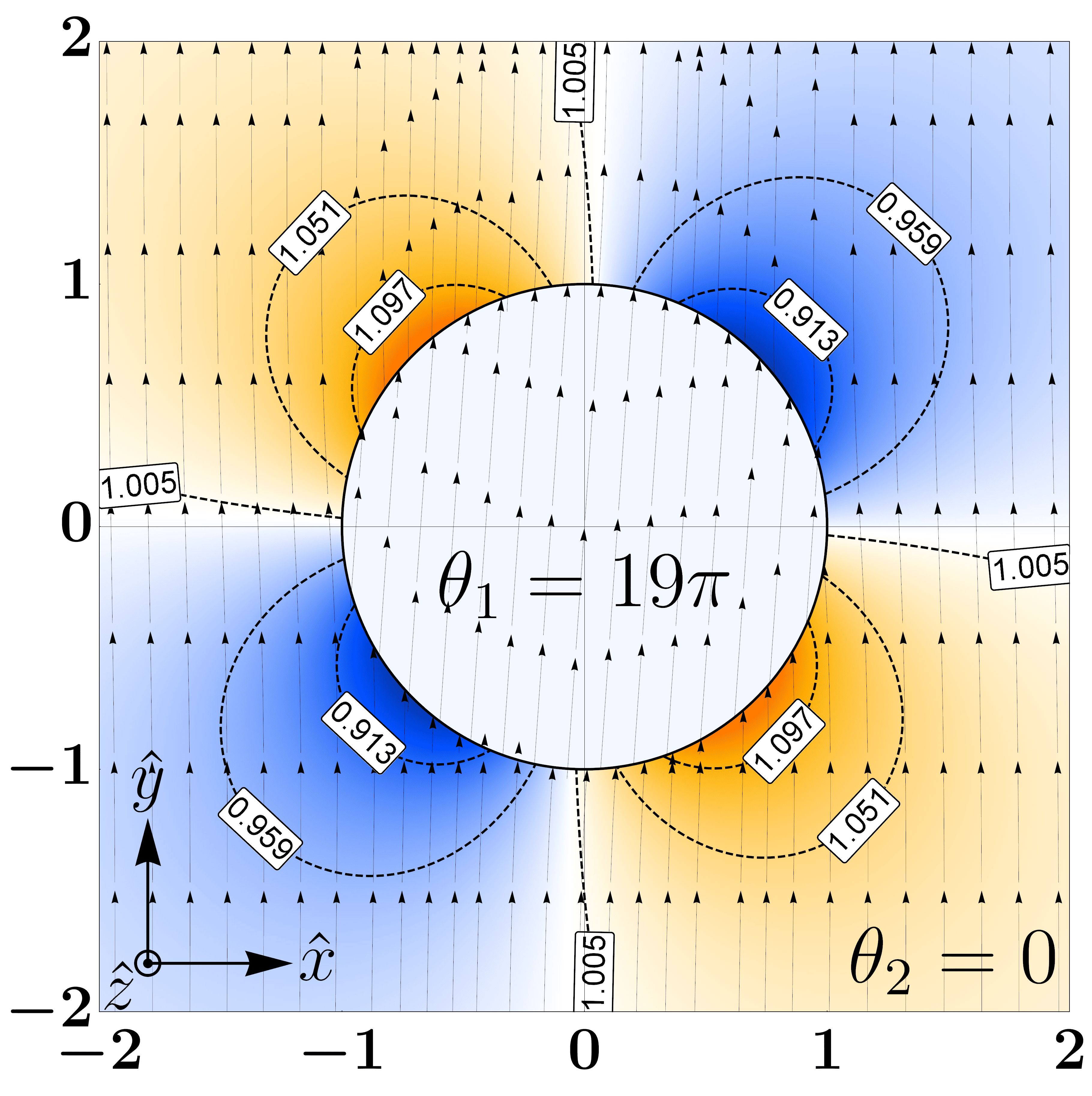}}
%\caption{}
\end{subfigure}
\begin{subfigure}{\textwidth}
\stackinset{l}{20pt}{t}{12pt}{(d)}{\includegraphics[width=0.9\textwidth]{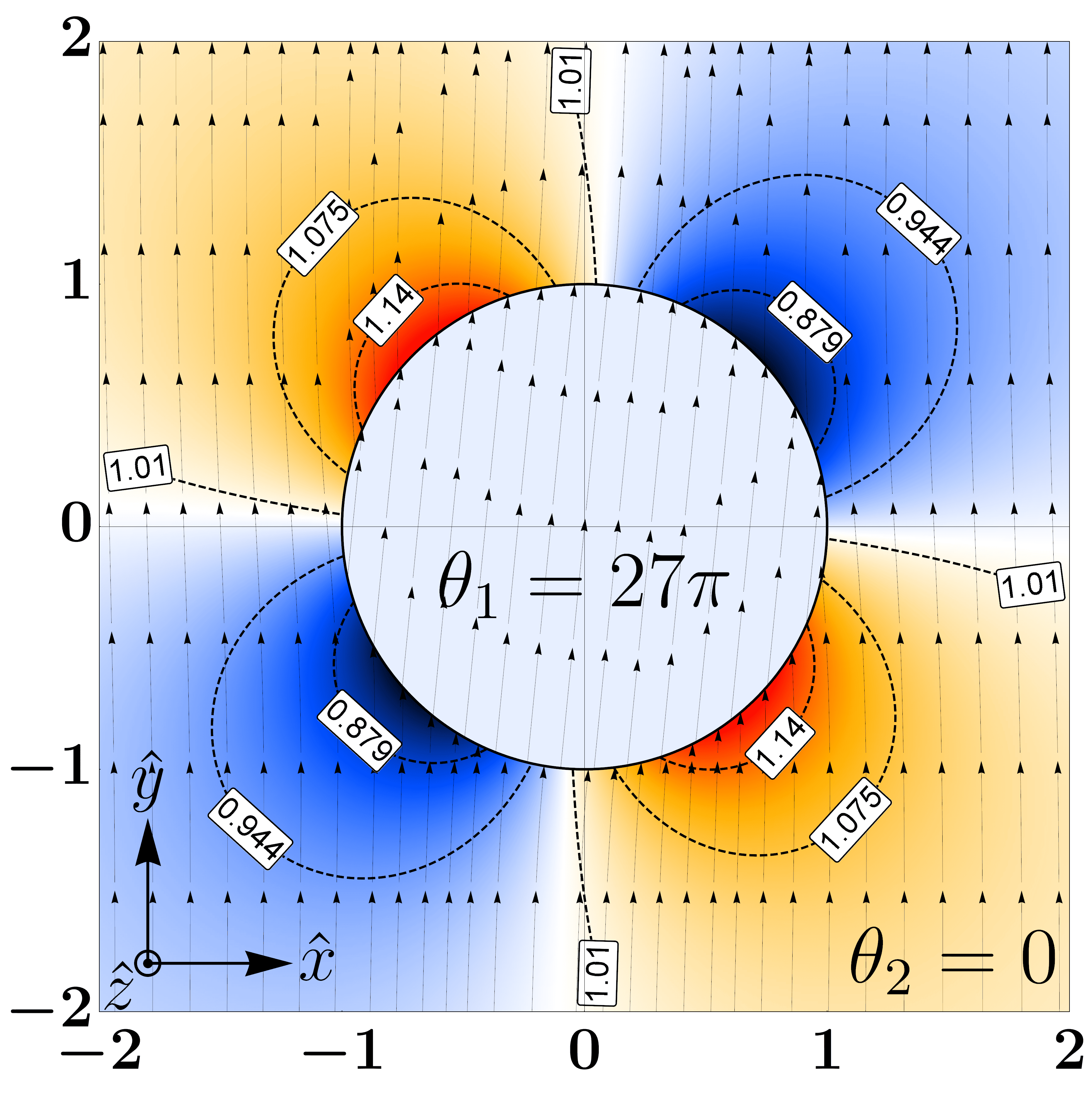}}
%\caption{}
\end{subfigure}
\end{minipage}
% \hspace{0.1mm}
% \hfill
\begin{minipage}{0.08\columnwidth}
\includegraphics[scale=0.35]{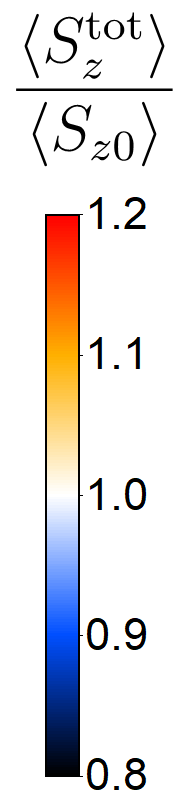}
\end{minipage}
\caption{Densidad del vector de Poynting relativo en las regiones interior y exterior en relación con el del campo EM de fondo. Las \textit{streamlines} son del campo $\mathbf{E}_{\perp}^{\textup{tot}}/E_{0}$. Los \textit{contourplot} muestran contornos de un vector de Poynting relativo constante. Los paneles (a-d) corresponden a $Z=1$ y $\theta_{1} =3\pi, 11\pi, 19\pi, 27\pi$, respectivamente.
}
\label{FIG:DensityPlotEn1} 
\end{figure}
%---------------------------------------------------------

Calcularemos ahora el promedio temporal del vector de Poynting, dada por la parte real de la Ec. \eqref{eq:Poyntingcomplejo}, en cada medio es $\langle \mathbf{S}_{m}^{\textup{tot}}\rangle=\langle S_{0}\rangle\hat{z}+\langle S_{0}\rangle\langle \mathbf{S}_{m}^{\theta}\rangle$, donde,
\begin{align}
    \langle \mathbf{S}_{1}^{\theta}\rangle&=-\frac{Z_{\theta}^{2}}{(1+Z_{\theta}^{2})}\tongo{z},\\
    \langle \mathbf{S}_{2}^{\theta}\rangle&=\frac{R^{2}Z_{\theta}}{(1+Z_{\theta}^{2})}\left [ \frac{R^{2}Z_{\theta}}{\rho^{4}}+\frac{2}{\rho^{2}}(\sin 2\phi-Z_{\theta}\cos 2\phi) \right ]\tongo{z},
\end{align}
donde $\langle S_{0}\rangle=c E_{0}^{2}/8\pi Z$ es el promedio del vector de Poynting de la OEM de fondo. En las Figs. (\ref{FIG:DensityPlotEn1}) mostramos las \textit{streamlines} de $\mathbf{E}_{m}^{\textup{tot}}/E_{0}$, la distribución espacial de $\langle \mathbf{S}_{m}^{\textup{tot}}\rangle/\langle S_{0}\rangle$ y contornos constantes del promedio temporal del vector de Poynting para diferentes valores de $\theta_{1}$. En todos lados ponemos $Z=1$ centrarnos en la contribución a las respuestas magnetoeléctricas de los TIs debida a $\theta$, frente a la de las ``propiedades ópticas'' ($\epsilon$ y $\mu$), establecer $\epsilon=1, \mu=1$ no es nuevo y ya ha sido considerado en la literatura, ver por ejemplo, comentarios antes de la Ec. (49) en \cite{crosse_electromagnetic_2015}. 

\begin{figure}[t]
    \centering
    \includegraphics[scale=0.6]{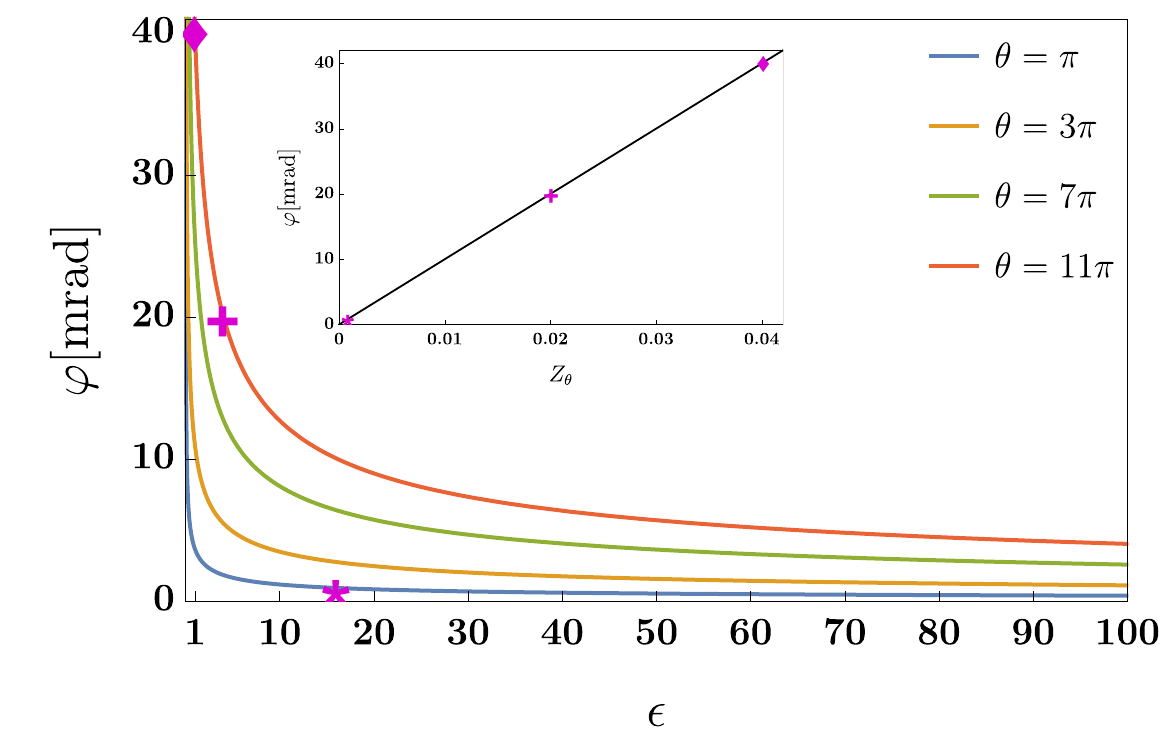}
    \caption{El asterisco ($\ast$) representa  Bi$_2$Se$_3$ con $\theta_{\textup{TI}} = \pi$ y $\epsilon_{\textup{TI}} =16$ según \cite{crosse_theory_2017}. 
    La suma ($+$) es TlBiSe$_2$ con $\theta_{\textup{TI}} = 11 \pi$ y $\epsilon_{\textup{TI}} =4$ según \cite{franca_modification_2022},
    y el diamante ($\blacklozenge$) representa uno de los casos de ``vacío'' con $\theta_{\textup{TI}}= 11 \pi$ y $\epsilon_{\textup{TI}} =1$.}
    \label{FIG:rotpol}
\end{figure}

El campo eléctrico dentro del TI, $\mathbf E^{\textup{tot}}_{1}$ es uniforme, y, debido a la Ec. (\ref{EQ:refracangle}), la polarización rota una cantidad fija $\varphi_{\textup{int}}$. Como era de esperar, la magnitud del efecto es mínima; sin embargo, para valores realistas del parámetro TMEP, la rotación del plano de polarización se puede medir con las técnicas actuales. Esta rotación esta dado por,
\begin{align}
    \cos \varphi \equiv \tongo{\mathbf{E}}_{1} \cdot \tongo{y} = (1 + Z^2_\theta )^{-1/2},
\end{align}
Para $Z=1$ y $\theta_{1}= 3 \pi, 11 \pi, 19 \pi$ y $27 \pi$ respectivamente, esto implica una rotación del plano de polarización de $0.63$, $2.30$, $3.97$ y $5.63$ grados, respectivamente, que: se debe enteramente al TMEP del TI, difiere de las rotaciones de Faraday o Kerr, y está dentro de la sensibilidad experimental actual. En la Fig. (\ref{FIG:rotpol}) se muestra la rotación de la polarización $\varphi$ en función de $\epsilon_{\textup{TI}}$ para diferentes valores de $\theta_{\textup{TI}}$. En la gráfica hemos marcado algunos valores de $\epsilon_{\textup{TI}}$ reportados en la literatura para algunos TIs. En el gráfico insertado de la Fig. (\ref{FIG:rotpol}) se muestra $\varphi$ en función de $Z_{\theta}$, el cual corresponde a una combinación de los parámetros de TI, como se define en la Ec. \eqref{eq:Zetatheta}.

Esta es una predicción novedosa que conduce a una nueva forma de observar el TME, y es una consecuencia de soluciones exactas de ondas TEM que no habían sido posibles hasta ahora con materiales dieléctricos convencionales. Además, son de relevancia en diversos contextos \cite{ibanescu_all-dielectric_2000, nordebo_dispersion_2014,catrysse_transverse_2011 ,shvedov_topological_2017, zhou_surface_2023}, por lo que son interesante buscarlas.
A continuación, analizaremos las cargas topológicas inducidas en la superficie. Esta perspectiva enriquecerá y facilitará la explicación de los efectos observados.
\subsection{Cargas topológicas inducidas}\label{Sec:CargasTopoInducidas}
La distribución de campo $\mathbf{E}_{m}^{\theta}$ y la respectiva rotación de la polarización puede entenderse de manera autoconsistente, orden por orden en $\theta$, en términos de las densidades de carga superficial topológica inducida. Este proceso infinito de ``auto-inducción'' se comentó en la Sec. (\ref{3.4}).
%La explicación física de la rotación del campo EM que reportamos, es debido a una compleja interacción entre las cargas y corrientes topológicas inducidas en la superficie del TI con los campos EM que se generan debido a las mismas en un proceso infinito de ``autoinduccion'' como se comentó en la sección (?). 
En esta sección demostraremos que este proceso converge a las Ecs. \eqref{eq:Etheta1} y \eqref{eq:Etheta2} a todo orden en $\theta$.

Consideremos el sistema estudiado en la sección anterior, donde inicialmente existe un campo eléctrico oscilante de orden cero en $\theta$, con una polarización definida en la dirección $\tongo{y}$\footnote{Podemos imaginar que este campo eléctrico es creado por placas paralela conectado a una diferencia de potencial variable en el tiempo y se encuentra orientada de tal forma que genera la polarización lineal/circular deseada.},
por consecuencia, el campo eléctrico es acompañado por un campo magnético siguiendo las Ecs. de Maxwell como en la Eq. \eqref{eq:Efondo}. 
%Por el bien de esta discusión, el campo electrico de fondo lo llamamos $\mathbf{E}^{\theta(n=0)}$. 
Si consideramos un cilindro TI embebido en este campo externo notamos que el campo magnético que es perpendicular a la superficie induce una carga topológica como sigue,
\begin{align}
    \sigma_{\theta}^{\theta(n=1)}=-\frac{1}{4\pi}\tilde{\theta}\,\mathbf{B}_{\textup{fondo}}\cdot\boldsymbol{\hat{\rho}}=\frac{\sqrt{\epsilon\mu}}{4\pi}\tilde{\theta}\,\mathbf{E}_{\textup{fondo}}\cdot\boldsymbol{\hat{\phi}}=\frac{\sqrt{\epsilon\mu}}{4\pi}\tilde{\theta}\,E_{0}\cos\phi e^{i(kz-\omega t)}
\end{align}
donde $\theta(n=1)$ significa que es a primer orden en $\theta$. Similarmente el campo eléctrico angular induce una corriente topológica de Hall $\mathbf{K}_{\theta}$ en la dirección $\tongo{z}$. De hecho la carga superficial y las densidades de corriente, $\mathbf{K}_{\theta}=c\sigma_{\theta}/\!\sqrt{\epsilon\mu}\,\tongo{z}$, satisfacen una ecuación de continuidad nos referiremos principalmente a $\sigma_{\theta}$.
%La $\sigma_{\theta}$ por su parte genera un campo eléctrico, que se determina con la expansión en polinomios del potencial antes utilizadas.
Como $\tilde{\theta}=(0-\theta_{1})<0$, las cargas positivas/negativas inducidas se generan en cerca de $\phi=\pi,0$ respectivamente. Dentro y fuera del TI, se induce un campo eléctrico de primer orden que lo llamaremos $\mathbf{E}_{m}^{\theta(n=1)}$. 

Ahora el enfoque es completamente distinto, tenemos una $\sigma_{\theta}$ en una la superficie y queremos encontrar el campo eléctrico $\mathbf{E}_{m}^{\theta(n=1)}$ que genera, dado que $\sigma_{\theta}$ es fuente de discontinuidad del campo eléctrico. Las soluciones del campo eléctrico dentro y fuera del cilindro son,
\begin{align}\label{eq:E_general_1}
\mathbf{E}_{1}^{\theta(n=1)}=&-\left (A\cos\phi+C\sin\phi\right )\boldsymbol{\hat{\rho}}-\left (C\cos\phi-A\sin\phi\right )\boldsymbol{\hat{\phi}}=-A\tongo{x}-C\tongo{y}\\\label{eq:E_general_2}
\mathbf{E}_{2}^{\theta(n=1)}=&\frac{R^{2}}{\rho^{2}}\left [(A\cos\phi +C\sin\phi)\boldsymbol{\hat{\rho}}-(C\cos\phi -A\sin\phi)\boldsymbol{\hat{\phi}}  \right ].
\end{align}
Estos campos satisfacen la condición de continuidad de la componente angular en la superficie y son soluciones generales sin especificar aún cuál es la distribución de carga en la superficie. La ecuación de la discontinuidad del campo eléctrico debido a la densidad de carga eléctrica superficial viene dada por,
\begin{align}
    \epsilon(\mathbf{E}_{2}^{\theta(n=1)}-\mathbf{E}_{1}^{\theta(n=1)})|_{R}\cdot \boldsymbol{\hat{\rho}}=4\pi\sigma_{\theta}^{\theta(n=1)}
\end{align}
Reemplazando, encontramos que los coeficientes son $A=Z_{\theta}E_{0}$ y $C=0$. Tenemos entonces que el campo total será el que existía de fondo más el generado por $\sigma_{\theta}$, \textit{i.e.}
\begin{align}
\mathbf{E}_{1}^{\textup{tot}}=&E_{0}\tongo{y}-Z_{\theta}E_{0}\tongo{x},\\
\mathbf{E}_{2}^{\textup{tot}}=&E_{0}\tongo{y}+\frac{R^{2}}{\rho^{2}}Z_{\theta}E_{0}\left [\cos\phi \boldsymbol{\hat{\rho}}+\sin\phi\boldsymbol{\hat{\phi}}  \right ].
\end{align}
Debido a que estamos considerando que el medio exterior es topológicamente trivial, $\theta_{2}=0$, entonces $Z_{\theta}<0$, y el campo inducido al interior del TI va en la dirección $\tongo{x}$. 

Los términos inducidos oscilan al unísono con el campo de fondo, por lo que el campo eléctrico de primer orden genera un campo magnético de primer orden $\mathbf{B}^{\theta(n=1)}_{m}$ que apunta en la dirección $\tongo{y} $. Hasta el primer orden en $\theta$ no hay más efectos. Sin embargo, el campo magnético de primer orden, induce densidades de carga superficial de segundo orden. Esto da lugar a un campo eléctrico de segundo orden $\mathbf{E}^{\theta(n=2)}_{m}$ paralelo a $-\tongo{y}$ y así sucesivamente.
%Consideraremos que este proceso de generar una densidad de carga y formar el correspondiente campo eléctrico ocurre instantáneamente, o podemos pensar en desplazamientos virtuales $\delta t$. Este cambio virtual del campo eléctrico induce el correspondiente campo magnético siguiendo las Ecs. de Maxwell, el cual nuevamente induce una segunda distribución de cargas eléctricas topológicas en la superficie. 
Este proceso continua indefinidamente y por lo tanto 
%el campo eléctrico total corresponde a la suma de todas las contribuciones de los campos eléctricos a todos los ordenes. 
el campo eléctrico total $\mathbf{E}^{\textup{tot}}_{m}$ resulta de la superposición del campo de fondo $E_{0} \tongo{y}$ más el campo eléctrico $\mathbf{E}_{m}^\theta$ atribuido a la densidad de carga superficial topológica, donde $m$ corresponde al campo en los medios $m=\{1,2\}$. Este campo eléctrico inducido se puede escribir como una suma orden por orden como $\mathbf{E}^{\theta}_{m} = \sum_{n=1}^{\infty} \mathbf{E}^{\theta (n)}_{m}$,
Esto se expresa como sigue,
\begin{align}
\mathbf{E}_{m}^{\textup{tot}}=E_{0}\tongo{y}+\sum_{n=1}^{\infty}\mathbf{E}_{m}^{\theta(n)},
\end{align}
Los campos a todos los ordenes tienen la forma genérica de las Ecs. \eqref{eq:E_general_1} y \eqref{eq:E_general_2}. Reemplazando en la ecuación anterior obtenemos,
\begin{align}\label{eq:CampoEGeneralPolares1}
\mathbf{E}_{1}^{\textup{tot}}=&E_{0}\tongo{y}-\sum_{n=1}^{\infty}(A_{n}\tongo{x}+C_{n}\tongo{y})\\\label{eq:CampoEGeneralPolares2}
\mathbf{E}_{2}^{\textup{tot}}=&E_{0}\tongo{y}+\frac{R^{2}}{\rho^{2}}\sum_{n=1}^{\infty}\left [(A_{n}\cos\phi +C_{n}\sin\phi)\boldsymbol{\hat{\rho}}-(C_{n}\cos\phi -A_{n}\sin\phi)\boldsymbol{\hat{\phi}}  \right ].
\end{align}
%
%Con un poco de suerte estas series deberían converger. Demostramos que las constantes $A_{n}$ y  $C_{n}$ dependen de las contantes del campo eléctrico a un orden menor $A_{n-1}$ y $C_{n-1}$ como sigue,
Debido a que los campos eléctricos de orden $n+1$ depende del campo magnético a orden $n$ a través de $\sigma_{\theta}^{\theta(n)}$, se puede demostrar que las constantes $A_{n}$ y  $C_{n}$ dependen de las contantes a un orden menor $A_{n-1}$ y $C_{n-1}$ como sigue,
\begin{align}
    A_{n}&=-Z_{\theta}C_{n-1}, & y,& & C_{n}&=Z_{\theta}A_{n-1}.
\end{align}
De esta forma las series quedan,
\begin{align}
\sum_{n=1}^{\infty}A_{n}&=A_{1}+A_{2}+A_{3}+\dots=A_{1}-Z_{\theta}C_{1}-Z_{\theta}C_{2}-\dots=A_{1}-Z_{\theta}\sum_{n=1}^{\infty}C_{n},\\
\sum_{n=1}^{\infty}C_{n}&=C_{1}+C_{2}+C_{3}+\dots=C_{1}+Z_{\theta}A_{1}+Z_{\theta}A_{2}+\dots=C_{1}+Z_{\theta}\sum_{i=1}^{\infty}A_{n}.
\end{align}
Notamos que las expresiones están acopladas y al desacoplar se obtiene,
\begin{align}
    \sum_{n=1}^{\infty}A_{n}&=\frac{A_{1}-Z_{\theta}C_{1}}{1+Z_{\theta}^{2}}=\frac{Z_{\theta}}{1+Z_{\theta}^{2}}E_{0},\\
    \sum_{n=1}^{\infty}C_{n}&=\frac{C_{1}+Z_{\theta}A_{1}}{1+Z_{\theta}^{2}}=\frac{Z_{\theta}^{2}}{1+Z_{\theta}^{2}}E_{0}.
\end{align}
Al reemplazar en las Ecs. \eqref{eq:CampoEGeneralPolares1} y \eqref{eq:CampoEGeneralPolares2} da como resultado los campos eléctricos totales,
\begin{align}
\mathbf{E}_{1}^{\textup{tot}}=&E_{0}\tongo{y}-E_{0}\frac{Z_{\theta}}{(1+Z_{\theta}^{2})}[\tongo{x}+Z_{\theta}\tongo{y}],
\label{eq:Ethetatot1}\\
\mathbf{E}_{2}^{\textup{tot}}=&E_{0}\tongo{y}+E_{0}\frac{Z_{\theta}}{(1+Z_{\theta}^{2})}\left ( \frac{R}{\rho} \right )^{2}\left [(Z_{\theta}\sin\phi+\cos\phi)\boldsymbol{\hat{\rho}}+(\sin\phi-Z_{\theta}\cos\phi)\boldsymbol{\hat{\phi}}  \right ].
\label{eq:Ethetatot2}
\end{align}
Estos campos son exactamente iguales a las ecuaciones \eqref{eq:Etheta1} y \eqref{eq:Etheta2}. La contribución de la densidad de carga topológica a todo orden corresponde a la suma infinita convergente siguiente,
\begin{align}\label{eq:Densidad_de_carga_inducida}
\sigma_{\theta}^{\textup{tot}}=\sum_{n=1}^{\infty}\sigma_{\theta}^{(n)}=&E_{0}\frac{\epsilon}{2\pi}\frac{Z_{\theta}}{(1+Z_{\theta}^{2})} (\cos\phi+Z_{\theta}\sin\phi),
\end{align}

\begin{figure}[t!]
\begin{minipage}{0.43\columnwidth}
\begin{subfigure}{\textwidth}
\stackinset{l}{20pt}{t}{12pt}{(a)}{\includegraphics[width=0.95\textwidth]{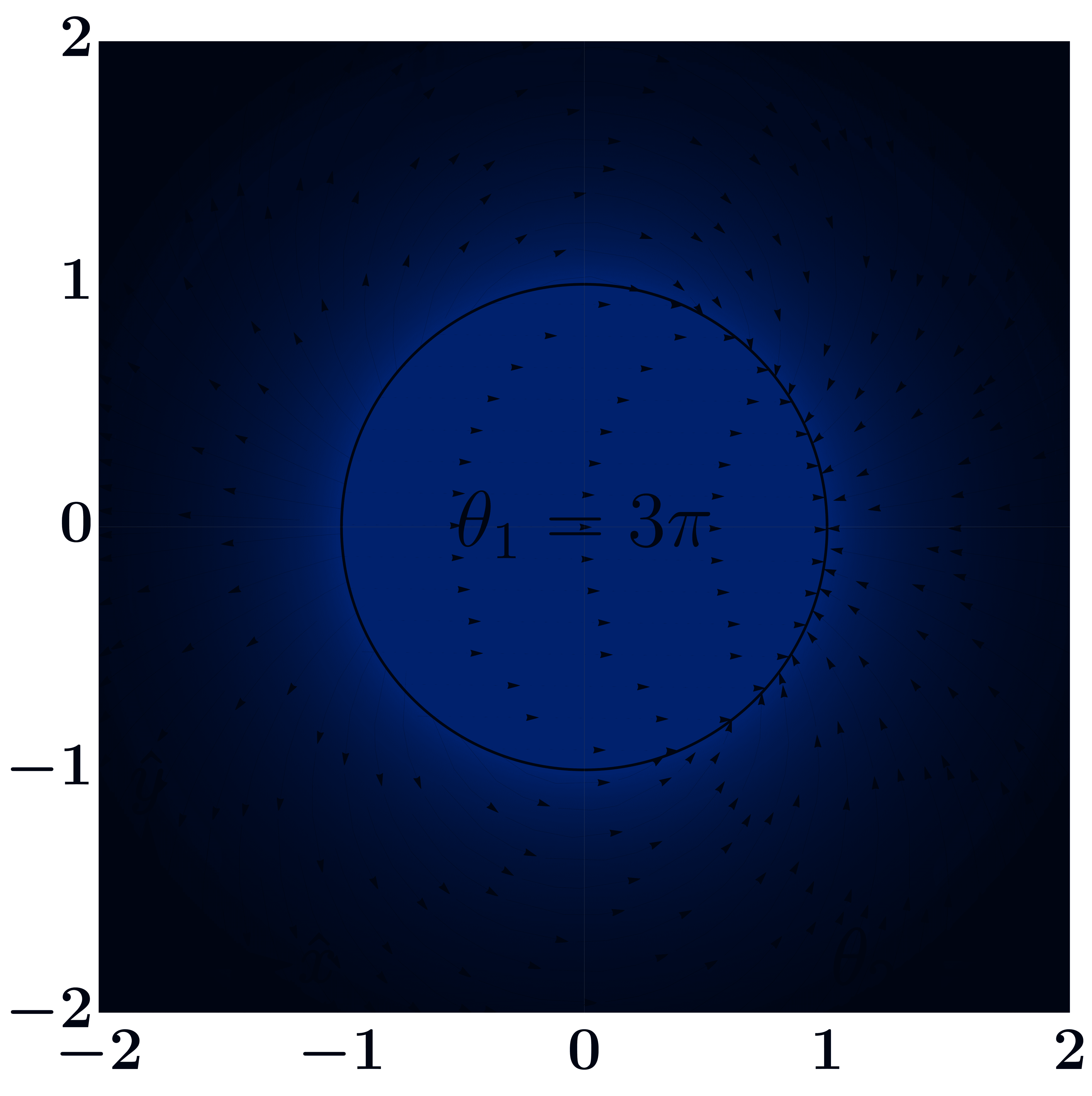}}
%\caption{}
\end{subfigure}
\begin{subfigure}{\textwidth}
\stackinset{l}{20pt}{t}{12pt}{(b)}{\includegraphics[width=0.95\textwidth]{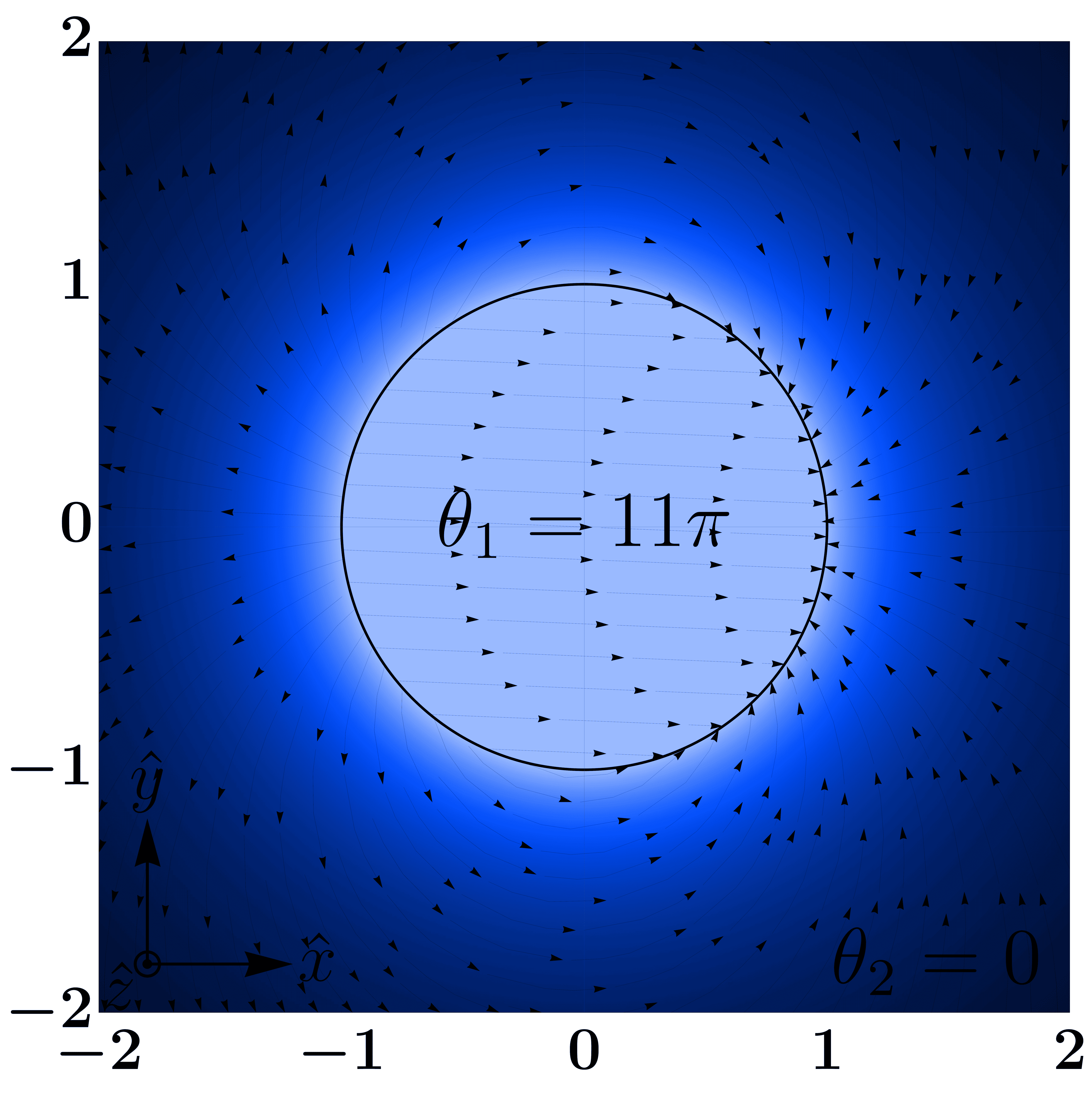}}
%\caption{}
\end{subfigure}
\end{minipage}
\hspace{0.1mm}
% \hfill
\begin{minipage}{0.43\columnwidth}
\begin{subfigure}{\textwidth}
\stackinset{l}{20pt}{t}{12pt}{(c)}{\includegraphics[width=0.95\textwidth]{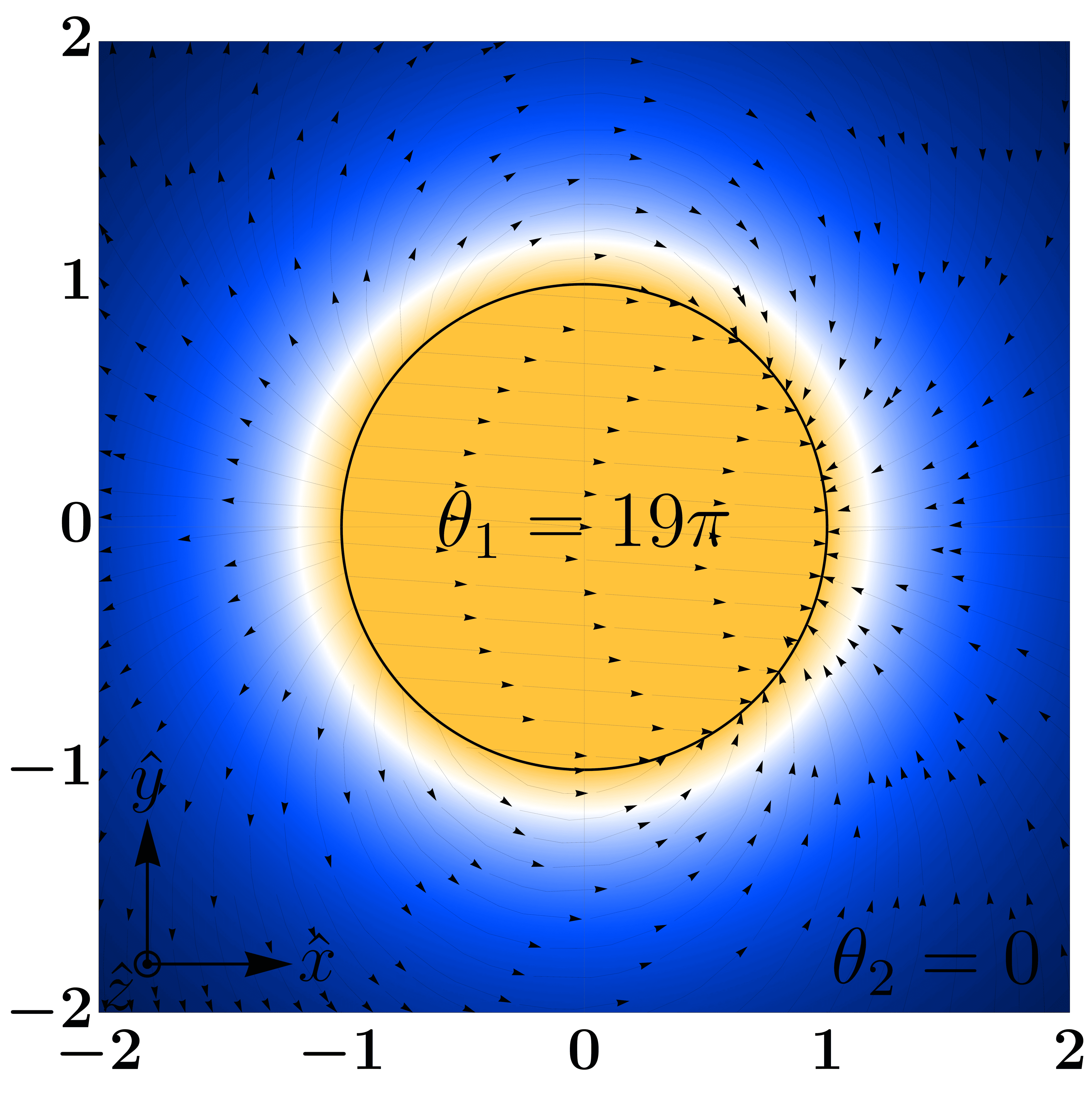}}
%\caption{}
\end{subfigure}
\begin{subfigure}{\textwidth}
\stackinset{l}{20pt}{t}{12pt}{(d)}{\includegraphics[width=0.95\textwidth]{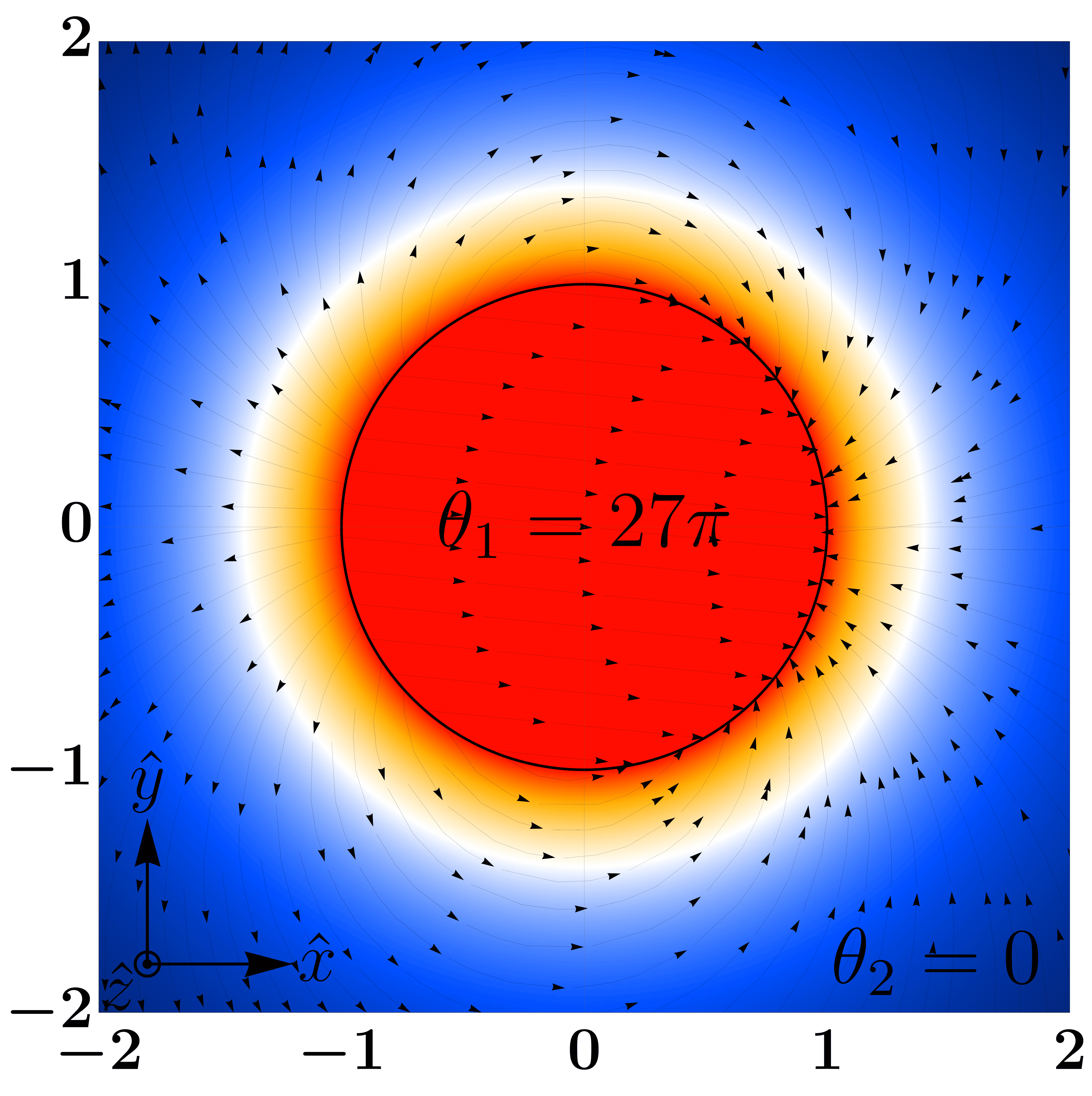}}
%\caption{}
\end{subfigure}
\end{minipage}
% \hspace{0.1mm}
% \hfill
\begin{minipage}{0.08\columnwidth}
\includegraphics[scale=0.12]{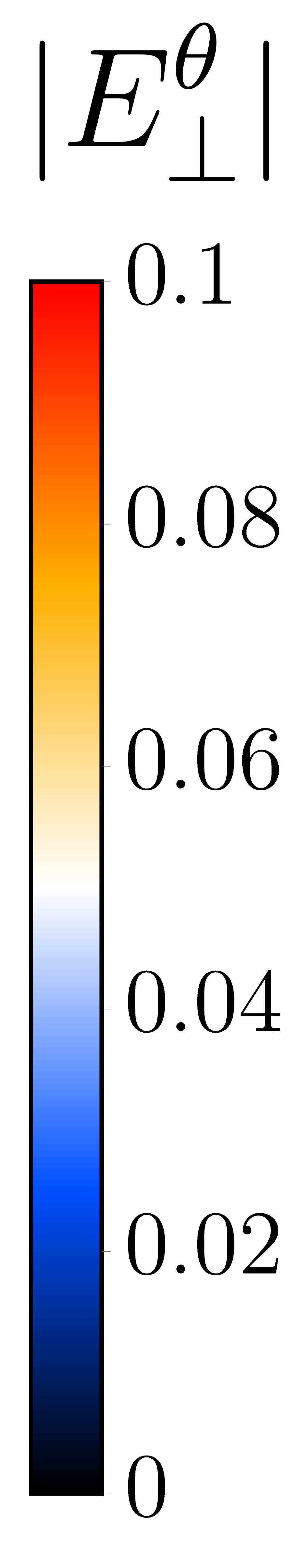}
\end{minipage}
\caption{Campos eléctricos generados por la densidad de cargas topológicas en la superficie del TI \eqref{eq:Densidad_de_carga_inducida}. Dentro del TI el efecto es constante y el campo eléctrico se dirige aproximadamente en la dirección del campo magnético de fondo. Lejos de la superficie el campo eléctrico se aproxima a la de un dipolo y decae como $\rho^{-2}$. Los paneles (a-d) corresponden a $Z=1$ y $\theta =3\pi, 11\pi, 19\pi, 27\pi$, respectivamente.
}
\label{FIG:Ethetan1} 
\end{figure}

En resumen, la rotación de la polarización de la OEM de fondo en un material TI surge de la compleja interacción entre la OEM de fondo y las cargas/corrientes eléctricas topológicas inducidas en la superficie del TI. En la Figs. (\ref{FIG:Ethetan1}) se muestra el campo eléctrico inducido por las carga topológica en la superficie de TI dada por la Eq. \eqref{eq:Densidad_de_carga_inducida}, que más bien, corresponde al incremento con respecto a la OEM de fondo. El campo $\mathbf{E}^{\theta}_{2}$ decae rápidamente como $\rho^{-2}$ en el medio exterior, aun así, se puede definir una \textit{skin depth} $\delta$, el cual corresponde a la distancia donde la amplitud promedio del campo eléctrico decae un factor de $1/e$ (un $37\%$ del valor de la superficie). El skin depth de este sistema esta dado por,
\begin{align}\label{eq:skindepth}
    \delta = \sqrt{e}R\approx 1.65 R,
\end{align}
el cual es independiente del valor de $\theta$ o de las propiedades del medio. Por ejemplo, si tenemos un cilindro TI de radio $R=1\,\mu m$, los efecto $\theta$ estarán \textit{confinados} alrededor de $1.65\,\mu m$ desde la superficie hacia afuera.
%---------------------------------------------------------
%---------------------------------------------------------

\subsection{Potencia transportada}
La potencia transportada por la OEM a lo largo de la guía es un parámetro físico fundamental que cuantifica la eficiencia de la guía de ondas. Para calcular esta potencia, utilizamos la definición integral de la potencia dada por la siguiente ecuación, 
\begin{align}\label{eq:potenciatheta}
    P=\int_{S}\left \langle \mathbf{S} \right \rangle\cdot d\mathbf{a}_{\perp}
\end{align}
que requiere una integración del promedio del vector de Poynting a lo largo de la sección transversal $S$ de la guía. Esto es debido a que $\left \langle \mathbf{S} \right \rangle$ tiene unidades de $[\frac{\textup{energía}}{\textup{área}\times \textup{tiempo}}]$, donde $\left \langle ~\right \rangle$ hace referencia al promedio temporal de la OEM debido a sus oscilaciones y se define como, $\left \langle \mathbf{S} \right \rangle=\textup{Re}[\frac{c}{8\pi\mu}\mathbf{E}\times\mathbf{B^{*}}]$. Además, el $\cdot d\mathbf{a}_{\perp}$ es porque nos interesa la energía que se propaga a lo largo de la dirección axial de la guía.

En ausencia del cilindro TI, la OEM de fondo tiene una potencia definida en una región transversal del espacio. A esta potencia la denominaremos ``potencia de fondo'', $P_{0m}$, la cual depende de la región integrada del medio $m$. Por ejemplo, $P_{01}=\int_{0}^{R}\langle S_{0}\rangle\rho d\rho d\phi=c E_{0}^{2}R^{2}/8Z$ y $P_{02}=c E_{0}^{2}(r^{2}-R^{2})/8Z$, donde $r$ es el radio de la sección transversal exterior. Como era de esperar, esta potencia escala con el área transversal, lo que implica que a mayor área transversal, mayor será el flujo del vector de Poynting. La potencia total tendrá entonces la contribución de fondo y una contención $\theta$ en cada región $m$ del espacio es $P_{m}^{\textup{tot}}=P_{0m}+P_{0m}P_{m}^{\theta}$, donde,
\begin{align}
    P_{1}^{\theta}&=-\frac{Z_{\theta}^{2}}{1+Z_{\theta}^{2}}, & &\textup{y}, & 
    P_{2}^{\theta}&=\frac{Z_{\theta}^{2}}{1+Z_{\theta}^{2}}\frac{R^2}{r^2}
\end{align}
$P_{m}^{\theta}$ representa la el incremento/disminución de la potencia de fondo y observamos que tiende a cero cuando $\tilde{\theta}\to 0$.  Dentro del material TI, la potencia siempre disminuye por un factor de $Z_{\theta}^{2}/(1+Z_{\theta}^{2})$ en comparación con la potencia de fondo. Por el contrario, la potencia en el medio exterior aumenta por el mismo factor, y a medida que nos alejamos de la superficie, la mejora se vuelve menos significativa. Dentro de la región de confinamiento, definida por la profundidad de penetración $r=\delta$ según la Ec. \eqref{eq:skindepth}, para $Z=1$ y $\theta_{1}=3\pi$, $11\pi$, $19\pi$ y $27\pi$, los valores de la potencia interior son $P_{1}^{\theta}=-0.01\%$, $-0.2\%$, $-0.5\%$, $-1\%$ y para la potencia exterior $P_{2}^{\theta}=0.004\%$ ,$0.06\%$, $0.2\%$, $0.4\%$. Como era de esperar, este efecto es extremadamente pequeño, del orden de $\alpha^{2}$. Si $Z_{\theta}\gg 1$, las potencias tienden a $P_{1}^{\theta}\to -100\%$ y $P_{2}^{\theta}\to 36.8\%$, lo que indica que la potencia, y en general la energía, se ``fuga'' hacia el medio exterior. Este comportamiento puede ser útil en dispositivos donde se necesite controlar la distribución de energía a lo largo de la guía de ondas o controlar cuánto de fuga queremos. 
%--------------------------------------------
\begin{figure}
\centering
\includegraphics[width=0.4\linewidth]{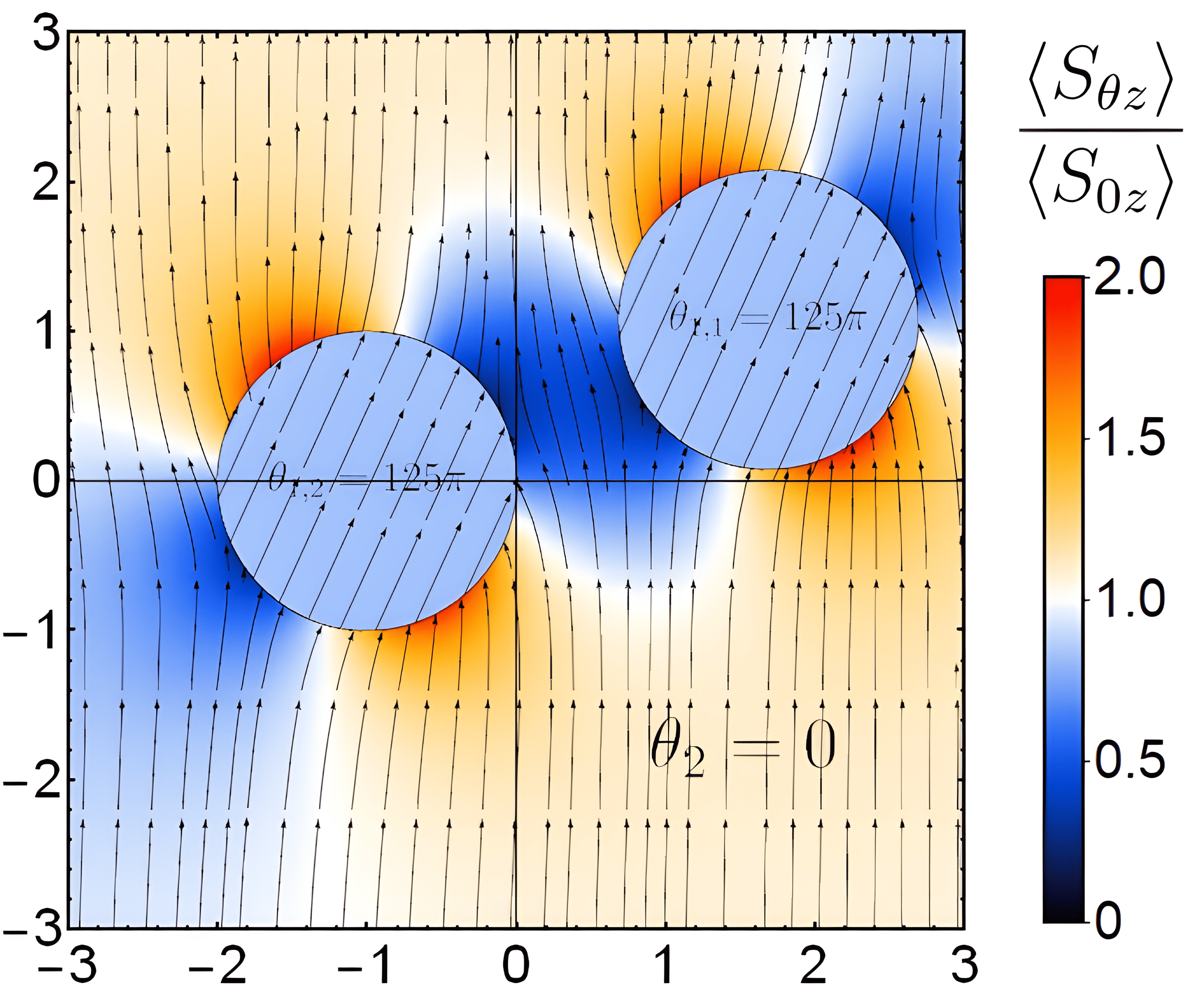}
%
%\hspace{-1mm}
%
\includegraphics[width=0.4\linewidth]{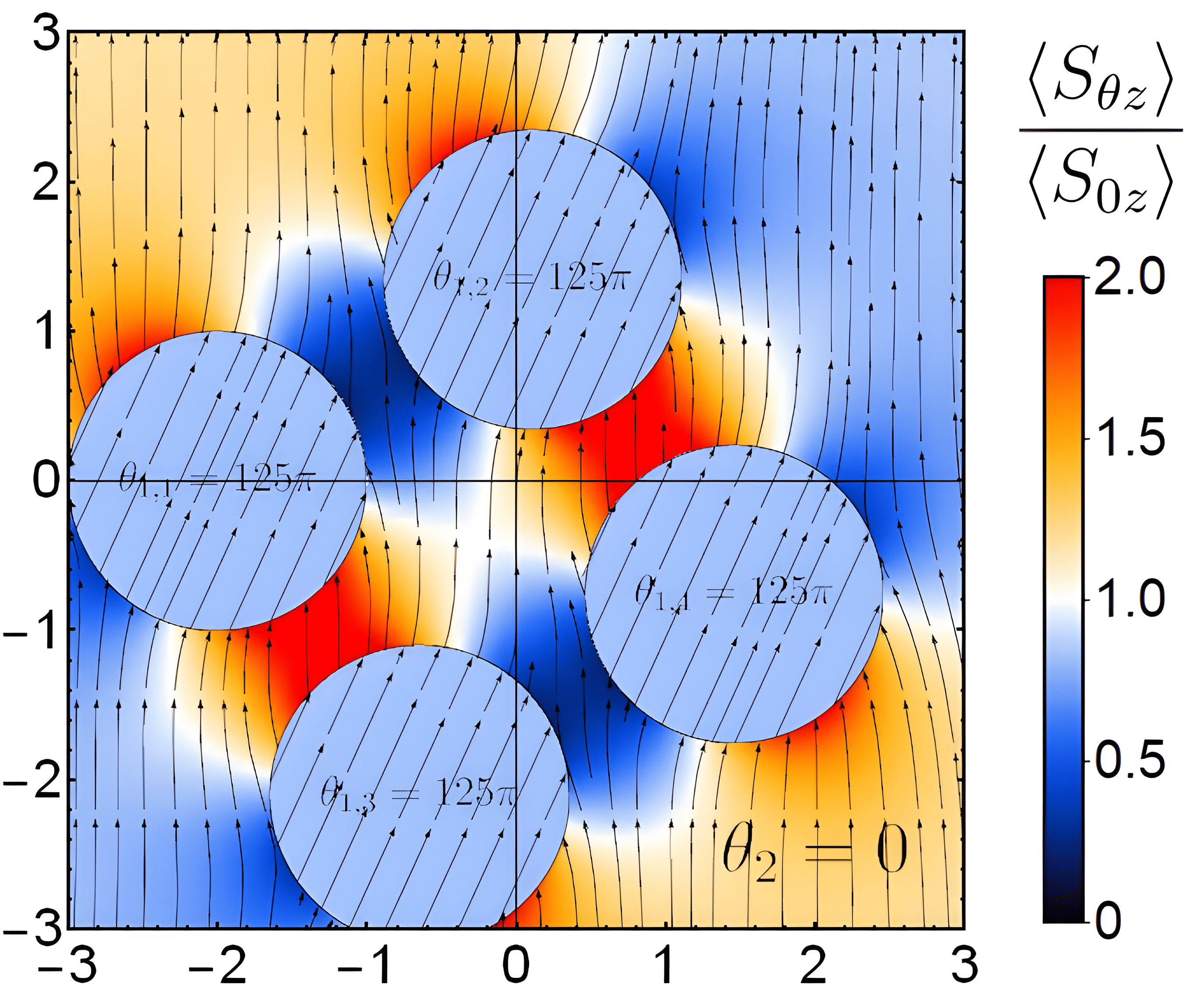}
%
%\hspace{-1mm}
%
\includegraphics[width=0.4\linewidth]{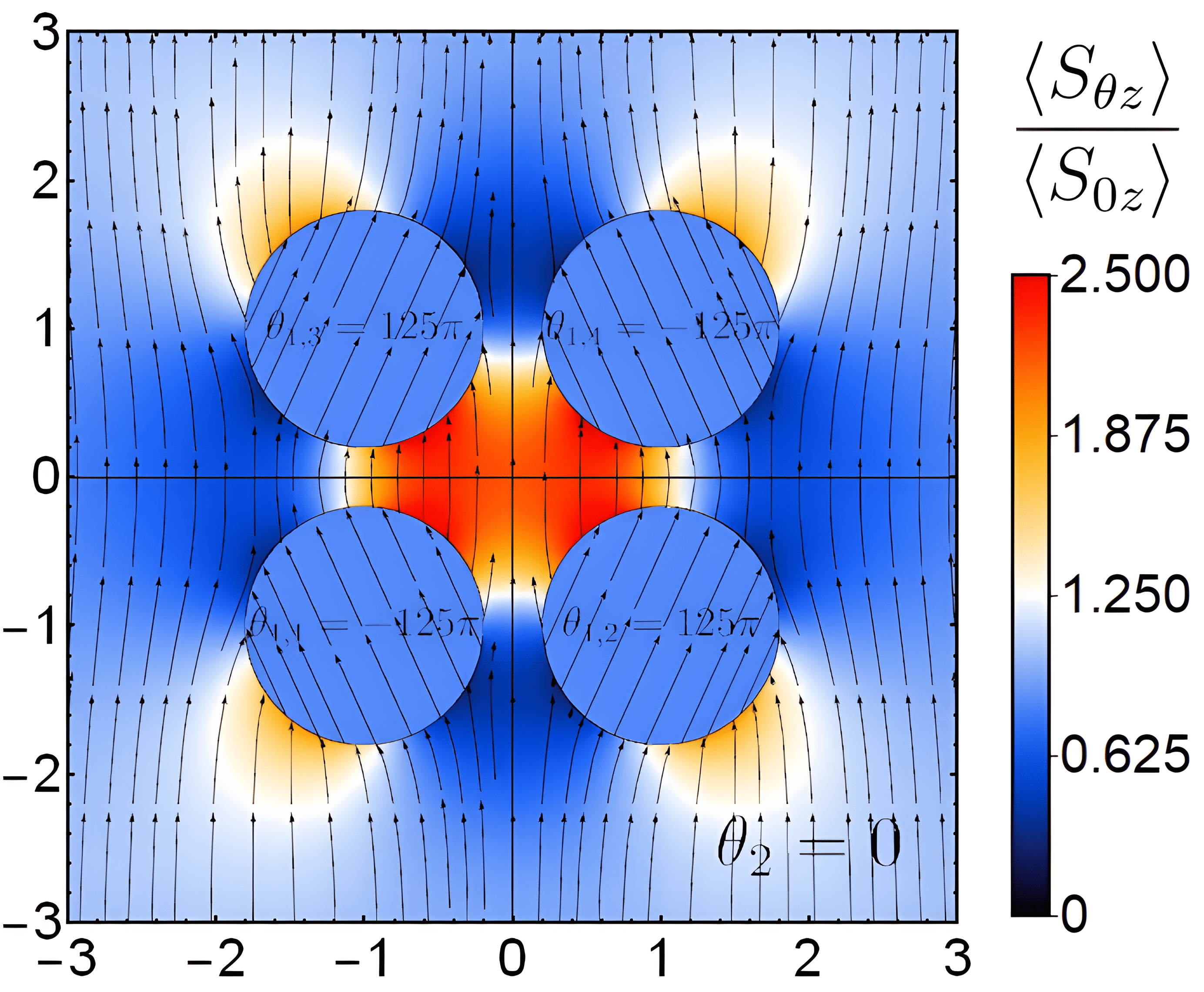}
\includegraphics[width=0.4\linewidth]{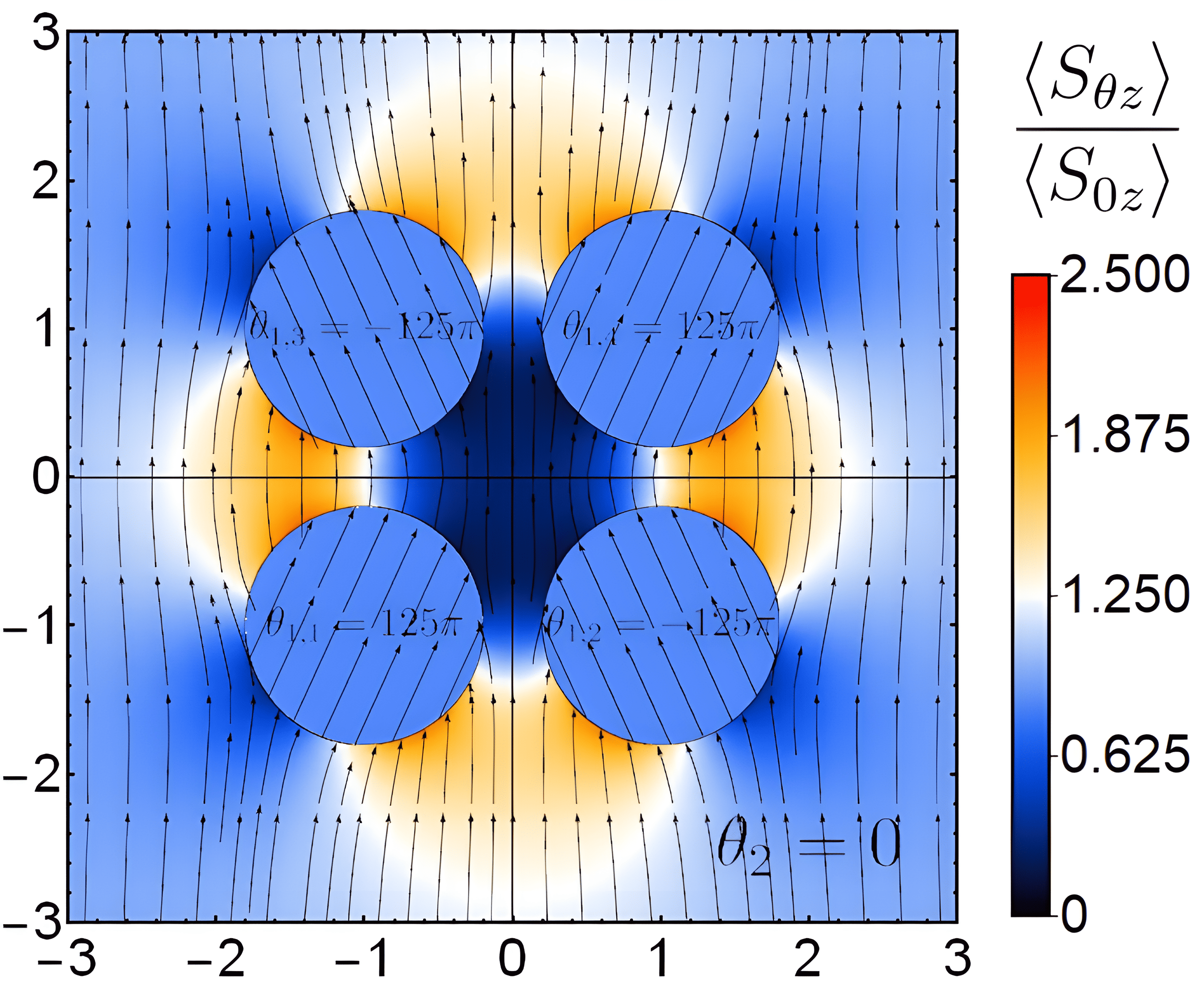}
\caption{Varios cilindros TI se colocan en diferentes posiciones de manera que sus campos EMs inducidos por las cargas topológicas se superpongan linealmente, generando distintas distribuciones del campo EM. Esto permite observar regiones del espacio donde los campos EMs se suman y la respuesta topológica se amplifica. Hemos seleccionado $Z=1$ y $\theta_{1,s}=\pm 125\pi$ para resaltar principalmente los efectos topológicos. Estos arreglos son útiles para enfocar o anular la energía EM en ciertas regiones del espacio que puedan ser de interés.}
\label{fig:plotmutiples}
\end{figure}
%--------------------------------------------
A lo largo de nuestro análisis, hemos considerado que el cilindro está inmerso en un medio con propiedades topológicas triviales, es decir, $\theta_{2}=0$. Sin embargo, dependiendo de cómo se rompa la TRS en la $\theta$-interfaz, la diferencia $\tilde{\theta}=\theta_{2}-\theta_{1}=\frac{\alpha}{\pi}2\pi p$ puede adquirir valores positivos, con $p\in\mathbb{Z}$ como es considerado en \cite{maciejko_topological_2010}. Este cambio de signo en $\tilde{\theta}$ está relacionado con la inversión de las cargas topológicas inducidas en la superficie. De este modo, podemos colocar múltiples cilindros formados por TIs con diferentes signos de rompimiento de la TRS, de manera que los campos EMs inducidos por las cargas topológicas en sus respectivas superficies se superpongan linealmente, creando patrones que controlan la distribución del campo EM total. En la Fig. (\ref{fig:plotmutiples}), se muestran diferentes configuraciones de cilindros TIs con valores del parámetro topológico $\theta_{m,s}$, donde $m=1$ corresponde al medio interior de cada cilindro $s$.

La inversión de cargas conlleva un cambio en la dirección del campo eléctrico $\mathbf{E}_{m}^{\theta}$ , que ahora sigue el mismo sentido que el campo magnético de fondo. A pesar de esta alteración en la dirección del campo eléctrico, no se observan cambios significativos en los efectos reportados. La única variación notable es el intercambio de intensidades entre los cuatro lóbulos observados en las Figs. (\ref{FIG:DensityPlotEn1}), donde los lóbulos más intensos se intercambian por los menos intensos. Este cambio se debe al cambio de signo del término de interacción $\mp2\mathbf{E}_{\textup{fondo}}\cdot(\mathbf{E}_{m}^{\theta})^{*}$ en la expresión del vector de Poynting $\left \langle \mathbf{S} \right \rangle\sim |\mathbf{E}_{\textup{fondo}}|^{2}\pm2\mathbf{E}_{\textup{fondo}}\cdot(\mathbf{E}_{m}^{\theta})^{*}+|\mathbf{E}_{m}^{\theta}|^{2}$, el cual es el responsable de formar los característicos lóbulos. Como mencionamos, este $\pm$ esta relacionado con el cambio de signos de las cargas topológicas.
\subsection{Influencia de la polarización de la onda electromagnética de fondo}
Uno puede preguntarse si la rotación de la polarización dentro del TI cambia o no al considerar que el campo EM de fondo tiene diferentes polarizaciones. Si el campo de fondo tiene polarizaciones circulares derecha o izquierda (RCP/LCP), la cantidad de rotación de polarización dentro del TI permanece constante y, en ambos casos, gira en la misma dirección que el campo de fondo. En un instante dado y para condiciones iniciales apropiadas, la estructura del campo de fondo con CP y los patrones del vector de Poynting son idénticos a los de la polarización lineal. Sin embargo, los promedios temporales difieren significativamente. Para el campo de fondo con CP, el patrón del vector de Poynting externo es isotrópico, con una dependencia $\sim \tilde \theta^2\rho^{-4}$. Aunque el campo EM de fondo esté rotando, las cargas responden a esta rotación manteniendo constante la estructura interna, de modo que el ángulo interno permanezca inalterado.

Para la polarización elíptica, se pueden obtener conclusiones similares a las obtenidas para la polarización lineal. Para comprender esto, podemos notar que el término de interacción del vector de Poynting, $2\mathbf{E}_{\textup{fondo}}\cdot(\mathbf{E}_{m}^{\theta})^{*}$, desaparece cuando las componentes de la OEM de fondo tienen la misma magnitud, $|E_{0x}|=|E_{0y}|$. Esto explica por qué, en el caso de la polarización lineal y elíptica, los lóbulos son robustos, mientras que en el caso de la polarización circular, los lóbulos desaparecen. Los resultados de este sistema ha sido reportado en \cite{Filipini2024polarization}.
\section{Cilindro TI hueco de grosor finito en un onda electromagnética de fondo}\label{5.5}
\begin{SCfigure}[1][t]
\caption{Guía de ondas compuesta por un único TI cilíndrico hueco de radio interior $R_1$ y exterior $R_2$, en presencia de una OEM externa (o de fondo) con el campo eléctrico orientado en la dirección $\tongo{y}$, \textit{i.e.}, $\mathbf{E}_{\textup{ext}}=E_{0}e^{i(k_{z}z-\omega t)}\hat{\mathbf{y}}$. El medio circundante ($\rho>R_2$) y el medio interior ($0<\rho<R_1$) son topológicamente trivial, con $\theta_1=\theta_3=0$. Para mantener las BCs a lo largo de toda la guía de ondas, todos los medios tienen el mismo índice de refracción, según la Ec.(\ref{eq:RIM}).
\label{FIG:GeoTEM2}}
\includegraphics[scale=0.6]{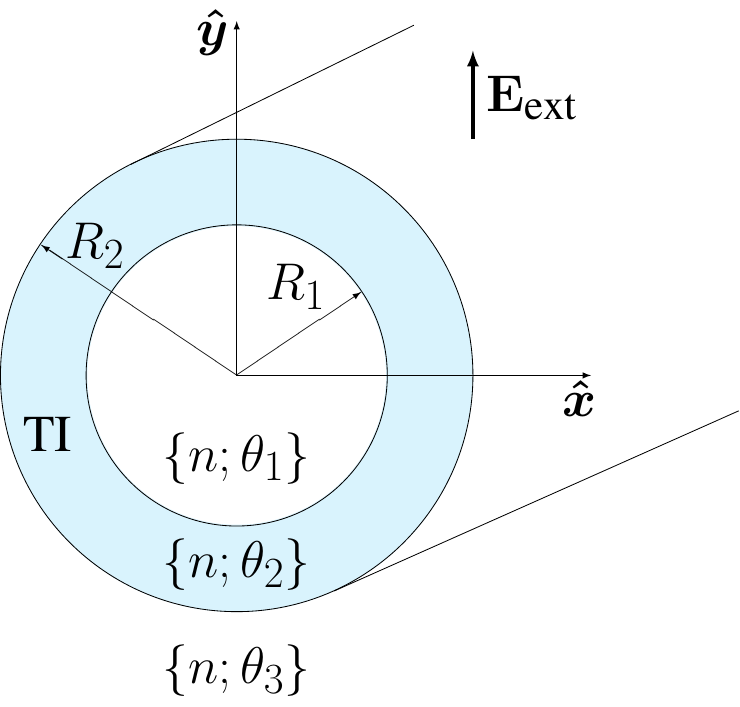}
\end{SCfigure}

Consideremos un cilindro TI hueco de radio interior $R_{1}$ y radio exterior $R_{2}$ como se muestra en la Fig (\ref{FIG:GeoTEM2}), embebido en la misma OEM definida en la Eq. \eqref{eq:Efondo}. En este sistema hay dos $\theta$-interfaces, en $R_{1}$ y $R_{2}$, el cual se debe aplicar las BCs \eqref{eq:CBcilindricas}. Nuevamente, resolvemos la ecuación de Laplace en 2D para el potencial eléctrico en una geometría cilíndrica con tres medios $m$, donde $m=1$ corresponde a la región $0<\rho<R_1$, $m=2$ a la región $R_1<\rho<R_2$, y $m=3$ a la región $R_2<\rho$. En este sistema, el campo EM debe satisfacer las BCs de la $\theta$-ED en las superficies $\rho=R_1$ y $\rho=R_2$. A medida que aumentamos el número de interfaces la expresión de los campos en los diferentes medios se vuelve mas engorrosa. Para un caso particular, consideraremos que el medio en el cual esta embebido es trivial, es decir $\theta_{1}=\theta_{3}=0$ y entonces $\theta_{2}\neq 0$. Con este procedimiento, encontramos que en las regiones $m=\{1,2,3\}$ el campo eléctrico total TEM, puede escribirse como $\mathbf{E}_{m}=E_{0} \tongo{y}+E_{0}\mathbf{E}_{m}^{\theta}$, donde las contribuciones $\theta$ son,
%---------------Figure------------
%---------------End Figure---------
\begin{align}
    \mathbf{E}_{1}^{\theta} &=-\Theta Z_{\theta_2}^{2}(\chi^{2}-1)\tongo{y}\,, 
    \label{EQ:E1theta}\\
    \mathbf{E}_{2}^{\theta} &= \Theta Z_{\theta_2}\Bigl[\Bigl(\frac{\chi R_2}{\rho}\Bigr)^{2}(\cos\phi\boldsymbol{\hat{\rho}}+\sin\phi\boldsymbol{\hat{\phi}})+\tongo{x}-Z_{\theta_2}(\chi^{2}-1)\tongo{y}\Bigr], \label{EQ:E2theta}\\
   \mathbf{E}_{3}^{\theta}&=\Theta Z_{\theta_2}(\chi^{2}-1)\Bigl(\frac{R_2}{\rho}\Bigr)^{2}[(\cos\phi+Z_{\theta_2}\sin\phi)\boldsymbol{\hat{\rho}}+(\sin\phi-Z_{\theta_2}\cos\phi)\boldsymbol{\hat{\phi}}],
   \label{EQ:E3theta}
\end{align}
donde $\Theta =1/(Z_{\theta_2}^{2}(\chi^{2}-1)-1)$, $Z_{\theta_2}=-Z\theta_{2}/2$ y $\chi = R_1/R_2$ determina el grosor del cilindro. A pesar del valor de $\theta_2$, el campo en la región $m=1$ es uniforme, con la misma polarización que el campo de fondo asintótico sin rotación. Esto se debe a que la configuración propuesta es antiparalela (ver Fig (\ref{FIG:rotacionTEM}.c)) y las rotaciones que ocurren en la interfaz $\rho=R_1$ y $\rho=R_2$ se cancelan. De manera bastante sorprendente, en el límite $R_1 \to R_2$, permanece un campo eléctrico radial y anisotrópico que reside solo en $\rho=R_2$, mientras que en las regiones $m=1$ y $m=3$ el campo eléctrico total es exactamente igual al campo de fondo. Explícitamente esto es, $ \mathbf{E}_{1}^{\theta}= \mathbf{E}_{3}^{\theta}=0$ y $\mathbf{E}_{2}^{\theta} =Z\theta_2\cos\phi\boldsymbol{\hat{\rho}}$.

\begin{figure}[t!]

\begin{subfigure}{\columnwidth}
  \stackinset{l}{18pt}{t}{10pt}{(a)}{\includegraphics[width=0.42\textwidth]{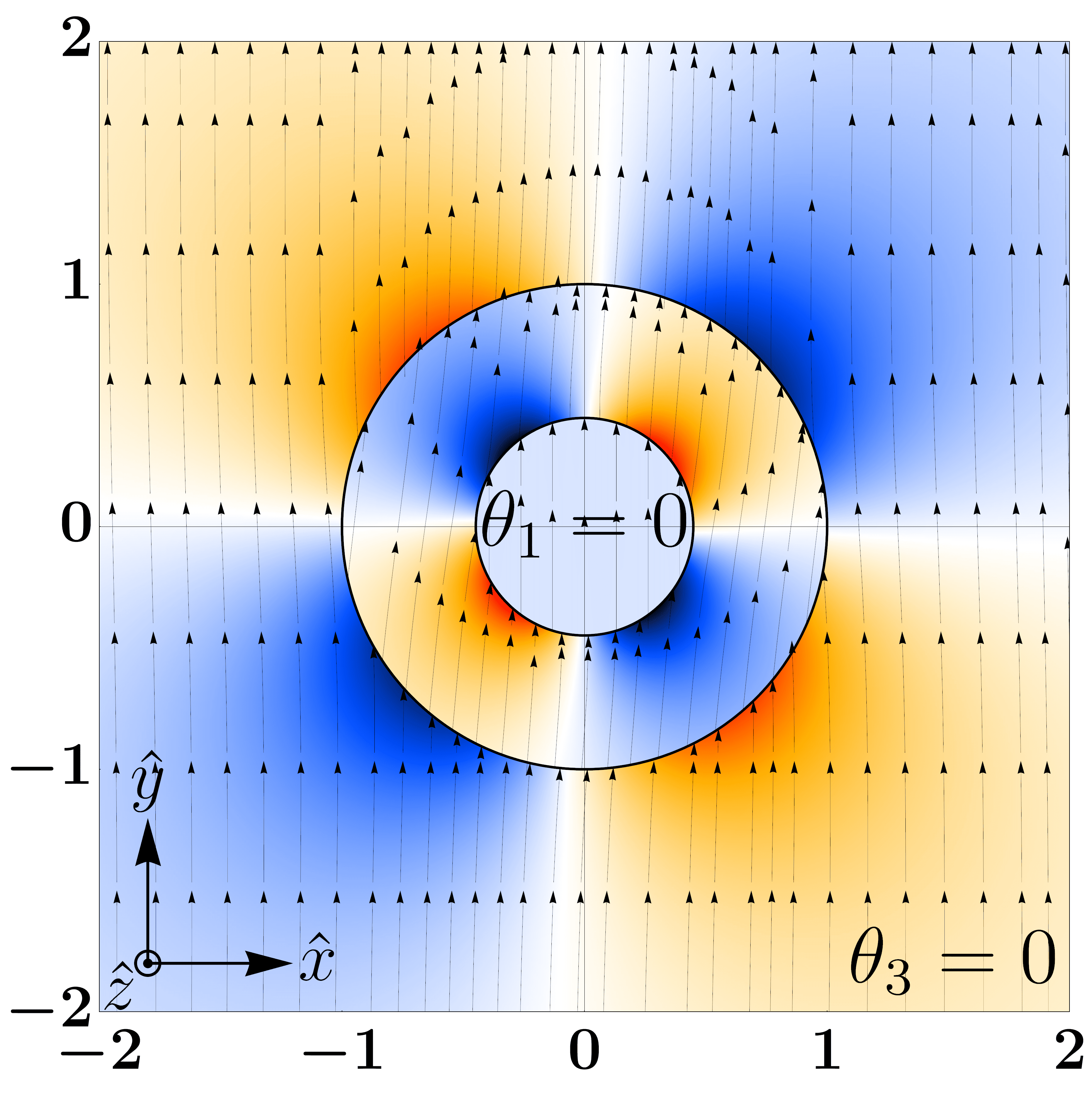}}
\hfill
\stackinset{l}{20pt}{t}{10pt}{(b)}{\includegraphics[width=0.48\textwidth]{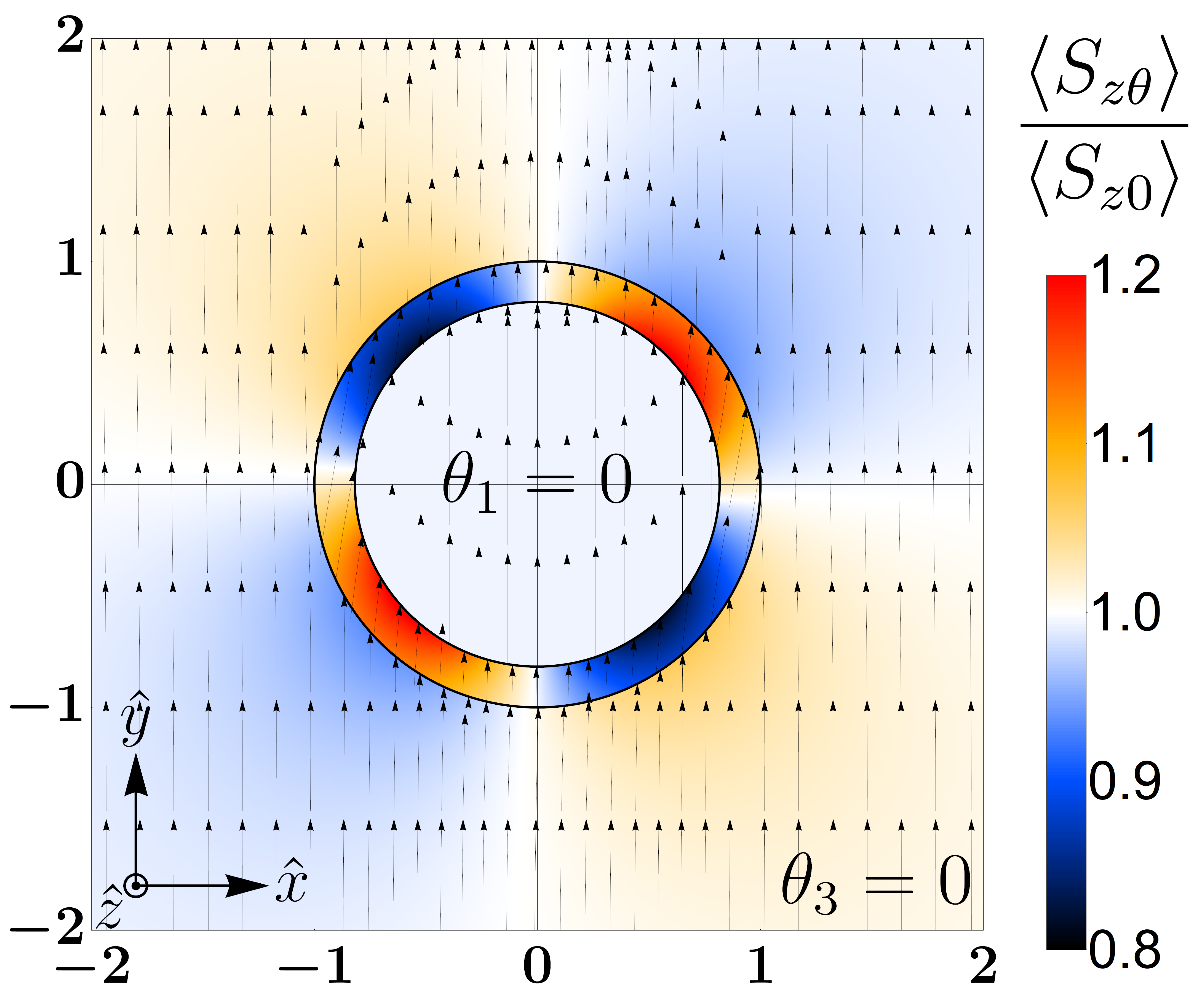}}
\end{subfigure}

\begin{subfigure}{\columnwidth}
  \stackinset{l}{18pt}{t}{10pt}{\textcolor{white}{(c)}}{\includegraphics[width=0.43\textwidth]{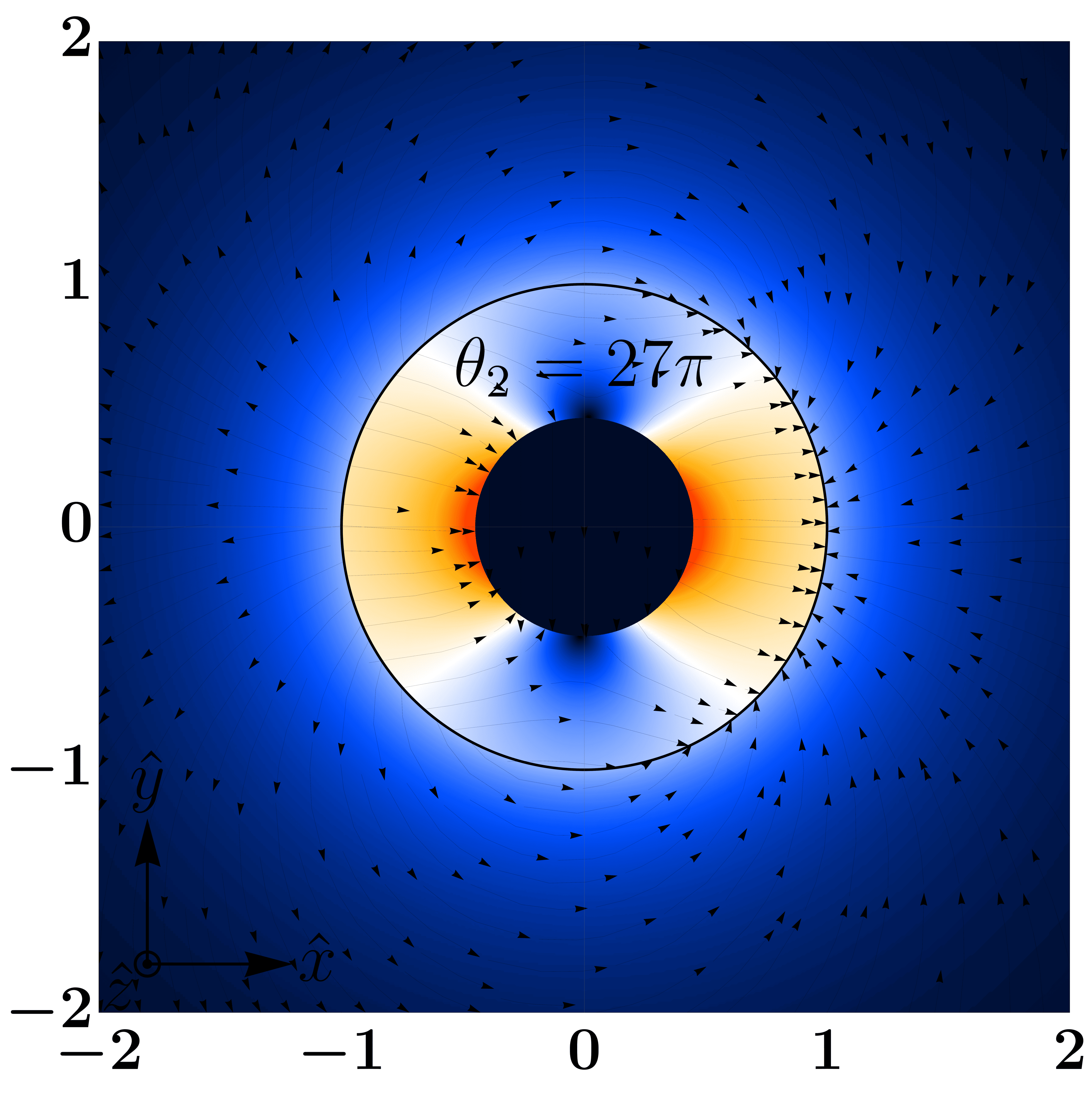}}
\hfill
\stackinset{l}{20pt}{t}{10pt}{\textcolor{white}{(d)}}{\includegraphics[width=0.5\textwidth]{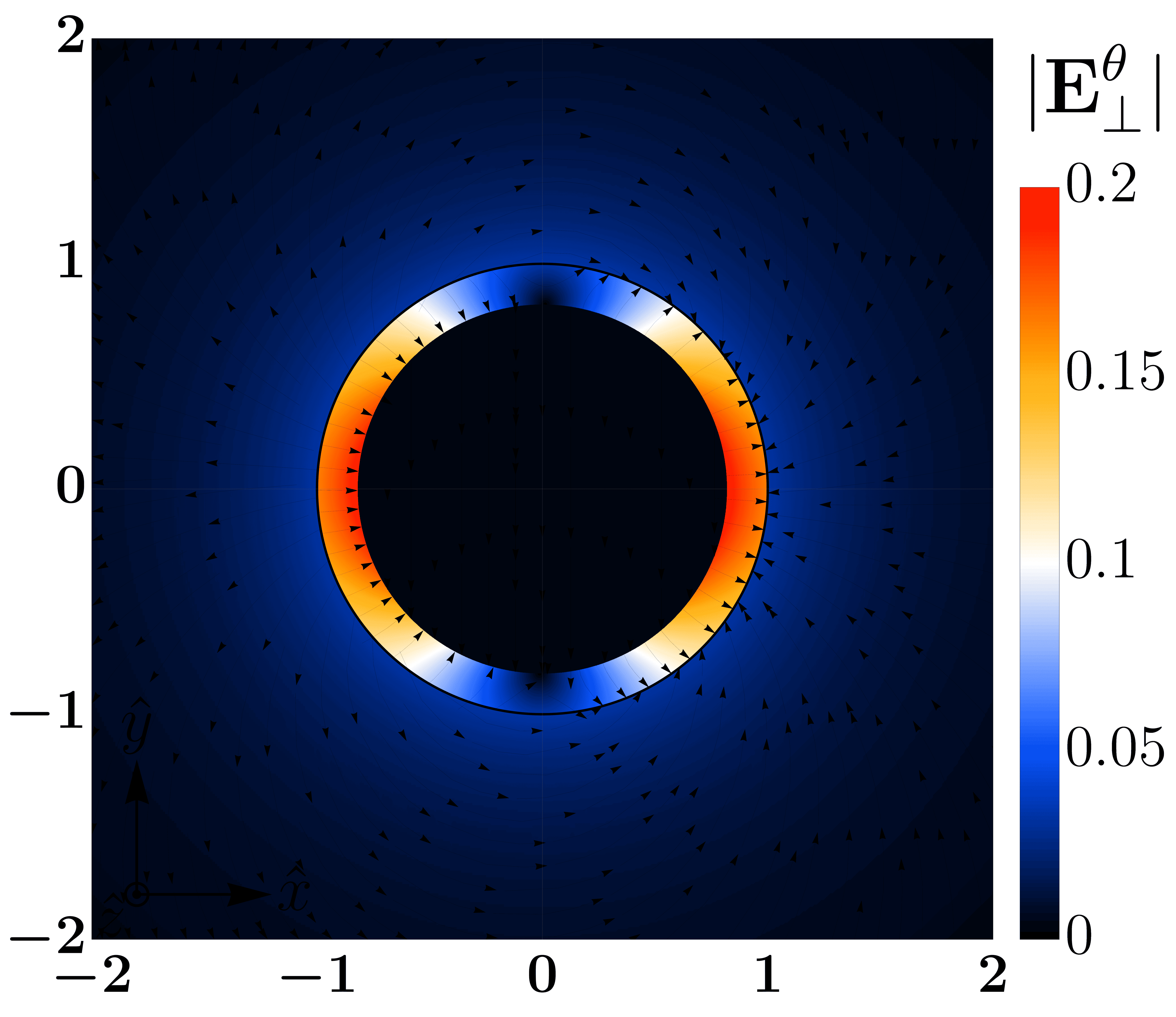}}
\end{subfigure}
%\begin{subfigure}{\columnwidth}
%\includegraphics[width=0.042\textwidth]{BarImage.png}
%\end{subfigure}
%--------------
\caption{\label{fig:DensitiyPlot2interfaces}(a,b) Campos eléctricos totales relativo al campo de fondo. (c,d) Campo eléctricos generados por la densidad de cargas topológicas en las superficies del TI. En todos los casos $Z=1$, $\theta_1=0=\theta_3$ y $\theta_2=27 \pi$ dentro del TI. En (a,c) $\chi=0.45$ y en (b,d) $\chi = 0.82$.}
\end{figure}

El promedio temporal del vector de Poynting en cada medio es $\langle\mathbf{S}_{m}^{\textup{tot}}\rangle=\langle S_{0}\rangle\tongo{z} +\langle S_{0}\rangle\langle \mathbf{S}_{m}^{\theta} \rangle$, donde,
\begin{align}
    \langle \mathbf{S}_{1}^{\theta} \rangle&=(\Theta^2-1)\tongo{z}\\
    \langle \mathbf{S}_{2}^{\theta} \rangle&=\Theta^2 Z_{\theta_2} \Bigl[Z_{\theta_2}\frac{R_1^4}{\rho ^4}+2\frac{R_1^2}{\rho ^2}(Z_{\theta_2}\cos 2\phi-\sin 2\phi)+Z_{\theta_2} (2 \chi ^2\!-\!(\chi^2\!-\!1)^2Z_{\theta_2}^2\!-\!1)\Bigr]\tongo{z}\\
    \langle \mathbf{S}_{3}^{\theta} \rangle&=\Theta ^2 Z_{\theta_2} \left(\frac{R_2^4 }{\rho ^4}(\chi ^2-1)^2 Z_{\theta_2}(Z_{\theta_2}^2+1)+2\frac{R_2^2 }{\Theta  \rho ^2}(\chi ^2-1) (\sin2 \phi-Z_{\theta_2} \cos2 \phi)\right)
\end{align}
En las Figs. (\ref{fig:DensitiyPlot2interfaces}.a.b) mostramos las \textit{streamlines} de $\mathbf{E}_{m}^{\textup{tot}}/E_{0}$ y la distribución espacial de $\langle\mathbf{S}_{m}^{\textup{tot}}\rangle/\langle S_{0}\rangle$ para $Z=1$ y $\theta_{2}=27\pi$. En (a) usamos $\chi_{a}=0.45$ y en (b) usamos $\chi_{b}=0.82$. En las Figs. (\ref{fig:DensitiyPlot2interfaces}.c.d) mostramos los campos $\mathbf{E}_{m}^{\theta}$ inducidos por las densidades de cargas topológicas de las superficie del TI. Las cargas topológicas depositadas en las superficies interna y externa del cilindro TI hueco respectivamente son,
%----------------
%--------------
\begin{align}\label{eq:cargasTopologicascilindrocongrosor1}
    \sigma_{\theta}(R_{1})&=\frac{\epsilon}{2\pi}Z_{\theta_2}\Theta(\chi^{2}-1)\cos\phi\\\label{eq:cargasTopologicascilindrocongrosor2}
    \sigma_{\theta}(R_{2})&=\frac{\epsilon}{2\pi}Z_{\theta_2}\Theta[Z_{\theta_2}(\chi^{2}-1)\sin\phi-\cos\phi]
\end{align}
En las Figuras (\ref{fig:DensitiyPlot2interfaces}.c.d), se muestra la distribución de las cargas topológicas en la superficie a través de las líneas de flujo (\textit{streamlines}) del campo eléctrico. En estas figuras, las cargas positivas actúan como fuentes del campo eléctrico, mientras que las cargas negativas funcionan como sumideros del mismo.

De las Ecs. \eqref{eq:cargasTopologicascilindrocongrosor1} y \eqref{eq:cargasTopologicascilindrocongrosor2}, notamos que las cargas positivas y negativas en la configuración antiparalela, para un ángulo fijo, no se intercalan perfectamente. Por ejemplo, para $\phi = \pi/2$, no hay cargas en la cara interior, pero sí en la cara exterior. Este ``desplazamiento'' angular varía en función de la distancia entre la cara interior y exterior: cuanto más delgado sea el grosor del cilindro hueco, menor será este desplazamiento. Además, las intensidades máximas de ambas densidades de carga son diferentes, con $\max \sigma_{\theta}(R_{2})>\max\sigma_{\theta}(R_{1})$. Sin embargo, a medida que se reduce el grosor, esta diferencia tiende a disminuir hasta desaparecer. Estas observaciones sugieren una influencia mutua entre las densidades de carga en ambas superficies. %Al igual que para el caso de una unica $\theta$-intefaz podemos definir 
Por último, la distancia a la cual se confinan los efectos de las cargas topológicas inducidas al considerar dos $\theta$-interfaces es la misma que la obtenida en la ecuación \eqref{eq:skindepth} para el caso de una sola $\theta$-interfaz, aun así, los efectos de la carga superficial interior en la región 1, $\rho<R_{1}$, son contantes y no dependen del tamaño del cilindro TI hueco. 
\subsection{Potencia y optimización por geometría}
\begin{figure}[t!]
%--------------
\stackinset{l}{20pt}{t}{12pt}{(a)}{\includegraphics[width=0.5\textwidth]{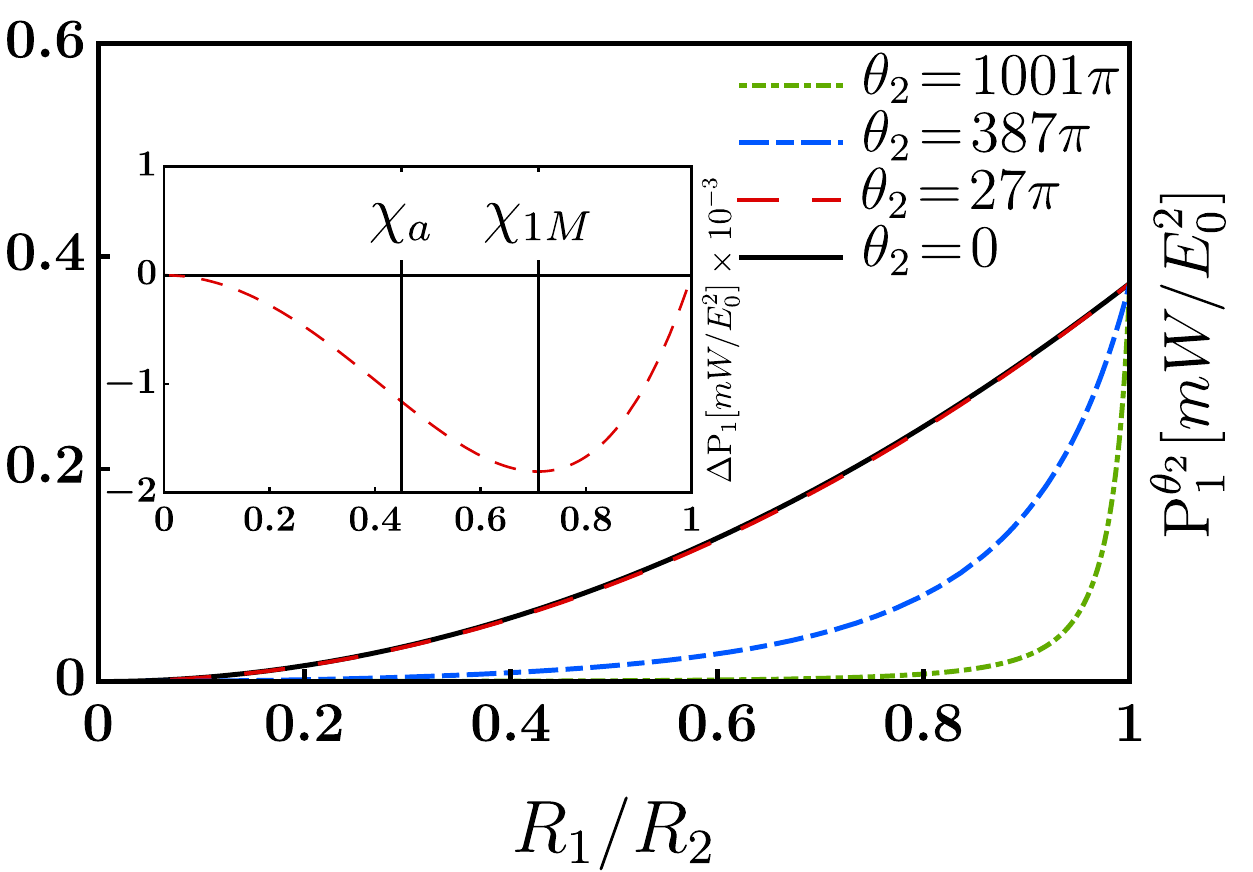}}
\hfill
\stackinset{l}{20pt}{t}{12pt}{(b)}{\includegraphics[width=0.5\textwidth]{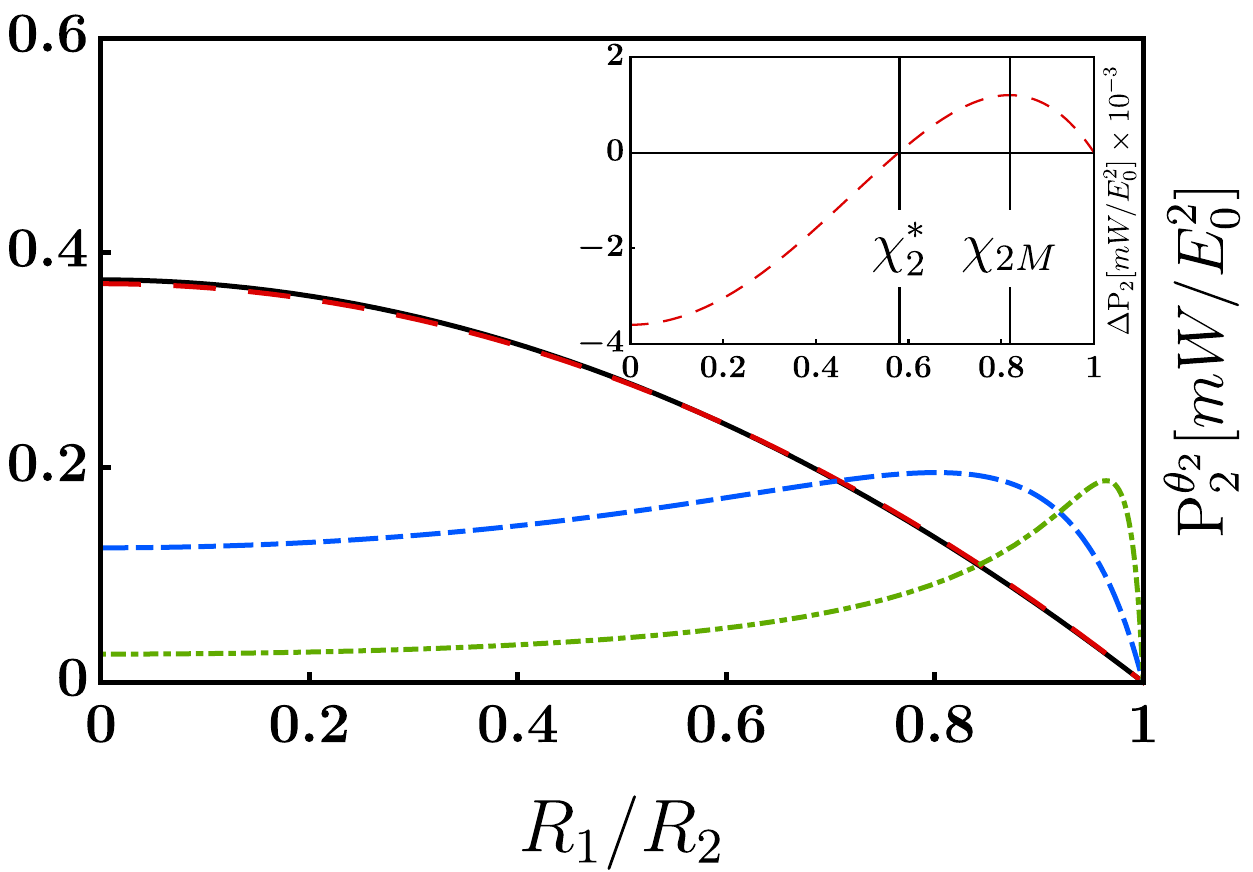}}
\caption{\label{fig:PotenciasTot_1_y_2} En (a,b) mostramos la potencia total transmitida a través de las regiones 1 y 2: $P_{1}^{\textup{tot}}$ y $P_{2}^{\textup{tot}}$, respectivamente para $R_2=10 \mu m$. Para $\theta_2=0$, las potencias correspondientes: $P_{01,2}$, se muestran en una línea negra continua. Para $\theta_2 = 27 \pi$, el recuadro de (a) muestra $\Delta P_{1}$. Las líneas verticales son $\chi_{a} = 0.45$ que define la geometría de la configuración en la Fig (\ref{fig:DensitiyPlot2interfaces}.a.c), y $\chi_{1M}$ que maximiza la diferencia. El recuadro de (b) muestra $\Delta P_{2}$ y los valores $\chi_2^\ast$ y $\chi_{b}=\chi_{2M}$ que define la geometría de la Fig (\ref{fig:DensitiyPlot2interfaces}.b.d).
}
\end{figure}

Recordemos que la potencia de define en la Ec. \eqref{eq:potenciatheta}. De esta forma, podemos definir la potencia total en cada region $m$: $P_1^{\textup{tot}}=\int_{0}^{2\pi}\int_{0}^{R_1}\left<S_{z,1} \right>\rho d\rho\phi$, $P_2^{\textup{tot}}=\int_{0}^{2\pi}\int_{R_1}^{R_2}\left<S_{z,2} \right>\rho d\rho\phi$ y $P_1^{\textup{tot}}=\int_{0}^{2\pi}\int_{R_2}^{r}\left<S_{z,3} \right>\rho d\rho\phi$. La potencia total en cada región, al igual que en el caso de una $\theta$-interfaz, se puede escribir como $P_{m}^{\textup{tot}}=P_{0m}+P_{0m}P_{m}^{\theta}$, donde,
\begin{align}
    P_{1}^{\theta}&=\Theta^{2}-1~,\\
    P_{2}^{\theta}&=\Theta^{2}Z_{\theta_2}^{2}(3\chi^{2}-1-Z_{\theta_2}^{2}(\chi^{2}-1)^{2})~,\\
    P_{3}^{\theta}&=\Theta^{2}Z_{\theta_2}^{2}(\chi^{2}-1)^{2}(1+Z_{\theta_2}^{2})\left (\frac{R_{2}}{r}\right )^{2}.
\end{align}
En la Fig. (\ref{fig:PotenciasTot_1_y_2}.a), para diferentes valores de $\theta_{2}$, comparamos la potencia transmitida en la región $m=1$: $P_{1}^{\textup{tot}}(\chi)$, con la potencia transmitida en esa misma región por el campo de fondo: $P_{01}$ (en línea negra continua). Para $\chi$ fijo, vemos que $P_1^{\textup{tot}}$ es más pequeño para valores más altos de $\theta_{2}$ y, para un $\theta_{2}$ dado, $P_1^{\textup{tot}}$ escala con $R_1 $ (lo cual es bastante trivial ya que cuanto mayor/menor $R_1$ mayor/menor será el área que se integra el vector de Poynting). Las diferencias,
\begin{align}
    \Delta P_m (\chi) \equiv P_m^{\textup{tot}}- P_{0m},
\end{align}
para $m=\{1,2\}$ cuantifican cuánto de la contribución de $\theta_{2}$ al campo EM está atrapado o ``confinado'' en la región $m$. La dependencia de $\chi$ en $\Delta P_m (\chi)$, permite encontrar configuraciones óptimas donde la contribución de $\theta_{2}$ sea importante en comparacion con la potencia de fondo. 

Por ejemplo, para cualquier $\theta_2$ dado, existe un $\chi_{1M}(\theta_2)$ crítico que minimiza $\Delta P_1$. El recuadro de la Fig. (\ref{fig:PotenciasTot_1_y_2}.a) muestra esta diferencia y los valores $\chi_a$ y $\chi_{1M}(27 \pi)$. De manera similar, en la Fig. (\ref{fig:PotenciasTot_1_y_2}.b), para diferentes valores de $\theta_{2}$, comparamos la potencia transmitida en la región $m=2$, $P_{2}^{\textup{tot}}(\chi)$, al transmitido en la misma región por el campo de fondo, $P_{02}$ (en línea negra continua). Para cada $\theta_2 \neq 0$ dado, existe un $\chi_2^\ast(\theta_2)$ tal que $\Delta P_2 (\chi_2^\ast)>0$, \textit{i.e.}, para los cuales la potencia transmitida en el conjunto del TI excede la de la misma región si el TI estuviera ausente. Cuanto mayor sea $\theta_2$, mayor será la ganancia; sin embargo, más cerca debe estar $\chi$ de 1, es decir, se producen mayores rendimientos para $\theta_{2}$ más altos y a través de fundas de TI más delgadas. Además, para $\theta_2$ dado, existe un $\chi_{2M}(\theta_2)$ con $\chi_2^\ast< \chi_{2M} < 1$ que maximiza $\Delta P_2$. En la Fig. (\ref{fig:DensitiyPlot2interfaces}.b.d), elegimos $\chi = \chi_{2M}$ para lo cual $P_{2}^{\textup{tot}}(\chi_{2M}) = \textup{max}(P_{2}^{\textup{tot}}(\chi)) = 1.01 P_{02}(\chi_{2M})$. 

Los resultados de este sistema ha sido reportado en \cite{Filipini2024polarization}.
\section{Cilindro TI de radio $R_1$ cubierto por un cilindro conductor en $R_2$}\label{5.6}
\begin{SCfigure}[1][t]
\caption{Guía de ondas compuesta por dos material cilíndricos de radio $R_1$ y radio $R_2$. El medio interior ($0<\rho<R_1$) es topológicamente trivial con $\theta_1=0$ y el medio que lo rodea es el TI con $\theta_2\neq 0$. Se agrega un conductor cilíndrico de radio $R_2$, que confina al campo electromagnético completamente. Para mantener las BCs a lo largo de toda la guía de ondas, todos los medios tienen el mismo índice de refracción, según la Ec.(\ref{eq:RIM}).
\label{FIG:GeoTEM3}}
\includegraphics[scale=0.6]{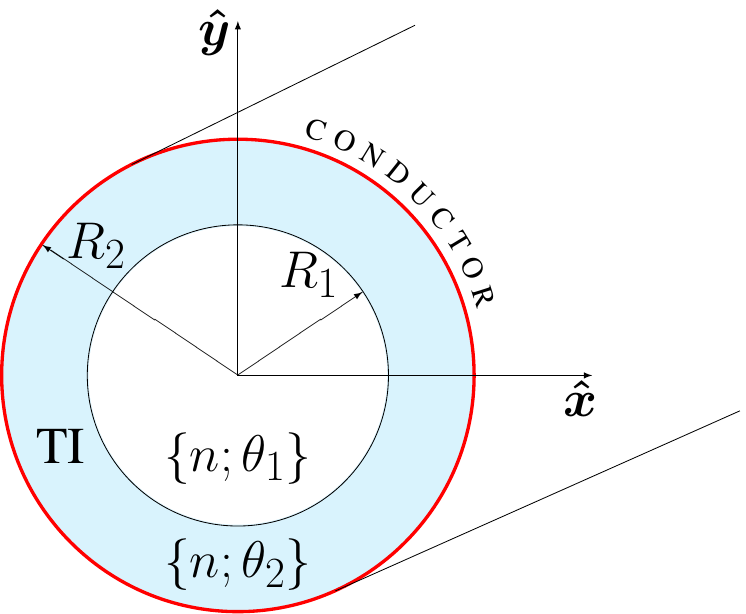}
\end{SCfigure}

Consideremos ahora otro sistema: un cilindro TI de radio $R_1$ y un cilindro conductor perfecto de radio $R_{2} > R_1$ que encierra completamente los campos EMs en la región $\rho < R_2$, ver Fig. (\ref{FIG:GeoTEM3}). El espacio entre el TI y el conductor está lleno de un dieléctrico de manera que su índice de refracción sea igual al del TI, es decir, $n_{\textup{TI}} = n_{\textup{dieléctrico}}$, condición necesaria para la propagación TEM. Dado que trabajamos en coordenadas polares, el potencial EM debe satisfacer la ecuación de Laplace en 2D tanto en el medio $m=1$ ($\rho < R_{1}$) como en el medio $m=2$ ($R_{1} < \rho < R_{2}$). En este sistema, debemos considerar las condiciones de contorno en la interfaz $\theta$ en $\rho = R_{1}$ y en la superficie del conductor en $\rho = R_{2}$.

La presencia de la superficie conductora rige el comportamiento de los campos eléctricos y establece ciertas BCs. Estas condiciones pueden tomar la forma de un campo eléctrico perpendicular justo fuera de la superficie, proporcional a la densidad de carga libre (BC de Neumann). Es decir, $\mathbf{E}\cdot \hat{n}=4\pi\sigma/\epsilon$, donde $\hat{n}$ es el vector normal a la superficie conductora. Alternativamente, podemos establecer que todo el conductor esté a un potencial constante (BC de Dirichlet), convirtiendo el conductor en una superficie equipotencial. Aunque estas dos condiciones parecen independientes, en realidad son consistentes, y podemos trabajar solo con una BC. Esto se debe a la relación entre la carga total en la superficie del conductor y el potencial asociado, expresada por $Q=C\Phi$, donde $C$ es la capacitancia del conductor, que depende de su geometría. En la práctica, la condición de ser una superficie equipotencial es relevante porque impone que la componente angular del campo eléctrico $E_{\phi}$ sea nula. Esta propiedad es esencial, ya que cualquier componente eléctrica angular implicaría un desequilibrio en las cargas superficiales del conductor. Dichas cargas responderían rápidamente al campo eléctrico angular para anularlo y restaurar el equilibrio inicial del sistema.

%---------------Figure------------
%---------------End Figure---------

Aplicar la BC en el conductor $\mathbf{E}_{2}|_{R_{2}}\cdot\boldsymbol{\hat{\phi}}=0$, obtenemos las siguientes relaciones para los coeficientes del campo EM en el medio $m=2$: $C_{\nu2}=-R_{2}^{-2\nu}D_{\nu2}$ y $A_{\nu2}=-R_{2}^{-2\nu}B_{\nu2}$. Para las BCs en $R_{1}$, se obtiene el siguiente sistema de ecuaciones algebraicas para los coeficientes restantes del campo EM,
\begin{align}\label{eq:Sistecuacionesthetacomplejo}
    \left(
\begin{array}{ccccc}
 1 & 2 Z_{\theta} & (1+\chi^{2\nu})R_{1}^{-2\nu} & 0  \\
 -2 Z_{\theta} & 1 & 0 &(1+\chi^{2\nu})R_{1}^{-2\nu}  \\
 -R_{1}^{2\nu} & 0 & (1-\chi^{2\nu}) & 0  \\
 0 & -R_{1}^{2\nu} & 0 & (1-\chi^{2\nu})  \\
\end{array}
\right).\left(
\begin{array}{c}
 A_{\nu 1} \\
 C_{\nu 1} \\
 B_{\nu 2} \\
 D_{\nu 2} \\
\end{array}
\right)=\left(
\begin{array}{c}
 0 \\
 0 \\
 0 \\
0  \\
\end{array}
\right),
\end{align}
con $B_{02}=0$, $\chi=R_{1}/R_{2}\in (0,1)$ y $Z_{\theta}=\tilde{\theta}Z/2$. Para obtener una solución no trivial de las constantes del campo EM, el determinante de la matriz debe ser igual a cero, es decir,
\begin{align}\label{eq:CondPropTEMComplejo}
    \det(\dots)=Z_{\theta}^2 (1-\chi ^{2 \nu })^2+1=0,
\end{align}
Sin embargo, debido a que $\chi$ es un parámetro geométrico real, la condición de propagación anterior no tiene solución para $Z_{\theta}\in \mathbb{R}$. Además, cuando  $\tilde{\theta}\to 0$, la condición de propagación conduce a una contradicción,  $1=0$. Esto implica que en las ecuaciones de Maxwell convencionales no es posible encontrar soluciones TEM con un único conductor, tal como lo establece el Teorema de Earnshaw.

No obstante, gracias al TMEP, es posible satisfacer la condición de propagación TEM \eqref{eq:CondPropTEMComplejo} en el conjunto de los números complejos. En general, los materiales presentan dispersión, lo que significa que las propiedades ópticas de un medio, como la permitividad eléctrica y la permeabilidad magnética, son dependientes de la frecuencia de oscilación del campo EM y se representan como números complejos, tal como mencionamos en la Sec. (\ref{4.2}). Considerando medios no magnéticos ($\mu = 1$) y dispersivos, donde $\epsilon(\omega) = \epsilon' + i \epsilon''$ y $\theta_{m}(\omega)=\theta'_{m}+i\theta''_{m}$, tenemos que en general $Z_{\theta} \in \mathbb{C}$, lo que conduce a la siguiente relación,
\begin{align}
    Z_{\theta}=\frac{\tilde{\theta}}{2\sqrt{\epsilon}}=\frac{\tilde{\theta}'+i\tilde{\theta}''}{2\sqrt{\epsilon'+i\epsilon''}}\equiv Z_{\theta}'+iZ_{\theta}''.
\end{align}
%
% donde,
% \begin{align}
%     Z_{\theta}'=&\frac{1}{2\sqrt{2}|\epsilon|}\sqrt{|\tilde{\theta}|^{2}|\epsilon|+2\epsilon''\tilde{\theta}''\tilde{\theta}'+\epsilon'(\tilde{\theta}'^{2}-\tilde{\theta}''^{2})}, &&& Z_{\theta}'
% \end{align}
%
donde $\{Z_{\theta}',Z_{\theta}''\}\in \mathbb{R}$. Por lo tanto, la condición de propagación TEM \eqref{eq:CondPropTEMComplejo} tiene como solución,
\begin{align}\label{eq:2CondicionesComplejo}
    Z_{\theta}'&=0, &&\textup{y,}&    \chi&=\left ( 1-\frac{1}{|Z_{\theta}''|} \right )^{\frac{1}{2\nu}}
\end{align}
donde,
\begin{align}\label{eq:ParteimaginariadeZtheta}
    Z_{\theta}''=\pm\frac{1}{2|\epsilon|}\sqrt{\epsilon'(\tilde{\theta}''^{2}-\tilde{\theta}'^{2})-2 \tilde{\theta}'\tilde{\theta}''\epsilon''}
\end{align}

La primera condición de las Ecs. \eqref{eq:2CondicionesComplejo} es una restricción de los parámetros complejos del medio, $Z_{\theta}'(\tilde{\theta},\epsilon)=0$, donde $Z_{\theta}'(\tilde{\theta},\epsilon) = 2 \tilde{\theta}'\tilde{\theta}''\epsilon'+\epsilon''(\tilde{\theta}''^{2}-\tilde{\theta}'^{2})$. Por ejemplo, si consideramos un medio con $\epsilon''=0$, implicaría que en un medio con $\tilde{\theta}'=0$ se satisface la condición de propagación TEM. 
%lo que lleva a $Z_{\theta}''=\pm \tilde{\theta}''/2\sqrt{\epsilon'}$, donde el signo $\pm$ está determinado por la diferencia de los TMEP de cada uno de los medios. 
La segunda relación de la Ec. \eqref{eq:2CondicionesComplejo} impone una restricción entre geometría, $\nu$ y los parámetros topo-ópticos, de tal forma que nos permite seleccionar el ``modo'' de propagación deseado.

\begin{figure}[t!]
\begin{minipage}{0.43\columnwidth}
\begin{subfigure}{\textwidth}
\stackinset{l}{15pt}{t}{12pt}{(a)}{\includegraphics[width=0.95\textwidth]{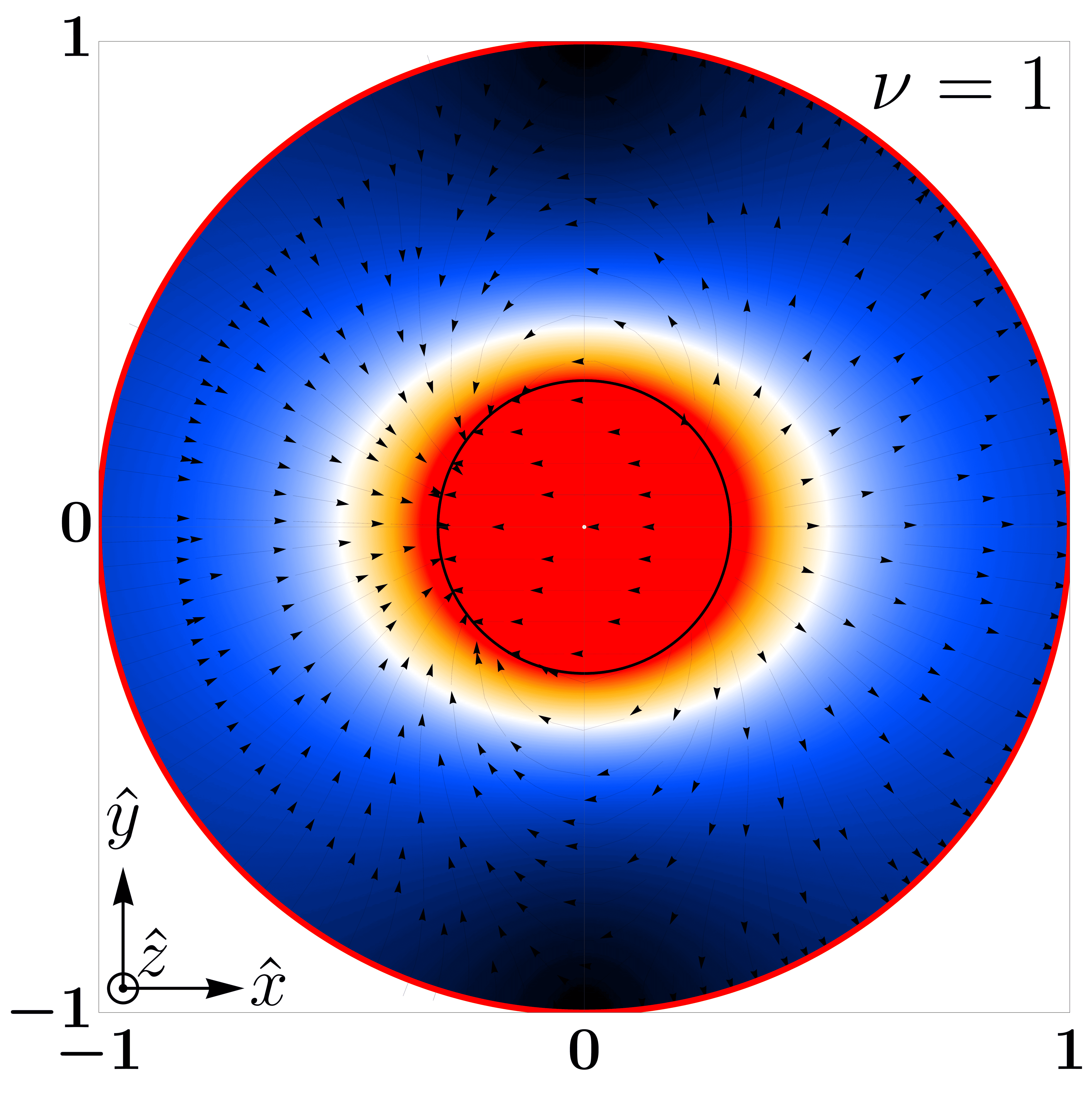}}
%\caption{}
\end{subfigure}
\begin{subfigure}{\textwidth}
\stackinset{l}{15pt}{t}{12pt}{(b)}{\includegraphics[width=0.95\textwidth]{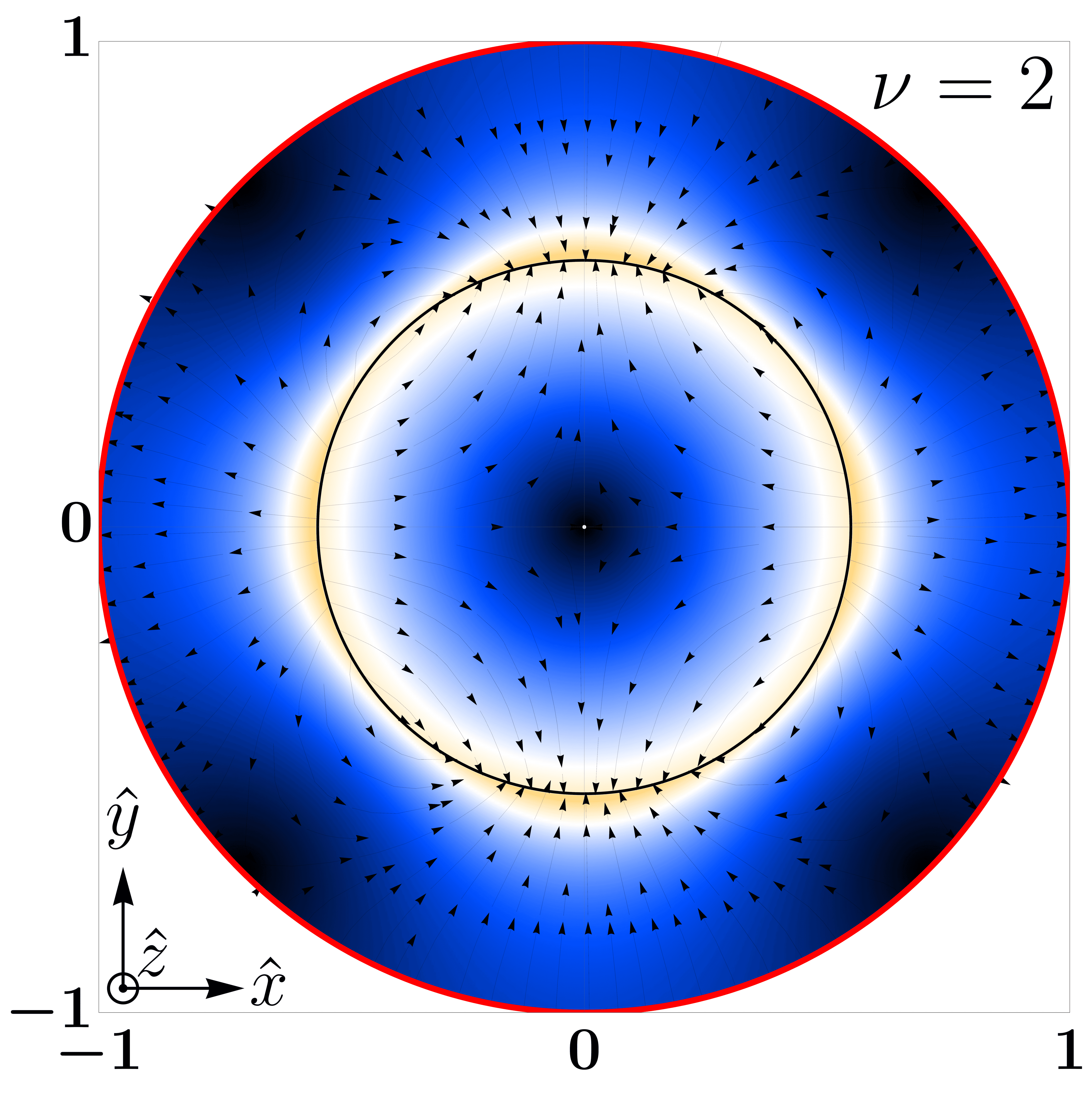}}
%\caption{}
\end{subfigure}
\end{minipage}
\hspace{0.1mm}
% \hfill
\begin{minipage}{0.43\columnwidth}
\begin{subfigure}{\textwidth}
\stackinset{l}{15pt}{t}{12pt}{(c)}{\includegraphics[width=0.95\textwidth]{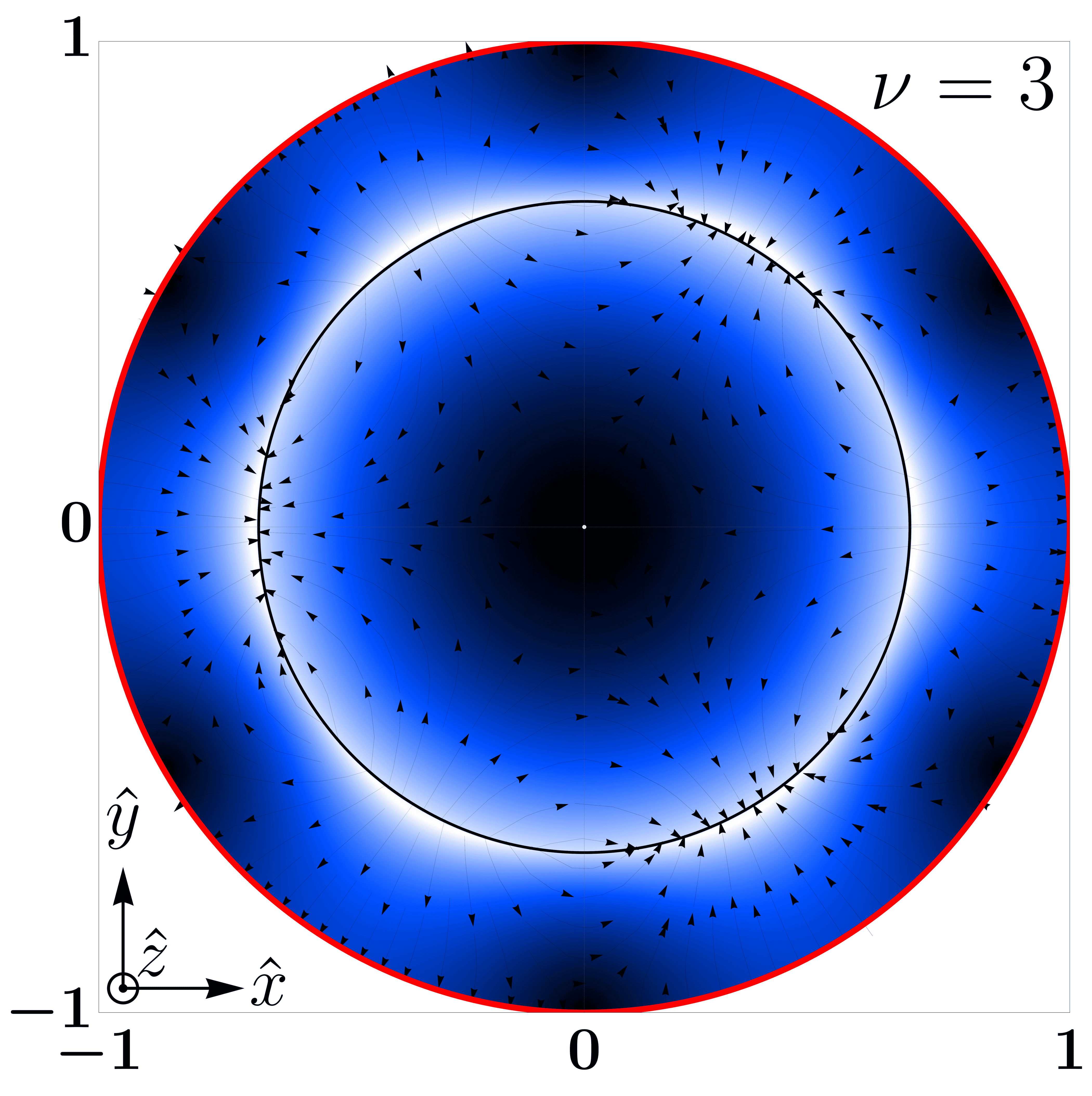}}
%\caption{}
\end{subfigure}
\begin{subfigure}{\textwidth}
\stackinset{l}{15pt}{t}{12pt}{(d)}{\includegraphics[width=0.95\textwidth]{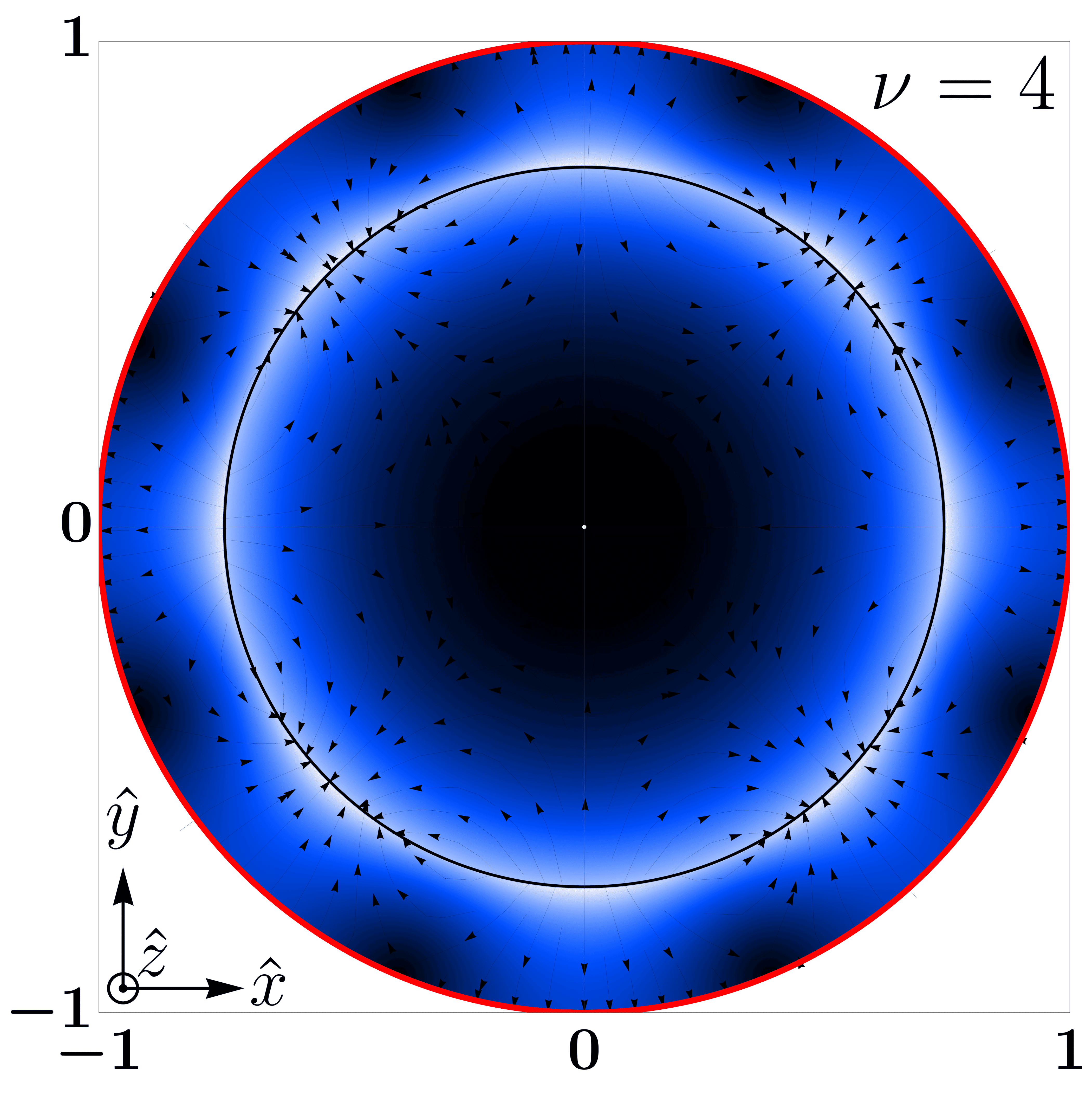}}
%\caption{}
\end{subfigure}
\end{minipage}
% \hspace{0.1mm}
% \hfill
\begin{minipage}{0.08\columnwidth}
\includegraphics[scale=0.12]{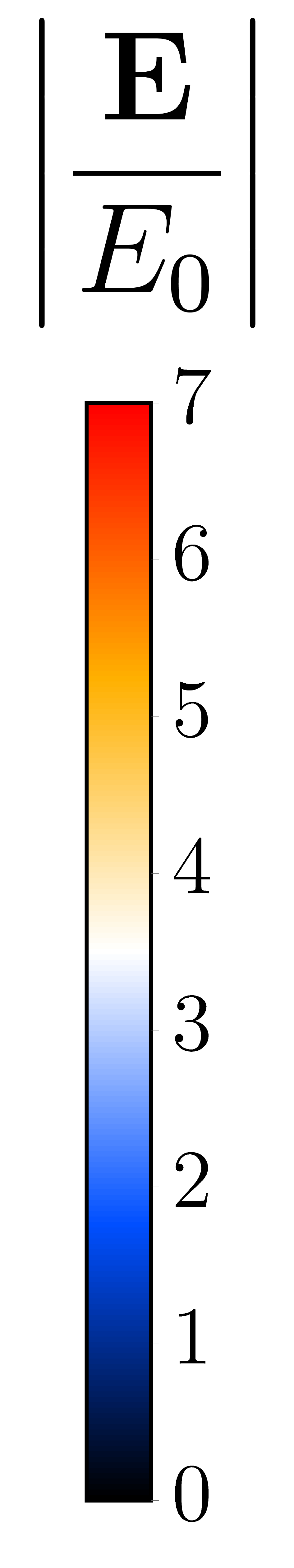}
\end{minipage}
\caption{\textit{Streamlines} de la parte real de los campos \eqref{eq:electricfieldconductor1}-\eqref{eq:electricfieldconductor2}, junto con la distribución de la magnitud. Hemos considerado $Z_{\theta}''=1.1$ con $\nu=1$ y $\chi=0.3$ en (a), $\nu=2$ y $\chi=0.55$ en (b), $\nu=3$ y $\chi=0.67$ en (c), y $\nu=4$ y $\chi=0.74$ en (d). En el caso de $\nu=1$, el campo eléctrico en el TI es $7$ veces más intenso que el campo eléctrico promedio en la superficie del conductor. Para $\nu>1$, la intensidad en el volumen del TI disminuye casi a cero, pero cerca de la superficie del TI reside campo eléctrico debido a las cargas superficiales topológicas. Para mantener las condiciones de contorno a lo largo de todo $z$ y $t$, es necesario que en ambas regiones la polarización sea circular pero de diferentes sentidos.}
\label{FIG:4modosTEMcomplejo} 
\end{figure}

Este último punto se puede clarificar siguiendo una serie de pasos:
\begin{enumerate}
    \item Consideremos que para cierto material en un rango de frecuencia $\Delta\omega$ \footnote{Recordemos que esta frecuencia no debe ser tal que la energía del fotón sea mayor que la brecha energética de los estados electrónicos superficiales y del bulk.}, cumple con $\epsilon''=0$ y $\tilde{\theta}'=0$, por lo que $Z_{\theta}'(\Delta\omega)=0$. Satisfaciendo la condición de propagación de la Ec. \eqref{eq:2CondicionesComplejo}.
    \item Por otro lado, para ese mismo rango consideremos que $\epsilon'=1$ y $\tilde{\theta}''=2.2$, lo que se obtiene $Z_{\theta}''(\Delta\omega)=1.1$. Para estos valores, determinados por el material y la frecuencia de operación, existen razones precisas únicas a las que debe ajustarse la geometría para un cierto $\nu$, es decir, $\chi(\nu=1)=0.3$, $\chi(\nu=2)=0.55$, $\chi(\nu=3)=0.67$, $\chi(\nu=4)=0.74$, \dots
    \item Una vez conocidos los posibles valores de $\chi$ (y solo esos), debemos construir la guía de onda eligiendo solo uno (a menos que podamos ajustar la razón $\chi$ con algún control).
    \item Por ejemplo, si ajustamos $\chi(\nu=4)=0.74$, significa que solo para $\nu=4$ los coeficientes del campo EM tienen valores no triviales, mientras que para $\nu\neq 4$ todos los coeficientes son iguales a cero.
\end{enumerate}
Siguiendo estos pasos, podemos construir una guía de onda que admita la propagación TEM con un solo conductor y seleccionar un único modo de propagación. Finalmente, el campo EM que satisface las BCs y las ecuaciones de la $\theta$-ED resulta,
\begin{align}\label{eq:electricfieldconductor1}
    \mathbf{E}_{1}^{\nu}&=-\frac{E_{0}}{\sqrt{2}(|Z_{\theta}''|-1)}\left ( \frac{\rho}{R_{2}} \right )^{\nu-1} [\boldsymbol{\hat{\rho}}+i\sgn(Z_{\theta}'')\boldsymbol{\hat{\phi}}]e^{i\sgn(Z_{\theta}'')\nu\phi},\\\label{eq:electricfieldconductor2}
    \mathbf{E}_{2}^{\nu}&=\frac{E_{0}}{\sqrt{2}}\left ( \frac{R_{2}}{\rho} \right )^{\nu+1}\left[\left(1+\frac{\rho^{2\nu}}{R_2^{2\nu}}\right)\boldsymbol{\hat{\rho}}-i\sgn(Z_{\theta}'')\left(1-\frac{\rho ^{2\nu}}{R_2^{2\nu}}\right )\boldsymbol{\hat{\phi}}\right]e^{i\sgn(Z_{\theta}'')\nu\phi},
\end{align}
donde $\sgn()$ es la función signo y $E_{0}$ es la promedio temporal de la amplitud del campo eléctrico en la superficie del conductor, \textit{i.e.}, $\langle|\mathbf{E}_{2}^{\nu}(R_{2})|^{2}\rangle=\frac{1}{2}\textup{Re}[\mathbf{E}_{2}^{\nu}(R_{2})^{*}\cdot \mathbf{E}_{2}^{\nu}(R_{2})]=|E_{0}|^{2}$. La Fig. (\ref{FIG:4modosTEMcomplejo}) muestra la parte real de los campos eléctricos $\textup{Re}(\mathbf{E}_{m}^{\nu}/E_{0})$ en $t=0$ y $z=0$, calculados mediante las Ecs. \eqref{eq:electricfieldconductor1}-\eqref{eq:electricfieldconductor2}, junto con su magnitud $|\textup{Re}(\mathbf{E}_{m}^{\nu}/E_{0})|$, donde hemos considerado $Z_{\theta}''=1.1$ y diferentes modos $\nu$ que determinan la geometría $\chi(\nu)$. 

Se observa que, para la mayoría de los valores de $\chi$, la parte real y el promedio temporal del campo eléctrico en la superficie del cilindro TI son más intensos que el campo eléctrico en la superficie del conductor, lo cual se expresa como,
\begin{align}
      \langle| \mathbf{E}_{1}^{\nu}(R_{1})|^2 \rangle&=\frac{\left(\chi ^{2 \nu }-1\right)^2}{2\chi^{2(1+\nu)}}\langle|\mathbf{E}_{2}^{\nu}(R_{2})|^2\rangle, &&\textup{y,}&  \langle |\mathbf{E}_{2}^{\nu}(R_{1})|^2\rangle&=\frac{\left(\chi ^{2 \nu }-1\right)^2+2 \chi ^{2 \nu }}{2\chi^{2(1+\nu)}}\langle|\mathbf{E}_{2}^{\nu}(R_{2})|^2\rangle,
\end{align}
Además, se destaca que en la superficie exterior del TI, la intensidad promedio del campo eléctrico es mayor que en la cara interna, es decir, $|\langle \mathbf{E}_{2}^{\nu}(R_{1})|^2 \rangle>\langle |\mathbf{E}_{1}^{\nu}(R_{1})|^2 \rangle>\langle |\mathbf{E}_{2}^{\nu}(R_{2})|^2 \rangle$, para un amplio rango de valores de $\chi$. Para los modos más altos, la intensidad del campo eléctrico disminuye principalmente en el volumen del cilindro TI, mientras que en su superficie reside un campo eléctrico debido a la carga topológica superficial.
%---------------------------------------------------------
%---------------------------------------------------------

Para mantener las BCs a lo largo de toda la guía, la propagación TEM presenta una polarización circular y es discontinua al cruzar de un medio a otro. Cuando $Z_{\theta}''>1$ ($Z_{\theta}''<-1$), la polarización de la OEM en el medio 1 es circular \textit{left-handed} (\textit{right-handed}), mientras que en el medio 2 es circular \textit{right-handed} (\textit{left-handed}). La polarización de la OEM se caracteriza mediante los parámetros de Stokes, definidos como sigue,
\begin{align}
    \boldsymbol{\mathcal{S}}_{m}\equiv\begin{pmatrix}
S_{0}\\ 
S_{1}\\ 
S_{2}\\ 
S_{3}
\end{pmatrix}_{m}=\begin{pmatrix}
|\mathbf{E}|^{2}\\ 
E_{x}^{*}E_{x}-E_{y}E_{y}^{*}\\ 
E_{x}^{*}E_{y}+E_{x}E_{y}^{*}\\ 
-i(E_{x}^{*}E_{y}-E_{x}E_{y}^{*})
\end{pmatrix}_{m},
\end{align}
donde $m$ es el índice del medio, $S_{0}$ es proporcional a la intensidad de la OEM, $S_{1}$ describe la cantidad de polarización lineal horizontal o vertical, $S_{2}$ describe la cantidad de polarización lineal de $+45^{o}$ o $-45^{o}$, y $S_3$ describe la cantidad de polarización circular derecha o izquierda contenida dentro del haz. El vector de Stokes para cada medio se expresa como,
\begin{align}
    \boldsymbol{\mathcal{S}}_{1}&=\frac{|E_{0}|^{2}}{(|Z_{\theta}''|-1)^{2}}\left ( \frac{\rho}{R_{2}} \right )^{2(\nu-1)}\begin{pmatrix}
1\\ 
0\\ 
0\\ 
\sgn(Z_{\theta}'')
\end{pmatrix},\\ 
\boldsymbol{\mathcal{S}}_{2}&=|E_{0}|^{2}\left ( \frac{R}{\rho} \right )^{2}\begin{pmatrix}
\frac{\rho ^{2 \nu }}{R_2^{2 \nu }}+\frac{R_2^{2 \nu }}{\rho ^{2 \nu }}\\ 
2 \cos (2 \phi )\\ 
2 \sin (2 \phi )\\ 
-\sgn(Z_{\theta}'')\left(\frac{R_2^{2 \nu }}{\rho ^{2 \nu }}-\frac{\rho ^{2 \nu }}{R_2^{2 \nu }}\right)
\end{pmatrix}.
\end{align}
Cuando $S_{3}>0$ ($S_{3}<0$), la OEM presenta una polarización circular \textit{left-handed} (\textit{right-handed}).
%\footnote{En algunos textos se define la polarización con respecto a la oscilación temporal de la onda electromagnética, el problema de esto es que depende del punto de vista del observador y trae consigo una ambigüedad en la definición. Nosotros hemos tomado la definición en función de la forma de la hélice en la coordenada espacial de propagación.}
Además, la polarización en el medio 2 depende de las coordenadas $\rho$ y $\phi$ en la contribución de la polarización lineal y circular, lo que determina la forma del campo eléctrico y da lugar a los patrones multipolares de los modos; para $\nu=1$, el patrón es dipolar, para $\nu=2$, el patrón es cuadrupolar, y así sucesivamente.

La densidad de cargas topológica $\sigma_{\theta}^{(\nu)}$, que reside en la superficie del TI para cada modo $\nu$ esta dada por,
\begin{align}\label{eq:CargasTopologicascomplejas}
    \sigma_{\theta}^{(\nu)}&=-\frac{1}{\chi^{\nu+1}}\sigma_{\textup{conductor}}^{(\nu)}, &&\textup{donde,}& \sigma_{\textup{conductor}}^{(\nu)}&=-\epsilon\sqrt{2} E_{0}e^{i \sgn Z_{\theta}''\nu \phi}/4\pi
\end{align}
es la carga en la cara interior del conductor. En la superficie del TI, las cargas son opuestas a las del conductor, y cuando $\chi\to 1$, $\sigma_{\theta}^{(\nu)}(\chi=1)=\sigma_{\textup{conductor}}^{(\nu)}$ , lo que indica la conservación de la carga. 
% De la relación de las densidades de cargas \eqref{eq:CargasTopologicascomplejas} podemos calcular la carga topológica, 
% %
% \begin{align}
%      \frac{q_{\theta}^{(\nu)}}{\pi R_{1}^2}&=-\frac{1}{\chi^{\nu+1}}\frac{q_{\textup{conductor}}^{(\nu)}}{\pi R_{2}^{2}}, &\to&&  q_{\theta}^{(\nu)}&=-\frac{1}{\chi^{\nu-1}}q_{\textup{conductor}}^{(\nu)}=-{\left ( \frac{|Z_{\theta}''|}{|Z_{\theta}''|-1} \right )^{\frac{\nu-1}{2\nu}}}q_{\textup{conductor}}^{(\nu)}
% \end{align}
% %
% A medida que se reduce el grosor del TI, la cantidad de carga topológica en un punto de la superficie, representada por $q_{\theta}$,
% $q_{\theta}=-\sqrt{Z_{\theta}''/(Z_{\theta}''-1)}q_{\textup{conductor}}$
% aumenta para mantener constante la carga inicial. Por ejemplo, para $Z_{\theta}''=1.1$, hay $3.32\,C$ más carga en la superficie del TI que en la superficie del conductor.

El sustento físico de la propagación TEM a lo largo de la guía radica en la diferencia de fase entre la carga topológica y el campo magnético. Si se induce una densidad de carga topológica que oscila en el tiempo, esta carga se convierte en fuente de un campo eléctrico $\mathbf{E}$ que también oscila en el tiempo, llenando todo el espacio dentro de la guía de ondas. A su vez, este campo eléctrico induce un campo magnético $\mathbf{B}$, perpendicular a la dirección de propagación y al campo eléctrico, como dicta la ley de Faraday. Ambos campos, $\mathbf{E}$ y $\mathbf{B}$, mantienen una fase $\delta_{B}$ debido a la dispersión del medio, con $n=|n|e^{i\delta_{B}}$ como el índice de refracción complejo. Entonces, el campo magnético queda, $\mathbf{B}=|n|e^{i\delta_{B}}\tongo{z}\times\mathbf{E}$. Para que la propagación TEM se mantenga, la componente normal del campo magnético en la superficie del TI debe inducir nuevamente una densidad de cargas topológicas $\sigma_{\theta}=-|\tilde{\theta}| e^{i\delta_{\theta}}\mathbf{B}\tongo{\rho}/4\pi$ en la superficie, comenzando así un nuevo ciclo. Al aplicar las BCs, llegamos a la condición de propagación TEM como muestra se muestra en la Ec. \eqref{eq:CondPropTEMComplejo}, que escrito en el lenguaje de fases queda,
\begin{align}
\frac{(|\tilde{\theta}|e^{i\delta_{\theta}})^2}{4(|n|e^{i\delta_{B}})^{2}} (1-\chi ^{2 \nu })^2+1&=0\\
    e^{2 i (\delta_{\theta}-\delta_{B})} |Z_{\theta}|^2 \left(1-\chi ^{2 \nu }\right)^2+1&=0.
\end{align}
Entonces, las BCs imponen una relación precisa entre las fases de ambos campos para que la condición de propagación TEM se satisfaga, \textit{i.e.}, $\delta_{\theta}-\delta_{B}=\pm \pi/2$. Por ejemplo, si $\delta_{B}=0$, entonces la oscilación de la densidad de cargas topológicas está desfasada o adelantada en 90 grados en comparación con el campo magnético. En general, esta diferencia de fase se satisface cuando se permite que el TMEP sea una numero complejo, lo cual ocurre cuando la respuesta del TI no es instantánea, como sucede en los materiales reales.

La componente $S_{0}$ de los parámetros de Stokes esta directamente relacionado con el vector de Poynting en cada medio, expresado como $\mathbf{S}_{m}=\tongo{z}c S_{0m}/8\pi Z$, donde $Z=\sqrt{\mu/\epsilon}$ es la impedancia del medio. Al integrar a lo largo de la sección transversal de la guía, obtenemos la potencia transmitida a lo largo de la dirección $\tongo{z}$ en cada medio $P^{\theta}_{m}$,
\begin{align}
    P^{\theta}_{1} &= \frac{c}{8Z}\frac{E_{0}^{2}R_{2}^{2}}{\nu}\frac{(1-\chi^{2\nu})^{2}}{\chi^{2\nu}}, &&\textup{y,}& P^{\theta}_{2} &=\frac{c}{8Z}\frac{E_{0}^{2}R_{2}^{2}}{\nu}\frac{(1-\chi^{4\nu})}{\chi^{2\nu}}.
\end{align}
Por consiguiente, la potencia total transmitida a lo largo de la guía es la suma, $P^{\theta}_{\textup{tot}}=P^{\theta}_{1}+P^{\theta}_{2}$, la cual se obtiene,
\begin{align}
    P^{\theta}_{\textup{tot}} = \frac{c}{4Z}\frac{E_{0}^{2}R_{2}^{2}}{\nu}\frac{(1-\chi^{2\nu})^{2}}{\chi^{2\nu}}
\end{align}
Tomemos como ejemplo medios no magnéticos con $\mu=1$, y tipo vacío con $\epsilon=1$, donde podemos escoger por ejemplo $Z_{\theta}''=1.1$ y un radio externo de $R_{2}=1\,\mu m$. La potencia total por unidad de campo eléctrico en la superficie conductora para diferentes valores de $\nu$ es de $P^{\theta}_{\textup{tot}} = \{25;12.5;8.3;6.2\}\times 10^{-3}\,[W/(V/m)]$ para $\nu=\{1;2;3;4\}$, respectivamente. Se observa que el modo más bajo, $\nu=1$, es dominante en la transmisión de energía EM, lo cual es favorable experimentalmente debido a que el cilindro TI para el modo más bajo tiene el valor de radio más pequeño en comparación con los modos más altos (ver Fig. (\ref{FIG:4modosTEMcomplejo}.a)). Los cálculos y análisis presentado de esta configuración aún esta en preparación \cite{underprep}.

\biblio %Se necesita para referenciar cuando se compilan subarchivos individuales - NO SACAR

%% file: Capitulos/06Slab.tex
Es bien conocido que una OEM que incide sobre la superficie de un TI induce una rotación observable del plano de polarización \cite{maciejko_topological_2010,crosse_electromagnetic_2015,crosse_optical_2016,tse_giant_2010}. En este contexto, anticipamos que una OEM que se propaga en una guía de ondas formada por TIs experimentará múltiples rotaciones de la polarización cuando la luz rebote en las paredes o se produzca una reflexión interna total. La acumulación de estas rotaciones en las interfaces podría dar lugar a efectos novedosos y detectables. En el capítulo anterior, discutimos las condiciones necesarias para que se produzca una onda TEM. Sin embargo, una onda TEM es un caso particular y muy interesante que difiere notablemente de otras soluciones permitidas en una guía de ondas\footnote{Difiere porque una onda TEM en una guía de ondas se propaga en la dirección del eje de la guía, es decir, no se propaga mediante reflexiones internas totales. Sorprendentemente, aunque la OEM no rebota en la superficie, la polarización de rota.}. La segunda parte de esta investigación se centra en desarrollar una solución completa para la propagación de OEM en un medio confinado constituido por TIs y materiales dieléctricos. Como se mencionó en el capítulo (\ref{Guías de Ondas}), las guías de ondas dieléctricas soportan tanto modos de propagación confinados como modos de radiación (ver Fig. (\ref{Guías de Ondas})). Dependiendo de la geometría transversal de la guía de ondas, esta soportará diferentes OEM confinadas, llamadas \textit{modos}. Se puede demostrar que estos modos forman una base completa que describe la OEM total \cite{jackson1999classical}.

Una guía de onda en forma de \textit{slab} es un tipo de guía de OEM que se utiliza en dispositivos fotónicos integrados para confinar, manipular, modular, acoplar y transmitir luz. Esta guía está formada por una estructura plana de capas delgadas de materiales dieléctricos con diferentes índices de refracción (ver Fig. (\ref{FIG:SlabGeo})). La diferencia en los índices de refracción ($n_2 > n_1 \geq n_3$, según laFig. anterior) asegura la propagación de la luz mediante reflexiones internas totales en las paredes, lo que confina la luz dentro de la guía de onda. En esta geometría, existen dos tipos de soluciones independientes: la polarización transversal eléctrica (TE) y la polarización transversal magnética (TM). Estas soluciones son independientes, lo que significa que es posible tener propagación TE o TM pura a lo largo de toda la guía.

En el contexto de guías de onda formadas con materiales topológicos, Crosse \cite{crosse_theory_2017} estudió la propagación de OEM en una guía de onda en forma de slab con paredes de TIs, informando sobre una hibridación de los modos TE y TM. Su enfoque fue considerar que, debido a que el efecto topológico es débil, el campo EM sería simplemente una combinación de los modos encontrados en la electrodinámica convencional de Maxwell, con amplitudes que cambian a lo largo de la dirección de la guía, $z$. Esto implica que si la OEM total en $z=0$ comienza con una polarización TE, a lo largo de una distancia, digamos en $z=l_{\theta}$, la OEM total tendrá una polarización TM. Durante todo su recorrido entre $0 < z < l_{\theta}$, la OEM es híbrida. En el contexto de materiales magnetoeléctricos generales, Talebi \cite{Talebi2016optical} encontró que la relación de dispersión de los campos EM cambia, indicando que los modos de propagación no son los mismos que los de la electrodinámica convencional, como sugirió Crosse. Sin embargo, Talebi no proporcionó detalles específicos sobre los modos de propagación alterados. Además, no se ha hecho un estudio de como son excitados estos nuevos modos permitidos en la slab. 

En esta tesis, focalizamos nuestros esfuerzos en el análisis semi-analítico de los modos confinados para establecer comparaciones entre las similitudes y diferencias con los modos permitidos derivados de las ecuaciones de Maxwell convencionales, y contrastar estos hallazgos con investigaciones previas, como las reportadas por Crosse \cite{crosse_theory_2017} y Talebi \cite{Talebi2016optical}. 

En la Sec. (\ref{6.1}) revisaremos la relación de dispersión que debe cumplir la OEM dentro de una guía de ondas con geometría arbitraria. Detallaremos la metodología a seguir para encontrar la solución completa del campo EM en el marco de la $\theta$-ED. En la Sec. (\ref{6.2}) consideraremos una guía de ondas en forma de slab dividida en tres medios con propiedades topo-ópticas diferentes y encontraremos la solución general del campo EM en cada uno de estos medios. Mostraremos específicamente las BCs que el campo EM debe satisfacer en una slab. En la Sec. (\ref{6.3}) aplicaremos las BCs y encontraremos que, para que se cumplan, los parámetros de propagación del campo EM deben satisfacer una ecuación característica. Mostraremos que la solución de esta ecuación característica discretiza los valores de las constantes de propagación, definiendo así los \textit{modos} del campo EM. La ecuación característica obtenida difiere de la derivada en la electrodinámica usual de Maxwell ($0$-ED), lo que implica que las relaciones de dispersión y los parámetros efectivos de la guía de ondas cambian. En la Sec. (\ref{6.5}) clasificaremos estos modos, que difieren significativamente de los modos en la $0$-ED. La diferencia principal radica en que todos los modos en la $\theta$-ED son híbridos, a diferencia de los modos TE y TM puros de la $0$-ED. En la Sec. (\ref{6.6}) definiremos un ``producto punto'' entre modos, demostrando que existe una relación de ortogonalización conocida como la reciprocidad de Lorentz, la cual nos permite normalizar dichos modos. Presentaremos gráficos de los campos EM dentro de la guía de ondas y definiremos un parámetro de respuesta topológica relativa a la respuesta de la $0$-ED, lo que nos permitirá identificar la configuración óptima para que esta respuesta sea significativamente más importante en comparación con los campos de la $0$-ED. En la Sec. (\ref{6.7}) definiremos una rotación efectiva del campo EM. Esta rotación es distinta de la que aparece cuando una OEM incide oblicuamente en una superficie TI; se trata de una configuración de ``equilibrio'' que el campo mantiene a lo largo de toda la guía de ondas. Finalmente, en la Sec. (\ref{6.8}) consideraremos un haz Gaussiano que incide en la guía de ondas en forma de slab para estudiar cuáles son los modos excitados que se propagan. Descubriremos que, en la $\theta$-ED, el haz Gaussiano excitará un número mayor de modos permitidos en comparación con lo que ocurre en la $0$-ED.

\section{Relación de dispersión y metodología}\label{6.1}
En general, consideramos que los campos EM oscilan temporalmente a una frecuencia angular $\omega$ y se propaga en la dirección $\tongo{z}$, de manera que $\mathbf{E}(\mathbf{r},t) =\mathbf{E}(\mathbf{r}_{\perp})e^{i(k_{z}z-\omega t)}$ y de forma análoga para el campo magnético. Consideramos que los medios que forman la guía de onda son homogéneos e isotrópicos, y asumimos que sus características no varían a lo largo de la dirección $z$. Por lo tanto, las propiedades $\epsilon$, $\mu$ y $\theta$ dependen únicamente de las coordenadas transversales y la frecuencia, es decir, $\epsilon = \epsilon(\mathbf{r}_{\perp}; \omega)$, $\mu = \mu(\mathbf{r}_{\perp}; \omega)$, y $\theta = \theta(\mathbf{r}_{\perp}; \omega)$.

Como describimos detalladamente en el capitulo de guías de ondas (\ref{Guías de Ondas}), los campos EM cumplen con las ecuaciones de Maxwell convencionales lejos de las interfaces entre los medios. Siguiendo el enfoque discutido en secciones anteriores, descomponemos los campos en componentes transversales y longitudinales, detallados en las Ecs. \eqref{EQ:T-L1}-\eqref{EQ:T-L6}. En las regiones alejadas de las interfaces, donde $\boldsymbol{\nabla}\theta = 0$, las componentes transversales de los campos se derivan a partir de las componentes longitudinales \cite{jackson1999classical},
\begin{align}
\mathbf{E}_{\perp}&=\frac{i}{k_{\perp}^{2}}(k_{z}\boldsymbol{\nabla}_{\perp}E_{z}-k_{0}\tongo{z}\times \boldsymbol{\nabla}_{\perp}B_{z}), &
\mathbf{B}_{\perp}&=\frac{i}{k_{\perp}^{2}}(k_{z}\boldsymbol{\nabla}_{\perp}B_{z}+k_{0}\epsilon\mu\tongo{z}\times \boldsymbol{\nabla}_{\perp}E_{z}),
\label{EQ:TransversalFields_UsualMaxwell}
\end{align}
las cuales son obtenidas de las Ecs. \eqref{EQ:T-L3} y \eqref{EQ:T-L5}. Se encuentra que la relación de dispersión del medio es,
\begin{align}    k_{\perp}^{2}=k_{0}^{2}\mu\epsilon-k_{z}^{2},
\label{EQ:DispersionRelation}
\end{align}
donde $k_{\perp} = \sqrt{k_{x}^2 + k_{y}^2}$ es el número de onda de corte, y $k_{0} = \omega/c = 2\pi/\lambda_{0}$ es el número de onda en el vacío. Al reemplazar estas expresiones en las Ecs. \eqref{EQ:T-L1}, \eqref{EQ:T-L2}, \eqref{EQ:T-L4}, y \eqref{EQ:T-L6}, derivamos las ecuaciones diferenciales para las componentes longitudinales de los campos,
\begin{align}
(\nabla_{\perp}^{2}+k_{\perp}^{2})B_{z}&=0, &&&
(\nabla_{\perp}^{2}+k_{\perp}^{2})E_{z}&=0,
\label{EQ:DifferentialEqLongitudinal_UsualMaxwell}
\end{align}
correspondientes a las ecuaciones de Helmholtz en las direcciones transversales. Las soluciones son ondas estacionarias en las direcciones transversales y una onda viajera en la dirección $\tongo{z}$. 

La metodología general para encontrar la solución completa en la $\theta$-ED en una geometría arbitraria consiste en determinar $E_{z}$ y $B_{z}$ utilizando las Ecs. \eqref{EQ:DifferentialEqLongitudinal_UsualMaxwell}, luego sustituirlas en las Ecs. \eqref{EQ:TransversalFields_UsualMaxwell} y aplicar las BCs impuestas por la $\theta$-ED. En la practica, la elección de la geometría y estructura de la guía de onda depende de factores como la banda de frecuencia operativa, la cantidad de energía EM que se desea transferir y las pérdidas de transmisión admisibles. 
%
% Se puede demostrar que todas las componente de los campos satisfacen las ecuaciones de Helmholtz  en un medio homogéneo,
% %
% \begin{align}
% (\nabla_{\perp}^{2}+k_{\perp}^{2})\begin{Bmatrix}
% \mathbf{E} \\ \mathbf{B}
% \end{Bmatrix}&=0.
% \end{align}
% %
\section{Componentes longitudinales en una guía de onda en forma de \textit{slab}\label{6.2}}
\begin{SCfigure}[][t!]
\caption{\small{Ilustración de un rayo EM propagándose a través del \textit{core} de una estructura en forma de \textit{slab}, delimitada en $x = \pm L$ por dos medios laterales con diferentes propiedades topo-ópticas, $\mathcal{M}_{m} = \{\epsilon_{m}, \mu_{m}, \theta_{m}\}$, donde $m = \{1, 2, 3\}$. El rayo se propaga en forma de zigzag con un ángulo $\phi_0$, y los índices de refracción de los medios laterales son menores que el del núcleo ($n_2 > n_1 \geq n_3$), lo que asegura la propagación mediante refracción interna total.}
\label{FIG:SlabGeo}}
\includegraphics[scale=0.55]{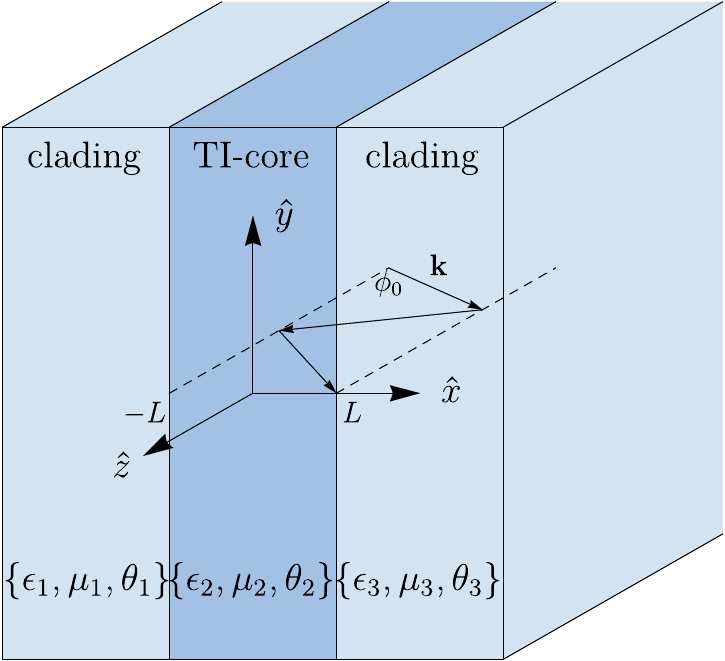}
\end{SCfigure}

Iniciamos nuestro análisis considerando una guía de ondas compuesta por tres medios en forma de slab infinita\footnote{La designación de ``infinita'' en la geometría se refiere a que una de las dimensiones características es extremadamente grande en comparación con las otras. En el caso específico de una guía de onda en forma de losa, dos de sus tres dimensiones son al menos dos órdenes de magnitud mayores que la tercera. }, dispuestos en la dirección $\tongo{x}$ y separados por una distancia de $2L$, como se ilustra en la Fig. (\ref{FIG:SlabGeo}). Cada uno de estos medios, denotados y caracterizados como $\mathcal{M}_{m}=\{\epsilon_{m},\mu_{m},\theta_{m}\}$ con $m = {1, 2, 3}$, presenta propiedades ópticas y topológicas distintivas. El medio $m=2$ actúa como el núcleo o \textit{core} de la guía, mientras que los medios $m=1$ y $m=3$ sirven como revestimientos o \textit{cladding} izquierdo y derecho, respectivamente. Un rayo de luz que incide, por ejemplo, en el plano $y=0$, forma un ángulo $\phi_{0}$ con las paredes de la guía (planos $x= \pm L$) de tal forma que $\phi_{0}<\cos^{-1}(n_1/n_2)\leq \cos^{-1}(n_3/n_2)$. El rayo se refleja y rebota entre estas superficies, propagándose a lo largo de la dirección $\tongo{z}$, por lo tanto, es guiado por la estructura de la guía de ondas.
%
%---------------Figure------------
%---------------End Figure---------
Debido a la simetría espacial en torno al eje $\tongo{y}$, los campos EMs no dependen de la coordenada $y$, lo que implica que $k_{y} = 0$. En esta configuración, las componentes longitudinales de los campos en cada medio satisfacen la ecuación de Helmholtz unidimensional, donde el número de onda de corte es $k_{\perp m}=k_{xm}$ en cada uno de los medios $m$. Esto es,
\begin{align}
(\partial_{x}^{2}+k_{xm}^{2})E_{z}(x)&=0, &&&
(\partial_{x}^{2}+k_{xm}^{2})B_{z}(x)&=0.
\label{EQ:DifferentialEqLongitudinal_UsualMaxwellenX}
\end{align}
Las soluciones a estas ecuaciones corresponden a combinaciones de exponenciales de la forma $\sim e^{\pm ik_{xm} x}$. El vector número de onda completo asociado a la OEM se expresa como $\mathbf{k}_{m\pm}=\pm k_{xm}\tongo{x}+k_{z}\tongo{z}$, manteniendo constante el valor de $k_{z}$ en todos los medios para asegurar la continuidad de las BCs a lo largo del eje $z$.

Para garantizar que las soluciones estén confinadas dentro del core, es crucial que la amplitud de la OEM decaiga rápidamente fuera del core. Para esto, buscamos soluciones con los siguiente número de ondas transversales: $k_{x1}=i \alpha_{1}$, $k_{x2}=\gamma$ y $k_{x3}=i\alpha_{3}$, donde $\{\alpha_{1},\gamma,\alpha_{3}\}\in \mathbb{R}^{+}$. Esto permite definir la profundidad de penetración en cada revestimiento como $\delta_{1,3} =1/ \alpha_{1,3}$, y la longitud de onda en la dirección $x$ como $\lambda_{x} = 2\pi/\gamma$.

La interacción entre las ondas que se propagan hacia $+\tongo{x}$ y $-\tongo{x}$ dentro del core genera patrones con paridad bien definida, la cual adoptaremos en esta tesis. Por lo tanto,
\begin{align}
    E_{z2}(x)&=E_{2+}e^{i\gamma x}+E_{2-}e^{-i\gamma x}\equiv E_{e}\sin\gamma x+E_{o}\cos\gamma x, \\
    B_{z2}(x)&=B_{2+}e^{i\gamma x}+B_{2-}e^{-i\gamma x}\equiv B_{e}\sin\gamma x+B_{o}\cos\gamma x
\end{align}
donde $C_{e} \equiv i(C_{2+} - C_{2-})$ y $C_{o} \equiv C_{2+} + C_{2-}$ ; con $C_{2\pm}=\{E_{2\pm},B_{2\pm}\}$, representan las soluciones pares e impares de la OEM, respectivamente. Es importante destacar que la paridad se refiere específicamente a la \textit{paridad del campo eléctrico transversal}, que depende de las derivadas espaciales de las componentes longitudinales. Con esto, las componentes longitudinales del campo en cada medio se expresan de la siguiente manera,
\begin{align}\label{EQ:ELongFieldTot}
E_{z}(x)&=\left\{\begin{matrix}
E_{1}e^{\alpha_{1}x}, &  & x\leq -L\\
E_{e}\sin\gamma x+E_{o}\cos\gamma x, &  & -L < x < L, \\ 
E_{3}e^{-\alpha_{3}x}, &  & x\geq L
\end{matrix}\right.\\\label{EQ:BLongFieldTot}
B_{z}(x)&=\left\{\begin{matrix}
B_{1}e^{\alpha_{1}x}, &  & x\leq -L\\
B_{e}\sin\gamma x+B_{o}\cos\gamma x, &  & -L < x < L, \\ 
B_{3}e^{-\alpha_{3}x}, &  & x\geq L
\end{matrix}\right.
\end{align}
Hemos excluido las soluciones no físicas, \textit{i.e.}, aquellas que divergen en $x=\pm\infty$. Al examinar \eqref{EQ:ELongFieldTot} y \eqref{EQ:BLongFieldTot}, identificamos un total de $12$ incógnitas, incluyendo $k_{z}$. Estas son $E_{1}$, $E_{e}$, $E_{o}$, $E_{3}$, $B_{1}$, $B_{e}$, $B_{o}$, $B_{3}$, $\alpha_{1}$, $\gamma$, $\alpha_{3}$, y $k_{z}$. Para encontrarlas, aplicamos todas las BCs en las interfaces de las slabs,
\begin{align}
 \Delta[E_{z}]|_{x=\pm L}&=0 & \Delta [\frac{1}{\mu}B_{z}]|_{x=\pm L},&=(\tilde{\theta}_{m}E_{z})|_{x=\pm L}, \label{EQ:BCZ}\\
\Delta [E_{y}]|_{x=\pm L}&=0 & \Delta [\frac{1}{\mu}B_{y}]|_{x=\pm L}&=(\tilde{\theta}_{m}E_{y})|_{x=\pm L} \label{EQ:BCY},\\
\Delta [\epsilon E_{x}]|_{x=\pm L}&=-(\tilde{\theta}_{m}B_{x})|_{x=\pm L}, & \Delta B_{x}|_{x=\pm L}&=0. \label{EQ:BCX}
\end{align}
Observamos que, inicialmente, contamos con $12$ BCs. Sin embargo, las BCs \eqref{EQ:BCY} y \eqref{EQ:BCX} son equivalentes, reduciendo el número efectivo de BCs a $8$. Aún necesitamos cuatro ecuaciones adicionales para determinar completamente el sistema. Las relaciones de dispersión \eqref{EQ:DispersionRelation} en cada medio nos proporcionan la conexión necesaria entre las incógnitas $\alpha_{1}$, $\gamma$, y $\alpha_{3}$. Estas relaciones resultan en,
\begin{align}
    \left.\begin{matrix}
        \alpha_{1}^{2}=k_{z}^{2}-k_{0}^{2}n^{2}_{1}\\
        \gamma^{2}=k_{0}^{2}n^{2}_{2}-k_{z}^{2}\\
        \alpha_{3}^{2}=k_{z}^{2}-k_{0}^{2}n^{2}_{3}
    \end{matrix}\right\}
    ~~~~\Rightarrow~~~~ 
    \begin{matrix}
    \gamma^{2}+\alpha_{1}^{2}=k_{0}^{2}(n^{2}_{2}-n^{2}_{1})\\
    \gamma^{2}+\alpha_{3}^{2}=k_{0}^{2}(n^{2}_{2}-n^{2}_{3})=k_{0}^{2}(n^{2}_{2}-n^{2}_{1})(1+\delta)\\
    \alpha_{3}^{2}-\alpha_{1}^{2}=k_{0}^{2}(n_{1}^{2}-n_{3}^{2})=k_{0}^{2}(n^{2}_{2}-n^{2}_{1})\delta
    \end{matrix}
\label{EQ:DispersionRelationInEachMedium}
\end{align}
donde $n_{m} = \sqrt{\epsilon_{m}\mu_{m}}$ representa el índice de refracción en cada medio, y $\delta$ es el parámetro de asimetría de la guía, definido como, 
%\textcolor{blue}{Pensar si cambiar la letra $\delta$ para esto u otras definiciones}
%
\begin{align}\label{eq:parámetrodeasimetría}
    \delta=\frac{n_{1}^{2}-n_{3}^{2}}{n_{2}^{2}-n_{1}^{2}}.
\end{align}
Es destacamos que $\delta \geq 0$, ya que asumimos que $n_{2} > n_{1} \geq n_{3}$ para asegurar la reflexión interna total de las OEMs que se propagan a través del core. Cuando $n_{1} = n_{3}$, se obtiene $\delta = 0$, indicando que la guía de onda es simétrica. Finalmente, tenemos 11 ecuaciones para las 12 incógnitas, y podremos expresar el sistema en términos de una única amplitud de onda, que será determinada normalizando el campo EM, como se discutirá más adelante. 
\section{Aplicación de las BCs: cálculo de las constantes de propagación}\label{6.3}
Primero consideramos las cuatro BCs de las componentes longitudinales de los campos, detalladas en las Ecs. \eqref{EQ:BCZ}, para determinar la amplitud en los medios $m=1$ y $m=3$ ,
\begin{align}
    E_{1}&=e^{v}[E_{o}\cos u-E_{e}\sin u], & B_{1}&=e^{v}\frac{\mu_{1}}{\mu_{2}}[(B_{o}-E_{o}\mu_{2}\tilde{\theta}_{1})\cos u-(B_{e}-E_{e}\mu_{2}\tilde{\theta}_{1})\sin u],\\
    E_{3}&=e^{w}[E_{o}\cos u+E_{e}\sin u], & B_{3}&=e^{w}\frac{\mu_{3}}{\mu_{2}}[(B_{o}+E_{o}\mu_{2}\tilde{\theta}_{2})\cos u+(B_{e}+E_{e}\mu_{2}\tilde{\theta}_{2})\sin u],
\end{align}
donde 
\begin{align}
    v = \alpha_{1}L,~~~~~~~u = \gamma L,~~~~~~~w = \alpha_{3}L,
\end{align}
representan los números de onda adimensionales en cada medio. Notamos que, dependiendo del signo del producto entre la amplitud del campo eléctrico y $\tilde{\theta}_{m}$, las amplitudes del campo magnético en los medios laterales pueden incrementarse o disminuirse. Al aplicar las siguientes cuatro BCs \eqref{EQ:BCX} o \eqref{EQ:BCY}, expresamos el sistema de ecuaciones algebraicas en forma matricial como sigue,
\begin{align}\label{EQ:BCsSystem}
    \mathbb{M}\cdot \mathbf{s}&=\mathbf{0}, &  &\textup{donde,}  & \mathbf{s}&=\begin{pmatrix}
E_e & E_o & B_e & B_o
\end{pmatrix}^{t},
\end{align}
y,
    \begin{align}
        \mathbb{M}&=\begin{pmatrix}
     \epsilon_3 u \sin u-\epsilon_2 w \cos u & \epsilon_3 u \cos u+\epsilon _2 w \sin u &  \tilde{\theta}_2 w\cos u &  -\tilde{\theta }_2 w\sin u \\
    \epsilon _2 v \cos u-\epsilon _1 u \sin u & \epsilon _1 u \cos u+\epsilon _2 v \sin u & \tilde{\theta }_1 v \cos u & \tilde{\theta }_1 v \sin u \\
     \mu _2 \mu _3 \tilde{\theta }_2 u \sin u & \mu _2 \mu _3 \tilde{\theta }_2 u \cos u & \mu _3 u \sin u-\mu _2 w \cos u & \mu _2 w \sin u+\mu _3 u \cos u \\
     \mu _2 \mu _1 \tilde{\theta }_1 u \sin u & -\mu _2 \mu _1 \tilde{\theta }_1 u \cos u & \mu _2 v \cos u-\mu _1 u \sin u & \mu _2 v \sin u+\mu _1 u \cos u 
    \end{pmatrix}.
    \label{EQ:BCsMatix}
    \end{align}
Para encontrar una solución no trivial del vector de los coeficientes $\mathbf{s}$, el determinante de la matriz $\mathbb{M}$ debe ser cero, \textit{i.e.,} $\det(\mathbb{M}) = 0$. Esta condición lleva a la \textit{ecuación característica} o \textit{condición de propagación}, que establece una relación entre $v$, $u$, y $w$, para ciertos parámetros topo-ópticos especificados. La ecuación característica, junto con las tres Ecs. \eqref{EQ:DispersionRelationInEachMedium}, determinan el conjunto completo de constantes de propagación $v$, $u$, $w$, y $k_{z}$ necesarias para que el campo EM se propague adecuadamente a lo largo de la guía de onda, cumpliendo con todas las BCs.

En la literatura es común representar la ecuación característica utilizando parámetros normalizados, lo que permite reducir de cinco ($n_1$, $n_2$, $n_3$, $L$ y $k_0$) a tres los parámetros en la ecuación característica. Estos parámetros normalizados incluyen la asimetría, definido anteriormente en la Ec.\eqref{eq:parámetrodeasimetría}, el \textit{índice guía normalizado} $b$ y el \textit{espesor normalizado} $R$ \cite{Kogelnik:74}, el cual estan dados por,
% El objetivo de este artículo es mostrar cómo puede reducirse el número de parámetros independientes mediante la introducción de parámetros debidamente normalizados y las reglas de escalado asociadas, y proporcionar gráficos universales a partir de los cuales puedan determinarse el índice de guía efectivo y el espesor de guía efectivo para cualquier configuración de guía de losa. Junto con las reglas de escalado, estos gráficos ofrecen una visión general compacta de las características de la guía de ondas y de las distintas posibilidades de diseño.
%
\begin{align}
    b\equiv\frac{v^{2}}{R^{2}}=\frac{k_{z}^{2}-k_{0}^{2}n^{2}_{1}}{k_{0}^{2}(n_{2}^{2}-n_{1}^{2})},
\end{align}
donde $R \equiv Lk_{0}\sqrt{n_{2}^{2} - n_{1}^{2}}$. De esta forma, si reemplazamos en las ecuaciones de dispersión definidas en \eqref{EQ:DispersionRelationInEachMedium} podemos notar que,
\begin{align}\label{eq:evwenfunciondebR}
    u&=R\sqrt{1-b}, &&& v&=R\sqrt{b}, &&& w&=R\sqrt{b+\delta}.
\end{align}
\begin{figure}[!ht]
%--------------
\stackinset{r}{28pt}{t}{35pt}{(a)}{\includegraphics[width=0.520\textwidth]{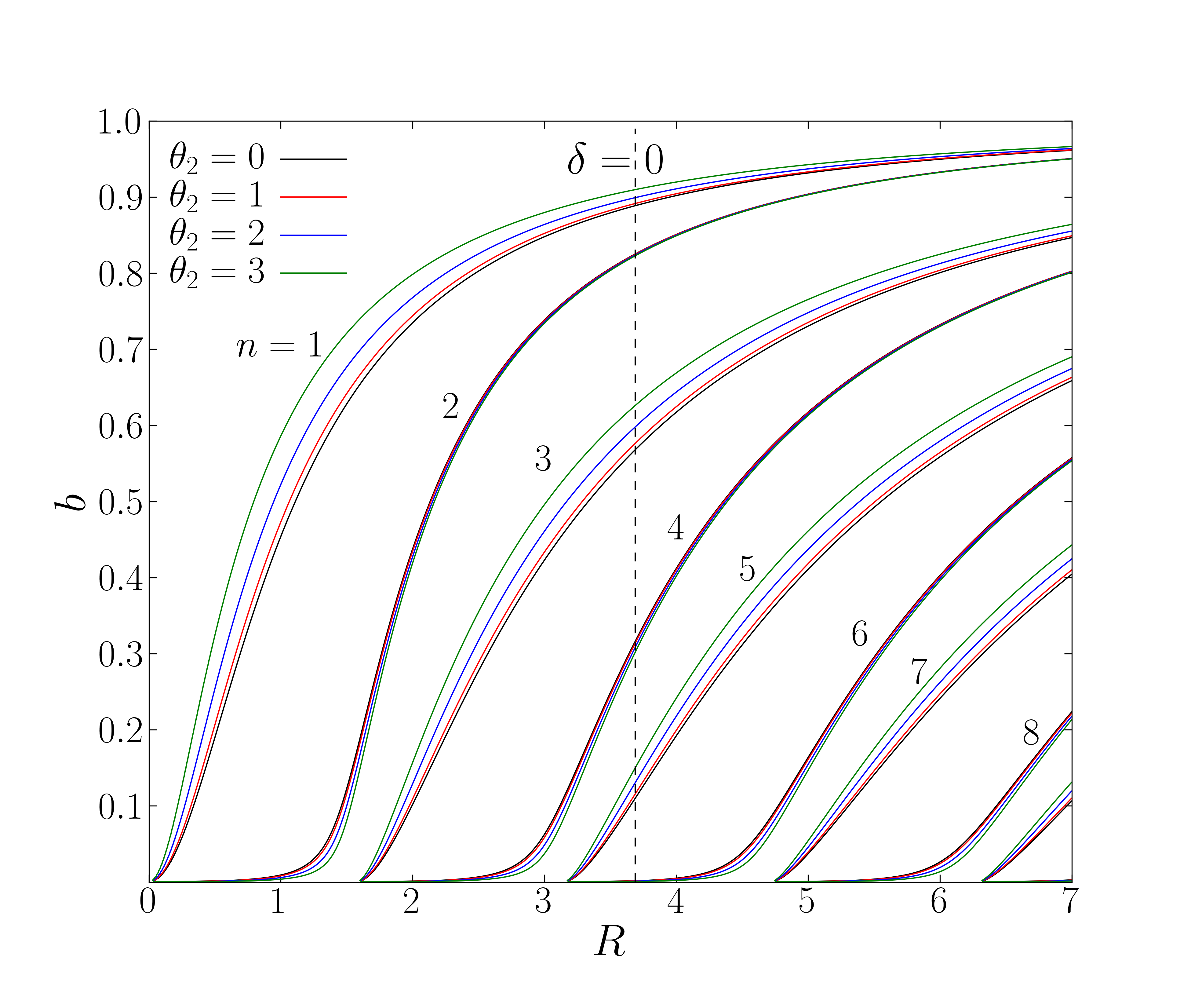}}
\stackinset{r}{28pt}{t}{35pt}{(b)}{\includegraphics[width=0.520\textwidth]{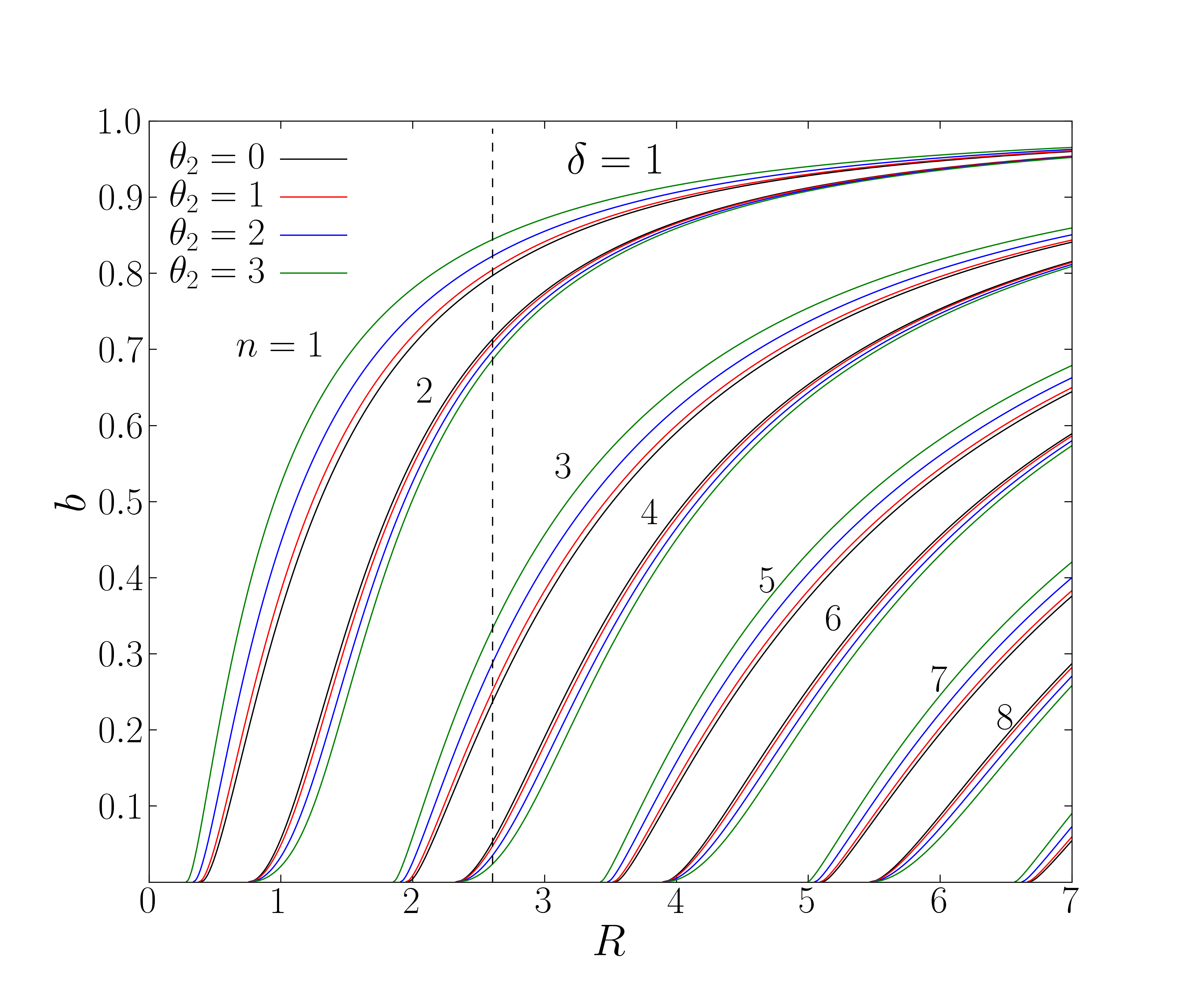}}
\stackinset{r}{28pt}{t}{35pt}{(c)}{\includegraphics[width=0.520\textwidth]{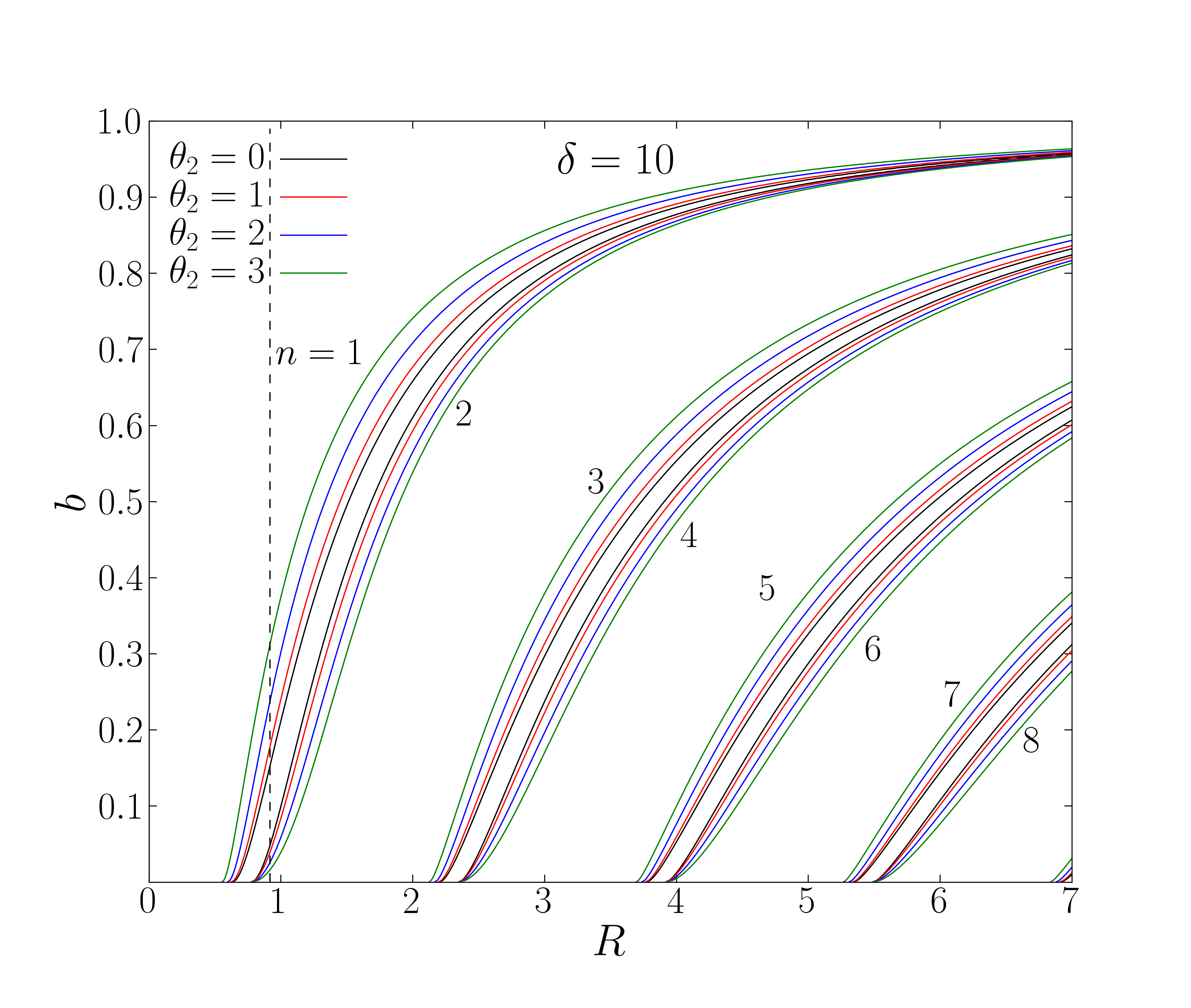}}
\stackinset{r}{28pt}{t}{35pt}{(d)}{\includegraphics[width=0.520\textwidth]{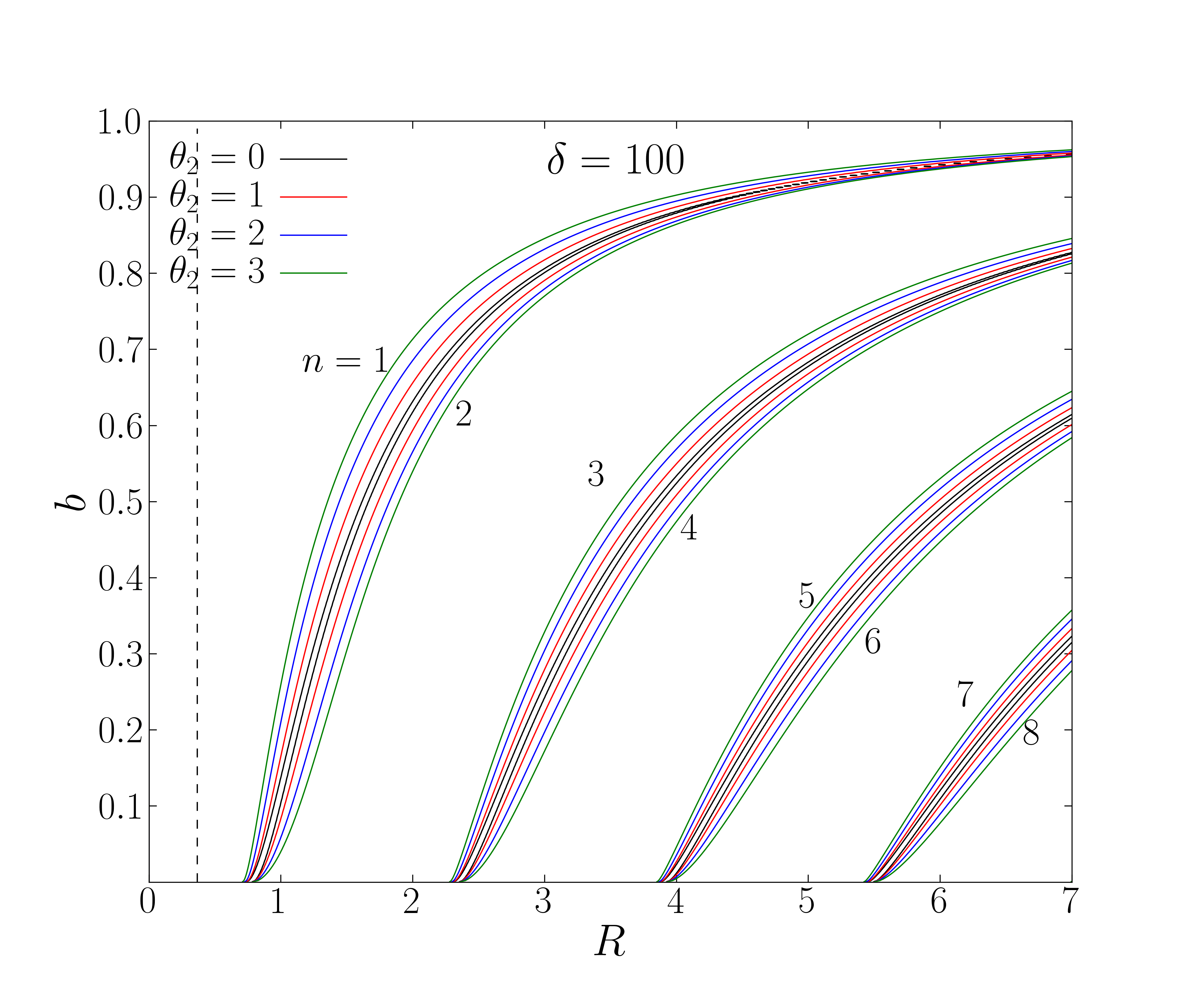}}
\caption{\label{fig:bvsR} Representación gráfica de la ecuación característica $\det(\mathbb{M})=0$ en función de $R$ y $b$ para un sistema con cladding triviales y un TI en el core, variando $\theta_{2}$ como se indica en la esquina superior izquierda de cada gráfico. En todos los casos, se fija $\mu_{1}=\mu_{2}=\mu_{3}=1$ y $\epsilon_{3}=1$, mientras que $\epsilon_{2}=16$ \cite{crosse_theory_2017}. Se exploran diferentes entornos para el TI, modificando así el factor de asimetría $\delta$ de la guía de onda: (a) $\epsilon_{1}=1$, (b) $\epsilon_{1}=8.5$, (c) $\epsilon_{1}=14.64$, y (d) $\epsilon_{1}=15.85$. En cada gráfico se considera una longitud de onda de operación $\lambda_{0}=3.3(2L)$, representada por una línea vertical en $R=\pi\sqrt{16-\epsilon_{1}}/3.3$. Este valor específico de $R$ define los valores permitidos de $b$ para la propagación de la OEM. Cada curva está etiquetada con un modo $n$, que puede ser interceptado o no por la línea vertical, dependiendo del valor de $\delta$. En el caso de $\delta=100$, no se observa modo de propagación.
}
\end{figure}
Para ilustrar los cambios en los parámetros de propagación cuando un TI está presente en el sistema, en la Fig. (\ref{fig:bvsR}) mostramos la ecuación característica considerando que la losa central es un TI y los medios laterales son topológicamente triviales (\textit{i.e.}, $\theta_{1} = \theta_{3} = 0$ y $\theta_{2} \neq 0$ en una configuración antiparalela). Además, hemos considerado que todos los medios son no magnéticos ($\mu_{1} = \mu_{2} = \mu_{3} = 1$) y que el medio 3 es aire ($\epsilon_{3} = 1$). Tomamos $\epsilon_{2} = 16$.  Al graficar, variamos $\epsilon_{1}$ para obtener valores de $\delta = \{0,1,10,100\}$ en (a), (b), (c) y (d) respectivamente. Cada gráfico en la Fig. (\ref{fig:bvsR}) muestra cuatro valores de $\theta_{2} = \{0,1,2,3\}$, que son altos comparados con los reportados para TIs o materiales Tellegen en general, pero útiles para visualizar los cambios de $b$ en función de $R$.

Es importante observar que, al definir una frecuencia específica del campo y los materiales utilizados, el parámetro $R$ queda determinado. Así, en la Fig. (\ref{fig:bvsR}), podemos ver que existen múltiples valores de $b$ que satisfacen la ecuación característica. Estos valores permitidos de $b$ corresponden a diferentes soluciones del campo EM, las cuales definiremos como distintos modos.

En cada gráfico, los modos están organizados de manera que, al incrementar progresivamente la frecuencia de oscilación del campo, la guía de onda admite modos $n=\{1, 2, 3, \dots\}$ cada vez más altos. Observamos que cuando $\delta \to 0$, principalmente los modos impares experimentan cambios significativos. A medida que el valor de $\delta$ aumenta, tanto los modos pares como los impares se modifican de manera similar. Para todos los valores de $\delta$ y $\theta_{2}$, notamos que $b_{\textup{n=impar}}(\theta_{2}) > b_{\textup{n=impar}}(0)$ y $b_{\textup{n=par}}(\theta_{2}) < b_{\textup{n=par}}(0)$. Estos cambios en el parámetro $b$ sugieren que, dado que $\lambda_{x} \sim 1/\sqrt{1-b}$ y $\delta_{x} \sim 1/\sqrt{b}$, los modos impares son más confinados y uniformes en comparación con los modos pares cuando hay un TI en la losa central. 

Es apropiado mencionar que los materiales que presentan valores mayores de la permeabilidad $\mu$, tanto de los medios laterales como la del TI, incrementa los efectos observables por el parámetro $\theta_2$. Esto último se menciona en \cite{Dvorquez2023mu}, donde se estudia la respuesta $\theta$ cuando hay $\mu$-metales y TIs en el sistema, pero no en el contexto de guías de ondas en forma de slab. 
\section{Solución semi-analítica para guía de onda simétrica} \label{6.4}
\begin{SCfigure}[1][t]
\caption{\small{Soluciones de la ecuación característica para el caso simétrico ($\delta=0$) con diferentes valores de $\theta_{2}=\alpha \theta_{\textup{TI}}/\pi$. Se consideran los valores $\mu_{1}=\mu_{2}=1$, $\epsilon_{1}=\epsilon_{3}=12$, $\epsilon_{2}=16$ y $\lambda_{0}=3.3\, (2L)$. Se muestran los primeros de cuatro modos de propagación en la guía de onda para la frecuencia de operación considerada $k_0=2\pi/\lambda_0$. Cada modo tiene un color diferente; $n=1$ azul, $n=2$ amarillo, $n=3$ verde y $n=4$ rojo.}
\label{FIG:v_vs_u}}
\includegraphics[scale=0.55]{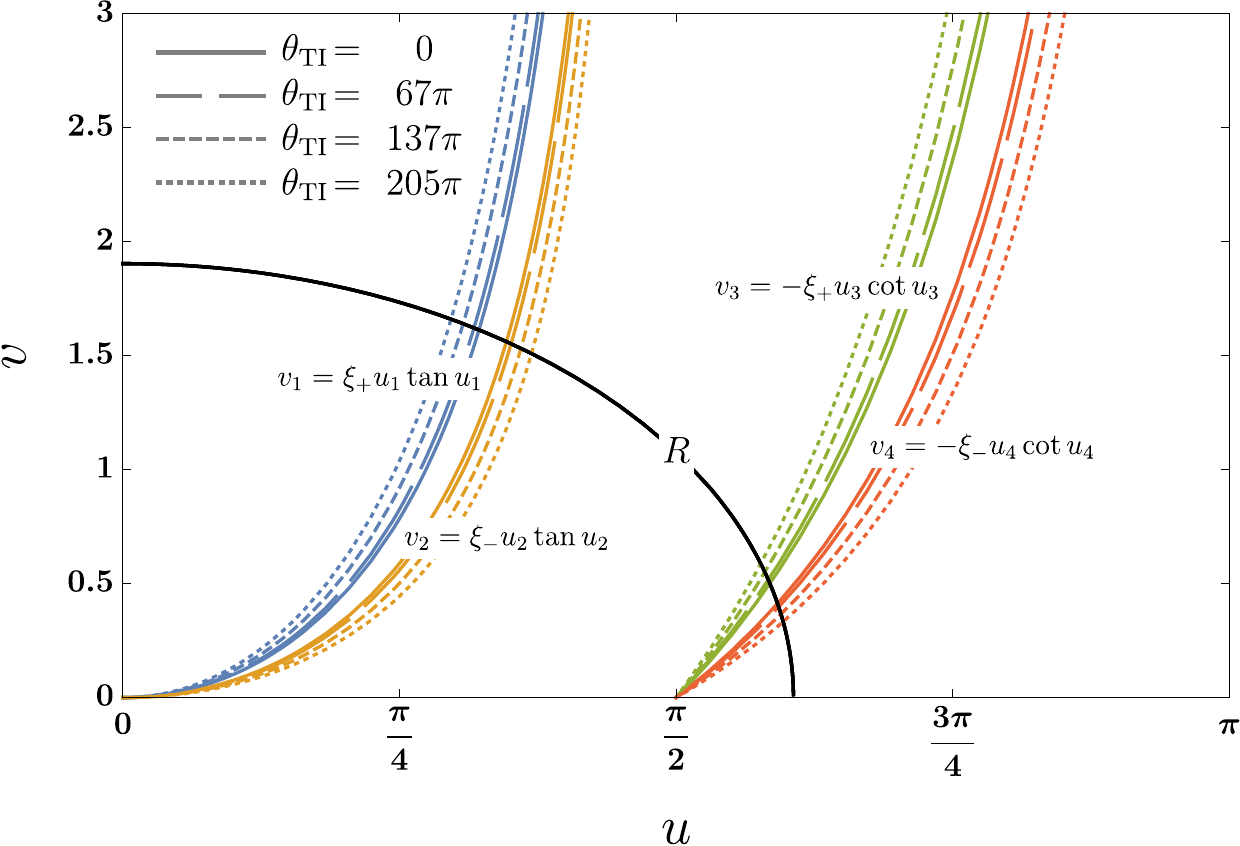}
\end{SCfigure}

Para investigar los efectos inducidos por el parámetro $\theta$ de manera más detallada, simplificaremos el sistema considerando que los cladding están hechos del mismo material, es decir,  $\epsilon_{3}=\epsilon_{1}$ y $\mu_{3}=\mu_{1}$, dando lugar a una slab simétrica con $\delta=0$, lo que implica $\alpha_{3}=\alpha_{1}\equiv\alpha$ o $w=v$. %El caso asimétrico es más complicado, pero el comportamiento básico no cambia \cite{Hu:09}. 
En este escenario, la relación entre los números de onda transversales, según la Ec. \eqref{EQ:DispersionRelationInEachMediumSymmetricCase}, se establece como sigue,
\begin{align}\label{EQ:DispersionRelationInEachMediumSymmetricCase}
    \left.\begin{matrix}
        v^{2}/L^{2}=k_{z}^{2}-k_{0}^{2}n^{2}_{1}\\
        u^{2}/L^{2}=k_{0}^{2}n^{2}_{2}-k_{z}^{2}
    \end{matrix}\right\}
    &&
    \Rightarrow 
    &&
    \begin{matrix}
    u^{2}+v^{2}=L^{2}k_{0}^{2}(n^{2}_{2}-n^{2}_{1})\equiv R^{2}.
    \end{matrix}
\end{align}
Además, consideramos que únicamente el core actúa es un TI con  $\theta_{2}\neq 0$, así las diferencias de los parámetros topológicos quedan  $\tilde{\theta}_{1}=\theta_{2}$ y $\tilde{\theta}_{2}=-\theta_{2}$. Con estas consideraciones, la ecuación característica queda,
\begin{align}
    \det (\mathbb{M})=(v-\xi_{+}u\tan u)(v-\xi_{-}u\tan u)(v+\xi_{+}u\cot u)(v+\xi_{-}u\cot u)=0, \label{EQ:AntiparallelBranches}
\end{align}
donde se ha definido la siguiente cantidad,
\begin{align}
    \xi_{\pm}&=\frac{1}{2} \left(\frac{\mu_{1}}{\mu_{2}}+\frac{\epsilon_{1}}{\epsilon_{2}}+\frac{\theta_{2}^{2} \mu_{1}}{\epsilon_{2}}\pm\sqrt{\left(\frac{\mu_{1}}{\mu_{2}}+\frac{\epsilon_{1}}{\epsilon_{2}}+\frac{\theta_{2}^{2} \mu_{1}}{\epsilon_{2}}\right)^2-4\frac{\mu_{1} \epsilon_{1}}{\mu_{2} \epsilon_{2}}}\right).
\end{align}
Hemos encontrado cuatro ecuaciones para $v$ en función de  $u$, $\epsilon_{1}$, $\epsilon_{2}$, $\mu_{1}$, $\mu_{2}$ y $\theta_{2}$, que satisfacen la condición de propagación \eqref{EQ:AntiparallelBranches}.
%y obtenemos la profundidad de la piel, $\delta = v^{-1}$, de la OEM en el revestimiento. 
Cada una de estas ecuaciones, denominadas ``ramas'', determina el comportamiento y la polarización de la OEM en la guía. %y define los modos de propagación permitidos. 
%Estas cuatro ecuaciones de $v(u)$ son independientes, en el sentido que un par modo de propagación satisface una y solo una de estas ``ramas''.  
En la Fig. (\ref{FIG:v_vs_u}) identificamos las primeras cuatro ramas como $n=1,2,3,4$, correspondientes a las parejas $(u_n, v_n)$ que anulan cada término de la ecuación característica. 
%
%---------------Figure------------
%---------------End Figure---------
Cada rama presenta una periodicidad debido a las funciones $\tan u$ y $\cot u$, lo que implica que los modos se repiten cada $u\to u+\pi$. Sin embargo, los modos en ramas repetidas no son idénticos, ya que las constantes de propagación son diferentes. Por tanto, existen infinitos pares de ($u_n,v_n$) con $n=1,2,3,4,5, 6, \dots$, donde las ramas se repiten en ciclos de cuatro, como se ilustra en la tabla (\ref{tab:etiquetaprimerosmodos}).
\begin{table}[h!]
    \centering
    \begin{tabular}{|c|c|}\hline
        Etiqueta del Modo & Condición de Propagación que satisface\\\hline
        $n=1$& $(v_1-\xi_{+}u_1\tan u_1)=0$\\
        $n=2$& $(v_2-\xi_{-}u_2\tan u_2)=0$\\
        $n=3$& $(v_3+\xi_{+}u_3\cot u_3)=0$\\
        $n=4$& $(v_4+\xi_{-}u_4\cot u_4)=0$\\\hline
        $n=5$& $(v_5-\xi_{+}u_5\tan u_5)=0$\\
        $n=6$& $(v_6-\xi_{-}u_6\tan u_6)=0$\\
        $n=7$& $(v_7+\xi_{+}u_7\cot u_7)=0$\\
        $n=8$& $(v_8+\xi_{-}u_8\cot u_8)=0$\\\hline
    \end{tabular}
    \caption{\label{tab:etiquetaprimerosmodos}Cada modo, o función propia de la ecuación de Helmholtz, satisface una única condición de propagación específica. Debido a la periodicidad de las funciones, existen infinitos modos de propagación que se repiten en ciclos de cuatro. Aunque los modos $n$ y $n+4$ cumplen con la misma condición de propagación, esto no significa que sean idénticos.}
    
\end{table}

Finalmente, para determinar los valores permitidos de $u_{n}$, es necesario encontrar la intersección entre la relación de dispersión $v_{n}^{2}=R^{2}-u_{n}^{2} $ y \eqref{EQ:AntiparallelBranches}, donde $R=Lk_{0}\sqrt{n_{2}^{2}-n_{1}^{2}}$ representa el radio del círculo. La frecuencia angular de operación $\omega=2\pi \nu_{0}$ o longitud de onda de vacío $k_{0}=2\pi/\lambda_{0}$ son necesarias para definir el radio del círculo y completar la definición de las ramas dados los atributos topo-ópticos de los materiales.

Por ejemplo, consideremos un sistema donde  $\mu_{1}=\mu_{2}=1$, $\epsilon_{1}=12$, $\epsilon_{2}=16$ y una longitud de onda de operación  $\lambda_{0}=3.3\, (2L)$. En la Fig. (\ref{FIG:v_vs_u}), presentamos las ramas correspondientes a diferentes valores de $\theta_{2}$ y observamos que, con la relación escogida entre $\lambda_{0}$ y $L$, existen cuatro modos guiados. Es crucial recordar que, para que la $\theta$-ED sea válida y describa correctamente a TIs, la frecuencia debe ser significativamente menor que la frecuencia asociada a la brecha de energía de los estados de superficie, \textit{i.e.}, $\nu_{0}\ll 73$ THz o $4.1~\mu m\ll \lambda_{0}$. 

Los puntos $(u_{n},v_{n})$ de intersección en la Fig. (\ref{FIG:v_vs_u}) están detallados en la tabla  (\ref{tab:Valoresnumericosthetagrande}), junto con los ángulos permitidos $\phi_{0}=\tan^{-1}(k_{z} L/u)$ en grados, la longitud de onda a lo largo de la guía  $\lambda_{z}=2\pi/k_{z}$, en la dirección transversal $\lambda_{x}=2\pi/\gamma$, y la longitud de penetración $\delta_{x}=1/\alpha$. Para valores de Bi$_2$Se$_3$, donde $\epsilon_{2}=16$ y $\theta_{\textup{TI}}=\pi$ \cite{crosse_theory_2017}, los cambios en las constantes de propagación son del orden del $\% 0.01$.
%$0.6~\mu m\ll L$

\begin{table}[H]
\begin{center}
\begin{tabular}{|c|c|c|c|c|c|c|}
\hline
\textbf{Modo $n$}    & $\theta_{\textup{TI}}$ & $(u_{n},v_{n})$& $\phi_{0}\,[\textup{deg}]$& $\lambda_{z}/L$&$\lambda_{x}/L$ &$\delta_{x}/L$\\ \hline
\multirow{5}{*}{$n=1$} 
& $ 0 $ & $(1.011,1.613)$ & $15.397$ & $1.711$ & $6.215$ & $0.62$\\%OK
& $11\pi $ & $(1.011,1.614)$ & $15.39$ & $1.711$ & $6.217$ & $0.62$\\%OK
& $67\pi $ & $(0.997,1.622)$ & $15.176$ & $1.71$ & $6.303$ & $0.616$\\%OK
& $137\pi $ & $(0.966,1.641)$ & $14.698$ & $1.706$ & $6.503$ & $0.61$\\%OK
& $205\pi $ & $(0.932,1.66)$ & $14.167$ & $1.702$ & $6.742$ & $0.602$\\ \hline%OK                                       

\multirow{5}{*}{$n=2$} 
& $ 0 $ & $(1.089,1.562)$ & $16.617$ & $1.722$ & $5.77$ & $0.64$\\%OK
& $11\pi $ & $(1.089,1.562)$ & $16.624$ & $1.722$ & $5.768$ & $0.64$\\%OK
& $67\pi $ & $(1.102,1.552)$ & $16.825$ & $1.724$ & $5.7$ & $0.644$\\%OK
& $137\pi $ & $(1.13,1.532)$ & $17.265$ & $1.728$ & $5.559$ & $0.653$\\%OK
& $205\pi $ & $(1.16,1.51)$ & $17.736$ & $1.732$ & $5.416$ & $0.662$\\ \hline%OK
                                                  
\multirow{5}{*}{$n=3$}     
& $ 0 $ & $(1.837,0.501)$ & $28.842$ & $1.884$ & $3.42$ & $1.997$\\%OK
& $11\pi $ & $(1.837,0.501)$ & $28.84$ & $1.884$ & $3.421$ & $1.995$\\%OK
& $67\pi $ & $(1.833,0.517)$ & $28.767$ & $1.882$ & $3.429$ & $1.936$\\%OK
& $137\pi $ & $(1.823,0.55)$ & $28.597$ & $1.879$ & $3.447$ & $1.817$\\%OK
& $205\pi $ & $(1.811,0.587)$ & $28.4$ & $1.876$ & $3.469$ & $1.703$\\ \hline%OK
                                                  
\multirow{5}{*}{$n=4$}
& $ 0 $ & $(1.859,0.413)$ & $29.216$ & $1.891$ & $3.38$ & $2.422$\\%OK
& $11\pi $ & $(1.859,0.412)$ & $29.218$ & $1.891$ & $3.38$ & $2.425$\\%OK
& $67\pi $ & $(1.862,0.398)$ & $29.273$ & $1.892$ & $3.374$ & $2.514$\\%OK
& $137\pi $ & $(1.868,0.366)$ & $29.384$ & $1.894$ & $3.363$ & $2.73$\\%OK
& $205\pi $ & $(1.875,0.333)$ & $29.491$ & $1.896$ & $3.352$ & $3.001$\\ \hline%OK
\end{tabular}
\caption{\small{\label{tab:Valoresnumericosthetagrande} Constantes de propagación de los primeros cuatro modos, determinadas por las intersecciones de las ramas de la ecuación característica y la ecuación circular de números de onda transversales, como se muestra en la Fig. (\ref{FIG:v_vs_u}). Hemos incluido los valores numéricos de las constantes de propagación para $\theta_{\textup{TI}}=11\pi$ debido a que mas adelante usaremos estos valores para clasificar los modos de propagación.}}
\end{center}
\end{table}
%En general, el cambio inducido por $\theta$ en los parámetros $(u_{n},v_{n})$ es pequeño. Para cuantificar el cambio, consideramos una expansión de Taylor en ambos parámetros de propagación. El valor de $u_{n}$
%
\section{Clasificación de modos electromagnéticos permitidos en la $\theta$-ED} \label{6.5}
Una vez obtenidos los valores de $u_{n}$ y $v_{n}$, es posible obtener el comportamiento del campo EM resultante. En el caso asimétrico general, tanto $u_{n}$ como $v_{n}$ son valores que deben determinarse numéricamente, debido a que no es posible encontrar una solución de la ecuación característica de forma analítica . Sin embargo, en el escenario simétrico-antiparalelo, $u_{n}$ requiere determinación numérica, mientras que $v_{n}$ se expresa analíticamente como función de $u_{n}$, permitiendo una solución semi-analítica.

Para cada ecuación independiente de $v_{n}$, proveniente de la Ec. \eqref{EQ:AntiparallelBranches}, la sustituimos en \eqref{EQ:BCsSystem} y resolvemos el sistema algebraico resultante. En la electrodinámica convencional ($0$-ED), donde $\theta_{m}=0$, \textit{i.e.}, no se consideran TIs, resulta que tres de las cuatro amplitudes desaparecen, como se resume en la siguiente tabla (\ref{tab:1}). 
%------------------------------------
\begin{table}[H]
\begin{tabular}{|c|c|c|c|}
\hline
\textbf{\begin{tabular}[c]{@{}c@{}}Condición de Propagación \\ que satisface\end{tabular}} & \textbf{\begin{tabular}[c]{@{}c@{}}Amplitudes \\ Nulas\end{tabular}} & \textbf{\begin{tabular}[c]{@{}c@{}}Amplitudes \\ Distintas de Cero\end{tabular}} & \textbf{\begin{tabular}[c]{@{}c@{}}Clasificación \\ del modo\end{tabular}} \\ \hline
$v_{1}=\frac{\mu _1 }{\mu _2}u_{1}\tan u_{1}$ & $E_{e1}=E_{o1}=B_{o1}=0$           & $B_{e1}\neq 0$ & TE-even-1 \\
$v_{2}=\frac{\epsilon_1}{\epsilon_2}u_{2}\tan u_{2}$& $E_{o2}=B_{e2}=B_{o2}=0$     & $E_{e2}\neq 0$ & TM-even-1 \\
$v_{3}=-\frac{\mu_1}{\mu_2}u_{2}\cot u_{3}$& $E_{e3}=E_{o3}=B_{e3}=0$              & $B_{o3}\neq 0$ & TE-odd-1  \\
$v_{4}=-\frac{\epsilon_1}{\epsilon_2}u_{4}\cot u_{4}$& $E_{e4}=B_{e4}=B_{o4}=0$    & $E_{o4}\neq 0$ & TM-odd-1  \\\hline
\end{tabular}
\caption{Para cada $v_n$ que satisface la ecuación característica cuando $\theta= 0$, se reemplaza en \eqref{EQ:BCsSystem} obteniendo los valores de las amplitudes. La amplitud $\neq 0$ nos permite clasificar los modos de la 0-ED.}
\label{tab:1}
\end{table}
%------------------------------------
En esta clasificación, cada solución de la ecuación característica define un tipo específico de modo, por ejemplo, TE-even-1 corresponde al primer modo donde el campo EM es transversal eléctrico (TE) y además, el campo eléctrico transversal es par. Debido a que las funciones $\tan u$ y $\cot u$ son periódicas, existen infinitas ``ramas'' donde el campo tiene una polarización TE-even, pero, aunque esta clasificación se repite en ciclos de cuatro, es importante destacar que cada ciclo representa modos distintos, es decir, TE-even-1 $\neq$ TE-even-2. Esto quiere decir que existe un par de números $\{u_5,v_5\}$ donde el campo EM tiene una polarización que es TE-even, pero como $u_1\neq u_5$ y $v_1\neq v_5$, se concluye que TE-even-1 $\neq$ TE-even-2.

La clasificación depende de la amplitud no nula resultante tras aplicar las BCs. En resumen, en la $0$-ED, la OEM se propaga con una polarización definida, TE o TM, manteniendo su perfil transversal a lo largo de toda la guía.

Al aplicar el mismo análisis en la $\theta$-ED, descubrimos que dos de las cuatro amplitudes desaparecen. Los resultados se resumen en la siguiente tabla (\ref{tab:2}), destacando los modos obtenidos para la $\theta$-ED:
%------------------------------------
\begin{table}[H]
\begin{tabular}{|c|c|c|c|}
\hline
\textbf{\begin{tabular}[c]{@{}c@{}}Condición de Propagación \\ que satisface\end{tabular}} & \textbf{\begin{tabular}[c]{@{}c@{}}Amplitudes \\ Nulas\end{tabular}} & \textbf{\begin{tabular}[c]{@{}c@{}}Amplitudes \\ Distintas de Cero\end{tabular}} & \textbf{\begin{tabular}[c]{@{}c@{}}Clasificación \\ del modo\end{tabular}} \\ \hline
$v_{1}=\xi_{+}u_{1}\tan u_{1}$& $E_{o1}=B_{o1}=0$& $E_{e1}=\frac{1}{\mathcal{F}_{+}}B_{e1},~~~B_{e1}\neq 0$& H-even-PTE-1\\
$v_{2}=\xi_{-}u_{2}\tan u_{2}$& $E_{o2}=B_{o2}=0$& $B_{e2}=\mathcal{F}_{-}E_{e2},~~~E_{e2}\neq 0$& H-even-PTM-1\\
$v_{3}=-\xi_{+}u_{3}\cot u_{3}$& $E_{e3}=B_{e3}=0$& $E_{o3}=\frac{1}{\mathcal{F}_{+}}B_{o3},~~~B_{o3}\neq 0$& H-odd-PTE-1\\
$v_{4}=-\xi_{-}u_{4}\cot u_{4}$& $E_{e4}=B_{e4}=0$& $B_{o4}=\mathcal{F}_{-}E_{o4},~~~E_{o4}\neq 0$& H-odd-PTM-1\\\hline
\end{tabular}
\caption{Para cada $v_n$ que satisface la ecuación característica cuando $\theta\neq 0$, se reemplaza en \eqref{EQ:BCsSystem} obteniendo los valores de las amplitudes. La amplitud $\neq 0$ nos permite clasificar los modos de la $\theta$-ED.}
\label{tab:2}
\end{table}
%------------------------------------
donde,
\begin{align}
    \mathcal{F}_{\pm}&=\frac{\mu_2\mu_1\theta_{2}}{\mu _1-\mu _2 \xi_{\pm}}.
\end{align}
En la $\theta$-ED, identificamos cuatro modos de propagación distintos, todos híbridos (H). Esto implica que una amplitud eléctrica longitudinal siempre se acompaña de una amplitud magnética longitudinal y viceversa. Debido a que el valor de $\theta_{2}$ es pequeño en comparación con los parámetros ópticos usuales, estos modos pueden pensarse como perturbaciones de los modos $0$-ED, indicando una polarización predominante (P) que proviene del $0$-ED. Por ejemplo, H-even-PTE-1 corresponde al primer modo híbrido con el campo eléctrico transversal completamente par (ambas componentes pares) y es predominantemente TE. 

La clasificación en ``modos'' no se debe únicamente a la discretización del espacio de soluciones al confinar la OEM. Se les denomina modos porque es posible definir un producto punto entre ellos, y de esta manera, el producto punto de dos modos distintos resulta ser ortogonal (ver Apéndice \ref{sec:ReciprocidaddeLorentz}).

Es posible expresar las amplitudes eléctricas y magnéticas inducidas únicamente en función de $\mathcal{F}_{+}$ o $\mathcal{F}_{-}$, ya que $\mathcal{F}_{+}\mathcal{F}_{-}=-\mu_{2}\epsilon_{2}$. Es importante notar que las amplitudes de los campos inducidos tienden a cero cuando el TI no está presente, es decir,
\begin{align}
    \lim_{\theta_{2}\to 0}\frac{1}{\mathcal{F}_{+}}&=0, &
    \lim_{\theta_{2}\to 0}\mathcal{F}_{-}&=0.
\end{align}
Por lo tanto, la hibridación es causada únicamente por la presencia de una TMEP no nulo en los TIs. En resumen, los campos EM se propagan en modos, de los cuales podemos identificar cuatro con comportamientos distintos. En la siguiente tabla se presentan los campos longitudinales para cada uno de estos cuatro modos diferentes:
%--------------------------
\begin{table}[H]
\centering
\begin{tabular}{|c|c|c|c|}
\hline
Modo                          & \multicolumn{1}{c|}{$E_{z}(x)$} & \multicolumn{1}{c|}{$B_{z}(x)$} & Dominio \\ \hline
\multirow{3}{*}{H-even-PTE-1}&                              $-B_{e1}\frac{\sin u_1 }{\mathcal{F}_{+}}e^{(x/L+1)v_{1}}$&                              $-B_{e1}\xi_{+}\sin u_1 e^{(x/L+1)v_{1}}$&         $x<-L$\\
                              &                              $B_{e1}\frac{1}{\mathcal{F}_{+}}\sin \left(\frac{u_1 x}{L}\right)$&                              $B_{e1}\sin \left(\frac{u_1 x}{L}\right)$&         $-L\leq x\leq L$\\
                              &                              $B_{e1}\frac{\sin u_1 }{\mathcal{F}_{+}}e^{(-x/L+1)v_{1}}$&                              $B_{e1}\xi_{+}\sin u_1 e^{(-x/L+1)v_{1}}$&         $L<x$\\ \hline
\multirow{3}{*}{H-even-PTM-1}&                              $-E_{e2}\sin u_2 e^{(x/L+1)v_{2}}$&                              $-E_{e2}\xi_{-}\mathcal{F}_{-}\sin u_2 e^{(x/L+1)v_{2}}$&         $x<-L$\\
                              &                              $E_{e2}\sin \left(\frac{u_2 x}{L}\right)$&                              $E_{e2}\mathcal{F}_{-}\sin \left(\frac{u_2 x}{L}\right)$&         $-L\leq x\leq L$\\
                              &                              $E_{e2}\sin u_2 e^{(-x/L+1)v_{2}}$&                              $E_{e2}\xi_{-}\mathcal{F}_{-}\sin u_2 e^{(-x/L+1)v_{2}}$&         $L<x$\\ \hline
\multirow{3}{*}{H-odd-PTE-1}&                              $B_{o3}\frac{\cos u_3 }{\mathcal{F}_{+}}e^{(x/L+1)v_{3}}$&                              $B_{o3}\xi_{+}\cos u_3 e^{(x/L+1)v_{3}}$&         $x<-L$\\
                              &                              $B_{o3}\frac{1}{\mathcal{F}_{+}}\cos \left(\frac{u_3 x}{L}\right)$&                              $B_{o3}\cos \left(\frac{u_3 x}{L}\right)$&         $-L\leq x\leq L$\\
                              &                              $B_{o3}\frac{\cos u_3 }{\mathcal{F}_{+}}e^{(-x/L+1)v_{3}}$&                              $B_{o3}\xi_{+}\cos u_3 e^{(-x/L+1)v_{3}}$&         $L<x$\\ \hline
\multirow{3}{*}{H-odd-PTM-1}&                              $E_{o4}\cos u_4 e^{(x/L+1)v_{4}}$&                              $E_{o4}\xi_{-}\mathcal{F}_{-}\cos u_4 e^{(x/L+1)v_{4}}$&         $x<-L$\\
                              &                              $E_{o4}\cos \left(\frac{u_4 x}{L}\right)$&                              $E_{o4}\mathcal{F}_{-}\cos \left(\frac{u_4 x}{L}\right)$&         $-L\leq x\leq L$\\
                              &                              $E_{o4}\sin u_4 e^{(-x/L+1)v_{4}}$&                              $E_{o4}\xi_{-}\mathcal{F}_{-}\sin u_4 e^{(-x/L+1)v_{4}}$&         $L<x$\\ \hline
\end{tabular}
\end{table}
%---------------------------
Para los modos de orden superior, las expresiones matemáticas de los campos se mantiene, aunque cambia el valor de  $u_{n}$, y por lo tanto, de $v_n$ también. Los campos EM transversales están descritos por las Ecs. \eqref{EQ:TransversalFields_UsualMaxwell}.
\section{Normalización y \textit{plot} de los modos permitidos en $\theta$-ED}\label{6.6}
Hasta este punto, hemos determinado 11 de las 12 incógnitas inicialmente identificadas en el problema. La incógnita restante corresponde a la amplitud de un modo específico. Para definirla, adoptamos una técnica de normalización ampliamente utilizada en guías de ondas EMs \cite{collin1990field}, que consiste en normalizar todos los modos de la OEM a una potencia unitaria. En regiones alejadas de fuentes EMs, todo par de modos $\ell$ y $s$ cumplen con la siguiente relación de ortogonalización (como se detalla en el Apéndice (\ref{sec:ReciprocidaddeLorentz})),
\begin{align}\label{eq:ortogonormalidaddemodos}
\frac{c}{4\pi}\int_{\mathbb{R}^{2}}\frac{1}{4\mu}\left(\mathbf{E}_{\perp\ell}\times\mathbf{B}^{*}_{\perp s}+\mathbf{E}^{*}_{\perp s}\times\mathbf{B}_{\perp\ell}\right )\cdot\mathbf{\hat{z}}da_{\perp}&=\delta_{\ell s}p_{\ell}.
\end{align}
Aquí, $\mathbf{E}_{\perp\ell}=\mathbf{E}_{\perp\ell}(\mathbf{r}_{\perp})$ representa el perfil del campo EM de un modo específico, sin incluir el factor $e^{i(k_{z}z-\omega t)}$, $\delta_{\ell s}$ corresponde a la delta de Kronecker y $p_\ell$ corresponde a la potencia del modo $\ell$. Si $p_\ell=1$, la relación \eqref{eq:ortogonormalidaddemodos} es de ortonormalización de los modos. 
% Para un par de modos generales híbridos, podemos expresar   
% %
% \begin{align}
%     \frac{1}{2}\int_{\mathbb{R}^{2}}\frac{1}{\mu k_{\perp\ell}^{2}k_{\perp s}^{2}}\left((k_{z\ell}\boldsymbol{\nabla}_{\perp}E_{z\ell}-k_{0}\tongo{z}\times \boldsymbol{\nabla}_{\perp}B_{z\ell})\times(k_{zs}\boldsymbol{\nabla}_{\perp}B_{zs}^{*}+k_{0}\epsilon\mu\tongo{z}\times \boldsymbol{\nabla}_{\perp}E_{zs}^{*})\right )\cdot\mathbf{\hat{z}}da_{\perp}&=\delta_{\ell s}\\
% \frac{1}{2}\int_{\mathbb{R}^{2}}\frac{1}{\mu k_{\perp\ell}^{2}k_{\perp s}^{2}}\left(k_{z\ell}k_{zs}\boldsymbol{\nabla}_{\perp}E_{z\ell}\times \boldsymbol{\nabla}_{\perp}B_{zs}^{*}+k_{z\ell}k_{0}\epsilon\mu\boldsymbol{\nabla}_{\perp}E_{z\ell}\times\tongo{z}\times \boldsymbol{\nabla}_{\perp}E_{zs}^{*}  
%   -k_{0}\tongo{z}\times \boldsymbol{\nabla}_{\perp}B_{z\ell}\times \right )\cdot\mathbf{\hat{z}}da_{\perp}&=\delta_{\ell s}
% \end{align}
% %
% %
% \begin{align}
% \mathbf{E}_{\perp}&=\frac{i}{k_{\perp}^{2}}(k_{z}\boldsymbol{\nabla}_{\perp}E_{z}-k_{0}\tongo{z}\times \boldsymbol{\nabla}_{\perp}B_{z}), &
% \mathbf{B}_{\perp}&=\frac{i}{k_{\perp}^{2}}(k_{z}\boldsymbol{\nabla}_{\perp}B_{z}+k_{0}\epsilon\mu\tongo{z}\times \boldsymbol{\nabla}_{\perp}E_{z}),
% \label{EQ:TransversalFields_UsualMaxwell}
% \end{align}
% %
Cada modo de propagación permitido requiere su propia normalización. Para cada modo, sustituimos las componentes transversales de los campos en la Ec. \eqref{eq:ortogonormalidaddemodos} y separamos la integral por cada uno de los medios $m={1,2,3}$. Por ejemplo, la amplitud que normaliza el modo $n=1$ es,
% %
% \begin{table}[H]
% \begin{tabular}{|c|c|}
% \hline
% Modo                          & \multicolumn{1}{c|}{Amplitud}  \\ \hline
% H-even-PTE-1&                              $B_{e1}=$\\ \hline
% H-even-PTM-1&                              $E_{e2}=$\\ \hline
% H-odd-PTE-1&                              $B_{o3}=$\\ \hline
% H-odd-PTM-1&                              $E_{o4}=$\\ \hline
% \end{tabular}
% \end{table}
%
%
\begin{align*}
    B_{e1}=\left [\frac{4 \xi_+^3\mathcal{F}_{+}^2 \mu _2^2 u_0^3}{L^3 k_{0}  k_{z1} \left(2 \cos ^2 u_0 \cot u_0 \left(\mu _1 \left(\mathcal{F}_{+}-\mu _2 \theta_{2}\right)^2+\mu _2^2 \epsilon _1\right)+\xi_+^3\mu _2 \left(2 u_0+\sin 2 u_0\right) \left(\mathcal{F}_{+}^2+\mu _2 \epsilon _2\right)\right)}  \right ]^{1/2}
\end{align*}
%
% %
% \begin{align}
%     B_{e1}=\left [\frac{4 \xi_+^3\mathcal{F}_{+}^2 \mu _2 u_0^3}{L^3 k_{0}  k_{z1} \left(2 \cos ^2 u_0 \cot u_0 \mu _2\left(\mu _1 \left(\mathcal{F}_{-}^{-1}\epsilon_{2}+ \theta_{2}\right)^2+\epsilon _1\right)+\xi_+^3\left(2 u_0+\sin 2 u_0\right) \left(\mathcal{F}_{+}^2+\mu _2 \epsilon _2\right)\right)}  \right ]^{1/2}
% \end{align}
% %
Este proceso se repite para cada uno de los modos.

%--------------
\begin{figure}[!ht]
\begin{center}
\stackinset{r}{120pt}{t}{35pt}{(a)}{\includegraphics[width=0.4\textwidth]{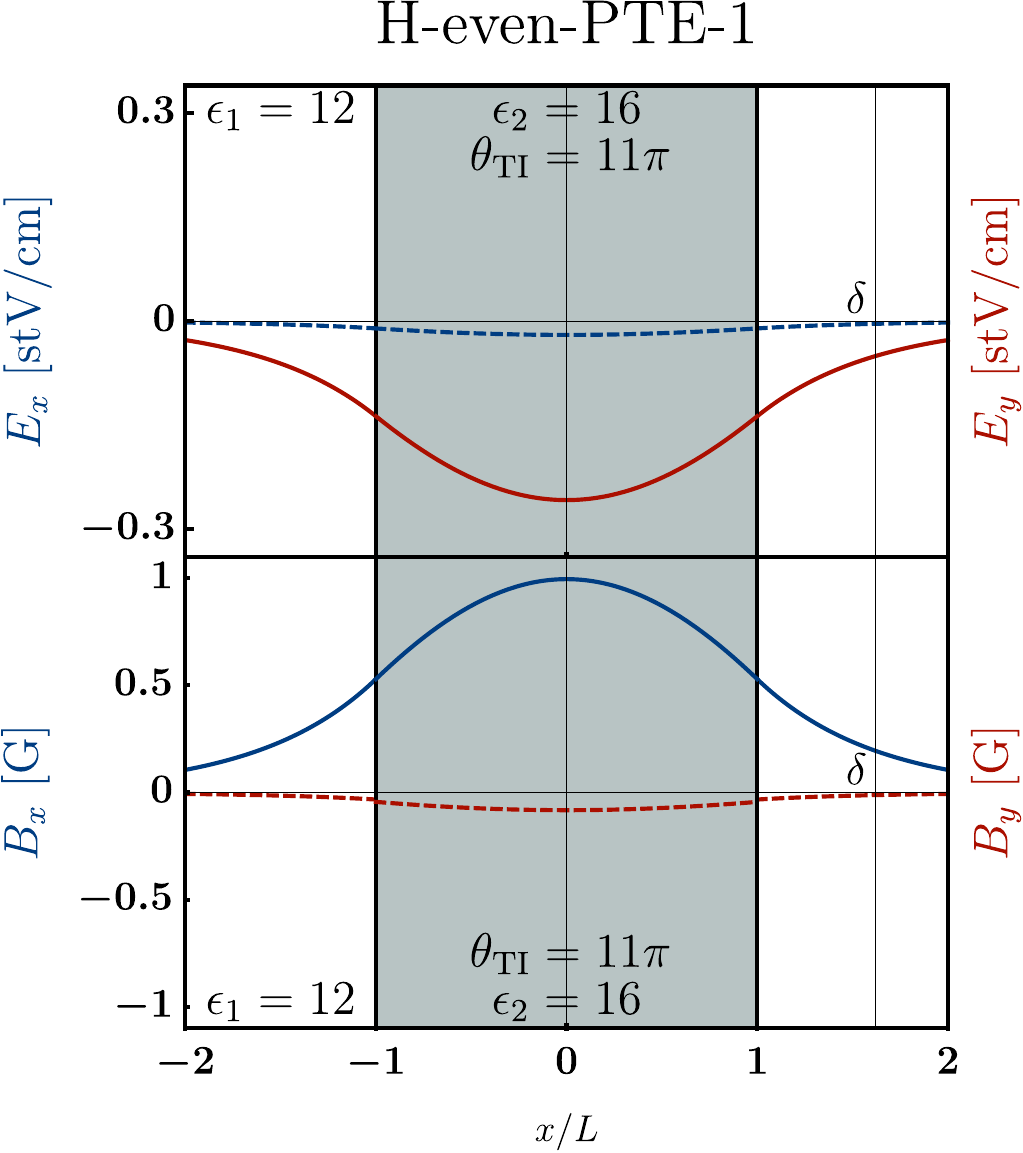}}
\includegraphics[width=0.471\textwidth]{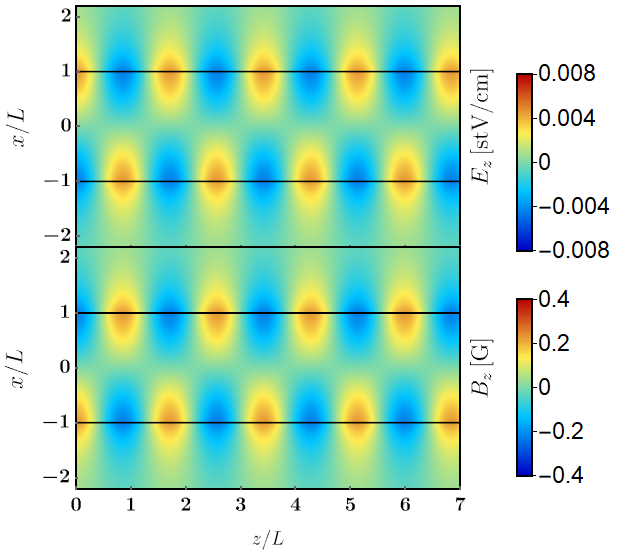}\\
\stackinset{r}{120pt}{t}{35pt}{(b)}{\includegraphics[width=0.4\textwidth]{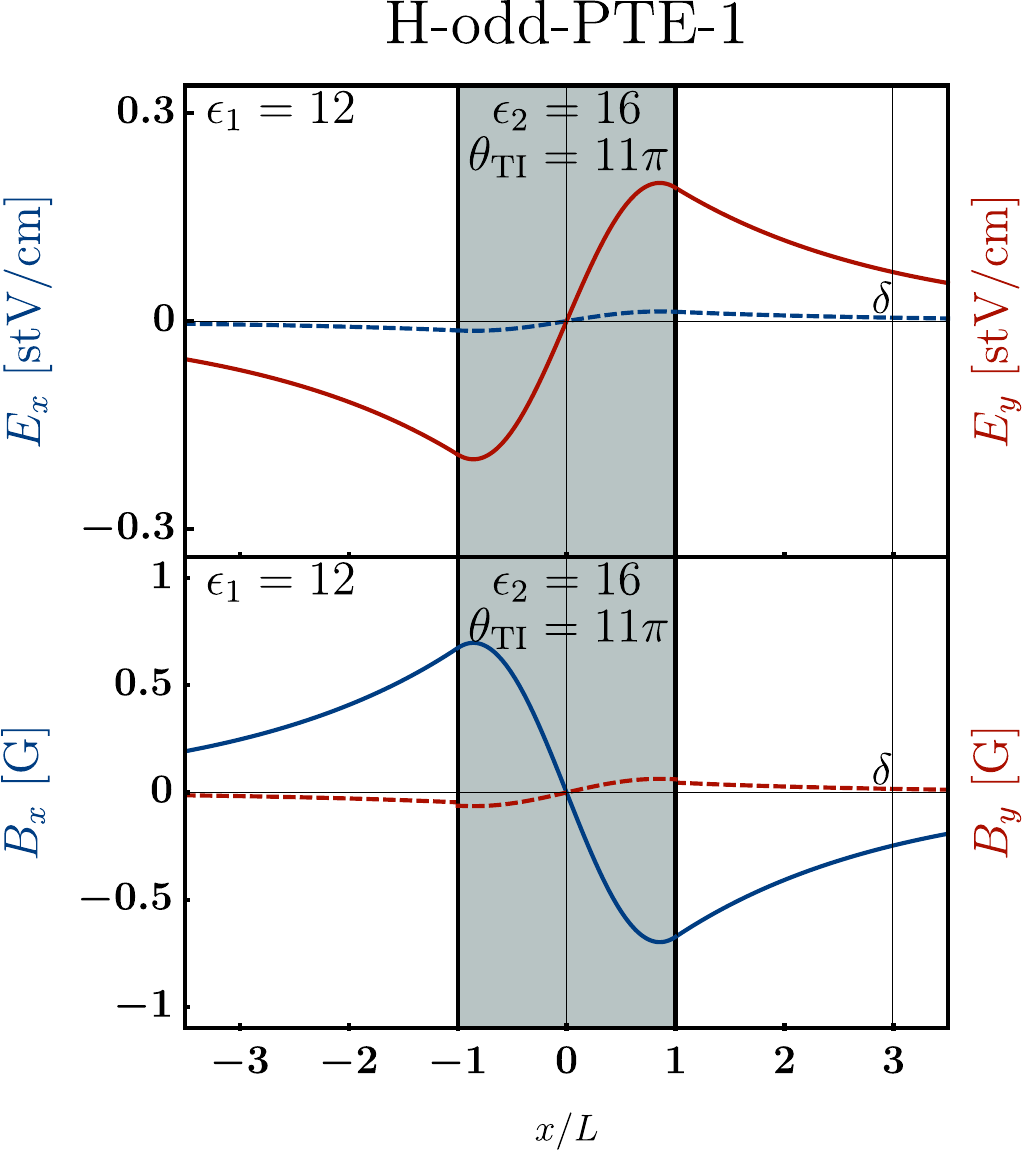}}
\includegraphics[width=0.471\textwidth]{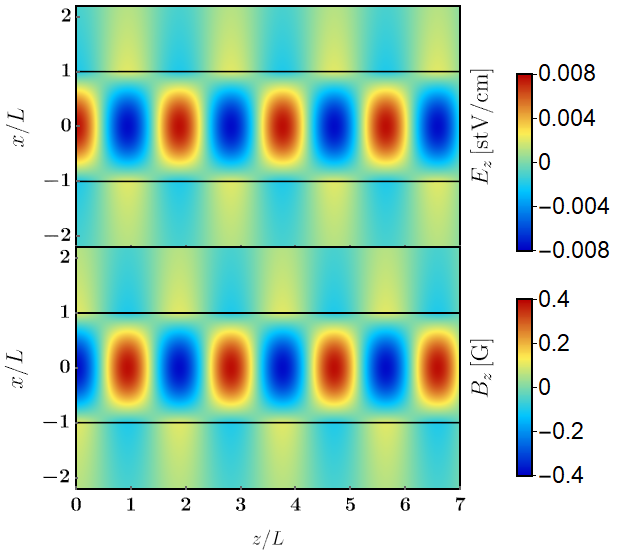}
\end{center}
\caption{\small{\label{fig:FieldsSlabTE} Campo EM de los dos primeros modos predominantemente transversales eléctricos (PTE), par (a) e impar (b), operando a una longitud de onda $\lambda_{0}=3.3(2L)$ en una guía de ondas simétrica con un núcleo TI de ancho $2L=40\,\mu m$. Las componentes $x$ e $y$ de los campos se muestran en azul y rojo, respectivamente. Las componentes longitudinales a lo largo de la dirección $z$ se ilustran con un \textit{density plot}. Los parámetros topo-ópticos utilizados son $\mu_{1}=\mu_{2}=1$, $\epsilon_{1}=12$, $\epsilon_{2}=16$ y $\theta_{2}=\alpha\theta_{\textup{TI}}/\pi=11\alpha$. Las curvas sólidas representan las componentes dominantes del modo TE, mientras que las curvas segmentadas indican las componentes inducidas únicamente por el TMEP $\theta_{2}$ del TI. Hemos marcado, con una linea vertical negra, el skin depth $\delta$ (no confundir con el parámetro de asimetría de la Ec. \eqref{eq:parámetrodeasimetría}). Notamos que el modo par es mas confinado el el modo impar.}
}
\end{figure}
%--------------
\begin{figure}[!ht]
\begin{center}
\stackinset{r}{225pt}{t}{35pt}{(a)}{\includegraphics[width=0.4\textwidth]{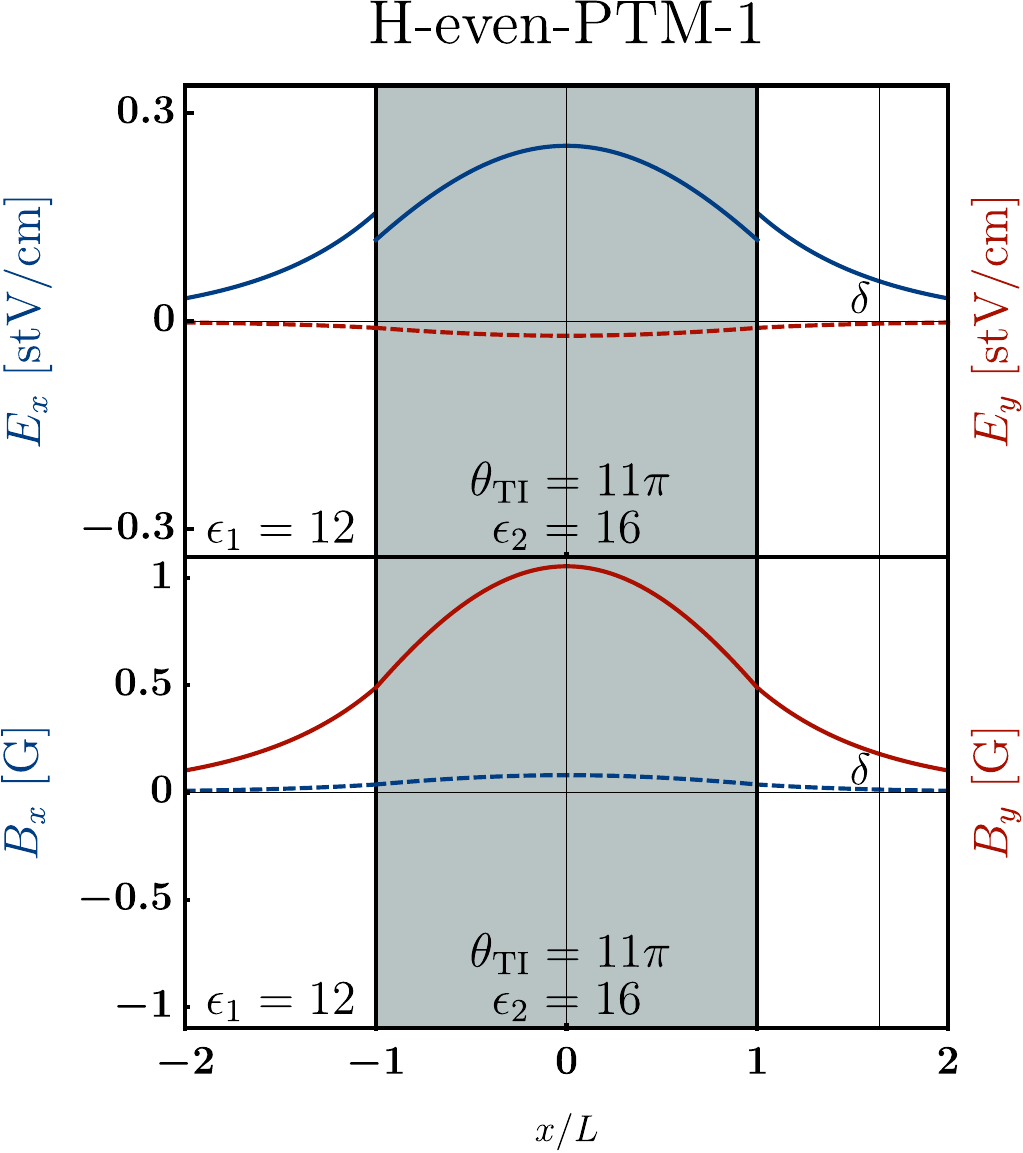}}
\includegraphics[width=0.471\textwidth]{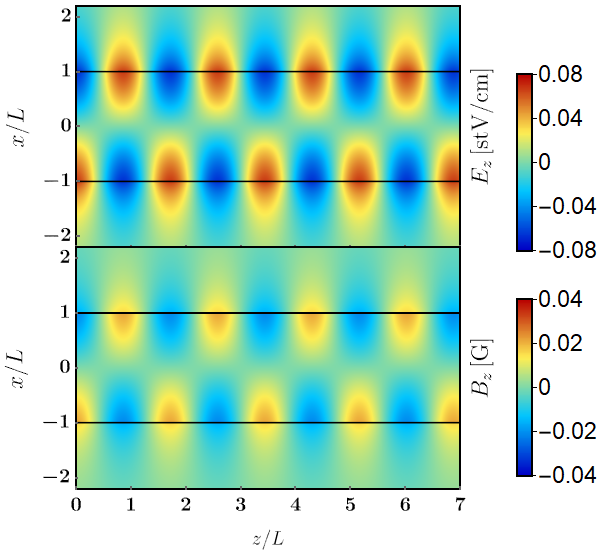}\\
\stackinset{r}{225pt}{t}{35pt}{(a)}{\includegraphics[width=0.4\textwidth]{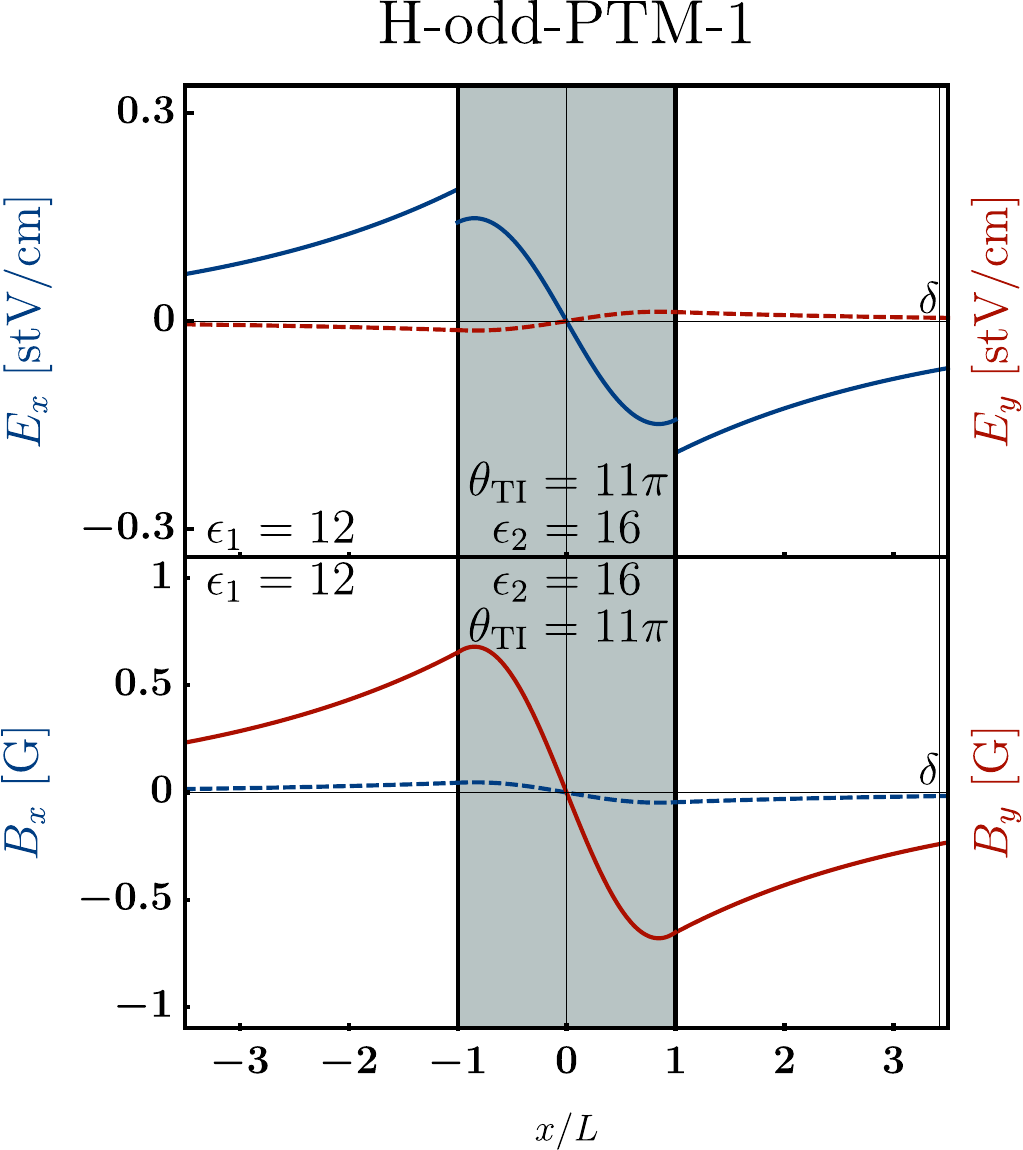}}
\includegraphics[width=0.471\textwidth]{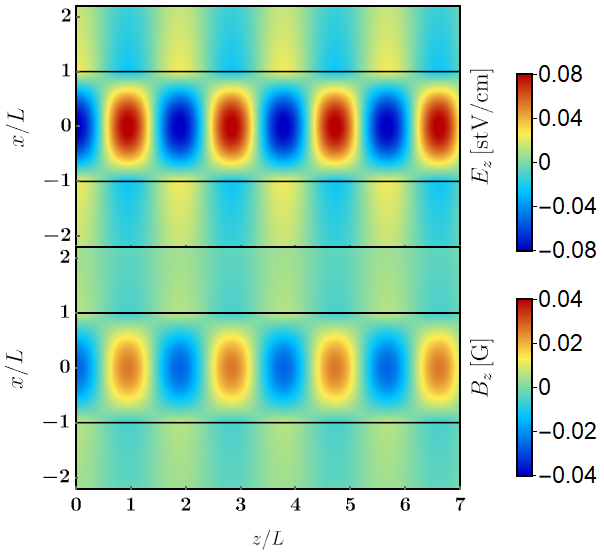}
\end{center}
\caption{\small{\label{fig:FieldsSlabTM} Campo EM de los dos primeros modos predominantemente transversales magnéticos (PTM), par (a) e impar (b), operando a una longitud de onda $\lambda_{0}=3.3(2L)$ en una guía de ondas simétrica con un núcleo TI de ancho $2L=40\,\mu m$. Las componentes $x$ e $y$ de los campos se muestran en azul y rojo, respectivamente. Las componentes longitudinales a lo largo de la dirección $z$ se ilustran con un \textit{density plot}. Los parámetros topo-ópticos utilizados son $\mu_{1}=\mu_{2}=1$, $\epsilon_{1}=12$, $\epsilon_{2}=16$ y $\theta_{2}=\alpha\theta_{\textup{TI}}/\pi=11\alpha$. Las curvas sólidas representan las componentes dominantes del modo TM, mientras que las curvas segmentadas indican las componentes inducidas únicamente por el TMEP $\theta_{2}$ del TI. Hemos marcado, con una linea vertical negra, el skin depth $\delta$ (no confundir con el parámetro de asimetría de la Ec. \eqref{eq:parámetrodeasimetría}). Notamos que el modo par es mas confinado el el modo impar.}
}
\end{figure}

En las Figs. (\ref{fig:FieldsSlabTE}) y (\ref{fig:FieldsSlabTM}) presentamos los campos EMs de los primeros cuatro modos de propagación permitidos, operando a una longitud de onda $\lambda_{0}=3.3(2L)$ en una guía de ondas simétrica con un núcleo-TI de ancho $2L=40\,\mu m$. Los parámetros topo-ópticos considerados son $\mu_{1}=\mu_{2}=1$, $\epsilon_{1}=12$, $\epsilon_{2}=16$ y $\theta_{2}=\alpha\theta_{\textup{TI}}/\pi=11\alpha$. Las curvas azules y rojas ilustran las componentes de los campos en las direcciones $x$ e $y$, respectivamente. Las componentes longitudinales a lo largo de la dirección $z$ se visualizan mediante un mapa de densidades (DensityPlot) en la columna derecha de las figuras. Cada una de estas componentes de los campos EM se presenta en sus respectivas unidades gaussianas.
%Cuando $\theta_{2}\to 0$ las curvas solidas no se anulan mientras que las segmentadas si. 

Los modos están normalizados conforme a la Ec. \eqref{eq:ortogonormalidaddemodos} a una potencia de $0.1\, W/cm$. Esto significa que las amplitudes de los campos EM mostradas en las Figs. (\ref{fig:FieldsSlabTE}) y (\ref{fig:FieldsSlabTM}) son las requeridas para que la potencia transmitida por un modo específico en una guía de onda de $2L=40\,\mu m$ de ancho sea de $0.1\, W$ por centímetro de longitud transversal.

Hasta ahora, hemos observado diferencias significativas en la amplitud de las componentes inducidas relativas a las componentes predominantes para cada uno de los modos H-PTE y H-PTM. Recordemos que en un modo TE y TM, las componentes transversales no nulas del campo eléctrico son $E_{y}$ y $E_{x}$ , respectivamente, descritas por las Ecs. \eqref{EQ:TransversalFields_UsualMaxwell}. Específicamente en la geometría en forma de slab tenemos,
\begin{align}
    E_{x}&=\frac{i}{k_{x}^{2}}k_{z}\partial_{x}E_{z}, &&\textup{y,} & E_{y}&=-\frac{i}{k_{x}^{2}}k_{0}\partial_{x}B_{z}.
\end{align}
Así, las componentes inducidas relativas a las predominantes se definen como,
%
% \begin{align}
% \mathbf{E}_{\perp}&=\frac{i}{k_{\perp}^{2}}(k_{z}\boldsymbol{\nabla}_{\perp}E_{z}-k_{0}\tongo{z}\times \boldsymbol{\nabla}_{\perp}B_{z}), &
% \mathbf{B}_{\perp}&=\frac{i}{k_{\perp}^{2}}(k_{z}\boldsymbol{\nabla}_{\perp}B_{z}+k_{0}\epsilon\mu\tongo{z}\times \boldsymbol{\nabla}_{\perp}E_{z}),
% \end{align}
%
\begin{align}\label{eq:thetaseñalgeneral}
    \textup{H-PTE:}~~~~~& \eta_{n}^{\textup{TE}}\equiv\left |\frac{E_{xn}^{\theta}}{E_{yn}}\right |=\frac{k_{zn}\mathcal{F}_{-}}{k_{0}\epsilon_{2}\mu_{2}}, & \textup{H-PTM:}~~~~~& \eta_{n}^{\textup{TM}}\equiv\left |\frac{E_{yn}^{\theta}}{E_{xn}}\right |=\frac{k_{0}\mathcal{F}_{-}}{k_{zn}}.
\end{align}
Donde, $n$ es el índice modal y $\eta_{n}$ es la intensidad del campo EM inducido por el parámetro $\theta$ relativa al campo EM predominante de la 0-ED. Aquí, la componente con el superíndice $\theta$ es la componente inducida. En el core TI de la guía se satisface $k_{zn}\leq k_{0}\sqrt{\epsilon_{2}\mu_{2}}$, como se ve en la relación de dispersión, entonces se obtiene que $\eta_{n}^{\textup{TE}}\leq\eta_{n}^{\textup{TM}}$, lo que indica que la $\theta$-señal es más prominente en un modo PTM que en un PTE. Para los campos EM en las Figs. (\ref{fig:FieldsSlabTE}) y (\ref{fig:FieldsSlabTM}), las intensidades relativas de la $\theta$-señal son $\eta_{1}^{\textup{TE}}=0.077$, $\eta_{2}^{\textup{TM}}=0.083$, $\eta_{3}^{\textup{TE}}=0.069$ y $\eta_{4}^{\textup{TM}}=0.091$.

\begin{figure}[!ht]
\begin{center}
\stackinset{r}{220pt}{t}{35pt}{(a)}{\includegraphics[width=0.4\textwidth]{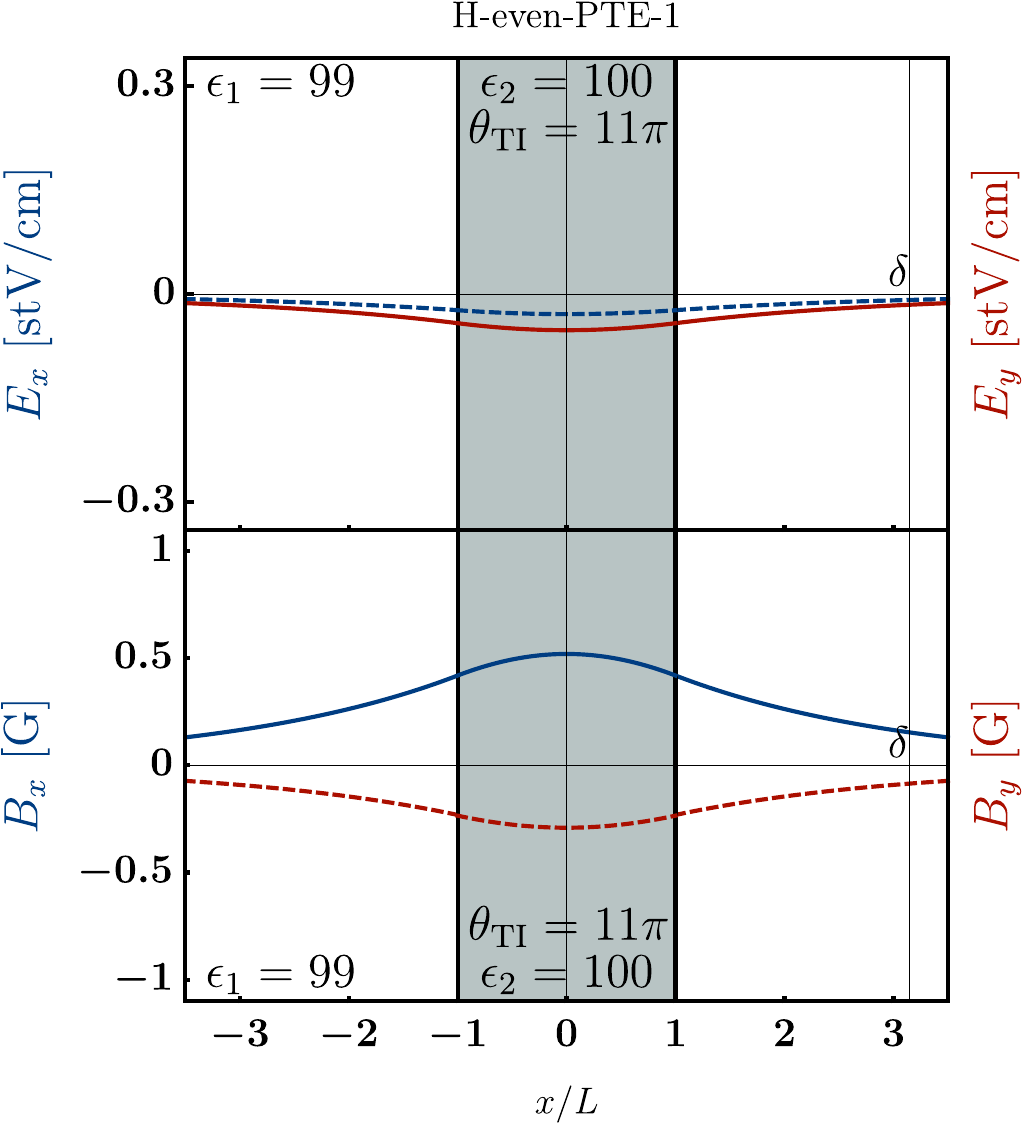}}
\includegraphics[width=0.471\textwidth]{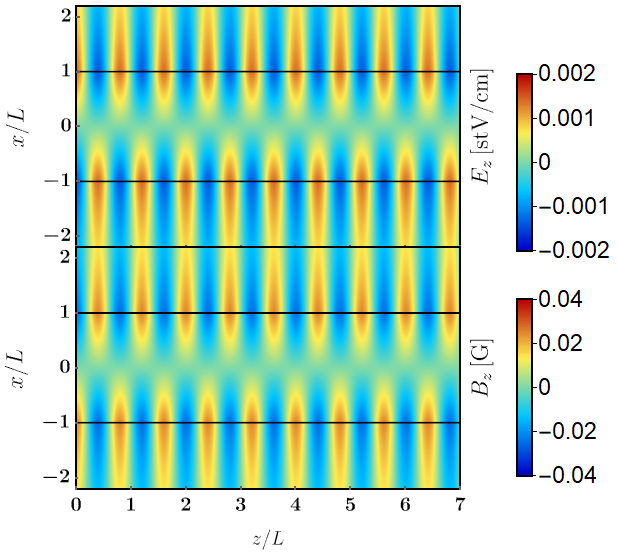}\\
\stackinset{r}{220pt}{t}{35pt}{(b)}{\includegraphics[width=0.4\textwidth]{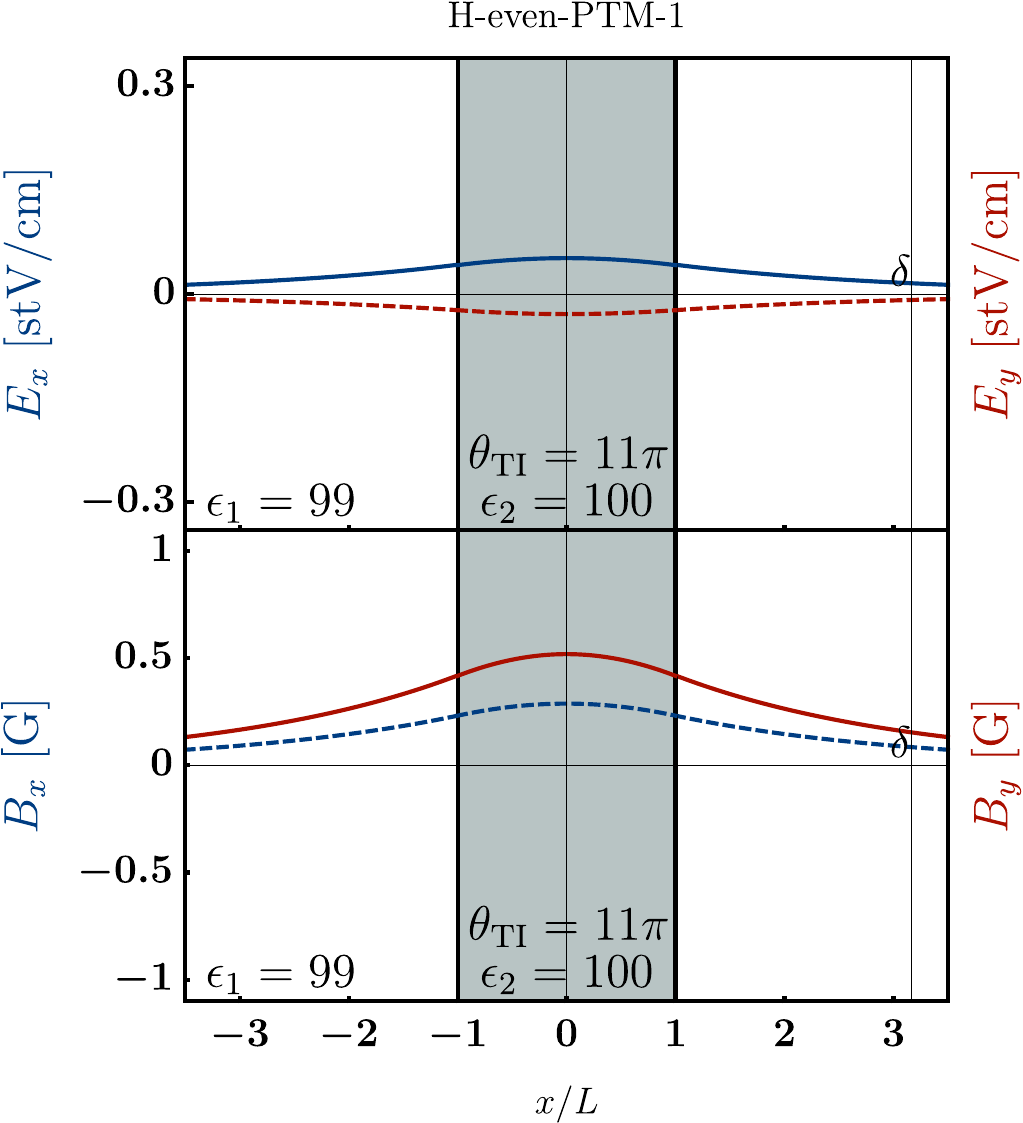}}
\includegraphics[width=0.471\textwidth]{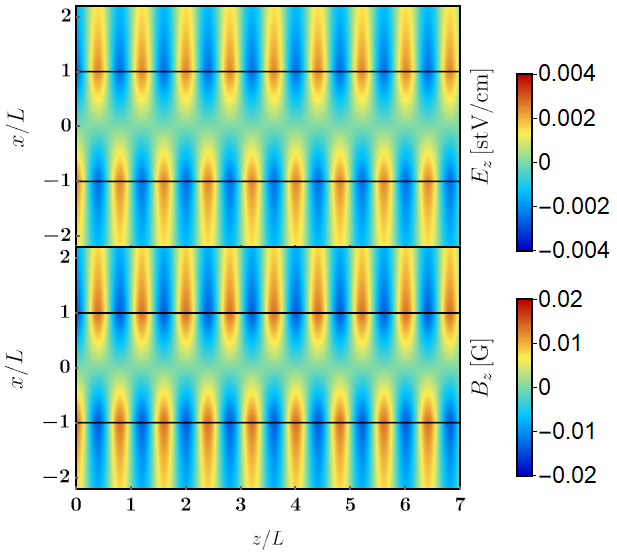}
\end{center}
\caption{\small{\label{fig:FieldsSlab100} Mejora de campos EMs inducidos debido al TMEP $\theta_{2}$. En (a) y (b) se utilizó una frecuencia de operación de $\nu_{0}=500\,[\textup{GHz}]$ y un TI con un ancho de $2L=\,150[\mu m]$, empleando los siguientes parámetros topo-ópticos: $\mu_{1}=\mu_{2}=1$, $\epsilon_{1}=99$, $\epsilon_{2}=100$ y $\theta_{2}=\alpha(11\pi)/\pi$ \cite{maciejko_topological_2010}. Se observa que los campos EM inducidos representan más del $50 \%$ en comparación con los campos predominantes en los dos modos únicos de propagación permitidos. Las curvas sólidas representan las componentes dominantes de los modos TE y TM, mientras que las curvas segmentadas ilustran las componentes inducidas específicamente por el TMEP $\theta_{2}$ del TI.}}
\end{figure}

Recordemos que la OEM se propaga haciendo múltiples reflexiones internas totales donde el rayo de luz y la pared de la guía forma un ángulo de incidencia $\phi_{0}$ (recordar Fig. (\ref{FIG:SlabGeo})). Independientemente de los parámetros topo-ópticos seleccionados, los modos con propagación más oblicua ($\phi_{0}$ cercano a $90°$) presentan $\eta_{n}^{\textup{TE}}\sim\eta_{n}^{\textup{TM}}$ , y para modos menos oblicuos (propagación más transversal, $\phi_{0}$ cercano a $0°$) $\eta_{n}^{\textup{TE}}\to 0$ y $0<\eta_{n}^{\textup{TM}}$. Planteamos la siguiente pregunta: ¿Es posible configurar un sistema para maximizar la $\theta$-señal en comparación con la señal predominante? La respuesta es afirmativa, pero requiere cumplir ciertos requisitos: 

Considerando que el valor de $\theta_{2}$ es pequeño, podemos expandir las funciones $\mathcal{F}_{\pm}$ a primer orden en $\theta_{2}$,
\begin{align}\label{eq:Fexpansion}
    \frac{1}{\mathcal{F}_{+}}&\approx -\frac{\mu_1\theta_{2}}{\epsilon_2\mu_1-\epsilon_1\mu_2}+\mathcal{O}(\theta_{2}^{3}), & \mathcal{F}_{-}&\approx \epsilon_2\mu_2\frac{\mu_1\theta_{2}}{\epsilon_2\mu_1-\epsilon_1\mu_2}+\mathcal{O}(\theta_{2}^{3}).
\end{align}
Además, podemos escribir $k_{zn}$ y $k_{0}$ en función de los parámetros ópticos,
\begin{align}\label{eq:kzyk0}
    k_{zn}=&\sqrt{\frac{u_{n}^{2}n_{1}^{2}+v_{n}^{2}n_{2}^{2}}{L^{2}(n_{2}^{2}-n_{1}^{2})}}, & k_{0}=&\sqrt{\frac{u_{n}^{2}+v_{n}^{2}}{L^{2}(n_{2}^{2}-n_{1}^{2})}}.
\end{align}
Recordando que $v_{n}$ es función de $u_{n}$, podemos generalizar su comportamiento como $v_{n}=\xi_{\pm}u_{n}f_{n}$, donde ``$+/-$'' e refiere a los modos PTE/PTM respectivamente, y $f_{n}$ es la función trigonométrica usual que define las ramas, por ejemplo, $\tan u_{n}$ y $-\cot u_{n}$ para los modos pares e impares respectivamente. Al remplazar las Ecs. \eqref{eq:kzyk0} y \eqref{eq:Fexpansion} en las Ecs. \eqref{eq:thetaseñalgeneral}, expandiéndolas hasta el primer orden en $\theta_{2}$ y considerando que los medios no son magnéticos ($\mu_{1}=\mu_{2}=1$), obtenemos,
\begin{align}\label{}
    \textup{H-PTE:}~~~~~& \left |\frac{E_{xn}^{\theta}}{E_{yn}}\right |\approx\theta_{2} \sqrt{\frac{\epsilon _2}{\tilde{\epsilon}^{2} }-\frac{1}{\tilde{\epsilon}(f_{n}^2+1)}},\\
    \textup{H-PTM:}~~~~~& \left |\frac{E_{yn}^{\theta}}{E_{xn}}\right |\approx\theta_{2} \sqrt{\frac{\epsilon _2}{\tilde{\epsilon}^{2} }+\frac{1}{\tilde{\epsilon}(f_{n}^2+1)}+\frac{1}{\epsilon_{2}-\tilde{\epsilon}}-\frac{f_{n}^4}{(f_{n}^2+1) \left(\epsilon_2\left(f_{n}^2+1\right)-\tilde{\epsilon}  f_{n}^2\right)}}.
\end{align}
donde $\tilde{\epsilon}=\epsilon_{2}-\epsilon_{1}$. En la situaciones donde $\tilde{\epsilon}$ es pequeño, el término dominante es $\epsilon_{2}/\tilde{\epsilon}^{2}$. Esto implica que cuanto mayor sea la permitividad del TI y más similar sea al material adyacente, más prominente será la $\theta$-señales en comparación con las señales predominantes de la electrodinámica usual de Maxwell. Considerando un TI con $\mu_{2}=1$, $\epsilon_{2}=100$ y $\theta_{2}=\alpha(11\pi)/\pi$ \cite{maciejko_topological_2010} apropiado para un Bi$_2$Se$_3$ topológico, junto con un material con $\mu_{1}=1$ y $\epsilon_{1}=99$, observamos un escenario donde, aunque es difícil confinar las OEM sin altas frecuencias y grandes guías de onda, es posible observar $\theta$-efectos significativos. Con una frecuencia de operación de $\nu_{0}=500\,[\textup{GHz}]$ y un ancho de guía de onda de $2L=150\,[\mu m]$ , los dos primeros modos más bajos que se propagan bajo estas condiciones muestran que $\eta_{1}^{\textup{TE}}=0.555$ y $\eta_{2}^{\textup{TM}}=0.559$, lo que representa una mejora de $\approx 78\%$ en comparación con las intensidades relativas obtenida de los cuatro modos  anteriores, como se muestra en la Fig. (\ref{fig:FieldsSlab100}).
\section{Rotación de los modos de propagación }\label{6.7}
En general, los modos de propagación son soluciones a las ecuaciones de Maxwell que cumplen todas las BCs, manteniendo constante la distribución transversal de los campos EM a lo largo del eje axial de la guía de ondas. Estos modos satisfacen una condición de auto-consistencia, es decir, la onda se reproduce idénticamente tras un recorrido zigzagueante de ida y vuelta entre las paredes de la guía. Esto indica que la polarización de la onda EM no rota con cada reflexión interna total, sino que conserva una configuración de equilibrio que permite su reproducción indefinida a lo largo de la dirección de propagación. 

Para analizar más profundamente este comportamiento, realizamos una rotación del sistema de coordenadas en el plano $x$-$z$, orientando el eje $z$ con la dirección de $\tongo{k}\equiv\tongo{z}'$, alineándolo con la dirección de propagación de la onda TEM\footnote{Recuerde que una OEM siempre tiene un campo EM transversal, es decir, es una onda TEM. Sin embargo, no debe confundirse con la solución TEM discutida en el capítulo (\ref{TEM}) ni con la onda TEM que se encuentra en cilindros coaxiales metálicos. En una guía de onda en forma de slab, la interacción de dos ondas TEM que viajan en las direcciones $+\tongo{x}$ y $-\tongo{x}$ da lugar a los patrones que hemos denominado "modos".} original. El campo EM original se expresa como,
\begin{align}
    \mathbf{E}^{\pm}(\mathbf{r})&=\begin{pmatrix}
\frac{i}{k_{x}^{2}}k_{z}\partial_{x}E_{z}^{\pm}\\ 
-\frac{i}{k_{x}^{2}}k_{0}\partial_{x}B_{z}^{\pm}\\ 
E_{z}^{\pm}
\end{pmatrix}e^{ik_{z}z}=\begin{pmatrix}
\mp\frac{1}{k_{x}}k_{z}E_{\pm}\\ 
\pm\frac{1}{k_{x}}k_{0}B_{\pm}\\ 
E_{\pm}
\end{pmatrix}e^{i(k_{z}z\pm k_{x}x)}
\end{align}

\begin{align}
\mathbf{B}^{\pm}(\mathbf{r})&=\begin{pmatrix}
\frac{i}{k_{x}^{2}}k_{z}\partial_{x}B_{z}^{\pm}\\ 
\frac{i}{k_{x}^{2}}k_{0}\mu\epsilon\partial_{x}E_{z}^{\pm}\\ 
B_{z}^{\pm}
\end{pmatrix}e^{ik_{z}z}=\begin{pmatrix}
\mp\frac{1}{k_{x}}k_{z}B_{\pm}\\ 
\mp\frac{1}{k_{x}}k_{0}\mu\epsilon E_{\pm}\\ 
B_{\pm}
\end{pmatrix}e^{i(k_{z}z\pm k_{x}x)}.
\end{align}
Aquí, el signo $\pm$ denota la dirección de propagación de la onda TEM en $\pm\tongo{x}$, y $A_{z}^{\pm}=A_{\pm}e^{\pm ik_{x}x}$ representa las componentes longitudinales de los campos para las dos ondas TEM. Implementamos el siguiente operador de rotación $\mathbb{R}^{\pm}$ del plano $x-z$ dejando el eje $y$ invariante,
\begin{align}
    \mathbb{R}^{\pm}=\begin{pmatrix}
\cos\phi_{0} & 0 & \mp\sin\phi_{0}\\ 
0 & 1 & 0\\ 
\pm\sin\phi_{0} & 0 & \cos\phi_{0}
\end{pmatrix}=\frac{1}{k_{0}\sqrt{\mu\epsilon}}\begin{pmatrix}
k_{z} & 0 & \mp k_{x}\\ 
0 & k_{0}\sqrt{\mu\epsilon} & 0\\ 
\pm k_{x} & 0 & k_{z}
\end{pmatrix}
\end{align}
donde $\cos\phi_{0}=\frac{k_{z}}{k_{0}\sqrt{\mu\epsilon}}$ y $\sin\phi_{0}=\frac{k_{x}}{k_{0}\sqrt{\mu\epsilon}}$, de acuerdo con la geometría del sistema. Aplicando esta rotación a los campos vectoriales, obtenemos:
\begin{align}
    \mathbf{E}^{\pm}(\mathbf{r}')&=\pm\frac{k_{0}\sqrt{\mu\epsilon}}{k_{x}}\begin{pmatrix}
- E_{\pm}\\ 
\frac{1}{\sqrt{\mu\epsilon}}B_{\pm}\\ 
0
\end{pmatrix}e^{ik_{0}\sqrt{\mu\epsilon}z'},\\
\mathbf{B}^{\pm}(\mathbf{r}')&=\mp \frac{k_{0}\sqrt{\mu\epsilon}}{k_{x}}\begin{pmatrix}
B_{\pm}\\ 
\sqrt{\mu\epsilon}E_{\pm}\\ 
0
\end{pmatrix}e^{ik_{0}\sqrt{\mu\epsilon}z'}.
\end{align}
En este contexto, $\mathbf{E}^{\pm}(\mathbf{r}')=\mathbb{R}^{\pm}\mathbf{E}(\mathbf{r}')$ y $\mathbf{r}'=\mathbb{R}^{\pm}\mathbf{r}$ definen los campos EM, que son perpendiculares entre sí, $\mathbf{E}^{\pm}(\mathbf{r}')\cdot\mathbf{B}^{\pm}(\mathbf{r}')=0$, como corresponde a una onda TEM. En la electrodinámica convencional de Maxwell, los modos en una slab son exclusivamente TE o TM, es decir, el campo eléctrico tiene componentes únicamente en la dirección $x'/y'$ respectivamente para los modos TM/TE. No obstante, en la $\theta$-ED, los modos son híbridos, indicando que ambas componentes eléctricas y magnéticas son significativas y distintas de cero.

Esto sugiere que un modo dado de la 0-ED rotó en el plano $x'$-$y'$ hasta alcanzar un equilibrio correspondiente al mismo modo de la $\theta$-ED. Podemos cuantificar esta rotación asumiendo que inicialmente que el campo eléctrico se encuentra en la polarización dominante de la 0-ED. Esto nos lleva a definir la rotación mediante la tangente del ángulo que relaciona la componente dominante con la inducida para cada polarización,
\begin{align}\label{eq:anglesTEyTM}
    \psi_{\textup{TE}}&\equiv\tan^{-1}\left (\frac{E^{\pm}_{x'}}{E^{\pm}_{y'}}\right )=\tan^{-1}\left (-\frac{\sqrt{\mu\epsilon}E_{\pm}}{B_{\pm}}\right )=\tan^{-1}\left (-\frac{\sqrt{\mu\epsilon}}{\mathcal{F}_{+}}\right )=\tan^{-1}\left (\frac{\mathcal{F}_{-}}{\sqrt{\mu\epsilon}}\right ),\\ 
    \psi_{\textup{TM}}&\equiv\tan^{-1}\left (\frac{E^{\pm}_{y'}}{E^{\pm}_{x'}}\right )=\tan^{-1}\left (-\frac{B_{\pm}}{\sqrt{\mu\epsilon}E_{\pm}}\right )=-\tan^{-1}\left (\frac{\mathcal{F}_{-}}{\sqrt{\mu\epsilon}}\right ),
\end{align}
\begin{figure}[!ht]
\begin{center}
\includegraphics[width=1\textwidth]{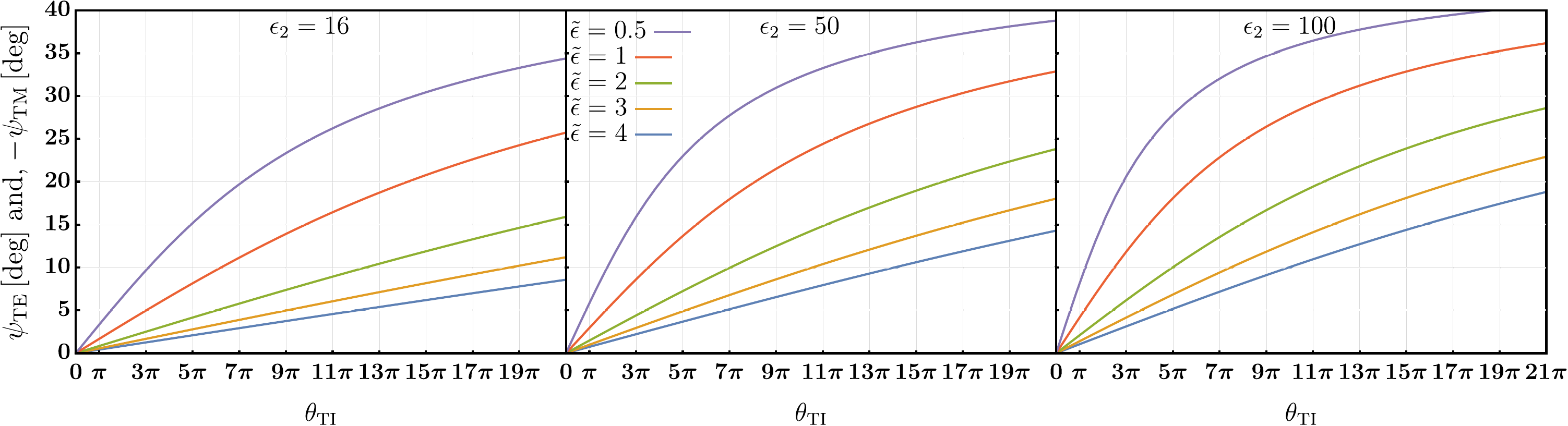}
\end{center}
\caption{\label{Fig:RotacionModes} Ángulo $\psi=\psi_{\textup{TE}}=-\psi_{\textup{TM}}$ que caracteriza la rotación de los modos TE/TM, medido en grados, en función del TMEP $\theta_{2}=\alpha\theta_{\textup{TI}}/\pi$. Se observa un incremento en el ángulo de rotación conforme aumenta $\epsilon_{2}$ y disminuye $\tilde{\epsilon}$, mostrando un crecimiento lineal para valores pequeños de $\theta_{\textup{TI}}$ según la relación $\psi \approx \sqrt{\epsilon_{2}}\theta_{2}/\tilde{\epsilon}$.}
\end{figure}
donde hemos utilizado que $\mathcal{F}_{+}\mathcal{F}_{-}=-\mu\epsilon$. Observamos que la polarización de la onda EM rota en sentido antihorario/horario para los modos TE/TM, respectivamente, y en la misma magnitud, $\psi_{\textup{TE}}=-\psi_{\textup{TM}}$. En la Fig. (\ref{Fig:RotacionModes}), se muestra la rotación de los modos en función del TMEP $\theta_{2}=\alpha\theta_{\textup{TI}}/\pi$ para una guía de ondas simétrica-antiparalela con diferentes valores de $\epsilon_{2}$ y $\tilde{\epsilon}=\epsilon_{2}-\epsilon_{1}$. A medida que la permitividad del TI, $\epsilon_{2}$, aumenta y la diferencia $\tilde{\epsilon}$ disminuye, se observan grandes rotaciones, llegando a decenas de grados. Para valores experimentales comunes de $\theta_{\textup{TI}}$, como $\theta_{\textup{TI}}=\pi$, se logran rotaciones de hasta $8.1\,[\textup{deg}]$ con $\epsilon_{2}=100$ y $\delta=0.5$. Estas rotaciones son comparables con las rotaciones reportadas en \cite{crosse_theory_2017,crosse_optical_2016} (de $\approx 6~[\textup{deg}]$) y de menor magnitud ($\approx 90~[\textup{deg}]$) con las reportadas en \cite{maciejko_topological_2010,tse_giant_2010}.
\section{Acoplamiento de un haz Gaussiano a la guía de ondas}\label{6.8}
Nos planteamos la siguiente pregunta: ¿Cómo se excitan estos nuevos modos y cómo interactúan al transportar energía electromagnética a lo largo de la guía de ondas? Para abordar esta cuestión, debemos considerar una OEM externa como fuente. En escenarios más realistas, se suelen utilizar pulsos Gaussianos generados por un láser. Un láser, situado en uno de los extremos de la guía de ondas, excitará los modos de propagación que viajarán a lo largo de la guía en la dirección $+\tongo{z}$. ``Excitar'' o ``encender'' un modo significa que, de todos los modos permitidos y posibles del modelo de slab, solo algunos pueden propagarse indefinidamente. La activación de ciertos modos permitidos depende de tres propiedades de la OEM externa: primero, la frecuencia con la que oscila, ya que, cuanto mayor sea la frecuencia y, por consecuencia, menor sea su longitud de onda, mayor será la posibilidad de activar múltiples modos permitidos; segundo, la paridad del perfil transversal de la OEM externa, donde, en una guía de ondas en forma de slab topológicamente trivial, si la OEM es par, solo activará modos pares, y lo mismo ocurrirá con los impares. Veremos que en la $\theta$-ED esto ya no se cumple completamente. Y tercero, la polarización del campo eléctrico externo.

La dinámica del campo EM cerca del láser es compleja y presenta desafíos para que este modelo de modos guiados capture con precisión todos los fenómenos que ocurren. Inicialmente, cualquier campo EM que ingrese a la guía en un ángulo no permitido verá cómo su amplitud se atenúa rápidamente a lo largo de la guía. Algunos rayos de luz escaparán, generando radiación, y con cada reflexión y emisión, perderán energía, reduciendo así su amplitud progresivamente. Únicamente los rayos que ingresan con los ángulos permitidos lograrán una propagación estable, lo que permite que la onda EM sea descrita por la superposición de los modos permitidos en la guía. Este modelo es preciso principalmente en regiones alejadas de la fuente láser.
\begin{figure}[t]
    \centering
    \includegraphics[width=0.9\linewidth]{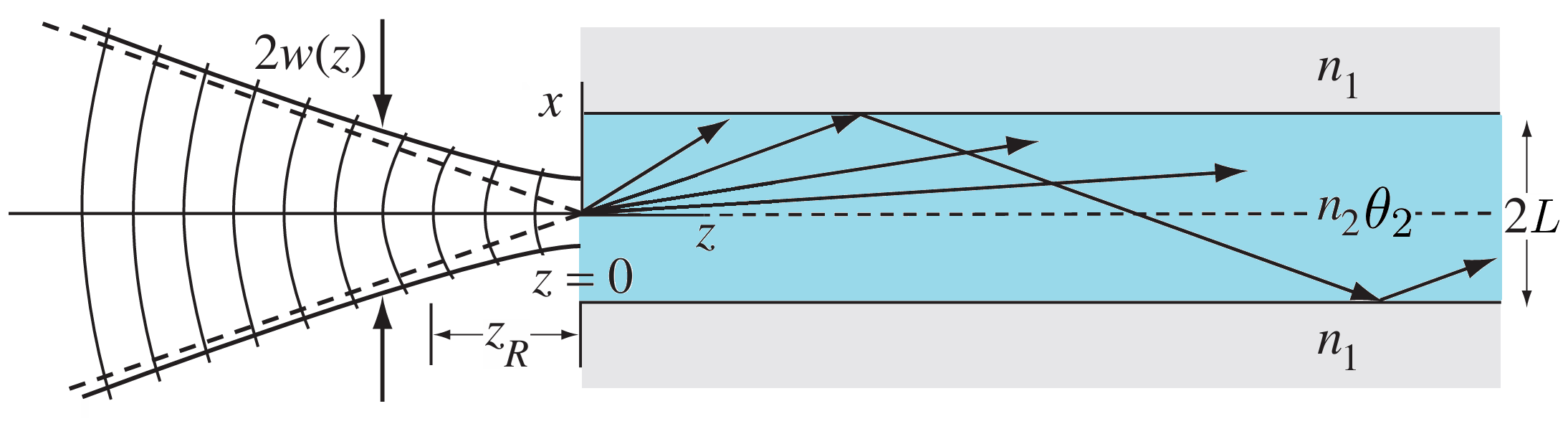}
    \caption{Un haz Gaussiano incide en la guía de onda en forma de slab en $z=0$. El ancho del haz en $z=0$ es $w_0$, sin embargo, siempre podemos hacer una traslación en $z\to z-z_0$ para aumentar el ancho del haz en $z=0$, es decir, $w_0\to w_0\sqrt{1+(z_0/z_R)^{2}}$. Solo algunos rayos de los que entran a la guía pueden propagarse indefinidamente.}
    \label{fig:gauss}
\end{figure}
Consideremos un campo EM arbitrario  $(\mathbf{E}_{0},\mathbf{B}_{0})$ situado en una posición específica de la guía de ondas, por ejemplo, en $z=0$, con una dependencia temporal de $e^{-i\omega t}$. El campo EM total dentro de la guía es la superposición lineal de todos los modos permitidos, que se ajustan al campo EM inicial en $z=0$ \cite{jackson1999classical}\footnote{ver Sec. \textit{8.12} de Jackson, tercera edición.},
\begin{align}
     \mathbf{E}_{\textup{total}}(\mathbf{r}_{\perp},z=0)&\equiv\mathbf{E}_ {0}(\mathbf{r}_{\perp})=\sum_{n}A_{n}\mathbf{E}_{n}, & \mathbf{B}_{\textup{total}}(\mathbf{r}_{\perp},z=0)&\equiv\mathbf{B}_{0}(\mathbf{r}_{\perp})=\sum_{n}A_{n}\mathbf{B}_{n}.\label{eq:InitialFields}
\end{align}
Aquí, $\mathbf{E}_{n}=\mathbf{E}_{n}(\mathbf{r_{\perp}})$ es el perfil del campo EM del modo $n$, que depende de las coordenadas transversales. $A_{n}$ representa la amplitud del modo $n$. Para determinar las amplitudes de los modos, aplicamos la relación de ortogonalidad de la Ec. \eqref{eq:ortogonormalidaddemodos} a los campos totales en la Ec. \eqref{eq:InitialFields} obteniendo,
\begin{align}\label{eq:Amplitudes_Modos}
    A_{n}=\frac{c}{16\pi}\int\frac{1}{\mu}\left [\mathbf{E}_{0\perp}\times\mathbf{B}_{n\perp}^{*}+\mathbf{E}_{n\perp}^{*}\times\mathbf{B}_{0\perp}  \right ]|_{z=0}\cdot\tongo{z}\,da_{\perp}.
\end{align}
Las exponenciales dependientes del tiempo se cancelan entre sí. Suponemos que la onda EM emitida por el láser se asemeja mucho a un haz Gaussiano (haz-G) que se propaga en la dirección $+\tongo{z}$. 

Un haz-G satisface las ecuaciones de Maxwell bajo la aproximación paraxial, esto es cuando el número de onda del haz, $\mathbf{k}_{\textup{haz-G}}=\mathbf{q}_{\perp}+k_{z}\tongo{z}$, es predominantemente longitudinal, es decir, $q_{\perp}^{2} \ll k_{z}^{2}$. En esta aproximación, el haz-G se modela como sigue \cite{zangwill2013modern},
\begin{align}
    \mathbf{E}_{\textup{haz-G}}(\mathbf{r},t)&=\mathbf{E}_{G\perp}(\mathbf{r},t)+\tongo{z}\frac{i}{k_{z}}\boldsymbol{\nabla}_{\perp}\cdot\mathbf{E}_{G\perp}(\mathbf{r},t),\\
    \mathbf{B}_{\textup{haz-G}}(\mathbf{r},t)&=-\frac{i}{k_{0}}\boldsymbol{\nabla}\times\mathbf{E}_{\textup{haz-G}}(\mathbf{r},t),
\end{align}
donde,
\begin{align}
   \mathbf{E}_{G\perp}(\mathbf{r},t)=\hat{\mathbf{E}}_{\perp}4\pi\left [\frac{w_{0}}{w(z)}  \right ]\exp{\left [-\frac{\rho^{2}}{w^{2}(z)}  \right ]}\exp{\left \{i\left [ \frac{k_{z}\rho^{2}}{2R(z)}+\alpha(z)+k_{z}z-\omega t\right ] \right \}}, 
\end{align}
Aquí, $\rho^{2} = x^{2} + y^{2}$ representa la coordenada transversal, $k_{z} = k_{0}\sqrt{\mu\epsilon}$ es la relación de dispersión del haz, $\mathbf{\hat{E}}_{\perp}=\bar{E}_{x}\hat{x}+\bar{E}_{y}\hat{y}$ describe la polarización unitaria del haz-G, y $w_{0}$ es la medida del ancho del haz en $z=0$. $R(z)$, $w(z)$, y $\alpha(z)$ son características definitorias del haz-G, ver Fig. (\ref{fig:gauss}), y que se definen como sigue,
\begin{align}
 R(z)&=z+\frac{z_{R}^{2}}{z},\\  w(z)&=w_{0}\sqrt{1+(z/z_{R})^{2}},\\ 
 \alpha(z)&=-\tan^{-1}\left ( \frac{z}{z_{R}} \right ),\\
z_{R}&=\frac{1}{2}k_{z}w_{0}^{2}. 
\end{align}
con $R(z)$ como el radio de curvatura de los frente de onda, $w(z)$ es el radio de la cintura, $\alpha(z)$ es la fase de Gouy y $z_R$ es la longitud de Rayleigh. El haz-G se propaga libremente fuera de la guía de ondas en el aire, donde $\mu = \epsilon = 1$, e ilumina en $z = 0$ el perfil transversal de la guía. Dentro de la guía de ondas, el haz-G solamente excitará los modos permitidos y será descrito como una superposición de tales modos. Nos interesa particularmente las componentes transversales del haz-G en $z = 0$, ya que son estas componentes las que activarán los modos de la guía, según se indica en la Ec. \eqref{eq:Amplitudes_Modos},
\begin{align}
    \mathbf{E}_{0\perp}|_{z=0}&=\mathbf{E}_{G\perp}(\mathbf{r}_{\perp})=\hat{\mathbf{E}}_{\perp}4\pi\exp{\left [-\frac{\rho^{2}}{w_{0}^{2}} \right ]},\\
    \mathbf{B}_{0\perp}|_{z=0}&=\mathbf{B}_{G\perp}(\mathbf{r}_{\perp})=\tongo{z}\times\left ( (k_z^2w_0^4+2\rho^2)\hat{\mathbf{E}}_{\perp}-4(\boldsymbol{\rho}\cdot\hat{\mathbf{E}}_{\perp})\boldsymbol{\rho} \right ) \frac{4\pi}{k_0k_zw_0^4}\exp{\left [-\frac{\rho^{2}}{w_{0}^{2}} \right ]},
\end{align}
donde $\boldsymbol{\rho} = \mathbf{r}_{\perp} = x\tongo{x} + y\tongo{y}$. Al insertar estas expresiones en la Ec. \eqref{eq:Amplitudes_Modos} e integrar la coordenada transversal $y$ desde $-\infty$ hasta $\infty$, obtenemos,
\begin{align*}
    A_{n}&=\frac{c}{16\pi}\int\frac{1}{\mu}\left [\mathbf{E}_{0\perp}\times\mathbf{B}_{n\perp}^{*}+\mathbf{E}_{n\perp}^{*}\times\mathbf{B}_{0\perp}  \right ]\cdot d\mathbf{a}_{\perp}\\
    %------------------------------
    %&=\frac{c}{4}\int\frac{dxdy}{\mu}\left [\hat{\mathbf{E}}_{\perp}\times\mathbf{B}_{n\perp}^{*}+\mathbf{E}_{n\perp}^{*}\times\tongo{z}\times\left ( (k_z^2w_0^4+2\rho^2)\hat{\mathbf{E}}_{\perp}-4(\boldsymbol{\rho}\cdot\hat{\mathbf{E}}_{\perp})\boldsymbol{\rho} \right ) \frac{1}{k_0k_zw_0^4}  \right ]\cdot \tongo{z}e^{-\frac{\rho^{2}}{w_{0}^{2}}}\\
    %------------------------------
    &=\frac{c}{4}\int\frac{dxdy}{\mu}\left [\hat{\mathbf{E}}_{\perp}\times\mathbf{B}_{n\perp}^{*}+\tongo{z}\left ( (k_z^2w_0^4+2\rho^2)\mathbf{E}_{n\perp}^{*}\cdot\hat{\mathbf{E}}_{\perp}-4(\boldsymbol{\rho}\cdot\hat{\mathbf{E}}_{\perp})(\boldsymbol{\rho}\cdot\mathbf{E}_{n\perp}^{*}) \right ) \frac{1}{k_0k_zw_0^4}  \right ]\cdot \tongo{z}e^{-\frac{\rho^{2}}{w_{0}^{2}}} \\
    %------------------------------
    %&=\frac{cw_0\sqrt{\pi}}{4}\int\frac{dx}{\mu}\left [(\hat{\mathbf{E}}_{\perp}\times\mathbf{B}_{n\perp}^{*})\cdot\tongo{z}+\left ( (k_z^2w_0^4+2x^2+w_0^2)\mathbf{E}_{n\perp}^{*}\cdot\hat{\mathbf{E}}_{\perp}-2\hat{\mathbf{E}}_{\perp}\cdot\begin{pmatrix} 2x^2 &0 \\ 0&w_0^2  \end{pmatrix}\mathbf{E}_{n\perp}^{*} \right ) \frac{1}{k_0k_zw_0^4}  \right ]e^{-\frac{x^{2}}{w_{0}^{2}}}  \\
    %------------------------------
    %&=\frac{cw_0\sqrt{\pi}}{4}\int\frac{dx}{\mu}\left [(\hat{\mathbf{E}}_{\perp}\times\mathbf{B}_{n\perp}^{*})_{z}+\frac{1}{k_0^2\sqrt{\mu\epsilon}w_0^4}\mathbf{E}_{n\perp}^{*}\cdot\begin{pmatrix} k_0^2\mu\epsilon w_0^4-2x^2+w_0^2 &0 \\   0&k_0^2\mu\epsilon w_0^4+2x^2-w_0^2  \end{pmatrix}\hat{\mathbf{E}}_{\perp}\right ]e^{-\frac{x^{2}}{w_{0}^{2}}}\\
    %------------------------------
    %&=\frac{cw_0\sqrt{\pi}}{4}\int\frac{dx}{\mu}\left [(\hat{\mathbf{E}}_{\perp}\times\mathbf{B}_{n\perp}^{*})_{z}+\frac{1}{k_0^2\sqrt{\mu\epsilon}w_0^2}\mathbf{E}_{n\perp}^{*}\cdot\begin{pmatrix}  k_0^2\mu\epsilon w_0^2-2\frac{x^2}{w_0^2}+1 &0 \\ 0&k_0^2\mu\epsilon w_0^2+2\frac{x^2}{w_0^2}-1 \end{pmatrix}\hat{\mathbf{E}}_{\perp}\right ]e^{-\frac{x^{2}}{w_{0}^{2}}}\\
    %------------------------------
    &=\frac{cw_0\sqrt{\pi}}{4}\int\frac{dx}{\mu}\left [(\hat{\mathbf{E}}_{\perp}\times\mathbf{B}_{n\perp}^{*})_{z}+\frac{1}{k_0^2w_0^2}\mathbf{E}_{n\perp}^{*}\cdot\begin{pmatrix}
    k_0^2w_0^2-2\frac{x^2}{w_0^2}+1 &0 \\ 0&k_0^2 w_0^2+2\frac{x^2}{w_0^2}-1 \end{pmatrix}\hat{\mathbf{E}}_{\perp}\right ]e^{-\frac{x^{2}}{w_{0}^{2}}}
    \end{align*}
donde se ha reemplazado la relación de dispersión del haz-G y se han utilizado las siguientes integrales,
\begin{align}
    \int_{-\infty}^{\infty}e^{-y^{2}/w_{0}^{2}}dy&=\sqrt{\pi}w_0, & \int_{-\infty}^{\infty}ye^{-y^{2}/w_{0}^{2}}dy&=0, &\textup{y,}~~~ \int_{-\infty}^{\infty}y^2e^{-y^{2}/w_{0}^{2}}dy&=\sqrt{\pi}w_0^3/2.
\end{align}
Ahora debemos integrar en $x$. Dado que los campos de los modos, 
\begin{align}
    \mathbf{E}_{n\perp}&=\frac{i}{k_{x}^2}[\tongo{x}k_z\partial_x E_{z}(x)-\tongo{y}k_0\partial_xB_z(x)],\\
    \mathbf{B}_{n\perp}&=\frac{i}{k_{x}^2}[\tongo{x}k_z\partial_x B_{z}(x)+\tongo{y}k_0\epsilon\mu\partial_xE_z(x)],
\end{align}
varían en función de $x$, los dividiremos en tres regiones para la integración: medio $m=1$: $x \in (-\infty, -L)$; medio $m=2$: $x \in [-L, L]$; y medio $m=3$: $x \in (L, \infty)$. 
Para obtener el resultado final de la amplitud, emplearemos las siguientes integrales que son fundamentales para resolver el problema,
\begin{align*}
    \int_{-\infty}^{-L}e^{v_n x/L}e^{-x^2/w_0^2}dx=&\frac{1}{2} \sqrt{\pi } w_0 e^{\frac{w_0^2 v_n^2}{4 L^2}} \text{erfc}\left[\frac{w_0 v_n}{2 L}+\frac{L}{w_0}\right],\\ 
    \int_{L}^{\infty}e^{-v_n x/L}e^{-x^2/w_0^2}dx=&\frac{1}{2} \sqrt{\pi } w_0 e^{\frac{w_0^2 v_n^2}{4 L^2}} \text{erfc}\left[\frac{w_0 v_n}{2 L}+\frac{L}{w_0}\right],\\
    \int_{-L}^{L}\cos(u_n x/L)e^{-x^2/w_0^2}dx=&\frac{1}{2} \sqrt{\pi } w_0 e^{-\frac{w_0^2 u_n^2}{4 L^2}} \left(\text{erf}\left[\frac{L}{w_0}+\frac{i w_0 u_n}{2 L}\right]+\text{erf}\left[\frac{L}{w_0}-\frac{i w_0 u_n}{2 L}\right]\right),\\
    \int_{-L}^{L}\sin(u_n x/L)e^{-x^2/w_0^2}dx=&0,
\end{align*}
donde utilizamos la función error de Gauss, $\textup{erf}(z)=\frac{2}{\sqrt{\pi}}\int_0^z e^{-t^2}dt$ y $\textup{erfc}(z)=1- \textup{erf}(z)$.
% %
% \begin{align}
%     \textup{erf}(z)&=\frac{2}{\sqrt{\pi}}\int_0^z e^{-t^2}dt, & \textup{erfc}(z)&=1- \textup{erf}(z)=\frac{2}{\sqrt{\pi}}\int_z^\infty e^{-t^2}dt,
% \end{align}
% %
%es la función error de Gauss. 
Consideraremos que todos los medios son no magnéticos, $\mu_{1} = \mu_{2} = \mu_{3} = 1$. Así, la amplitud $A_{n}$ se puede escribir como,
\begin{align}\notag
    A_{n} &= \frac{c\sqrt{\pi} e^{-\frac{L^2}{w_0^2}}}{8k_0^2}\sin u_n \Biggl\{ \mathcal{A}_{2n}\left [ i(\textup{w}(z_{2n})-\textup{w}(z_{2n}^*))\cot u_n-\textup{w}(z_{2n})-\textup{w}(z_{2n}^*) \right ] +2 \mathcal{A}_{1n}\textup{w}(z_{1n}) \\\notag
    &+i\frac{4L^2}{u_nv_nw_0}(k_{zn}\bar{E}_xE_{e}+k_0\bar{E}_yB_{e})\left (u_n-v_n\cot u_n\right ) \\\label{eq:AmpltudnGauss}
    &+i\frac{Lk_0}{v_n}\bar{E}_{y}\left ( 2iz_{1n}^*-\zeta_{01n}\textup{w}(z_{1n})\right )\left ( E_o(\tilde{\theta}_{2}+\tilde{\theta}_{1})\cot u_n + E_e(\tilde{\theta}_{2}-\tilde{\theta}_{1}) \right )  \Biggl\}
\end{align}
donde hemos definido las siguientes constantes,
\begin{align}
    \mathcal{A}_{mn}&=\frac{1}{k_{xm}}(\zeta_{3mn}k_{zn}\bar{E}_{x}E_e+\zeta_{0mn}k_{0}\bar{E}_{y}B_e ),\\
    \zeta_{0mn}&=\sqrt{\pi}w_0^2\left (\frac{k_{xm}^2}{2}-k_0^2-k_{zn}k_0  \right )\\
    \zeta_{3mn}&=\sqrt{\pi}w_0^2\left (\frac{k_{xm}^2}{2}+k_0^2\left (1+\frac{k_0\epsilon_{m}}{k_{zm}}  \right )  \right )\\
    z_{mn}&=\frac{w_0 k_{xm}}{2}+i\frac{L}{w_0}
\end{align}
con $m$ representando el índice del medio y $\textup{w}(z)=e^{-z^2}\textup{erfc}[-iz]$ corresponde a la función de Faddeeva \cite{al1967vlf}. La Ec.\eqref{eq:AmpltudnGauss} determina la amplitud del modo $n$ permitido en la guía de ondas. Si ignoramos el término topológico de la tercera fila de la Ec. \eqref{eq:AmpltudnGauss}, observamos que las amplitudes que se acoplan con la polarización del haz-G son únicamente pares, es decir, términos como $E_e$ y $B_e$, indicando que un haz-G típicamente solo excita a modos pares, pues su perfil transversal es par. Sin embargo, en la $\theta$-ED, es posible excitar modos impares si $\tilde{\theta}_{2}+\tilde{\theta}_{1}\neq 0$ y modos pares si $\tilde{\theta}_{2}-\tilde{\theta}_{1}\neq 0$. La primera aseveración es interesante debido a que un haz-G, que su perfil transversal es par, polarizado en la dirección $\tongo{y}$ puede excitar una modo impar. El análisis detallado de la presencia de este modo impar con un haz-G par se presentará en un artículo en preparación \cite{underprep2}. 

Para las consideraciones que hemos hecho en este estudio, solamente el core de la guía es un TI y los cladding son dieléctricos topológicamente triviales, \textit{i.e.}, $\theta_{1} = \theta_{3} = 0$ y $\theta_{2} \neq 0$, esto resulta en un término lineal en $\theta_{2}$: $\sim (\tilde{\theta}_{2}-\tilde{\theta}_{1})\bar{E}_{y}E_{e}=-2\theta_{2}\bar{E}_{y}E_{e}$, recordando que $\tilde{\theta}_{m}=\theta_{m+1}-\theta_{m}$.
\begin{figure}[!ht]
\begin{center}
    \includegraphics[width=0.46\textwidth]{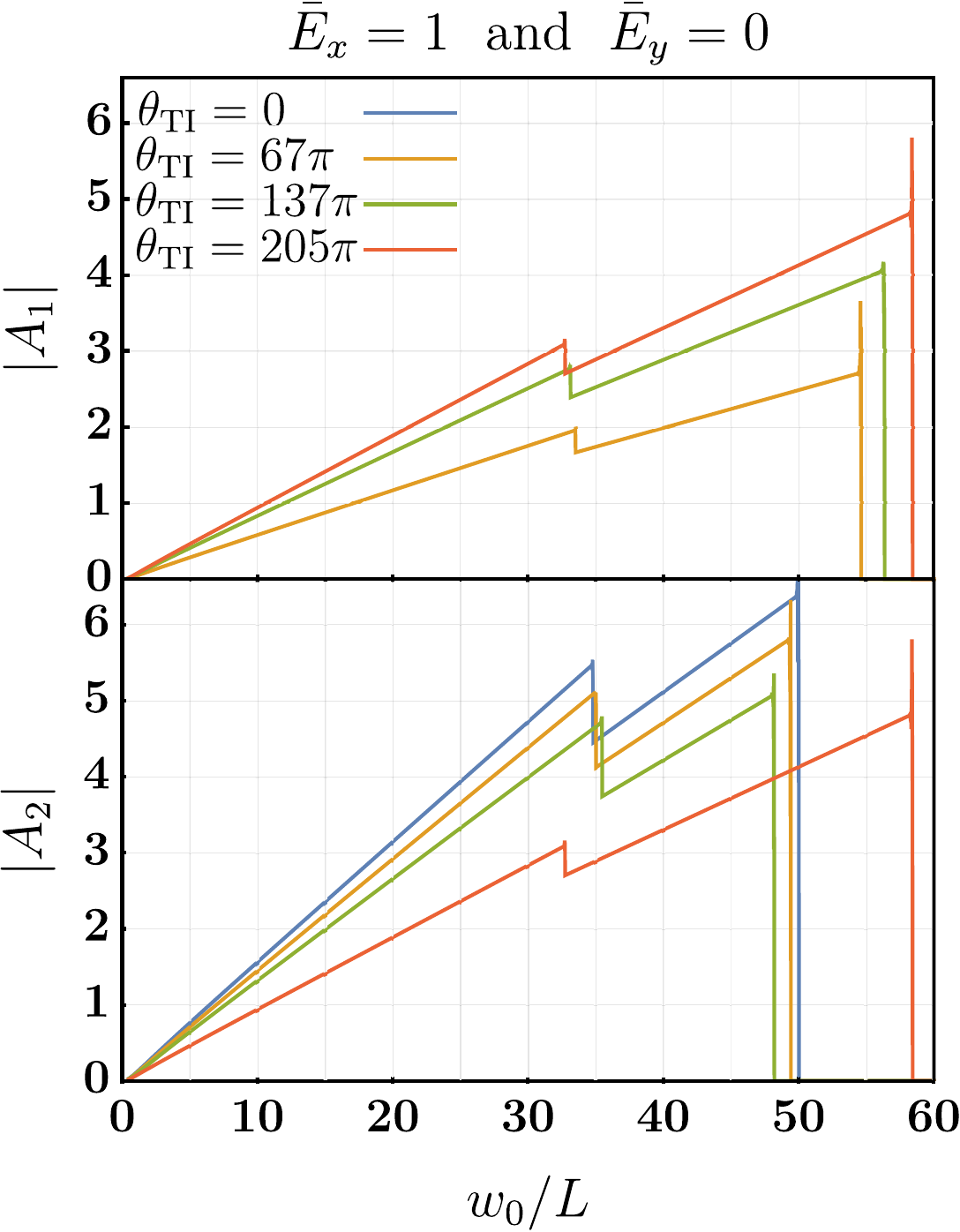}
    \includegraphics[width=0.46\textwidth]{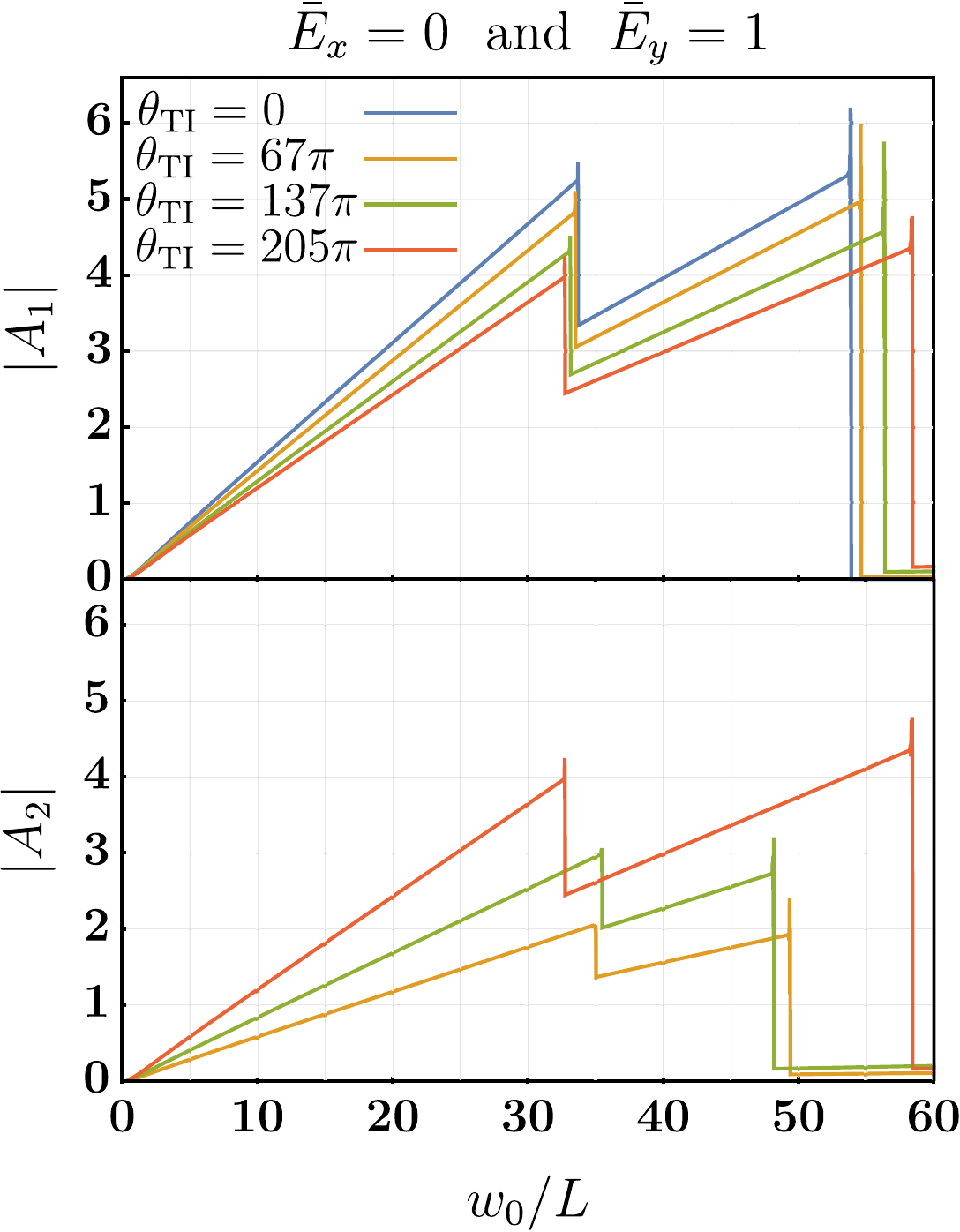}
\end{center}
\caption{\small{\label{Fig:Amplitudesn} 
Amplitudes de los dos primeros modos excitados cuando el haz-G tiene una polarización en el eje $\tongo{x}$ (columna izquierda) y en el eje $\tongo{y}$ (columna derecha). Cuando $|A_1|\neq 0$ y $|A_2|=0$ el campo total es un TE puro y cuando $|A_1|= 0$ y $|A_2|\neq0$ el campo total es TM puro. Cuando tenemos $|A_1|\neq0$ y $|A_2|\neq0$ el campo total es una combinación de TE y TM (no confundir con un campo total híbrido). Se ha considerado un haz-G oscilando a $k_0=2\pi/\lambda_0$, con $\lambda_0=3.3(2L)$ y un ancho de guía de $2L=40\,\mu m$. Los medios son no magnéticos, con $\mu_1=\mu_2=1$, y las constantes dieléctricas son $\epsilon_2=16$ y $\epsilon_1=12$. Se observa que para un medio no topológico (línea azul), un haz-G con una polarización en la dirección $\tongo{x}$ solamente excitará un modo TM y cuando la polarización esta en la dirección $\tongo{y}$ solamente excitará un modo TE. Sin embargo, en presencia de un TI, un haz-G en cualquiera de las dos polarizaciones excita simultáneamente los dos primeros modos. Los picos observados se deben a que, matemáticamente, las funciones $\textup{erf}(z)$ tienden a cero más rápido que las funciones $e^{z^2}$ en los puntos donde se encuentran los picos. No estamos seguros si estos picos son de origen físico o se deben a una falta de precisión numérica.}}
\end{figure}
¿Qué significan estos acoplamientos? Cuando decimos que $\bar{E}_{y}$ se acopla con un $B_{e}$, estamos indicando que un haz-G polarizado en $\tongo{y}$ excita un campo magnético longitudinal (a la dirección de propagación) \textit{even} con amplitud $B_{e}$. Esto tiene sentido físico, ya que dentro de la guía de ondas, las ecuaciones de Maxwell permiten simultáneamente las tríadas de componentes de campos: $\{B_{x},E_{y},B_{z}\}\neq 0$ para el caso TE y $\{E_{x},B_{y},E_{z}\}\neq 0$ para el caso TM. Por lo tanto, cuando el haz-G tiene una polarización en la dirección $\tongo{y}$, excita la configuración TE, logrando un campo magnético longitudinal. Este análisis también se aplica cuando decimos que $\bar{E}_{x}$ se acopla con un $E_{e}$: un haz-G polarizado en $\tongo{x}$ excita la configuración TM con un campo eléctrico longitudinal. En la $\theta$-ED, observamos un fenómeno no permitido en la 0-ED, donde un $\bar{E}_{y}$ se acopla con un $E_{e}$, un fenómeno que ocurre porque un campo eléctrico paralelo a la superficie, en la dirección $\tongo{y}$, induce un campo magnético en la misma dirección $\tongo{y}$, que a su vez excita la configuración TM de orden $\theta_{2} \sim \alpha$.

En la Fig. (\ref{Fig:Amplitudesn}) se muestra la amplitud de los modos $n=1$ y $n=2$ en función de $w_{0}/L$ para diferentes valores de $\theta_{2}=\alpha\theta_{\textup{TI}}/\pi$, con $\theta_{\textup{TI}}=\{0,67\pi,137\pi,205\pi\}$ y dos configuraciones de la polarización del haz-G. Para medios triviales, es decir, $\theta_{\textup{TI}}=0$ (línea azul), cada polarización del haz-G excita únicamente un modo a la vez. Sin embargo, cuando se introduce un TI en el sistema, ambos modos son excitados con amplitudes comparables.

Finalmente, cuando utilizamos una fuente externa arbitraria que ``ilumina'' a la guía de onda en $z=0$, de tal forma que solamente excitamos los dos primeros modos, se puede ver de la Ec. \eqref{eq:InitialFields} que el campo EM total dentro de la guía de onda se puede expresar como,
\begin{align}\label{eq:TotalE}
    \mathbf{E}_{\textup{tot}}(\mathbf{r})&=A_{1}(\theta)\mathbf{E}_{1}(x;\theta)e^{i k_{z1}(\theta)z}+A_{2}(\theta)\mathbf{E}_{2}(x;\theta)e^{i k_{z2}(\theta)z},\\\label{eq:TotalB}
    \mathbf{B}_{\textup{tot}}(\mathbf{r})&=A_{1}(\theta)\mathbf{B}_{1}(x;\theta)e^{i k_{z1}(\theta)z}+A_{2}(\theta)\mathbf{B}_{2}(x;\theta)e^{i k_{z2}(\theta)z}, 
\end{align}
donde la dependencia explícita en $\theta$ indica que la amplitud, las constantes de propagación y los modos EM permitidos dependen del TMEP. Aquí, el sub-índice $1$ y $2$ corresponde a los dos primeros modos permitidos. Esto implica que tanto el modo de vibración del campo EM, el modo de acoplamiento y la forma de propagación cambian en función de $\theta$. 

En las Figs. (\ref{fig:EMTotal10}) y (\ref{fig:EMTotal01}) se muestra el campo EM total inducido por un haz-G cuando la polarización está en el eje $\tongo{x}$ y $\tongo{y}$, respectivamente. Se observa que se induce un campo EM predominantemente con la polarización permitida por la 0-ED, descrita anteriormente. Sin embargo, cuando hay un TI en el sistema, se induce un campo EM en la polarización contraria. El TMEP $\theta_{2}$ no solo induce un campo EM en la polarización contraria a la predominante, sino que también modifica el campo EM de la polarización dominante. 

El campo inducido de la polarización contraria tiene una envolvente simétrica con una longitud de onda modulada de $\lambda_{\textup{mod}}\approx 540 L$. Estos pulsos se pueden medir y se acortan a medida que $\theta_{2}$ aumenta, lo cual es una señal inequívoca de las corrientes topológicas en la superficie del TI. Además, las oscilaciones de la OEM inducida con una longitud de onda mediana de $\lambda_{\textup{mediana}}\sim 20 L$ tienen un pequeño desajuste con respecto a la OEM dominante. Antes del máximo del pulso (donde el pulso inducido es más ancho), la onda se atrasa o adelanta en comparación con la dominante, y después del máximo del pulso la onda se adelanta o atrasa, respectivamente. Aún no está claro por qué existe este desajuste, pero se intuye que puede ser debido a las corrientes y cargas topológicas de la superficie, que se retrasan o adelantan a medida que la OEM dominante se propaga.

En el trabajo de Crosse \cite{crosse_theory_2017}, solo se reportó un cambio en las amplitudes del modo cuando hay un TI en el sistema, dotándolas de una dependencia no trivial en la dirección $z$. Esto se justifica porque el tamaño relativo de $\alpha$ y $\epsilon$ es tal que las modificaciones son pequeñas y los nuevos modos deben tratarse como perturbaciones de los modos permitidos usualmente. Según Crosse, los campos totales son,
\begin{align}
    \mathbf{E}_{\textup{tot}}^{\textup{Crosse}}(\mathbf{r})&=A_{1}(z;\theta)\mathbf{E}_{1}(x;0)e^{i k_{z1}(0)z}+A_{2}(z;\theta)\mathbf{E}_{2}(x;0)e^{i k_{z2}(0)z},
\end{align}
y similar para el campo magnético. Esta dependencia en $z$ de las amplitudes provoca que el campo EM total oscile de TM a TE y viceversa a lo largo de $z$. Esta conclusión se debe a que $\mathbf{E}_{1}(x;0)$ y $\mathbf{E}_{2}(x;0)$ de la electrodinámica usual son modos puramente TE y TM, respectivamente. Por lo tanto, cuando $A_{1}(z;\theta)=0$ y $A_{2}(z;\theta)\neq 0$, decimos que el campo es puramente TM, y cuando $A_{1}(z;\theta)\neq 0$ y $A_{2}(z;\theta)=0$, decimos que el campo es puramente TE. Dado que nuestro modelo es exacto, en algún límite debemos recuperar lo reportado por Crosse y observar estas oscilaciones TE/TM. Podemos considerar que a primer orden en $\theta$, nuestros modos y constantes de propagación se pueden separar en una contribución de la 0-ED más una de la $\theta$-ED. De hecho,
\begin{align}\notag
    \mathbf{E}_{\textup{tot}}^{\textup{exacto}}(\mathbf{r})&=A_{1}(\theta)[\mathbf{E}_{1}(x;0)+\mathbf{E}_{1}^{\theta}(x)]e^{i k_{z1}(0)z}e^{i k_{z1}^{\theta}z}+A_{2}(\theta)[\mathbf{E}_{2}(x;0)+\mathbf{E}_{2}^{\theta}(x)]e^{i k_{z2}(0)z}e^{i k_{z2}^{\theta}z}\\\notag
    &=A_{1}(z;\theta)[\mathbf{E}_{1}(x;0)+\mathbf{E}_{1}^{\theta}(x)]e^{i k_{z1}(0)z}+A_{2}(z;\theta)[\mathbf{E}_{2}(x;0)+\mathbf{E}_{2}^{\theta}(x)]e^{i k_{z2}(0)z}\\
    &=\mathbf{E}_{\textup{tot}}^{\textup{Crosse}}(\mathbf{r})+A_{1}(z;\theta)\mathbf{E}_{1}^{\theta}(x)e^{i k_{z1}(0)z}+A_{2}(z;\theta)\mathbf{E}_{2}^{\theta}(x)e^{i k_{z2}(0)z},
\end{align}
donde $A_{i}(z;\theta)\equiv A_{i}(\theta)e^{i k_{zi}^{\theta}z}$. Por lo tanto, la diferencia entre lo obtenido por Crosse \cite{crosse_theory_2017} corresponde a dos términos, uno por cada modo, que tienen contribuciones en $\theta$ a todos los órdenes. Este nuevo término es el que contrarresta la perfecta oscilación entre un modo TE y un modo TM, o viceversa, reportada. La implicación física de este nuevo término se discute en un artículo en preparación \cite{underprep2}.
\begin{landscape} % Inicia la página rotada
\thispagestyle{empty}
    \begin{figure}[h]
    \centering
    \includegraphics[width=1.4\textwidth]{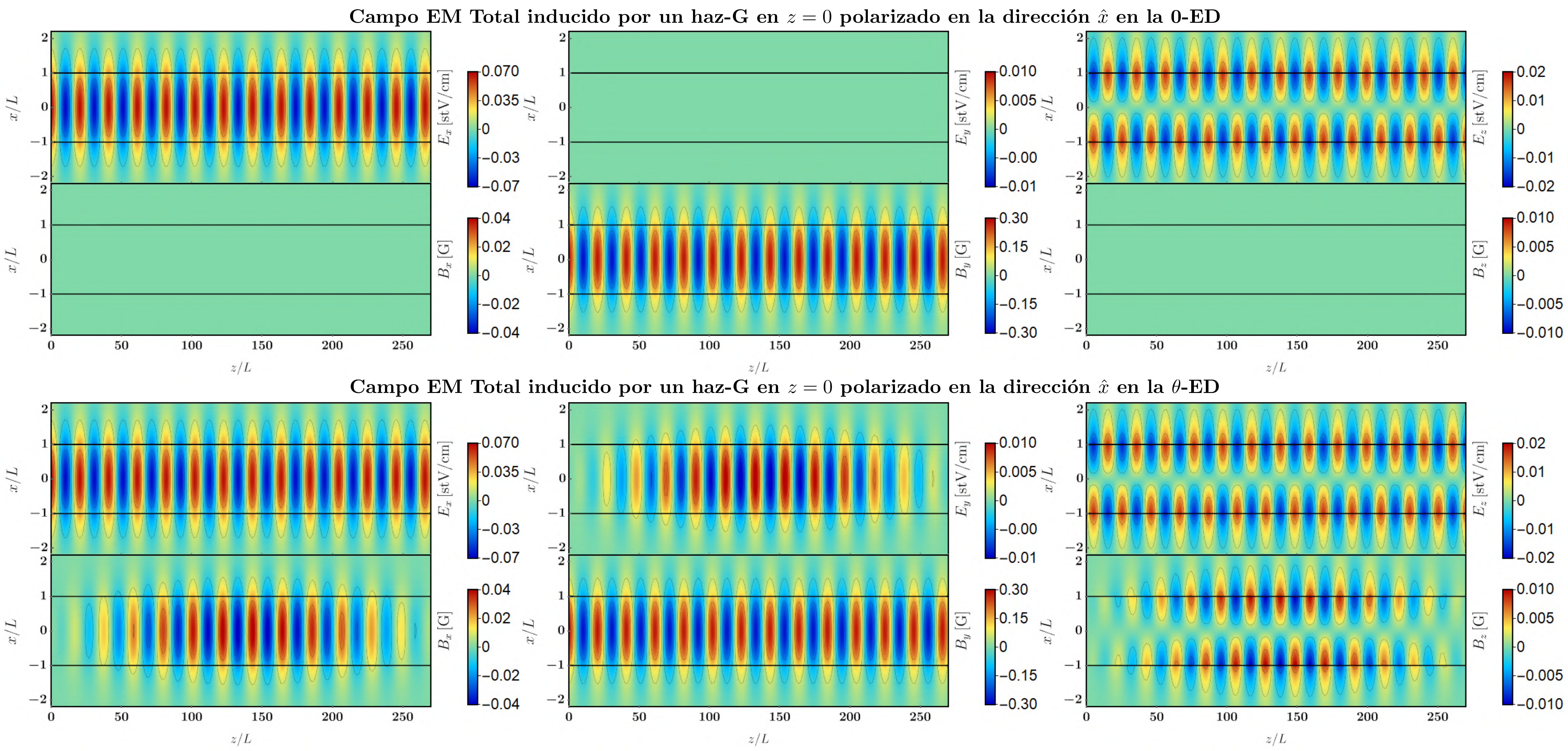}
    \caption{\small{\label{fig:EMTotal10}\textit{DensityPlot} en $t=0$ de la parte real de las componentes del campo EM total \eqref{eq:TotalE} y \eqref{eq:TotalB} inducido por un haz-G polarizado en la dirección $\tongo{x}$. En la fila superior se muestra el campo EM total para un sistema no topológico, \textit{es decir}, $\theta_2=0$, como punto de referencia, y en la fila inferior para un TI con $\theta_2=\alpha\theta_{\textup{TI}}/\pi=11\alpha$. Mostramos contornos de amplitud constantes del campo EM total. Hemos considerado un haz-G oscilando a $k_0=2\pi/\lambda_0$, con $\lambda_0=3.3(2L)$ y un ancho del haz-G de $w_0=40,\mu m$, igual que el ancho de la guía $2L=40,\mu m$. Los medios son no magnéticos y tienen una constante dieléctrica de $\epsilon_{1}=\epsilon_{3}=12$ y $\epsilon_{2}=16$. Observamos que, cuando hay un TI presente en el sistema, aparece una nueva contribución en todas las componentes donde en la 0-ED están prohibidas. La nueva contribución presenta dos tipos de oscilaciones: una rápida del orden de $\lambda_{\textup{corta}}\sim L$ y una lenta envolvente del orden de $\lambda_{\textup{larga}}=\lambda_{\textup{mod}}\sim 540L$. Las oscilaciones de frecuencia mediana que se observan en el plot, del orden de $\lambda_{\textup{mediana}}\sim 20 L$, son en realidad una contribución promedio del conjunto de oscilaciones rápidas.}}
    \end{figure}
\end{landscape}
\begin{landscape} % Inicia la página rotada
\thispagestyle{empty}
\begin{figure}[h] % Usar el entorno figure para permitir el uso de \caption
    \centering
    \includegraphics[width=1.4\textwidth]{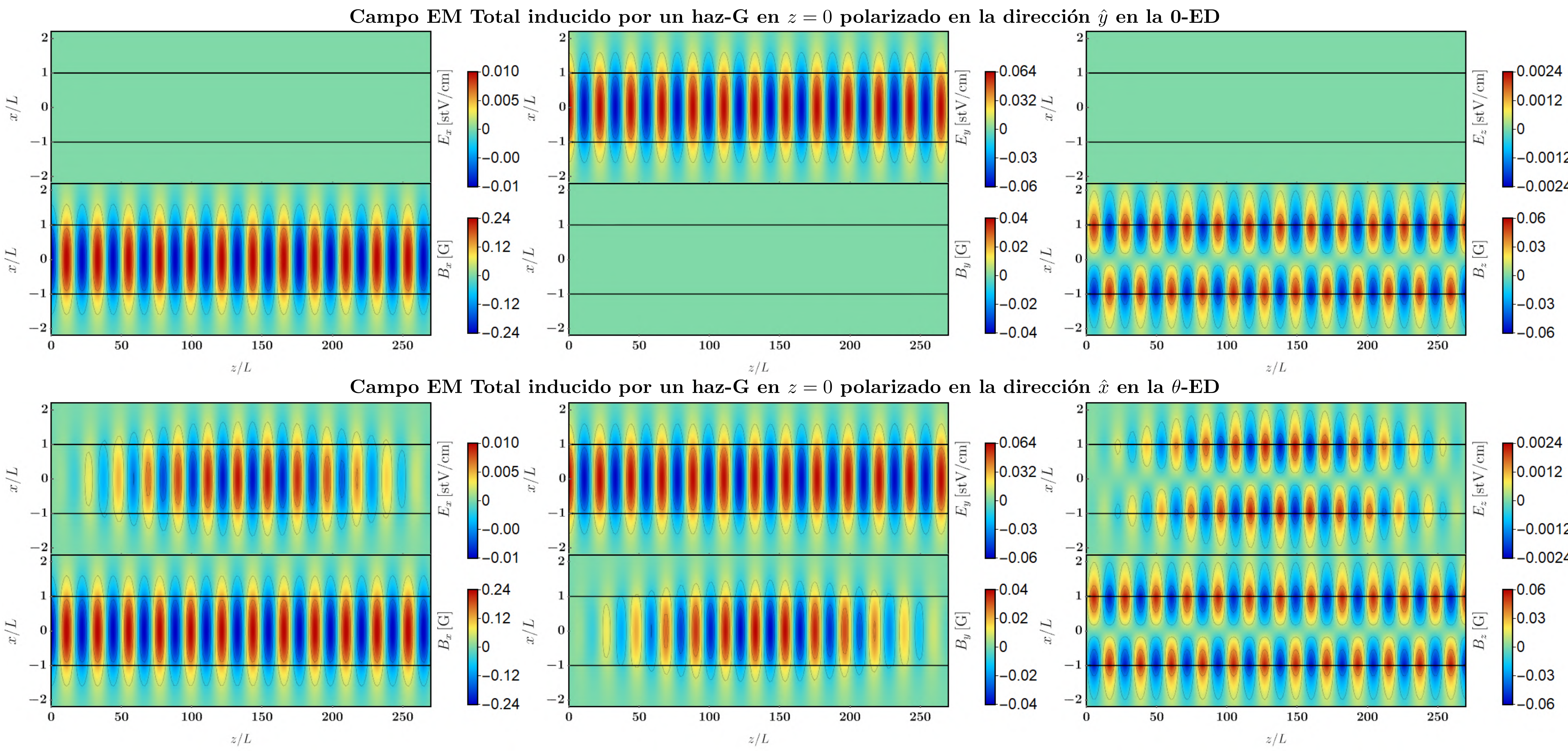}
    \caption{\small{\label{fig:EMTotal01} Similar a la Fig. (\ref{fig:EMTotal10});\textit{DensityPlot} en $t=0$ de la parte real de las componentes del campo EM total \eqref{eq:TotalE} y \eqref{eq:TotalB} inducido por un haz-G polarizado en la dirección $\tongo{y}$. En la fila superior, se muestra el campo EM total para un sistema no topológico, \textit{es decir}, $\theta_2=0$, como punto de referencia, y en la fila inferior para un TI con $\theta_2=\alpha\theta_{\textup{TI}}/\pi=11\alpha$. Mostramos contornos de amplitud constantes del campo EM total. Hemos considerado un haz-G oscilando a $k_0=2\pi/\lambda_0$, con $\lambda_0=3.3(2L)$ y un ancho del haz-G de $w_0=40,\mu m$, igual que el ancho de la guía $2L=40,\mu m$. Los medios son no magnéticos y tienen una constante dieléctrica de $\epsilon_{1}=\epsilon_{3}=12$ y $\epsilon_{2}=16$. Observamos que, cuando hay un TI presente en el sistema, aparece una nueva contribución en todas las componentes donde en la 0-ED están prohibidas. La nueva contribución presenta dos tipos de oscilaciones: una rápida del orden de $\lambda_{\textup{corta}}\sim L$ y una lenta envolvente del orden de $\lambda_{\textup{larga}}=\lambda_{\textup{mod}}\sim 540L$. Las oscilaciones de frecuencia mediana que se observan en el plot, del orden de $\lambda_{\textup{mediana}}\sim 20 L$, son en realidad una contribución promedio del conjunto de oscilaciones rápidas.}}
\end{figure}
\end{landscape}

%Poner polarizacion circular 

 %--------------

\biblio %Se necesita para referenciar cuando se compilan subarchivos individuales - NO SACAR

%% file: Capitulos/07Conclusion.tex
En esta tesis, hemos presentado una revisión del marco teórico y de algunos aspectos experimentales que, en conjunto, culminaron en el desarrollo de los aislantes topológicos (TI). En el capítulo (\ref{Fases Topológicas}), analizamos cómo algunas propiedades físicas de los materiales, como la conductividad y la polarización, están relacionadas con invariantes topológicos. En particular, exploramos la conexión entre la fase de Berry, que está asociada con la curvatura del espacio de parámetros del sistema, y los invariantes topológicos que distinguen las diferentes fases topológicas.
También, repasamos brevemente las principales características de los modelos para el efecto Hall cuántico entero y anómalo, así como los TIs 2D y 3D, destacando la naturaleza robusta de los estados de borde (o superficiales) y su carácter topológico. 

En el capítulo (\ref{Efecto Magnetoeléctrico Topológico}), nos enfocamos en el modelo efectivo, conocido como electrodinámica $\theta$ ($\theta$-ED), que describe la respuesta electromagnética de los TIs 3D cuando se abre una brecha energética en su superficie. Principalmente, se destaca el efecto magnetoeléctrico topológico (TME) que se manifiesta en los TIs. En un material que exhibe el TME, un campo eléctrico induce una magnetización y un campo magnético induce una polarización eléctrica, estableciendo una interrelación entre ambos campos, incluso en condiciones estáticas. Este fenómeno surge debido a las modificaciones en las condiciones de contorno que deben satisfacer los campos en la $\theta$-ED. Estudios previos han investigado cómo las fuentes electromagnéticas estáticas interactúan con un TI \cite{qi_monopole_2009,martin2015green,martin2016electro,martin2016electromagnetic,martin2016green}, así como la interacción de una onda electromagnética incidente en un TI \cite{PhysRevB.80.113304,maciejko_topological_2010,tse_giant_2010,wu_quantized_2016,crosse_electromagnetic_2015,crosse_optical_2016} han dado los primeros pasos para el estudio de la respuesta electromagnética de estos materiales. En esta tesis, exploramos sistemas altamente sensibles a las condiciones de contorno para investigar el TME, centrándonos principalmente en la propagación de ondas electromagnéticas en medios confinados que incluyen TIs en su estructura.

En el capítulo (\ref{Guías de Ondas}) comenzamos definiendo las ecuaciones diferenciales en el espacio de frecuencia, lo que impone una respuesta causal de los parámetros topo-ópticos que caracterizan al material. Hemos reportado nuevos términos topológicos que modifican la ley de conservación de la energía en medios donde el campo electromagnético oscila armónicamente. Estos nuevos términos son proporcionales a $\sim \boldsymbol{\nabla}\theta\cdot(\mathbf{E}\times\mathbf{E}^{*})$, lo cual es no trivial cuando la onda incide en un material topológico con polarización circular o elíptica.

En el capítulo (\ref{TEM}), investigamos soluciones de ondas electromagnéticas transversales (TEM) en guías de ondas. Actualmente, existen dos tipos principales de guías de ondas: metálicas y dieléctricas, que se utilizan en diferentes regímenes del espectro electromagnético. Para las radiofrecuencias, el cable coaxial metálico es el más utilizado, ya que confina todo el campo electromagnético entre dos cilindros metálicos coaxiales. En las guías de ondas metálicas, el modo TEM se destaca debido a su relación lineal entre la frecuencia y el vector de onda, lo que garantiza que un pulso de diferentes frecuencias conserve su forma al propagarse por la guía. Sin embargo, una desventaja crucial de las guías de ondas coaxiales metálicas es que resultan ineficaces en longitudes de onda ópticas debido a las elevadas pérdidas por absorción en el metal. Por esta razón, las guías de ondas ópticas se limitan al uso de materiales dieléctricos. Se destaca que, debido al teorema de Earnshaw \cite{landau2013electrodynamics}, hasta ahora no ha sido posible recrear un modo TEM utilizando únicamente materiales dieléctricos \cite{ibanescu_all-dielectric_2000}. En \cite{martin2016electro} se demostró que es posible eludir el teorema de Earnshaw con TIs. En esta tesis, hemos explorado diversas configuraciones que evaden este teorema, demostrando así la existencia de modos TEM mediante una combinación de materiales dieléctricos y topológicos que no son permitidas en la electrodinámica convencional de Maxwell.

La primera configuración involucra la inducción de cargas y corrientes de Hall en la superficie de un cilindro de TI mediante una onda electromagnética de fondo. Las cargas topológicas inducidas en la superficie son fuentes de campos EM que interactúan con el campo de fondo, provocando una rotación de la polarización de este. Esta rotación es una señal clara y observable de las propiedades superficiales topológicas de los TI. La segunda configuración corresponde a un cilindro hueco TI de cierto grosor en el mismo campo de fondo, el cual se logró confinar la potencia EM que se propaga a lo largo de la dirección de la guía. Específicamente, para un valor dado de la razón entre el radio interno y externo del cilindro, $R_{1}/R_{2}$, la potencia transmitida es máxima. El salto en $\theta$ junto al campo magnético normal al cilindro genera una discontinuidad del campo eléctrico que actúa como una densidad de carga superficial topológica. Junto a ella, existe una densidad de corriente de Hall superficial inducida. Demostramos que el campo eléctrico total puede entenderse como una superposición infinita de los campos inducidos por estas densidades de carga y corrientes superficiales topológicas. La suma infinita, de hecho converge y conduce precisamente a los campos electromagnéticos correctos.

La tercera configuración aprovecha la naturaleza compleja del parámetro magnetoeléctrico topológico $\theta(\omega)$, permitiendo la propagación de ondas TEM en medios con menos de dos conductores en un rango puramente imaginario de este parámetro, lo cual ocurre físicamente cuando la respuesta topológica del material es causal. La propagación TEM es robusta debido a la diferencia de fase que hay entre la carga topológica y el campo magnético. El mantenimiento de las condiciones de contorno a lo largo de $z$ y $t$ requiere que en ambas regiones la polarización sea circular pero de diferentes sentidos, lo cual puede lograrse mediante una adecuada configuración y ajuste de los parámetros ópticos complejos del material.

Nuestros hallazgos proporcionan evidencia del efecto magnetoeléctrico topológico, manifestándose a través de la rotación de la polarización de la onda TEM. Esta rotación, distinta de los fenómenos de Faraday y Kerr, no está condicionada por la presencia de una componente magnética en la dirección de propagación, ni asociada con la birrefringencia del material o con un campo reflejado. Su origen radica en la discontinuidad del campo EM en la dirección transversal de propagación. Además, la propagación TEM con menos de dos conductores es viable en guías de onda formadas por materiales descritos por la $\theta$-ED, siempre que se cumplan ciertas condiciones específicas, como la coincidencia de índices de refracción para todos los medios y las requeridas para satisfacer le ecuación característica dictadas en cada configuración.

Finalmente, en el capítulo (\ref{Slab}), presentamos un estudio detallado de la solución completa del campo electromagnético en una guía de ondas en forma de slab, compuesta por un TI en el núcleo (core) y dieléctricos como revestimiento (cladding). Mostramos que, aunque el parámetro topológico $\theta$ es pequeño en comparación con la permitividad, los modos permitidos en la guía de ondas se ven significativamente alterados. Estos cambios incluyen una inevitable hibridación de los modos transversales eléctricos (TE) y magnéticos (TM), así como modificaciones en los parámetros de propagación, como las longitudes de onda transversales y longitudinales a lo largo del eje de la guía. Además, reportamos variaciones en el parámetro de confinamiento de la onda electromagnética y en los ángulos permitidos, lo que influye directamente en la velocidad de propagación de la energía de la onda.

Se determinó que los efectos observables del parámetro topológico $\theta$ se amplifican cuando la permitividad relativa del TI es grande y cuando el revestimiento tiene una permitividad relativa similar a la del TI. Esta mejora se cuantifica y para TIs reales se puede observar un aumento de hasta un $50\%$ en comparación con el modo dominante permitido por la electrodinámica usual. Además, se propone que los efectos $\theta$ podrían mejorar al considerar revestimientos con permitividades relativas más altas. Hemos cuantificado la rotación de los modos de propagación en el sistema de referencia donde la onda es TEM. Además, reportamos rotaciones del orden de la docena de grados, lo cual es considerable en comparación con las típicas rotaciones reportadas en la literatura. La rotación del modo TE y TM de la electrodinámica usual ocurre en la misma magnitud pero en diferentes sentidos.

Por último, acoplamos un haz Gaussiano en el borde de la guía de onda que modifica las constantes de acoplamiento de los modos híbridos permitidos en la guía. El perfil simétrico del haz Gaussiano excita principalmente a los modos pares permitidos, pero hemos reportado que es posible excitar un modo impar gracias a la diferencia de los parámetros $\theta$ de los medios. Las amplitudes de los modos, el mismo campo electromagnético y las constantes de propagación son dependientes del parámetro $\theta$, lo que refleja la compleja interacción entre la onda electromagnética inicial y las cargas topológicas inducidas. La modificación de los modos reportadas enriquece la contribución $\theta$, influyendo en nuevos efectos que no habían sido reportadas por Crosse \cite{crosse_theory_2017} y Talebi \cite{Talebi2016optical}.

Estos resultados amplían nuestra comprensión de las ondas electromagnéticas en medios confinados formados por TIs y dieléctricos, destacando tanto la existencia de modos TEM en guías de ondas no metálicas como la inevitable hibridación de los modos permitidos. Los modos híbridos en guías de ondas ofrecen una flexibilidad superior en comparación con los modos TE y TM puros, permitiendo un mejor acoplamiento con fuentes externas y un control más preciso sobre la distribución del campo electromagnético. Además, estos modos proporcionan un confinamiento más eficiente de la energía, reduciendo las pérdidas por radiación, y minimizan la dispersión de las señales. Estos efectos, inducidos por el efecto magnetoeléctrico topológico, son fundamentales para medir y explorar las propiedades topológicas de las corrientes robustas que se manifiestan en la superficie de los TIs.

\biblio %Se necesita para referenciar cuando se compilan subarchivos individuales - NO SACAR

%% file: Capitulos/Apendice.tex
% Las siguientes líneas están para que la numeración en el apéndice sea correcta - NO CAMBIAR.
% Usar estas en caso de utilizar la clase book, de lo contrario comentar
\renewcommand{\thesection}{A\arabic{section}}
\renewcommand{\thetable}{A\arabic{section}.\arabic{table}}
\counterwithin{table}{section}
\counterwithin{figure}{section}

% Usar estas en caso de utilizar cualquier clase que NO SEA book.
%\renewcommand{\thesubsection}{A\arabic{subsection}}
%\renewcommand{\thetable}{A\arabic{subsection}.\arabic{table}}
%\counterwithin{table}{subsection}
%\counterwithin{figure}{subsection}

%---------- Escribir desde aquí en adelante

\section{Conductividad cuántica de Hall}\label{ConductividadHall}

El estado propio de Landau $|u_{n\mathbf{k}}\rangle$ perturbado a primer orden por un campo eléctrico es,
\begin{align}
    |u_{n\mathbf{k}}\rangle_{E}=|u_{n\mathbf{k}}\rangle+\sum_{m(\neq n)}\frac{\langle u_{m\mathbf{k}}|(-qEy)|u_{n\mathbf{k}}\rangle}{\varepsilon_{n\mathbf{k}}-\varepsilon_{m\mathbf{k}}}|u_{m\mathbf{k}}\rangle
\end{align}
La densidad de corriente superficial a lo largo del eje $\hat{x}$, $K_x$, en presencia del campo $E$ en la dirección $\hat{y}$ es,
\begin{align}\notag
    \langle K_{x} \rangle_{E}&=\sum_{n\mathbf{k}}f(\varepsilon_{n\mathbf{k}})\langle u_{n\mathbf{k}}|_{E}\Big{(}\frac{qv_{x}}{L^{2}}\Big{)}|u_{n\mathbf{k}}\rangle_{E}\\\notag
    &=\langle K_{x} \rangle_{E=0}-\frac{q^{2}E}{L^{2}}\sum_{n\mathbf{k}}\sum_{m(\neq n)}\frac{f(\varepsilon_{n\mathbf{k}})}{\varepsilon_{n\mathbf{k}}-\varepsilon_{m\mathbf{k}}}\Big(\langle u_{m\mathbf{k}}|y|u_{n\mathbf{k}}\rangle\langle u_{n\mathbf{k}}|v_{x}|u_{m\mathbf{k}}\rangle\\
    &~~~~~~~~~~~~~~~~~~~~~~~~~~~~~~~~~+\langle u_{n\mathbf{k}}|y|u_{m\mathbf{k}}\rangle\langle u_{m\mathbf{k}}|v_{x}|u_{n\mathbf{k}}\rangle\Big)+\cancel{\mathcal{O}(E^2)}^{\approx 0}
\end{align}
donde $v_{x}$ es la velocidad de la carga a lo largo del eje $\hat{x}$ y $f(\varepsilon_{n\mathbf{k}})$ es la función de distribución de Fermi. 
El primer término es la densidad de corriente en equilibrio, que obviamente es cero. El segundo término contribuye a la corriente Hall, es decir, $\sigma_{xy}=\langle K_{x} \rangle_{E}/E$. Para continuar con el cálculo debemos usar la ecuación de movimiento de Heisenberg $v_{x}=\frac{1}{i\hbar}[x,H_{\mathbf{k}}]$ lo que conduce a,
\begin{align}\notag
    \langle u_{n\mathbf{k}}|v_{x}|u_{m\mathbf{k}}\rangle &=\frac{1}{i\hbar}\langle u_{n\mathbf{k}}|(xH_{\mathbf{k}}-H_{\mathbf{k}}x)|u_{m\mathbf{k}}\rangle\\
    &=-\frac{1}{i\hbar} (\varepsilon_{n\mathbf{k}}-\varepsilon_{m\mathbf{k}})\langle u_{n\mathbf{k}}|x|u_{m\mathbf{k}}\rangle.
\end{align}
Además, como $\langle u_{n\mathbf{k}}|u_{m\mathbf{k}}\rangle=\delta_{nm}$, usaremos la identidad $\langle u_{n\mathbf{k}}|\partial_{k_{j}}u_{m\mathbf{k}}\rangle=-\langle \partial_{k_{j}}u_{n\mathbf{k}}|u_{m\mathbf{k}}\rangle$. Al reemplazar obtenemos la corriente de Hall,

\begin{align}\notag
    \sigma_{xy}&=\frac{i q^{2}}{\hbar L^{2}}\sum_{n\mathbf{k}}\sum_{m(\neq n)}f(\varepsilon_{n\mathbf{k}})\Big(\langle u_{m\mathbf{k}}|\partial_{k_{y}}u_{n\mathbf{k}}\rangle\langle u_{n\mathbf{k}}|\partial_{k_{x}}u_{m\mathbf{k}}\rangle-\langle u_{n\mathbf{k}}|\partial_{k_{y}}u_{m\mathbf{k}}\rangle\langle u_{m\mathbf{k}}|\partial_{k_{x}}u_{n\mathbf{k}}\rangle\Big)\\\notag
    &=\frac{i q^{2}}{\hbar L^{2}}\sum_{n\mathbf{k}}\sum_{m(\neq n)}f(\varepsilon_{n\mathbf{k}})\Big(-\langle \partial_{k_{x}}u_{n\mathbf{k}}|u_{m\mathbf{k}}\rangle\langle u_{m\mathbf{k}}|\partial_{k_{y}}u_{n\mathbf{k}}\rangle+\langle \partial_{k_{y}}u_{n\mathbf{k}}|u_{m\mathbf{k}}\rangle\langle u_{m\mathbf{k}}|\partial_{k_{x}}u_{n\mathbf{k}}\rangle\Big)\\\notag
    &=\frac{i q^{2}}{\hbar L^{2}}\sum_{n\mathbf{k}}f(\varepsilon_{n\mathbf{k}})\Big(-\langle \partial_{k_{x}}u_{n\mathbf{k}}|(\mathbb{I}-\cancel{|u_{n\mathbf{k}}\rangle\langle u_{n\mathbf{k}}|})|\partial_{k_{y}}u_{n\mathbf{k}}\rangle+\langle \partial_{k_{y}}u_{n\mathbf{k}}|(\mathbb{I}-\cancel{|u_{n\mathbf{k}}\rangle\langle u_{n\mathbf{k}}|})|\partial_{k_{x}}u_{n\mathbf{k}}\rangle\Big)\\\notag
    &=\frac{i q^{2}}{\hbar L^{2}}\sum_{n\mathbf{k}}f(\varepsilon_{n\mathbf{k}})\Big(-\partial_{k_{x}}\langle u_{n\mathbf{k}}|\partial_{k_{y}}u_{n\mathbf{k}}\rangle+\partial_{k_{y}}\langle u_{n\mathbf{k}}|\partial_{k_{x}}u_{n\mathbf{k}}\rangle\Big)\\\notag
    &=-\frac{q^{2}}{\hbar L^{2}}\sum_{n\mathbf{k}}f(\varepsilon_{n\mathbf{k}})\Big(\partial_{k_{x}}A_{ny}-\partial_{k_{y}}A_{nx}\Big)\\
    &=-\frac{q^{2}}{\hbar L^{2}}\sum_{n\mathbf{k}}f(\varepsilon_{n\mathbf{k}})F_{nz}
\end{align}
donde hemos usado $x_{i}\to i\partial_{k_{i}}$ y corresponde a la Ec. \eqref{eq:ConductividadHallconFermi}.
\section{Reciprocidad de Lorentz en la $\theta$-ED }\label{sec:ReciprocidaddeLorentz}
Consideremos un modo $\ell$ de la EMW que oscila armónicamente en el tiempo a una frecuencia $\omega$,
\begin{align}
    \mathbf{E}_{\ell}&=\mathbf{E}_{0\ell}(\mathbf{r})e^{-i\omega t}, & \mathbf{B}_{\ell}&=\mathbf{B}_{0\ell}(\mathbf{r})e^{-i\omega t}.
\end{align}
Al reemplazar en las ecuaciones vectoriales de sin fuentes de la $\theta$-ED obtenemos,
\begin{align}
\boldsymbol{\nabla}\times \mathbf{E}_{0\ell}&=ik_{0}\mathbf{B}_{0\ell} &
\boldsymbol{\nabla}\times(\frac{1}{\mu}\mathbf{B}_{0\ell})&=-ik_{0}\epsilon\mathbf{E}_{0\ell}+\boldsymbol{\nabla}\theta\times\mathbf{E}_{0\ell}
\end{align}
Aplicamos productos escalares convenientes (con otro modo $s$) y conjugamos la ecuación de la derecha. Luego restamos las ecuaciones de la siguiente manera,
\begin{align}\notag
    \frac{1}{\mu}\mathbf{B}^{*}_{0s}\cdot\boldsymbol{\nabla}\times\mathbf{E}_{0\ell}-\mathbf{E}_{0\ell}\cdot\boldsymbol{\nabla}\times\frac{1}{\mu}\mathbf{B}^{*}_{0s}&=ik_{0}\left (\frac{1}{\mu}\mathbf{B}_{0\ell}\cdot\mathbf{B}^{*}_{0s}-\epsilon\mathbf{E}_{0\ell}\cdot\mathbf{E}^{*}_{0s}\right )-\mathbf{E}_{0\ell}\cdot(\boldsymbol{\nabla}\theta\times\mathbf{E}^{*}_{0s})\\
    \boldsymbol{\nabla}\cdot\left (\mathbf{E}_{0\ell}\times\frac{1}{\mu}\mathbf{B}^{*}_{0s} \right )&=ik_{0}\left (\frac{1}{\mu}\mathbf{B}_{0\ell}\cdot\mathbf{B}^{*}_{0s}-\epsilon\mathbf{E}_{0\ell}\cdot\mathbf{E}^{*}_{0s}\right )+\boldsymbol{\nabla}\theta\cdot(\mathbf{E}_{0\ell}\times\mathbf{E}^{*}_{0s}) \label{EQ:S_in_LorTeo}
\end{align}
donde hemos considerado que todos los parámetros topo-ópticos son reales. Consideremos el conjugado de la ecuación \eqref{EQ:S_in_LorTeo} y un cambio en los nombres de los modos $\ell\leftrightarrow s $,
\begin{align}
    \boldsymbol{\nabla}\cdot\left ( \mathbf{E}^{*}_{0s}\times\frac{1}{\mu}\mathbf{B}_{0\ell} \right )&=-ik_{0}\left ( \frac{1}{\mu}\mathbf{B}_{0\ell}\cdot\mathbf{B}^{*}_{0s}-\epsilon\mathbf{E}_{0\ell}\cdot\mathbf{E}^{*}_{0s}\right )-\boldsymbol{\nabla}\theta\cdot(\mathbf{E}_{0\ell}\times\mathbf{E}_{0s}^{*})
\end{align}
y sumamos las ecuaciones,
\begin{align}
    \boldsymbol{\nabla}\cdot\left ( \mathbf{E}_{0\ell}\times\frac{1}{\mu}\mathbf{B}^{*}_{0s}+\mathbf{E}^{*}_{0s}\times\frac{1}{\mu}\mathbf{B}_{0\ell} \right )&=0.
\end{align}
Ahora expandimos el operador diferencial $\boldsymbol{\nabla}=\boldsymbol{\nabla}_{\perp}+\tongo{z}\partial_{z}$ , y asumimos la dependencia armónica en $z$, es decir $\mathbf{E}_{0\ell}\propto e^{ik_{z\ell}z}$,
\begin{align}
    \boldsymbol{\nabla}_{\perp}\!\cdot\!\left ( \mathbf{E}_{0\ell}\times\frac{1}{\mu}\mathbf{B}^{*}_{0s}+\mathbf{E}^{*}_{0s}\times\frac{1}{\mu}\mathbf{B}_{0\ell} \right)_{\perp}\!+\!i(k_{z\ell}-k_{zs})\!\left(\mathbf{E}_{0\ell}\times\frac{1}{\mu}\mathbf{B}^{*}_{0s}+\mathbf{E}^{*}_{0s}\times\frac{1}{\mu}\mathbf{B}_{0\ell}\right )\cdot\mathbf{\hat{z}}&=0\label{eq22}
\end{align}
Integramos en la superficie transversal sobre todo $\mathbb{R}^{2}$. El primer término desaparece, porque los campos tienden a cero en el infinito. Por otro lado, en el segundo termino, los campos EM transversales son los que están en la dirección $\tongo{z}$ al hacer el producto cruz, por lo tanto,
\begin{align}
    i(k_{z\ell}-k_{zs})\int_{\mathbb{R}^{2}}\frac{1}{\mu}\left(\mathbf{E}_{\perp\ell}\times\mathbf{B}^{*}_{\perp s}+\mathbf{E}^{*}_{\perp s}\times\mathbf{B}_{\perp\ell}\right )\cdot\mathbf{\hat{z}}da_{\perp}&=0.
\end{align}
como $(k_{z\ell}-k_{zs})\neq0$, la integral debe ser cero. 
% Además, para un modo $\ell$ que se propaga hacia atrás tenemos, 
% %
% \begin{align}
%     i(-k_{z\ell}-k_{zs})\int_{\mathbb{R}^{2}}\frac{1}{\mu}\left(\mathbf{E}_{\perp\ell}\times\mathbf{B}^{*}_{\perp s}-\mathbf{E}^{*}_{\perp s}\times\mathbf{B}_{\perp\ell}\right )\cdot\mathbf{\hat{z}}da_{\perp}&=0.
% \end{align}
% %
% Sumando las dos ecuaciones anteriores obtenemos,
% %
% \begin{align}
%     \int_{\mathbb{R}^{2}}\frac{1}{\mu}\left(\mathbf{E}_{\perp\ell}\times\mathbf{B}^{*}_{\perp s}\right )\cdot\mathbf{\hat{z}}da_{\perp}&=0. & &\textup{para}~~k_{z\ell}\neq k_{z\ell}
% \end{align}
% %
Por lo tanto, tenemos la relación de ortogonormalidad de los modos,
\begin{align}\label{eq:ortogonormalidad}
\frac{c}{4\pi}\int_{\mathbb{R}^{2}}\frac{1}{4\mu}\left(\mathbf{E}_{\perp\ell}\times\mathbf{B}^{*}_{\perp s}+\mathbf{E}^{*}_{\perp s}\times\mathbf{B}_{\perp\ell}\right )\cdot\mathbf{\hat{z}}da_{\perp}&=\delta_{\ell s}p_{\ell}.
\end{align}
donde hemos introducido el factor $c/4\pi$ para que $p_{\ell}$ tenga unidades de potencia [\textit{erg/sec}] en unidades Gaussianas,  $\mathbf{E}_{\perp \ell}=\mathbf{E}_{\perp \ell}(\mathbf{r}_{\perp})$ es el perfil EM del modo $\ell$ sin el factor $e^{i(k_{z\ell}z-\omega t)}$ y $\delta_{\ell s}$ corresponde a la delta de Kronecker. Cuando $p_{\ell}=1$, decimos que los modos son ortonormales. 

\biblio %Se necesita para referenciar cuando se compilan subarchivos individuales - NO SACAR